\documentclass[12pt]{article}
\usepackage{amsmath,amssymb}
\usepackage{graphicx}
\usepackage{color}
\allowdisplaybreaks[1]

\makeatletter
 
  \@addtoreset{equation}{section}
 \makeatother

\newcommand{\bfp}{{\mathbf p}}

\newcommand{\bfpsi}{\bar{\psi}}

\newcommand{\hq}{\hat{q}}

\newcommand{\hL}{\widehat{L}}

\newcommand{\tp}{\widetilde{p}}
\newcommand{\tq}{\widetilde{q}}

\newcommand{\tA}{{\widetilde{A}}}
\newcommand{\tB}{{\widetilde{B}}}
\newcommand{\tC}{{\widetilde{C}}}
\newcommand{\tD}{{\widetilde{D}}}
\newcommand{\tE}{{\widetilde{E}}}
\newcommand{\tF}{{\widetilde{F}}}
\newcommand{\tP}{{\widetilde{P}}}
\newcommand{\tQ}{{\widetilde{Q}}}
\newcommand{\tX}{{\widetilde{X}}}

\newcommand{\tDelta}{\widetilde{\Delta}}
\newcommand{\tLambda}{{\widetilde{\Lambda}}}

\newcommand{\cA}{{\cal A}}
\newcommand{\cB}{{\cal B}}

\newcommand{\cD}{{\cal D}}
\newcommand{\cE}{{\cal E}}
\newcommand{\cF}{{\cal F}}
\newcommand{\cG}{{\cal G}}
\newcommand{\cK}{{\cal K}}
\newcommand{\cY}{{\cal Y}}

\newcommand{\cN}{{\cal N}}
\newcommand{\cO}{{\cal O}}
\newcommand{\cR}{{\cal R}}

\newcommand{\Z}{\mathbb{Z}}
\newcommand{\R}{\mathbb{R}}

\newcommand{\nn}{\nonumber}
\newcommand{\ket}{\rangle}
\newcommand{\bfra}{\langle}
\newcommand{\kett}{\rangle\!\rangle}
\newcommand{\bfraa}{\langle\!\langle}

\newcommand{\id}{{1\!\!1}} 
\newcommand {\be}{\begin{equation}}
\newcommand {\ee}{\end{equation}}
\newcommand {\bea}{\begin{eqnarray}}
\newcommand {\eea}{\end{eqnarray}}

\newcommand{\Tr}{{\rm Tr}\,}
\newcommand{\tr}{{\rm tr}\,}
\newcommand{\del}{\partial}

\newcommand{\ve}{\varepsilon}

\newcommand{\Wsj}{{\left\{\begin{matrix}J'&J''&J\\j&j&j\end{matrix}\right\}}}

\newcommand{\bA}{{\tt A}}
\newcommand{\bB}{{\tt B}}
\newcommand{\bC}{{\tt C}}
\newcommand{\ui}{{\underline{i}}}
\newcommand{\vev}[1]{\left\langle #1 \right\rangle}

\allowdisplaybreaks
\begin{document}
\thispagestyle{empty} \addtocounter{page}{-1}
\begin{flushright}
OIQP-15-7
%
\end{flushright} 
\vspace*{7mm}

\begin{center}
{\large \bf A one-loop test for construction of 4D $\cN=4$ SYM }\\
{\large \bf from 2D SYM via fuzzy sphere geometry}\\
\vspace*{2cm}
So Matsuura$^*$ and Fumihiko Sugino$^\dagger$\\
\vskip0.7cm
{}$^*${\it Department of Physics, and Research and Education Center for Natural Science, }\\
\vspace*{1mm}
{\it Keio University, 4-1-1 Hiyoshi, Yokohama 223-8521, Japan}\\
\vspace*{0.2cm}
{\tt s.matsu@phys-h.keio.ac.jp}\\
\vskip0.4cm
{}$^\dagger${\it Okayama Institute for Quantum Physics, } \\
\vspace*{1mm}
{\it Kyoyama 1-9-1, Kita-ku, Okayama 700-0015, Japan}\\
\vspace*{0.2cm}
{\tt fusugino@gmail.com}\\
\end{center}
\vskip1cm
\centerline{\bf Abstract}
\vspace*{0.3cm}
{\small 
As a perturbative check of the construction of four-dimensional (4D) $\cN=4$ supersymmetric Yang-Mills theory (SYM) 
from mass deformed $\cN=(8,8)$ SYM on the two-dimensional (2D) lattice, 
the one-loop effective action for scalar kinetic terms is computed in $\cN=4$ $U(k)$ SYM on $\R^2 \times$ (fuzzy $S^2$), 
which is obtained by expanding 2D $\cN=(8,8)$ $U(N)$ SYM with mass deformation around its fuzzy sphere classical solution. 
The radius of the fuzzy sphere is proportional to the inverse of the mass.  
We consider two successive limits: (1) decompactify the fuzzy sphere to a noncommutative (Moyal) plane 
and (2) turn off the noncommutativity of the Moyal plane. 
It is straightforward at the classical level to obtain the ordinary $\cN=4$ SYM on $\R^4$ in the limits, while it is nontrivial 
at the quantum level.    
The one-loop effective action for $SU(k)$ sector of the gauge group $U(k)$ coincides with that of 
the ordinary 4D $\cN=4$ SYM in the above limits. 
Although ``noncommutative anomaly'' appears in the overall $U(1)$ sector of the $U(k)$ gauge group, this can be 
expected to be a gauge artifact not affecting gauge invariant observables.    
}



\newpage

\section{Introduction}
\label{sec:intro}

The correspondence between four-dimensional 
(4D) ${\cal N}=4$ $U(N)$ supersymmetric Yang-Mills theory (SYM) 
and type IIB superstring theory on $AdS_5\times S^5$ 
is one of the most typical examples of the AdS/CFT duality conjecture 
\cite{Maldacena:1997re,Gubser:1998bc,Witten:1998qj}. 
The correspondence between the 4D ${\cal N}=4$ $U(N)$ SYM 
in the large $N$ and large 't Hooft coupling limit 
and the classical gravity limit of the superstring
has been supported by numerous pieces of evidence and is almost established. 
On the other hand, the strong claim of this correspondence 
between the gauge theory with finite $N$ and quantum superstring theory 
has been poorly explored and still remains conjectural. 
This is partly because of a lack of numerical as well as analytical tools beyond perturbative treatment 
on the gauge theory side. 
A possible way to overcome this situation is to construct 
a lattice formulation of 4D ${\cal N}=4$ SYM as its nonperturbative framework. 
Indeed, in the case of a lower-dimensional version 
of the duality in the D0-brane system \cite{Itzhaki:1998dd}, numerical simulations 
\cite{Anagnostopoulos:2007fw,Catterall:2008yz,Hanada:2008gy,Hanada:2008ez,Hanada:2009ne,Hanada:2013rga,Kadoh:2015mka}
have been developed to provide quantitative nonperturbative tests of the duality 
beyond the supergravity approximation and 
bring new aspects into black hole physics. 
A similar numerical study for the 4D theory will enable  
an explicit check of this conjecture.
Furthermore, 
if the strong duality conjecture is true, 
the discretized formulation
will provide a nonperturbative description of type IIB superstrings. 

Constructing lattice formulations for 4D supersymmetric gauge theories is, 
however, not straightforward. 
As an obstacle, it seems impossible to maintain all the supersymmetry 
on the lattice because of the breakdown of 
the Leibniz rule~\cite{Drell:1976mj,Dondi:1976tx,Bouguenaya:1985ym}.
Indeed, 
a no-go theorem has been proved for constructing
a lattice theory 
with keeping translational invariance, locality, and the Leibniz rule~\cite{Kato:2008sp}. 
In order to circumvent this problem, 
several lattice formulations that keep nilpotent supersymmetries (up to gauge transformations), 
which do not generate space-time translations, 
are constructed
by applying the so-called orbifolding 
procedure~\cite{Kaplan:2002wv,Cohen:2003xe,Cohen:2003qw,Kaplan:2005ta,Endres:2006ic,Giedt:2011zza,Matsuura:2008cfa,Joseph:2013jya} 
or topological 
twists~\cite{Catterall:2003wd,Sugino:2003yb,Sugino:2004qd,Sugino:2004uv,Sugino:2006uf,Sugino:2008yp,Kikukawa:2008xw,Kadoh:2009yf}  
to the discretization. 
For one- and two-dimensional theories, we can see by a perturbative argument 
that the continuum limit gives the target theories 
without any fine-tuning as a result of the exact supersymmetries on the lattice. 
This has been nonperturbatively shown by numerical simulations as 
well~\cite{Suzuki:2007jt,Kanamori:2007ye,Kanamori:2007yx,Kanamori:2008bk,Hanada:2009hq,Catterall:2010fx,Giguere:2015cga}. 
However, 
for 4D supersymmetric gauge theories including $\cN=4$ SYM, 
the nilpotent supersymmetries are not sufficient 
to forbid all the relevant operators that prevent the full 4D supersymmetry from restoring in the continuum limit.
Therefore, we normally need to tune a number of parameters in taking the continuum limit, 
which makes it almost impossible to carry out a numerical simulation%
~\footnote{
The exceptions are ${\cal N}=1$ pure SYM theories 
in three and four dimensions. 
The exact parity or chiral symmetry rather than supersymmetry on the lattice 
plays a key role in restoring the supersymmetry and all the other symmetries 
in the continuum limit~\cite{Maru:1997kh,Giedt:2009yd}. 
}. 
For some progress in 4D $\cN=4$ SYM from 4D lattice, see~\cite{Catterall:2011pd,Catterall:2013roa,Catterall:2015ira}. 
Regarding the planar part of the 4D $\cN=4$ SYM, its nonperturbative construction has been given by the plane-wave matrix 
model~\cite{Ishii:2008ib,Ishiki:2011ct}.  

In~\cite{Hanada:2010kt,Hanada:2010gs,Hanada:2011qx}, 
a new approach to circumvent this issue is proposed for 4D $\cN=2, \,4$ SYM theories, 
where two different discretizations by lattice 
and matrix~\cite{Berenstein:2002jq,Das:2003yq,Myers:1999ps} are combined. 
The strategy is as follows. 
We first construct lattice formulations reminiscent of 
the ``plane-wave matrix string''~\cite{Das:2003yq,Bonelli:2002mb}, 
which are mass deformations of the lattice theories of 
two-dimensional (2D) $\cN=(4,4)$ and ${\cal N}=(8,8)$ SYM with $U(N)$ gauge group 
given in~\cite{Sugino:2003yb,Sugino:2004qd,Sugino:2004uv}.  
As a result of this deformation, fuzzy sphere configurations realize
as classical solutions of the theories. 
In particular, if we expand field variables around a solution representing $k$-coincident fuzzy spheres, 
fluctuations can be regarded as fields of the supersymmetric 
$U(k)$ SYM 
on the direct product space of the 2D lattice 
and the fuzzy sphere. The degrees of freedom of the fuzzy sphere are $n^2\equiv N^2/k^2$. 
As mentioned above, the continuum limit of the 2D lattice 
can be taken without any fine-tuning. 
In addition, by taking a large-$N$ limit of the original 2D 
lattice theories with scaling the deformation parameter appropriately, 
the fuzzy sphere becomes the 2D noncommutative (Moyal) plane $\R^2_\Theta$. 
In the rest of this paper, we call this limit the ``Moyal limit''. 
Therefore, if we first take the continuum limit of the lattice theory
followed by the large-$N$ limit, we obtain the 4D SYM theories 
on $\R^2 \times \R^2_\Theta$. 
In particular, it is argued~\cite{Matusis:2000jf} and shown in the light-cone gauge~\cite{Hanada:2014ima} that 
the commutative limit ($\R^2_\Theta\to \R^2$) of the noncommutative $\cN=4$ SYM theory is continuous. 
Therefore, numerical simulation of 
the ordinary 4D $\cN=4$ SYM theory is expected to be possible by this hybrid formulation.

In~\cite{Hanada:2010kt,Hanada:2010gs}, 
it is discussed that there appears no radiative correction preventing 
restoration of the full supersymmetry of the 4D $\cN=4$ theory based on 
power counting. 
This argument relies on an assumption that the mass deformation 
of the 2D theory is soft in 4D theory as well as 2D theory. 
The deformation parameter $M$ has a positive mass dimension 
and the deformation is indeed soft in 2D theory, which 
does not ruin the feature of the 2D lattice theory 
that no fine-tuning is needed in taking the continuum limit. 
On the other hand, the situation is more subtle 
from the viewpoint of the 4D theory, 
since the same parameter $M$ comes in both the radius of 
the fuzzy sphere and the noncommutative parameter. Furthermore, it is related to the UV cutoff in the fuzzy sphere directions.  
Thus, we cannot say that $M$ is an IR deformation in the usual sense. 
There is still the possibility 
of unexpected divergences radiatively generated by the so-called UV/IR mixing due to the noncommutativity~\cite{Minwalla:1999px} 
or via the relation to the UV cutoff of the fuzzy sphere, which may spoil some symmetries to be restored in 
the 4D continuum theory.

The purpose of this paper is to check
if there appear such unexpected divergences by perturbative calculation 
in Feynman-type gauge fixing at the one-loop order. 
As mentioned above, the lattice continuum limit can be safely taken even after the deformation, 
which allows us to start with the deformed theory on the continuum 2D space-time; 
namely, the deformed 2D $\cN=(8,8)$ SYM theory on $\R^2$. 
We explicitly calculate the one-loop radiative corrections to 
the one- and two-point functions of bosonic fields,
which have larger superficial degrees of UV divergences and are more nontrivial compared to higher-point functions. 
For the $SU(k)$ part, we will see that there is no unexpected divergence and the 4D rotational ($SO(4)$) symmetry 
is restored at this order. 
On the other hand, for the overall $U(1)$ part, 
there appears a non-trivial correction regarded as 
the so-called ``noncommutative anomaly''~\cite{Chu:2001xi}, 
which breaks the $SO(4)$ symmetry. 
However, according to the results for theories on $\R^2\times \R^2_\Theta$ as well as $\R^4$ 
in the light-cone gauge~\cite{Hanada:2014ima,Mandelstam:1982cb}, 
it is considered that this anomaly arises only accompanied with 
wave function renormalizations. 
This implies that the anomaly is a gauge artifact and does 
not affect correlation functions among gauge invariant observables, in which the $SO(4)$ symmetry is expected to be restored.   

The organization of this paper is as follows. 
We present a brief review of the continuum deformed 2D $\cN=(8,8)$ SYM theory in the next section, 
and expand fields around a classical solution of $k$-coincident 
fuzzy spheres by using the so-called 
fuzzy spherical harmonics in section~\ref{sec:FS sln}. 
In section~\ref{sec:loop}, the successive limits leading to the 4D target theory are presented, and 
propagators are explicitly given for perturbative calculations. 
In section~\ref{sec:1pt}, the one-point functions of bosonic fields are computed at the one-loop order. 
As a warm up exercise, calculations are presented in some detail.  
In section~\ref{sec:2pt}, we calculate the radiative corrections 
to the scalar two-point functions at the one-loop level. 
Since we are interested in the Moyal limit, 
each Feynman diagram is evaluated in this limit. 
The use of three theorems proved in appendix~\ref{app:proof} considerably simplifies the computations. 
In section~\ref{sec:decpt limit}, 
the one-loop effective action for the scalar kinetic terms is obtained in the successive limits.  
Section~\ref{sec:conclusion} is devoted to a summary of the results obtained so far and discussions of future subjects. 
In appendix~\ref{app:action} we give the explicit form 
of the deformed action in the balanced topological field theory (BTFT) description. 
In order for this paper to be as self-contained as possible, 
we derive various useful properties of the fuzzy spherical harmonics in appendix~\ref{app:FSH}. 
Computational details of the one-point functions are collected in appendix~\ref{app:one-point_func}. 
Appendix~\ref{app:proof} is devoted to proofs of the theorems given in section~\ref{sec:2pt}. 
We give precise relations between the fuzzy sphere and the Moyal plane 
by taking the Moyal limit of the fuzzy spherical harmonics in appendix~\ref{app:FSH flat limit}. 
To examine the nonperturbative stability of the $k$-coincident fuzzy sphere solution, tunneling amplitudes to some other 
vacua are evaluated in appendix~\ref{app:tunneling}. 

\section{Review of the deformed 2D $\cN=(8,8)$ SYM}
\label{sec:review}
In this section we briefly review the deformed 2D $\cN=(8,8)$ SYM given in \cite{Hanada:2010kt}. 
Let us start with $U(N)$ $\cN=(8,8)$ SYM on $\R^2$ in a Euclidean signature.
This theory has gauge fields $A_\mu$ ($\mu=1,2$), eight scalar fields, and sixteen fermionic fields. 
For convenience, we express the scalar fields as 
$(X_i, X_a)$ $(i=3,\cdots,7,\, a=8,9,10)$ and the fermionic fields as
$\psi_{r\,\alpha}$ $(r=1,\cdots,8,\, \alpha=\pm \frac12)$. 
Each field is represented by an $N\times  N$ matrix. 
Note that the scalars $(X_8, X_9, X_{10})$ and eight pairs of the fermions $\psi_r\equiv (\psi_{r\,+\frac12}, \psi_{r\,-\frac12})$ 
form a triplet and doublets of an $SU(2)_R$ subgroup of the $SO(8)$ R-symmetry, 
respectively, 
and the other fields are $SU(2)_R$ singlets. 
In this expression, the action can be written as 
\begin{align}
S_0=\frac{1}{g_{2d}^2}\int d^2x\, \Tr \Bigl\{
&F_{12}^2
+\left(\cD_\mu X_i\right)^2
+\left(\cD_\mu X_a\right)^2
-\frac{1}{2}\left[X_i, X_j \right]^2 
-\frac{1}{2}\left[X_a, X_b\right]^2 
-\left[X_a, X_i \right]^2 \nn \\
& +i \bfpsi_{r\,\alpha}(\hat\gamma_\mu)_{rs}\cD_\mu \psi_{s\,\alpha}
 -\bar\psi_{r\,\alpha} (\hat\gamma_i)_{rs}[X_i,\psi_{s\,\alpha}]
 -\bar\psi_{r\,\alpha} (\hat\sigma_a)_{\alpha\beta}[X_a,\psi_{r\,\beta}]
\Bigr\}, 
\label{S0}
\end{align}
where 
$\cD_\mu=\partial_\mu+i[A_\mu, \ \cdot \ ]$ are covariant derivatives for adjoint fields, 
$F_{12}=\partial_1A_2-\partial_2A_1+i[A_1, A_2]$ is the gauge field strength, 
$\hat\gamma_I = (\hat\gamma_\mu,\hat\gamma_i)$ 
$(I=(\mu,i)=1,\cdots,7)$ are $8\times 8$ matrices satisfying 
$\{\hat\gamma_I, \hat\gamma_J\}=-2\delta_{IJ}$, $\hat\sigma_a$ 
are Pauli matrices, 
\begin{align}
 \hat\sigma_8=\sigma_1, \quad 
 \hat\sigma_9=\sigma_2, \quad 
 \hat\sigma_{10}=\sigma_3, 
\end{align}
and  $\bar\psi_{r\alpha} \equiv \psi_{r\beta}(i\sigma_2)_{\beta\alpha}$. 
In this paper, we use the following convention 
for $\hat{\gamma}_I$ matrices:
\bea
& & \hat{\gamma}_1  \equiv  \begin{bmatrix}   &       &  \id_2 &        \\
                                              &        &        &  \id_2  \\
                                       -\id_2  &        &       &          \\ 
                                               & -\id_2 &    &         \end{bmatrix}, \qquad                                
\hat{\gamma}_2 \equiv \begin{bmatrix}   &      & -i\sigma_2  &       \\
                                             &     &         &  i\sigma_2 \\ 
                                   -i\sigma_2  &      &     &       \\
                                              & i\sigma_2  &        &  \end{bmatrix},  \nn \\
& &  \hat{\gamma}_3 \equiv \begin{bmatrix}   &         &        & i\sigma_2    \\
                                             &       & i\sigma_2   &   \\ 
                                             & i\sigma_2 &     &       \\
                                      i\sigma_2  &      &        &  \end{bmatrix}, \qquad
\hat{\gamma}_4 \equiv \begin{bmatrix}   &         &        & \id_2    \\
                                             &       & -\id_2   &   \\ 
                                             & \id_2 &     &       \\
                                      -\id_2  &      &        &  \end{bmatrix},  \nn \\
& &  \hat{\gamma}_5 \equiv \begin{bmatrix}   & \sigma_1  &        &     \\
                                      -\sigma_1 &       &         &   \\ 
                                             &     &     &  -\sigma_1  \\
                                             &      & \sigma_1  &  \end{bmatrix}, \qquad 
\hat{\gamma}_6 \equiv \begin{bmatrix}   &  \sigma_3  &        &     \\
                                -\sigma_3 &          &        &   \\ 
                                             &       &       & -\sigma_3  \\
                                              &      & \sigma_3  &  \end{bmatrix},  \nn \\     
& & \hat{\gamma}_7  \equiv  \begin{bmatrix} -i\sigma_2 &      &         &       \\
                                                 & -i\sigma_2 &         &      \\ 
                                                   &      & i\sigma_2 &       \\
                                                  &         &        & i\sigma_2 \end{bmatrix} .                           
\eea
Here and in what follows, blank elements in matrices mean null. 

Next we deform the action by adding 
mass terms to scalars and fermions as well as 
the so-called Myers term \cite{Myers:1999ps}: 
\begin{equation}
 S=S_0+S_M 
 \label{action}
\end{equation}
with
\begin{align}
S_M=\frac{1}{g_{2d}^2}\int d^2x\, \Tr \Bigl\{
&\frac{M^2}{9} X_a^2 +i \frac{M}{3} \epsilon_{abc} X_a[X_b,X_c]
+\frac{2M^2}{81}X_i^2 
-m_r \bfpsi_{r\kappa}\psi_{r\kappa} \nn \\ 
&-i\frac{4M}{9}X_7\left(F_{12}+i[X_3,X_4]\right)
\Bigr\}, 
\label{mass action}
\end{align}
where $\epsilon_{abc}$ is a third-rank antisymmetric tensor defined by  $\epsilon_{8,9,10}=1$ 
and  
\begin{align}
m_r=\left(
\frac{M}{9},\frac{M}{9},\frac{M}{3},\frac{M}{3},
\frac{2M}{9},\frac{2M}{9},\frac{2M}{9},\frac{M}{3}
\right). 
\end{align}
We have also introduced the pure imaginary term $\Tr\left(-i\frac{4M}{9}X_7\left(F_{12}+i[X_3,X_4]\right)\right)$ in (\ref{mass action}). 
This is motivated by the deformation of the BTFT description of 
2D $\cN=(8,8)$ SYM, which manifestly preserves two supercharges~\cite{Hanada:2010kt}~\footnote{
The other fourteen supercharges are softly broken by the deformation. 
}.  
In the BTFT description, we rename the scalar fields as 
\begin{align}
& B_1(x)\equiv -X_5(x), \quad B_2(x)\equiv X_6(x), \quad B_3(x)\equiv X_7(x), \nn \\
& C(x)\equiv 2X_8(x), \quad 
\phi_\pm(x) \equiv X_9(x) \pm i X_{10}(x), 
\label{rename_scalar}
\end{align}
and write the fermions as 
\begin{align}
 \psi_1&\equiv \left(\begin{matrix}
 \rho_{-3} \\ \rho_{+3}
 \end{matrix}\right),\
 \psi_2\equiv \left(\begin{matrix}
 \rho_{-4} \\ \rho_{+4}
 \end{matrix}\right),\
 \psi_3\equiv \left(\begin{matrix}
 \psi_{-2} \\ \psi_{+2}
 \end{matrix}\right),\
 \psi_4\equiv \left(\begin{matrix}
 \psi_{-1} \\ \psi_{+1}
 \end{matrix}\right),\nn \\ 
 \psi_5&\equiv \left(\begin{matrix}
 \chi_{-1} \\ \chi_{+1}
 \end{matrix}\right),\
 \psi_6\equiv \left(\begin{matrix}
 \chi_{-2} \\ \chi_{+2}
 \end{matrix}\right),\ 
 \psi_7\equiv \left(\begin{matrix}
 \chi_{-3} \\ \chi_{+3}
 \end{matrix}\right),\ 
 \psi_8\equiv \left(\begin{matrix}
 -\frac{\eta_{-}}{2} \\ \frac{\eta_+}{2}
 \end{matrix}\right). 
\end{align}
Then we can show that the deformed action (\ref{action}) is invariant 
under the transformation: 
\begin{eqnarray}
& &
Q_{\pm}A_\mu=\psi_{\pm\mu},
\quad
Q_\pm\psi_{\pm\mu}
=
\pm iD_\mu\phi_{\pm}, 
\quad Q_\mp\psi_{\pm\mu}
=
\frac{i}{2}D_\mu C\mp\tilde{H}_\mu, 
\nonumber\\
& &
Q_{\pm}\tilde{H}_\mu
=
\left[\phi_{\pm},\psi_{\mp\mu}\right]
\mp\frac{1}{2}\left[C,\psi_{\pm\mu}\right]
\mp\frac{i}{2}D_\mu\eta_{\pm} + \frac{M}{3} \psi_{\pm \mu},
\nonumber\\ 
& &
Q_{\pm}X_{\ui}=\rho_{\pm \ui},
\quad
Q_\pm\rho_{\pm \ui}
=
\mp \left[X_{\ui},\phi_{\pm}\right], 
\quad 
Q_\mp\rho_{\pm \ui}
=
-\frac{1}{2}[X_\ui, C]\mp\tilde{h}_{\ui}, 
\nonumber\\
& &
Q_{\pm}\tilde{h}_{\ui}
=
\left[\phi_{\pm},\rho_{\mp \ui}\right]
\mp\frac{1}{2}\left[C,\rho_{\pm \ui}\right]
\pm\frac{1}{2}\left[X_{\ui},\eta_{\pm}\right]
+\frac{M}{3} \rho_{\pm \ui}, 
\nonumber\\
& &
Q_{\pm}B_{\bA}
=
\chi_{\pm \bA}, 
\quad
Q_{\pm}\chi_{\pm \bA}
=
\pm[\phi_\pm,B_{\bA}], 
\quad 
Q_\mp\chi_{\pm \bA}
=
-\frac{1}{2}[B_{\bA},C]
\mp H_{\bA}, 
\nonumber\\
& &
Q_\pm H_{\bA}
=
[\phi_\pm,\chi_{\mp \bA}]
\pm\frac{1}{2}\left[B_{\bA},\eta_\pm\right]
\mp\frac{1}{2}\left[C,\chi_{\pm \bA}\right]
+\frac{M}{3} \chi_{\pm \bA}, 
\nonumber\\ 
& &
Q_\pm C
=
\eta_\pm, 
\quad
Q_\pm\eta_\pm
=
\pm\left[\phi_\pm,C\right]
+\frac{2M}{3} \phi_\pm, 
\quad
Q_\mp\eta_\pm
=
\mp\left[\phi_+,\phi_-\right]
\pm\frac{M}{3} C, 
\nonumber\\
& &
Q_\pm\phi_\pm=0, 
\quad
Q_{\mp}\phi_\pm=\mp\eta_\pm. 
\label{SUSY algebra}
\end{eqnarray}
Here, the range of the index $i$ ($=3,\cdots, 7$) has been decomposed into $\ui=3, 4$ and $\bA=1, 2, 3$. 
The supercharges $Q_\pm$ satisfy the anti-commutation relations,  
\begin{equation}
Q_+^2
=
\frac{M}{3}J_{++}, 
\quad
Q_-^2
=
-\frac{M}{3}J_{--}, 
\quad 
\{Q_+,Q_-\}
=
-\frac{M}{3}J_0,  
\label{Q-anticommutator}
\end{equation}
up to gauge transformations, where $J_0$, $J_{++}$ and $J_{--}$ are 
generators of $SU(2)_R$ symmetry acting on fields as 
\bea
J_{++} & = & \int d^2x \left[\psi_{+\mu}^{\alpha}(x)\frac{\delta}{\delta\psi_{-\mu}^{\alpha}(x)} 
+\rho_{+\ui}^{\alpha}(x)\frac{\delta}{\delta\rho_{-\ui}^{\alpha}(x)} 
+\chi_{+\bA}^{\alpha}(x)\frac{\delta}{\delta\chi_{-\bA}^{\alpha}(x)} 
-\eta_+^{\alpha}(x)\frac{\delta}{\delta\eta_-^{\alpha}(x)} \right. \nn \\
 & & \hspace{14mm}
\left. +2\phi_+^{\alpha}(x) \frac{\delta}{\delta C^{\alpha}(x)} 
-C^{\alpha}(x) \frac{\delta}{\delta\phi_-^{\alpha}(x)}\right], \nn \\
J_{--} & = & \int d^2x \left[\psi_{-\mu}^{\alpha}(x)\frac{\delta}{\delta\psi_{+\mu}^{\alpha}(x)} 
+\rho_{-\ui}^{\alpha}(x)\frac{\delta}{\delta\rho_{+\ui}^{\alpha}(x)} 
+\chi_{-\bA}^{\alpha}(x)\frac{\delta}{\delta\chi_{+\bA}^{\alpha}(x)} 
-\eta_-^{\alpha}(x)\frac{\delta}{\delta\eta_+^{\alpha}(x)} \right. \nn \\
 & & \hspace{14mm}
\left. -2\phi_-^{\alpha}(x) \frac{\delta}{\delta C^{\alpha}(x)} 
+C^{\alpha}(x) \frac{\delta}{\delta\phi_+^{\alpha}(x)}\right], \nn \\
J_0 & = & \int d^2x \left[ \psi_{+\mu}^{\alpha}(x)\frac{\delta}{\delta\psi_{+\mu}^{\alpha}(x)} 
-\psi_{-\mu}^{\alpha}(x)\frac{\delta}{\delta\psi_{-\mu}^{\alpha}(x)}
+\rho_{+\ui}^{\alpha}(x)\frac{\delta}{\delta\rho_{+\ui}^{\alpha}(x)} 
-\rho_{-\ui}^{\alpha}(x)\frac{\delta}{\delta\rho_{-\ui}^{\alpha}(x)}
 \right. \nn \\
 & & \hspace{14mm}
+\chi_{+\bA}^{\alpha}(x)\frac{\delta}{\delta\chi_{+\bA}^{\alpha}(x)} 
-\chi_{-\bA}^{\alpha}(x)\frac{\delta}{\delta\chi_{-\bA}^{\alpha}(x)}  
+\eta_+^{\alpha}(x)\frac{\delta}{\delta\eta_+^{\alpha}(x)} 
-\eta_-^{\alpha}(x)\frac{\delta}{\delta\eta_-^{\alpha}(x)}   \nn \\
 & & \hspace{14mm}
\left. 
 +2\phi_+^{\alpha}(x) \frac{\delta}{\delta \phi_+^{\alpha}(x)} 
-2\phi_-^{\alpha}(x) \frac{\delta}{\delta\phi_-^{\alpha}(x)}\right]  
\label{SU2R_gen}
\eea  
($\alpha$ is an index for the gauge group generators), which satisfy
\be
[J_{++},\,J_{--}]=J_0,\qquad [J_0,\,J_{\pm\pm}]=\pm 2J_{\pm\pm}. 
\ee  
The eigenvalues of $J_0$ are $\pm 1$ for the fermions with the index $\pm$, 
$\pm 2$ for $\phi_\pm$, and zero for the other bosonic fields. 
As mentioned above, 
$\phi_\pm$ and $C$ form an $SU(2)_R$ triplet 
and each pair of 
$(\psi_{+\mu},\psi_{-\mu})$, 
$(\rho_{+\ui}, \rho_{-\ui})$, 
$(\chi_{+\bA},\chi_{-\bA})$, 
$(\eta_+,-\eta_-)$, and $(Q_+,Q_-)$ forms a doublet.   

The invariance of the action under $Q_\pm$-transformations is most easily 
seen by the fact that the deformed action (\ref{action}) is recast as
\begin{eqnarray}
S=
\left(
Q_+Q_- - \frac{M}{3}
\right)
{\cal F}, 
\label{Q-closed form}
\end{eqnarray}
where 
\begin{align}
{\cal F}
=
\frac{1}{g^2_{2d}}\int d^2x\, {\rm Tr}\Bigl\{
&-iB_{\bA}\Phi_{\bA}
-
\frac{1}{3}\epsilon_{\bA\bB\bC}B_{\bA}[B_{\bB},B_{\bC}]
-\frac{M}{9} B_\bA^2 
-\frac{2M}{9} X_{\ui}^2
\nonumber\\
&-
\psi_{+\mu}\psi_{-\mu}
-
\rho_{+\ui}\rho_{-\ui}
-
\chi_{+\bA}\chi_{-\bA}
-
\frac{1}{4}\eta_+\eta_-
\Bigl\}  
\label{F}
\end{align}
with $\Phi_1=2(-D_1X_3-D_2X_4)$, 
$\Phi_2=2(-D_1X_4+D_2X_3)$,  
$\Phi_3=2(-F_{12}+i[X_3,X_4])$. 
In fact, the invariance is shown by (\ref{Q-anticommutator}) 
and the commutation relation between $(J_{\pm\pm},J_0)$ and $Q_\pm$: 
\begin{equation}
[J_{\pm\pm}, Q_\pm]=0, \qquad [J_{\pm\pm}, Q_\mp]=Q_\pm, \qquad [J_0,Q_\pm] =\pm Q_\pm . 
\label{Q_doublet}
\end{equation}
In appendix \ref{app:action} we give the explicit form of the action in the BTFT description.

\section{Fuzzy sphere solution and mode expansion}
\label{sec:FS sln}

In this section, we expand the action around a particular supersymmetry preserving solution 
(a $k$-coincident fuzzy sphere solution)
and explicitly give its mode expansion, which is convenient to carry out perturbative calculations 
in the subsequent sections. 

\subsection{Action around fuzzy sphere solution}

As a result of the deformation by (\ref{mass action}), 
the theory has fuzzy sphere solutions as minima of the action
($S=0$) preserving $Q_\pm$ supersymmetries: 
\begin{equation}
X_a(x)=\frac{M}{3} L_a, \quad
(\text{other fields}) = 0, 
\label{FS sln0}
\end{equation}
where $L_a$ belong to an $N$-dimensional (not necessary irreducible) representation 
of the $SU(2)$-algebra satisfying 
\begin{equation}
[L_a, L_b]=i \epsilon_{abc} L_c. 
\end{equation}
Among a lot of possible solutions, 
we consider $k$-coincident fuzzy $S^2$ with the size $n$ described by 
\begin{equation}
L_a=L_a^{(n)} \otimes \id_k, 
\label{FS sln}
\end{equation}
where $L_a^{(n)}$ are the $n$-dimensional irreducible representation of $su(2)$
with $N=nk$. 
The fields $X_a$ are expanded around the solution (\ref{FS sln}) as  
\begin{align}
X_a(x)=\frac{M}{3}L_a +  \tilde{X}_{a}(x). 
\end{align}
Introducing the ``field strength"  
\begin{align}
F_{ab}&\equiv \frac{M}{3} \left( i[L_a, \tX_b] - i [L_b, \tX_a] + \epsilon_{abc} \tX_c \right) 
+ i[\tX_a, \tX_b], \nn \\
F_{\mu a} &\equiv \del_\mu \tX_a -i \frac{M}{3} [L_a, A_\mu] + i [A_\mu, \tX_a],
\end{align}
and the ``covariant derivatives"
\begin{align}
\cD_a &\equiv i \frac{M}{3}\left[ L_a, \ \cdot\  \right] + i [ \tX_a, \ \cdot\  ] 
\end{align}
recasts the bosonic and fermionic parts of the action (\ref{action}) : $S=S_b+S_f$ as 
\begin{align}
\label{boson action}
S_b=\frac{1}{g_{2d}^2}\int d^2x \,\Tr \Bigl\{
&F_{12}^2
+\frac{1}{2} F_{ab}^2
+F_{\mu a}^2
+\left(\cD_\mu X_i\right)^2 
+\left(\cD_a X_i\right)^2
-\frac{1}{2}\left[X_i, X_j\right]^2 \nn \\
&+\frac{2M^2}{81}X_i^2 
-i\frac{4M}{9}X_7\left(F_{12}+i[X_3,X_4]\right)
\Bigr\}, \\
S_f=\frac{1}{g_{2d}^2}\int d^2x\, \Tr \Bigl\{
& i \bfpsi_{r\alpha}(\hat\gamma_\mu)_{rs}\cD_\mu \psi_{s\alpha}
 +i\bar\psi_{r\alpha} (\hat\sigma_a)_{\alpha\beta} \cD_a\psi_{r\beta} \nn \\
&-\bfpsi_{r\alpha}(\hat\gamma_i)_{rs}[X_i,\psi_{s\alpha}] 
-m_r \bfpsi_{r\alpha}\psi_{r\alpha}, 
\Bigr\}. 
\label{fermion action}
\end{align}

Upon perturbative calculations, we fix the gauge to a Feynman-type gauge: 
\begin{equation}
F(A,\tX)\equiv \del_\mu A_\mu + i \frac{M}{3} [L_a,\tX_a] =0. 
\end{equation}
Under the gauge transformation 
\be
\delta A_\nu=\cD_\mu c, \qquad  \delta \tilde{X}_a = i\frac{M}{3}[L_a, c] +i[\tilde{X}_a,c] \label{gaugetr_tildeX}
\ee
(the latter is obtained from $\delta X_a=-i[c,X_a]$), $F(A,\tX)$ changes as 
\bea
\delta F(A,\tilde{X})  =  \del_\mu \cD_\mu c -\frac{M^2}{9}[L_a,[L_a,c]] 
-\frac{M}{3}\left([L_a,[\tilde{X}_a,c]] \right).
\eea 
Thus, the gauge fixing terms to the Feynman-type gauge 
and the associated Faddeev-Popov ghost terms 
are given by 
\begin{align}
S_{\rm GF}  =  \frac{2}{g_{2d}^2}\int d^2x \,\Tr\biggl\{ &\frac12 F(A,\tilde{X})^2 
+\bar{c}\,\delta F(A,\tilde{X})\biggr\} \nn \\
=\frac{1}{g_{2d}^2}\int d^2x \, \Tr\biggl\{
&\left(\del_\mu A_\mu \right)^2 - \left(\frac{M}{3}\right)^2  [L_a,\tX_a]^2
+i \frac{2M}{3}\left(\del_\mu A_\mu\right) [ L_a, \tX_a ] \nn \\
&-2\left(\del_\mu \bar{c}\right)\left( \cD_\mu c\right)
-i\frac{2M}{3} [ L_a, \bar{c} ] \left( \cD_a c\right)
\biggr\}, 
\label{gf action}
\end{align}
where $c$ and $\bar{c}$ are ghost and anti-ghost fields, respectively. 
The action after the gauge fixing is given by the summation of 
(\ref{boson action}), (\ref{fermion action}), and (\ref{gf action}): 
\begin{equation}
 S_{\rm tot}=S_b + S_f + S_{\rm GF}. 
 \label{total action}
\end{equation}

\subsection{Fuzzy spherical harmonics and mode expansion}

As discussed in \cite{Hanada:2010kt}, 
the derivatives and the gauge fields along directions of the fuzzy $S^2$ with the radius $R=\frac{3}{M}$ are given by 
two linearly independent combinations of $i\frac{M}{3} [L_a, \ \cdot \ ]$ ($a=8,9,10$) 
and by the two corresponding combinations of $\tX_a$, respectively. 
Integration over the fuzzy sphere corresponds to taking a partial trace over the $n$ dimensions in the total trace ``Tr''. 
This means that the action (\ref{boson action}) and (\ref{fermion action}) can be 
regarded as the action of mass-deformed 4D $\cN=4$ SYM theory 
on $\R^2\times ({\rm fuzzy}\ S^2)$. 
In doing perturbative calculations, 
it is convenient to expand all the fields in the action (\ref{total action}) by the momentum basis 
on $\R^2$ and by the basis of fuzzy spherical harmonics on the fuzzy $S^2$. 
The fuzzy spherical harmonics are $n\times n$ matrices, and their definitions and relevant properties are presented  
in appendix~\ref{app:FSH}. 

The fields $A_\mu(x)$, $X_i(x)$, $c(x)$, and $\bar{c}(x)$ 
are expanded by using the scalar fuzzy spherical harmonics 
$\hat{Y}^{(jj)}_{Jm}$ as
\begin{align}
 A_\mu(x) &= \sum_{J=0}^{2j} \sum_{m=-J}^{J} \int \frac{d^2 p}{(2\pi)^2} \,e^{ip\cdot x} \ 
 \hat{Y}^{(jj)}_{J\, m} \otimes a_{\mu,\,J \,m}(p), \nn \\ 
 X_i(x) &= \sum_{J=0}^{2j} \sum_{m=-J}^{J} \int \frac{d^2 p}{(2\pi)^2} \,e^{ip\cdot x} \ 
 \hat{Y}^{(jj)}_{J\, m} \otimes x_{i,\, J \, m}(p), \nn \\
 c(x) &= \sum_{J=0}^{2j} \sum_{m=-J}^{J} \int \frac{d^2 p}{(2\pi)^2}\, e^{ip\cdot x} \ 
 \hat{Y}^{(jj)}_{J\, m} \otimes c_{J\, m}(p), \nn \\
 \bar{c}(x) &= \sum_{J=0}^{2j} \sum_{m=-J}^{J} \int \frac{d^2 p}{(2\pi)^2}\, e^{ip\cdot x} \ 
 \hat{Y}^{(jj)}_{J\, m} \otimes \bar{c}_{J\, m}(p),
 \label{expand scalar}
\end{align}
where $n=2j+1$, and $a_{\mu,\, J\, m}(p)$, $x_{i,\, J\, m}(p)$, $c_{J\, m}(p)$, and $\bar{c}_{J\, m}(p)$ are 
$k\times k$ matrices. 

$\vec{\tilde{Y}}(x) \equiv \left( \tilde{X}_9(x),\tilde{X}_{10}(x),\tilde{X}_8(x)  \right)^T$ 
is expanded by the vector fuzzy spherical harmonics $\vec{\hat{Y}}^{\rho}_{J\, m\, (jj)}$ as
\begin{align}
 \vec{\tilde{Y}}(x) = \sum_{\rho=-1}^{1} \sum_{J=\delta_{\rho, 0}}^{2j-\delta_{\rho,-1}} 
 \sum_{m=-Q}^{Q} \int \frac{d^2 p}{(2\pi)^2} \,e^{ip\cdot x} \ 
 \vec{\hat{Y}}^{\rho}_{J\, m\, (jj)} \otimes y_{J\, m,\, \rho}(p), 
 \label{expand vector}
\end{align}
where $Q\equiv J+\delta_{\rho,\,1}$, and $y_{J\, m,\, \rho}(p)$ are $k \times k$ matrices. 
Note that $m$ runs from $-J-1$ to $J+1$ for $\rho=1$, and 
$J$ runs from $0$ to $2j-1$ for $\rho=-1$. Also, $J=1,\cdots, 2j$ for $\rho=0$.  

The fermions $\psi_{r\alpha}(x)$ are expanded by the spinor fuzzy spherical harmonics, 
\begin{align}
 \hat{Y}^{\kappa}_{J\, m\, (jj)}=\left(\begin{matrix}
  \hat{Y}^{\kappa}_{J\, m\, (jj) \, \alpha=\frac{1}{2}}\\
  \hat{Y}^{\kappa}_{J\, m\, (jj) \, \alpha=-\frac{1}{2}}
\end{matrix}\right) , 
\end{align}
as 
\begin{equation}
 \left(\sigma_1\right)_{\alpha\beta} \psi_{r\,\beta}(x)=
 \sum_{\kappa=\pm 1} \sum_{J=\frac{1}{2} \delta_{\kappa,-1}}^{2j-\frac{1}{2}\delta_{\kappa,-1}} 
 \sum_{m=-U}^U
 \int \frac{d^2 p}{(2\pi)^2} \, e^{ip\cdot x} \ 
 \hat{Y}^\kappa_{J \, m \,(jj)\,\alpha} \otimes \psi_{J \, m,\, \kappa}(p)_r, 
 \label{expand spinor}
\end{equation}
where $U\equiv J+\frac{1}{2} \delta_{\kappa,\,1}$, and $\psi_{J\,m,\,\kappa}(p)_r$ are $k\times k$ matrices. 
Note that $m=-J-\frac12,\cdots, J+\frac12$ and $J=0,\cdots, 2j$ for $\kappa=1$, while 
$m=-J, \cdots, J$ and $J=\frac12, \cdots, 2j-\frac12$ for $\kappa=-1$. 

We also define the vertex coefficients
\bea
& & \hat{C}^{J_1\,m_1\,(jj)}_{J_2\,m_2\,(jj)\, J_3\,m_3\,(jj)} \equiv 
\frac{1}{n}\,\tr_n\left\{ \left(\hat{Y}^{(jj)}_{J_1\,m_1}\right)^\dagger \hat{Y}^{(jj)}_{J_2\,m_2}
\hat{Y}^{(jj)}_{J_3\,m_3}\right\},  
\label{VC_FSH_def} \\
& & \hat{\cD}^{J\,m\,(jj)}_{J_1\,m_1\,(jj)\,\rho_1\,J_2\,m_2\,(jj)\,\rho_2} 
\equiv \sum_{i=1}^3\frac{1}{n}\,\tr_n\left\{\left(\hat{Y}^{(jj)}_{J\,m}\right)^\dagger 
\hat{Y}^{\rho_1}_{J_1\,m_1\,(jj)\,i} \hat{Y}^{\rho_2}_{J_2\,m_2\,(jj)\,i}\right\}, 
\label{VC_D_def} \\
& & \hat{\cE}_{J_1\,m_1\,(jj)\,\rho_1\,J_2\,m_2\,(jj)\,\rho_2\,J_3\,m_3\,(jj)\,\rho_3} \nn \\
& & \hspace{4mm}\equiv \sum_{i',j',k'=1}^3\epsilon_{i'j'k'}\frac{1}{n}\,\tr_n \left\{
\hat{Y}^{\rho_1}_{J_1\,m_1\,(jj)\,i'} \hat{Y}^{\rho_2}_{J_2\,m_2\,(jj)\,j'}
 \hat{Y}^{\rho_3}_{J_3\,m_3\,(jj)\,k'}\right\}, 
 \label{VC_E_def} \\
& & \hat{\cF}^{J_1\,m_1\,(jj)\,\kappa_1}_{J_2\,m_2\,(jj)\,\kappa_2\,J\,m\,(jj)} 
\equiv \sum_{\alpha=-\frac12}^{\frac12}\frac{1}{n} \,\tr_n\left\{ 
\left(\hat{Y}^{\kappa_1}_{J_1\,m_1\,(jj)\,\alpha}\right)^\dagger \hat{Y}^{\kappa_2}_{J_2\,m_2\,(jj)\,\alpha} 
\hat{Y}^{(jj)}_{J\,m}\right\}, 
\label{VC_F_def} \\
& & \hat{\cG}^{J_1\,m_1\,(jj)\,\kappa_1}_{J_2\,m_2\,(jj)\,\kappa_2\,J\,m\,(jj)\,\rho} \nn \\
& & \hspace{4mm}\equiv \sum_{\alpha,\beta=-\frac12}^{\frac12}\sum_{i=1}^3 \sigma^i_{\alpha\beta}\frac{1}{n} \,\tr_n\left\{
\left(\hat{Y}^{\kappa_1}_{J_1\,m_1\,(jj)\,\alpha}\right)^\dagger 
\hat{Y}^{\kappa_2}_{J_2\,m_2\,(jj)\,\beta} \hat{Y}^\rho_{J\,m\,(jj)\,i}\right\}. 
\label{VC_G_def}
\eea
Substituting (\ref{expand scalar}), (\ref{expand vector}), and (\ref{expand spinor}) into 
the action (\ref{total action}) and using these vertices, 
we obtain the action with respect to the modes 
both for the momentum in the 2D flat directions and the angular momentum 
in the fuzzy $S^2$ directions. 
We write it as 
\begin{equation}
S_{\rm tot}= S_{2,B}+S_{2,F}+S_{3,B}+S_{3,F}+S_4,
\end{equation}
where 
$S_{2,B}$ and $S_{2,F}$ are bosonic and fermionic kinetic terms, 
$S_{3,B}$ and  $S_{3,F}$ denote bosonic and fermionic 3-point 
interaction terms, 
and $S_4$ represents (bosonic) 4-point interaction terms. 
The interaction terms are further decomposed as
\begin{align}
 S_{3,B}&=S_{3,B}^{C} + S_{3,B}^{\cD} + S_{3,B}^{\cE} , \nn \\
 S_{3,F}&=S_{3,F}^{\cF}+S_{3,F}^{\cG}, \nn \\
 S_4&=S_4^{CC}+S_4^{C\cD}+S_4^{\cD\cD}+S_4^{\cE\cE}, 
\end{align} 
corresponding the vertex coefficients that appear. 
The explicit form of the kinetic terms reads 
\begin{align}
\label{S2B}
S_{2, B} 
 =&  \frac{n}{g_{2d}^2}\int\frac{d^2q}{(2\pi)^2}\sum_{J=0}^{2j}\sum_{m=-J}^J(-1)^m \nn \\
&  \times\tr_k\left[
\left(q^2 +\frac{M^2}{9}J(J+1)\right)\, a_{\nu,\,J\,-m}(-q)a_{\nu,\,J\,m}(q) \right. \nn \\
&  \hspace{12mm}+\left(q^2+ \frac{2M^2}{81}+\frac{M^2}{9}J(J+1)\right) 
x_{i,\,J\,-m}(-q) x_{i,\,J\,m}(q) \nn \\
&  \hspace{12mm}+\frac{4M}{9}x_{7,\,J\,-m}(-q)\left(q_1 \,a_{2,\,J\,m}(q) -q_2 \,a_{1,\,J\,m}(q)\right)\nn \\
&  \hspace{12mm}\left. -2\left(q^2 +\frac{M^2}{9}J(J+1)\right)\,\bar{c}_{J\,-m}(-q)\,c_{J\,m}(q)\right] \nn \\
& \hspace{-7mm} + \frac{n}{g_{2d}^2}\int\frac{d^2q}{(2\pi)^2}\sum_{J=0}^{2j}\sum_{m=-J-1}^{J+1}(-1)^{m-1}\,\tr_k\left[
\left(q^2 +\frac{M^2}{9}(J+1)^2\right) 
y_{J\,-m,\,\rho=1}(-q)\,y_{J\,m,\,\rho=1}(q)\right] \nn \\
& \hspace{-7mm} + \frac{n}{g_{2d}^2}\int\frac{d^2q}{(2\pi)^2}\sum_{J=0}^{2j-1}\sum_{m=-J}^{J}(-1)^{m-1}\,\tr_k\left[
\left(q^2 +\frac{M^2}{9}(J+1)^2\right) 
y_{J\,-m,\,\rho=-1}(-q)\,y_{J\,m,\,\rho=-1}(q) \right] \nn \\
& \hspace{-7mm} + \frac{n}{g_{2d}^2}\int\frac{d^2q}{(2\pi)^2}\sum_{J=1}^{2j}\sum_{m=-J}^{J}(-1)^{m-1}\,\tr_k\left[
\left(q^2 + \frac{M^2}{9}J(J+1)\right) 
 y_{J\,-m,\,\rho=0}(-q) \,y_{J\,m,\,\rho=0}(q) \right], \\ 
S_{2,F} =&  \frac{n}{g_{2d}^2}\int\frac{d^2q}{(2\pi)^2}\sum_{J=0}^{2j}\sum_{m=-J-\frac12}^{J+\frac12}
(-1)^{-m-\frac12}\,\tr_k\left[\Psi_{J\,-m,\,\kappa=1}(-q)^T\hat{D}_{J,\,\kappa=1}(q) \Psi_{J\,m,\,\kappa=1}(q)
\right] \nn \\
&+ \frac{n}{g_{2d}^2}\int\frac{d^2q}{(2\pi)^2}\sum_{J=\frac12}^{2j-\frac12}\sum_{m=-J}^{J}
(-1)^{-m+\frac12}\,\tr_k\left[\Psi_{J\,-m,\,\kappa=-1}(-q)^T\hat{D}_{J,\,\kappa=-1}(q) \Psi_{J\,m,\,\kappa=-1}(q)
\right] ,
\label{S2F}
\end{align}
with $q^2\equiv q_\mu q_\mu$, 
\be
\Psi_{J\, m,\,\kappa}(q)\equiv (\psi_{J\, m,\,\kappa}(q)_1, \cdots , \,\psi_{J\, m,\,\kappa}(q)_8)^T, 
\label{Psi_Jm}
\ee  
\bea
& & \hspace{-7mm}\hat{D}_{J,\,\kappa=1}(q) \equiv  \begin{bmatrix} 
   &             &              &              &   -q_1   &    q_2     &         &     \\
                &    &          &             &    -q_2   &    -q_1     &       &     \\
                 &            &    &      &        &          &   -q_1   & -q_2  \\
             &            &            &      &      &         &   q_2   &  -q_1   \\
    q_1     &    q_2     &           &        &        &       &         &       \\
   -q_2     &    q_1      &          &          &        &      &        &        \\
          &           &    q_1   &  -q_2     &        &            &      &        \\
          &           &   q_2    &   q_1   &          &           &            &     
\end{bmatrix} \nn \\
& & \hspace{-7mm}-\frac{M}{3}\,{\rm diag}\left(J+\frac13,\,J+\frac13,\,J+1,\,J+1,\,J+\frac23,\,J+\frac23,\,J+\frac23,\,J+1\right) 
\eea
and 
\bea
& & \hspace{-7mm}\hat{D}_{J,\,\kappa=-1}(q)  \equiv  \begin{bmatrix} 
   &             &              &              &   -q_1   &    q_2     &         &     \\
                &    &          &             &    -q_2   &    -q_1     &       &     \\
                 &            &    &      &        &          &   -q_1   & -q_2  \\
             &            &            &      &      &         &   q_2   &  -q_1   \\
    q_1     &    q_2     &           &        &        &       &         &       \\
   -q_2     &    q_1      &          &          &        &      &        &        \\
          &           &    q_1   &  -q_2     &        &            &      &        \\
          &           &   q_2    &   q_1   &          &           &            &            
\end{bmatrix} \nn \\
& & \hspace{-7mm}
+\frac{M}{3}\,{\rm diag}\left(J+\frac76,\,J+\frac76,\,J+\frac12,\,J+\frac12,\,J+\frac56,\,J+\frac56,\,J+\frac56,\,J+\frac12\right).
\eea 
Here ``$\tr_k$'' denotes the $k$-dimensional trace acting on the modes.     
The 3-point interaction terms are expressed as 
\begin{align}
S_{3, B}^{C} 
 = & \frac{n}{g_{2d}^2}\int \frac{d^2p}{(2\pi)^2} \frac{d^2q}{(2\pi)^2}\frac{d^2r}{(2\pi)^2}
\,(2\pi)^2\delta^2(p+q+r)\nn \\
 & \times \sum_{J,J',J''=0}^{2j}\sum_{m=-J}^J\sum_{m'=-J'}^{J'} \sum_{m''=-J''}^{J''} (-1)^m\,
\hat{C}^{J\,-m\,(jj)}_{J'\,m'\,(jj)\,J''\,m''\,(jj)} \nn \\
 & \hspace{7mm}\times \tr_k\Biggl[2(p_1-r_1)\,a_{1,\,J\,m}(q) a_{2,\,J'\,m'}(p)a_{2,\,J''\,m''}(r)  \nn \\
 & \hspace{18mm} +2(p_2-r_2)\,a_{2,\,J\,m}(q) a_{1,\,J'\,m'}(p)a_{1,\,J''\,m''}(r) \nn \\
 & \hspace{18mm} +2(p_\mu-r_\mu)\,a_{\mu,\,J\,m}(q) x_{i,\,J'\,m'}(p)x_{i,\,J''\,m''}(r) \nn \\
 & \hspace{18mm} +\frac{4M}{9}x_{7,\,J\,m}(p)a_{1,\,J'\,m'}(q)a_{2,\,J''\,m''}(r) 
-\frac{4M}{9}x_{7,\,J\,m}(p)a_{2,\,J'\,m'}(q)a_{1,\,J''\,m''}(r) \nn \\
 & \hspace{18mm} +\frac{4M}{9}x_{7,\,J\,m}(p)x_{3,\,J'\,m'}(q)x_{4,\,J''\,m''}(r) 
-\frac{4M}{9}x_{7,\,J\,m}(p)x_{4,\,J'\,m'}(q)x_{3,\,J''\,m''}(r) \nn \\
 & \hspace{18mm}  +2p_\mu\,\bar{c}_{J\,m}(p)\,a_{\mu,\,J'\,m'}(q)\,c_{J''\,m''}(r) 
-2p_\mu\,\bar{c}_{J\,m}(p)\,c_{J'\,m'}(q)\,a_{\mu,\,J''\,m''}(r)\Biggr],
\label{S3B_C}
\end{align}
\begin{align}
S_{3, B}^{\cD} 
 = &  \frac{n}{g_{2d}^2}\int \frac{d^2p}{(2\pi)^2} \frac{d^2q}{(2\pi)^2}\frac{d^2r}{(2\pi)^2}
\,(2\pi)^2\delta^2(p+q+r)\nn \\
& \times \sum_{J=0}^{2j}\sum_{m=-J}^J\sum_{\rho',\rho''=-1}^1\sum_{J'=\delta_{\rho', 0}}^{2j-\delta_{\rho',-1}} 
\sum_{J''=\delta_{\rho'', 0}}^{2j-\delta_{\rho'',-1}} \sum_{m'=-Q'}^{Q'} \sum_{m''=-Q''}^{Q''} 
(-1)^m\,\hat{\cD}^{J\,-m\,(jj)}_{J'\,m'\,(jj)\,\rho'\,J''\,m''\,(jj)\,\rho''} \nn \\
& \hspace{7mm}\times 
\tr_k\left[2(p_\mu-r_\mu)\,a_{\mu,\,J\,m}(q)\,y_{J'\,m',\,\rho'}(p)\,y_{J''\,m'',\,\rho''}(r)\right] 
\nn \\
& \hspace{-4mm}+ \frac{n}{g_{2d}^2}\int \frac{d^2p}{(2\pi)^2} \frac{d^2q}{(2\pi)^2}\frac{d^2r}{(2\pi)^2}
\,(2\pi)^2\delta^2(p+q+r)\nn \\
& \times \sum_{J,J'=0}^{2j} \sum_{m=-J}^J\sum_{m'=-J'}^{J'} \sum_{\rho''=-1}^1 
\sum_{J''=\delta_{\rho'', 0}}^{2j-\delta_{\rho'',-1}} \sum_{m''=-Q''}^{Q''}(-1)^{m'}\,\frac{M}{3}\sqrt{J(J+1)} \nn \\
& \times\Bigl\{ \hat{\cD}^{J'\,-m'\,(jj)}_{J''\,m''\,(jj)\,\rho''\,J\,m\,(jj)\,0} \,
\tr_k\left[2a_{\mu,\,J\,m}(p)\,a_{\mu,\,J'\,m'}(q)\,y_{J''\,m'',\,\rho''}(r) \right.\nn \\
& \hspace{50mm} +2x_{i,\,J\,m}(p)\,x_{i,\,J'\,m'}(q)\,y_{J''\,m'',\,\rho''}(r) \nn \\
& \hspace{50mm}\left. -2\bar{c}_{J\,m}(p)\,c_{J'\,m'}(q)\,y_{J''\,m'',\,\rho''}(r)\right] \nn \\
& \hspace{7mm} +\hat{\cD}^{J'\,-m'\,(jj)}_{J\,m\,(jj)\,0\,J''\,m''\,(jj)\,\rho''} \,
\tr_k\left[-2a_{\mu,\,J'\,m'}(p)\,a_{\mu,\,J\,m}(q)\,y_{J''\,m'',\,\rho''}(r)  \right.\nn \\
 & \hspace{53mm} -2x_{i,\,J'\,m'}(p)\,x_{i,\,J\,m}(q)\,y_{J''\,m'',\,\rho''}(r) \nn \\
 & \hspace{53mm} \left. +2\bar{c}_{J\,m}(p)\,y_{J''\,m'',\,\rho''}(q)\,c_{J'\,m'}(r)\right]
\Bigr\}, 
\label{S3B_D}
\end{align}
\begin{align}
S_{3,B}^{\cE} = & \frac{n}{g_{2d}^2}\int \frac{d^2p}{(2\pi)^2} \frac{d^2q}{(2\pi)^2}\frac{d^2r}{(2\pi)^2}
\,(2\pi)^2\delta^2(p+q+r)\nn \\
& \times \sum_{\rho,\rho',\rho''=-1}^1 \sum_{J=\delta_{\rho, 0}}^{2j-\delta_{\rho, -1}}
\sum_{J'=\delta_{\rho', 0}}^{2j-\delta_{\rho', -1}}\sum_{J''=\delta_{\rho'', 0}}^{2j-\delta_{\rho'', -1}}
\sum_{m=-Q}^Q\sum_{m'=-Q'}^{Q'} \sum_{m''=-Q''}^{Q''} i\frac{2M}{3}\rho (J+1) \nn \\
 & \hspace{7mm}\times \hat{\cE}_{J\,m\,(jj)\,\rho\,J'\,m'\,(jj)\,\rho'\,J''\,m''\,(jj)\,\rho''} \,
 \tr_k\left[y_{J\,m,\,\rho}(p)\,y_{J'\,m',\,\rho'}(q)\,y_{J''\,m'',\,\rho''}(r)
\right], 
\label{S3B_E}
\end{align}
\begin{align}
S_{3,F}^{\cF} = & \frac{n}{g_{2d}^2}\int \frac{d^2p}{(2\pi)^2} \frac{d^2q}{(2\pi)^2}\frac{d^2r}{(2\pi)^2}
\,(2\pi)^2\delta^2(p+q+r)\nn \\
 & \hspace{-7mm}\times \sum_{\kappa,\kappa''=\pm 1}\sum_{J=\frac12\delta_{\kappa,-1}}^{2j-\frac12\delta_{\kappa,-1}} 
\sum_{J''=\frac12\delta_{\kappa'',-1}}^{2j-\frac12\delta_{\kappa'',-1}} 
\sum_{m=-U}^U\sum_{m''=-U''}^{U''}\sum_{J'=0}^{2j}\sum_{m'=-J'}^{J'} 
(-1)^{m-\frac12\kappa}\,\hat{\cF}^{J\,-m\,(jj)\,\kappa}_{J''\,m''\,(jj)\,\kappa''\,J'\,m'\,(jj)} \nn \\
 & \hspace{-4mm}\times \tr_k\left[2\Psi_{J\,m,\,\kappa}(p)^T\tilde{\gamma}_\mu \Psi_{J''\,m'',\,\kappa''}(r)\,
a_{\mu,\,J'\,m'}(q)  + 2\Psi_{J\,m,\,\kappa}(p)^T\tilde{\gamma}_i \Psi_{J''\,m'',\,\kappa''}(r)\,
x_{i,\,J'\,m'}(q)\right], 
\label{S3F_F}
\end{align}
\begin{align}
S_{3,F}^{\cG}  = & \frac{n}{g_{2d}^2}\int \frac{d^2p}{(2\pi)^2} \frac{d^2q}{(2\pi)^2}\frac{d^2r}{(2\pi)^2}
\,(2\pi)^2\delta^2(p+q+r)\nn \\
 & \times \sum_{\kappa,\kappa''=\pm 1}\sum_{J=\frac12\delta_{\kappa,-1}}^{2j-\frac12\delta_{\kappa,-1}} 
\sum_{J''=\frac12\delta_{\kappa'',-1}}^{2j-\frac12\delta_{\kappa'',-1}} 
\sum_{m=-U}^U\sum_{m''=-U''}^{U''}\sum_{\rho'=-1}^1\sum_{J'=\delta_{\rho', 0}}^{2j-\delta_{\rho',-1}}\sum_{m'=-Q'}^{Q'} 
(-1)^{m-\frac12\kappa}\,\nn \\
 & \hspace{7mm}\times 
\hat{\cG}^{J\,-m\,(jj)\,\kappa}_{J''\,m''\,(jj)\,\kappa''\,J'\,m'\,(jj)\,\rho'}\, 
\tr_k\left[2\Psi_{J\,m,\,\kappa}(p)^T\Psi_{J''\,m'',\,\kappa''}(r)\,
y_{J'\,m',\,\rho'}(q) \right] . 
\label{S3F_G}
\end{align}
Finally, the 4-point interaction terms are given by 
\begin{align}
S_4^{CC} = & \frac{n}{g_{2d}^2}\int\frac{d^2p}{(2\pi)^2} \frac{d^2q}{(2\pi)^2}\frac{d^2r}{(2\pi)^2}
\frac{d^2\ell}{(2\pi)^2}
\,(2\pi)^2\delta^2(p+q+r+\ell)\nn \\   
&  \times \sum_{J,J',J'',J''',J_1=0}^{2j} \sum_{m=-J}^J \sum_{m'=-J'}^{J'}  \sum_{m''=-J''}^{J''}
\sum_{m'''=-J'''}^{J'''}\sum_{m_1=-J_1}^{J_1} \nn \\
&  \hspace{7mm}\times (-1)^{m_1} \,
\hat{C}^{J_1\,m_1\,(jj)}_{J\,m\,(jj)\,J'\,m'\,(jj)} \,
\hat{C}^{J_1\,-m_1\,(jj)}_{J''\,m''\,(jj)\,J'''\,m'''\,(jj)} \nn \\
&  \hspace{7mm}\times \tr_k\left[ 
-2a_{1,\,J\,m}(p)a_{2,\,J'\,m'}(q)a_{1,\,J''\,m''}(r)a_{2,\,J'''\,m'''}(\ell) \right. \nn \\
&  \hspace{18mm}+2a_{1,\,J\,m}(p)a_{1,\,J'\,m'}(q)a_{2,\,J''\,m''}(r)a_{2,\,J'''\,m'''}(\ell) \nn \\
&  \hspace{18mm}-2a_{\mu,\,J\,m}(p) x_{i,\,J'\,m'}(q)a_{\mu,\,J''\,m''}(r) x_{i,\,J'''\,m'''}(\ell) \nn \\
&  \hspace{18mm}+2a_{\mu,\,J\,m}(p)a_{\mu,\,J'\,m'}(q) x_{i,\,J''\,m''}(r)x_{i,\,J'''\,m'''}(\ell) \nn \\
&  \hspace{18mm}-x_{i,\,J\,m}(p) x_{j,\,J'\,m'}(q)x_{i,\,J''\,m''}(r) x_{j,\,J'''\,m'''}(\ell) \nn \\
&  \hspace{18mm}\left.+x_{i,\,J\,m}(p)x_{i,\,J'\,m'}(q) x_{j,\,J''\,m''}(r)x_{j,\,J'''\,m'''}(\ell) \right],
\label{S4_CC}
\end{align}
\begin{align}
S_4^{C\cD}  = & \frac{n}{g_{2d}^2}\int\frac{d^2p}{(2\pi)^2} \frac{d^2q}{(2\pi)^2}\frac{d^2r}{(2\pi)^2}
\frac{d^2\ell}{(2\pi)^2}
\,(2\pi)^2\delta^2(p+q+r+\ell)\nn \\   
 & \times \sum_{J,J'',J_1=0}^{2j}\sum_{m=-J}^J\sum_{m''=-J''}^{J''} \sum_{m_1=-J_1}^{J_1} 
\sum_{\rho',\rho'''=-1}^1 \sum_{J'=\delta_{\rho', 0}}^{2j-\delta_{\rho',-1}}\sum_{J'''=\delta_{\rho''', 0}}^{2j-\delta_{\rho''',-1}}
\sum_{m'=-Q'}^{Q'}\sum_{m'''=-Q'''}^{Q'''} \nn \\
 & \hspace{7mm}\times (-1)^{m_1}\,\hat{C}^{J_1\,m_1\,(jj)}_{J\,m\,(jj)\,J''\,m''\,(jj)}\,
\hat{\cD}^{J_1\,-m_1\,(jj)}_{J'\,m'\,(jj)\,\rho'\,J'''\,m'''\,(jj)\,\rho'''} \nn \\
 & \hspace{7mm}\times \tr_k\left[
2a_{\mu,\,J\,m}(p)\,a_{\mu,\,J''\,m''}(r)\,y_{J'\,m',\,\rho'}(q)\,y_{J'''\,m''',\,\rho'''}(\ell) \right. \nn \\
 & \hspace{17mm}
 \left. +2x_{i,\,J\,m}(p)\,x_{i,\,J''\,m''}(r)\,
 y_{J'\,m',\,\rho'}(q)\,y_{J'''\,m''',\,\rho'''}(\ell) \right], 
\label{S4_CD}
\end{align}
\begin{align}
S_4^{\cD\cD} = & \frac{n}{g_{2d}^2}\int\frac{d^2p}{(2\pi)^2} \frac{d^2q}{(2\pi)^2}\frac{d^2r}{(2\pi)^2}
\frac{d^2\ell}{(2\pi)^2}
\,(2\pi)^2\delta^2(p+q+r+\ell)\nn \\   
 & \hspace{-7mm}\times \sum_{J,J''=0}^{2j}\sum_{m=-J}^J\sum_{m''=-J''}^{J''} \sum_{\rho',\rho''',\rho_1=-1}^1 
\sum_{J'=\delta_{\rho', 0}}^{2j-\delta_{\rho',-1}} \sum_{J'''=\delta_{\rho''', 0}}^{2j-\delta_{\rho''',-1}}\sum_{J_1=\delta_{\rho_1, 0}}^{2j-\delta_{\rho_1,-1}} 
\sum_{m'=-Q'}^{Q'}\sum_{m'''=-Q'''}^{Q'''}\sum_{m_1=-Q_1}^{Q_1} \nn \\
 & \times (-1)^{m+m''+m_1} 
\hat{\cD}^{J\,-m\,(jj)}_{J'\,m'\,(jj)\,\rho'\,J_1\,m_1\,(jj)\,\rho_1}\, 
\hat{\cD}^{J''\,-m''\,(jj)}_{J'''\,m'''\,(jj)\,\rho'''\,J_1\,-m_1\,(jj)\,\rho_1} \nn \\
 & \times \tr_k\left[2a_{\mu,\,J\,m}(p)\,y_{J'\,m',\,\rho'}(q)\, a_{\mu,\,J''\,m''}(r)\,y_{J'''\,m''',\,\rho'''}(\ell) \right. \nn \\
 & \hspace{10mm}\left. +2x_{i,\,J\,m}(p)\,y_{J'\,m',\,\rho'}(q)\, x_{i,\,J''\,m''}(r)\,y_{J'''\,m''',\,\rho'''}(\ell)  
\right],
\label{S4_DD}
\end{align}
\begin{align}
S_4^{\cE\cE} = & \frac{n}{g_{2d}^2}\int\frac{d^2p}{(2\pi)^2} \frac{d^2q}{(2\pi)^2}\frac{d^2r}{(2\pi)^2}
\frac{d^2\ell}{(2\pi)^2}
\,(2\pi)^2\delta^2(p+q+r+\ell)\sum_{\rho,\rho',\rho'',\rho''',\rho_1=-1}^1 \nn \\   
&  \hspace{-7mm}\times  \sum_{J=\delta_{\rho, 0}}^{2j-\delta_{\rho,-1}} 
\sum_{J'=\delta_{\rho', 0}}^{2j-\delta_{\rho',-1}} \sum_{J''=\delta_{\rho'', 0}}^{2j-\delta_{\rho'',-1}} \sum_{J'''=\delta_{\rho''', 0}}^{2j-\delta_{\rho''',-1}} 
\sum_{J_1=\delta_{\rho_1, 0}}^{2j-\delta_{\rho_1,-1}} 
\sum_{m=-Q}^{Q}  \sum_{m'=-Q'}^{Q'}\sum_{m''=-Q''}^{Q''}\sum_{m'''=-Q'''}^{Q'''}
\sum_{m_1=-Q_1}^{Q_1}  \nn \\
&  \times (-1)^{-m_1}
\hat{\cE}_{J\,m\,(jj)\,\rho\,J'\,m'\,(jj)\,\rho'\,J_1\,m_1\,(jj)\,\rho_1}\,
\hat{\cE}_{J''\,m''\,(jj)\,\rho''\,J'''\,m'''\,(jj)\,\rho'''\,J_1\,-m_1\,(jj)\,\rho_1} \nn \\
&  \times
\tr_k\left[y_{J\,m,\,\rho}(p)\,y_{J'\,m',\,\rho'}(q)\,y_{J''\,m'',\,\rho''}(r)\,y_{J'''\,m''',\,\rho'''}(\ell)
\right]. 
\label{S4_EE}
\end{align}
For later convenience, 
we present another expression for $S_4^{C\cD}$ in terms of only the $\cD$ coefficients rather than the $C$ and $\cD$ coefficients: 
\begin{align}
S_4^{C\cD}  = & \frac{n}{g_{2d}^2}\int\frac{d^2p}{(2\pi)^2} \frac{d^2q}{(2\pi)^2}\frac{d^2r}{(2\pi)^2}
\frac{d^2\ell}{(2\pi)^2}
\,(2\pi)^2\delta^2(p+q+r+\ell)\nn \\   
 & \times \sum_{J,J''=0}^{2j}\sum_{m=-J}^J\sum_{m''=-J''}^{J''} 
\sum_{\rho',\rho''',\rho_1=-1}^1 \sum_{J'=\delta_{\rho', 0}}^{2j-\delta_{\rho',-1}}\sum_{J'''=\delta_{\rho''', 0}}^{2j-\delta_{\rho''',-1}}
\sum_{J_1=\delta_{\rho_1, 0}}^{2j-\delta_{\rho_1,-1}}
\sum_{m'=-Q'}^{Q'}\sum_{m'''=-Q'''}^{Q'''} \sum_{m_1=-Q_1}^{Q_1}\nn \\
 & \hspace{7mm}\times (-1)^{m-1+m''+m_1}\,\hat{\cD}^{J\,-m\,(jj)}_{J_1\,m_1\,(jj)\, \rho_1 \,J'''\,m'''\,(jj)\,\rho'''}\,
\hat{\cD}^{J''\,-m''\,(jj)}_{J'\,m'\,(jj)\,\rho'\,J_1\,-m_1\,(jj)\,\rho_1} \nn \\
&  \hspace{7mm}\times \tr_k\left[
2a_{\mu,\,J\,m}(p)\,a_{\mu,\,J''\,m''}(r)\,y_{J'\,m',\,\rho'}(q)\,y_{J'''\,m''',\,\rho'''}(\ell) \right. \nn \\
 & \hspace{17mm}\left. +2x_{i,\,J\,m}(p)\,x_{i,\,J''\,m''}(r)\,y_{J'\,m',\,\rho'}(q)\,y_{J'''\,m''',\,\rho'''}(\ell) \right].
\label{S4_CD2}
\end{align}

\section{Loop corrections}
\label{sec:loop}

The action obtained in the previous section is that of a 4D theory 
on $\R^2\times ({\rm fuzzy}\ S^2)$. 
We consider the following successive limits: 
\begin{itemize}
\item {\bf Step 1} (Moyal limit): Take $n=2j+1 \to \infty$ with the fuzziness $\Theta\equiv \frac{18}{M^2 n}$ and $k$ fixed. 
Then, the IR and UV cutoffs on $S^2$ become $M\propto n^{-1/2}\to 0$ and 
$\Lambda_j\equiv \frac{M}{3}\cdot 2j\propto n^{1/2}\to \infty$, respectively.
Namely, the fuzzy $S^2$ is decompactified to the noncommutative (Moyal) plane $\R^2_\Theta$.
\item {\bf Step 2} (commutative limit): Send the noncommutativity parameter $\Theta$ to zero.
\end{itemize}
At the level of the classical action or tree level amplitudes at least, 
the theory coincides with the ordinary 
$\cN=4$ $U(k)$ SYM on $\R^4$ after these steps. 
One might further expect that Step 1 would be safe even quantum mechanically since $4D$
$\cN=4$ SYM is UV finite and the deformation by the mass $M$ would be soft. 
However, the situation is not so simple. 
Although the deformation parameter $M$ giving masses naively seems soft, 
the softness is not clear in the sense that $M$ gives not only the IR cutoff but also the UV cutoff.
In addition, the obtained theory at Step 1 is a noncommutative field theory; i.e, there might 
appear nontrivial divergence through the so-called UV/IR mixing in non-planar diagrams~\cite{Minwalla:1999px}. 
In fact, we have to consider non-planar contributions as well as planar ones 
and take care of both the UV and IR divergences. 
In the following, we first compute the one-point functions 
at the one-loop order in the next section (section~\ref{sec:1pt}),  
and then explicitly calculate 
the two-point functions of the scalar fields $X_i$ in section~\ref{sec:2pt}. 
This will give a check of whether we can take the limits safely 
even in the quantum mechanical sense. 

Upon the perturbative calculation, we rescale all the fields as
\be
(\mbox{field})\to \frac{g_{2d}}{\sqrt{2n}}\,(\mbox{field})
\label{field_rescale}
\ee
so that the kinetic terms take the canonical form. 
The kinetic terms of gauge fields $a_\mu$ and a scalar $x_7$ are written as  
\be
\int \frac{d^2q}{(2\pi)^2}\,\sum_{J=0}^{2j}\sum_{m=-J}^J\,\frac12 (-1)^m\,\tr_k\left[\varphi_{K,\,J\,-m}(-q) 
\, \tilde{\Delta}_J(q)_{K\,K'}\, \varphi_{K',\,J\,m}(q)\right], 
\ee
where $K, \,K'=1,2,3$, 
\be
\varphi_{K=1,\,J\,m} \equiv a_{1,\,J\,m}, \qquad 
\varphi_{K=2,\,J\,m} \equiv a_{2,\,J\,m}, \qquad 
\varphi_{K=3,\,J\,m} \equiv x_{7,\,J\,m}, 
\ee
and the kinetic kernel is a $3\times 3$ matrix: 
\be
\tilde{\Delta}_J(q)\equiv \begin{bmatrix}q^2+\frac{M^2}{9}J(J+1) & 0   & \frac{2M}{9}q_2 \\
0  & q^2 +\frac{M^2}{9}J(J+1) & -\frac{2M}{9}q_1 \\ 
-\frac{2M}{9}q_2 & \frac{2M}{9}q_1 & q^2+\frac{2M^2}{81} +\frac{M^2}{9}J(J+1) \end{bmatrix}.
\ee
Then, we obtain the propagator 
\bea
& & \vev{\varphi_{K,\,J\,m}(q)_{s\,t}\,\varphi_{K',\,J'\,m'}(q')_{s'\,t'}}_0 \nn \\
& & \hspace{7mm}= \delta_{s\,t'}\,\delta_{t\,s'}\,\delta_{J\,J'}\, \delta_{m+m',\,0}\,(-1)^{m'}\,(2\pi)^2\,\delta^2(q+q') 
\left(\tilde{\Delta}_J(q)^{-1}\right)_{K\,K'}.
\label{propa_ax7}
\eea
Here, $s, \,t, \,s', \,t'(=1, \cdots, k)$ denote the color indices. 
The inverse of the kinetic kernel is given by 
\begin{align}
&\hspace{-4mm}\tilde{\Delta}_J(q)^{-1} = \frac{1}{q^2+\frac{M^2}{9}J(J+1)} \,\id_3 \nn \\
& +\frac{\frac{2M}{9}}{[q^2+\frac{M^2}{9}J(J+\frac13)][q^2+\frac{M^2}{9}(J+1)(J+\frac23)]}
\begin{bmatrix}     &    & -q_2 \\
                    &    &  q_1 \\
                q_2 & -q_1 & -\frac{M}{9} \end{bmatrix} \nn \\
& + \frac{\frac{4M^2}{81}}{[q^2 + \frac{M^2}{9}J(J+1)][q^2+\frac{M^2}{9}J(J+\frac13)][q^2+\frac{M^2}{9}(J+1)(J+\frac23)]} 
\begin{bmatrix} -q_2^2 & q_1q_2 &    \\
                q_1q_2 & -q_1^2 &    \\
                       &        & -q^2 \end{bmatrix}.
\end{align}       

The propagators for the other bosons and ghosts can be easily read off.  
For scalars, 
\bea
& & \vev{x_{i,\,J\,m}(q)_{s\,t}\, x_{i',\,J'\,m'}(q')_{s'\,t'}}_0 \nn \\
& & \hspace{7mm}= \delta_{i\,i'}\, \delta_{s\,t'}\,\delta_{t\,s'}\,\delta_{J\,J'}\, \delta_{m+m',\,0}\,(-1)^{m'}\,(2\pi)^2\,\delta^2(q+q') 
\, \frac{1}{q^2+\frac{M^2}{9}(J+\frac13)(J+\frac23)}  \nn \\
\eea
with $i,i'=3,4,5,6$. 
For ghosts and $y$-fields, 
\bea
& & \vev{c_{J\,m}(q)_{s\,t}\,\bar{c}_{J'\,m'}(q')_{s'\,t'}}_0 \nn \\
& & \hspace{7mm}=  \delta_{s\,t'}\,\delta_{t\,s'}\,\delta_{J\,J'}\, \delta_{m+m',\,0}\,(-1)^{m'}\,(2\pi)^2\,\delta^2(q+q') 
\, \frac{-1}{q^2+\frac{M^2}{9}J(J+1)}, \\
& & \vev{y_{J\,m,\,\rho}(q)_{s\,t}\,y_{J'\,m',\,\rho'}(q')_{s'\,t'}}_0 \nn \\
& & \hspace{7mm}=  \delta_{\rho\,\rho'}\, \delta_{s\,t'}\,\delta_{t\,s'}\,\delta_{J\,J'}\, \delta_{m+m',\,0}\,(-1)^{m'+1}\,
(2\pi)^2\,\delta^2(q+q') \,\tilde{\Delta}_{(\rho, J)}(q)^{-1}
\eea
with 
\be
\tilde{\Delta}_{(\rho, J)}(q)^{-1}\equiv\frac{\delta_{\rho, 0}}{q^2+\frac{M^2}{9}J(J+1)} + 
\frac{\delta_{\rho, 1} + \delta_{\rho, -1}}{q^2+\frac{M^2}{9}(J+1)^2}. 
\label{prop_y}
\ee

Finally, the fermion propagators are obtained as 
\bea
& & \vev{\hat{\Psi}_{J\,m,\,\kappa}(q)_{r,\,s\,t}\,\hat{\Psi}_{J'\,m',\kappa'}(q')_{r',\,s'\,t'}}_0 \nn \\
& & \hspace{7mm}= \delta_{\kappa,\,\kappa'}\,\delta_{s\,t'}\,\delta_{t\,s'}\,\delta_{J\,J'}\, \delta_{m+m',\,0}\,
(-1)^{m+\frac12\kappa}\,(2\pi)^2\,\delta^2(q+q') \, \left(\hat{D}_{J,\,\kappa}(q)^{-1}\right)_{r\,r'},
\eea
where $r,\,r'(=1, \cdots, 8)$ label the spinor components of (\ref{Psi_Jm}),
and the inverse of the kernel is given by
\bea
\hspace{-7mm}\hat{D}_{J,\,\kappa=1}(q)^{-1} 
& = & \frac{1}{q^2+\frac{M^2}{9}(J+1)^2} \nn \\
& & \hspace{3mm}\times \begin{bmatrix} 0        &        &        &        &         &         &        &       \\
                         &   0    &        &        &         &         &        &        \\
                         &         & -\frac{M}{3}(J+1) &   &  &         &        &  q_2   \\
                         &       &      &  -\frac{M}{3}(J+1) &      &   &        &  q_1   \\
                         &        &        &         &     0  &        &        &         \\
                         &        &        &        &         &  0    &         &         \\
                        &        &         &         &        &         &  0    &         \\
                        &        &  -q_2   & -q_1    &       &          &       & -\frac{M}{3}(J+1) \end{bmatrix} 
\nn 
\eea  
\bea
& & +\frac{1}{q^2+\frac{M^2}{9}(J+1)(J+\frac23)} 
\begin{bmatrix}   0     &         &        &         &          &        &          &         \\
                        &   0     &        &         &          &        &           &        \\
                        &          &  0    &         &          &        &  q_1      &        \\
                        &          &        &   0    &           &       &  -q_2     &         \\
                        &          &        &          &    0   &        &           &         \\
                        &         &        &         &           &   0   &           &         \\
                        &         & -q_1   & q_2     &          &        & -\frac{M}{3}(J+1) &    \\
                        &         &        &         &          &        &           &    0    \end{bmatrix}  
\nn \\
& & +\frac{1}{q^2 + \frac{M^2}{9}(J+\frac13)(J+\frac23)} \nn \\
& & \hspace{7mm} \times 
\begin{bmatrix} -\frac{M}{3}(J+\frac23) &          &        &      &   q_1    &    -q_2      &    &    \\
                              & -\frac{M}{3}(J+\frac23) &   &     &   q_2    &   q_1       &     &   \\
                                     &               &   0  &      &         &             &     &    \\
                                     &              &        & 0   &         &            &      &    \\
                          -q_1     &   -q_2     &        &      & -\frac{M}{3}(J+\frac13) &    &   &   \\
                            q_2    &  -q_1      &        &      &     & -\frac{M}{3}(J+\frac13) &  &   \\
                                   &             &         &    &               &          &  0   &    \\
                                   &             &       &       &          &          &        &  0  \end{bmatrix}
\nn \\
& & + \frac{\frac{M}{9}}{[q^2+\frac{M^2}{9}(J+1)^2][q^2+\frac{M^2}{9}(J+1)(J+\frac23)]}
\begin{bmatrix}  0  &   &        &           &     &      &    &     \\
                    & 0 &        &            &    &      &    &     \\
                    &   & q_1^2 &  -q_1q_2  &     &       &     &     \\
                    &    & -q_1q_2 & q_2^2  &     &      &      &     \\
                    &    &       &         &  0   &      &      &     \\
                    &    &       &         &     &  0    &     &      \\
                    &    &        &        &     &      &  0  &       \\
                    &    &        &        &     &      &     &   0    \end{bmatrix}   \nn \\
& &          
\eea       
and 
\bea
\hspace{-7mm}\hat{D}_{J,\,\kappa=-1}(q)^{-1} 
& = & \frac{1}{q^2+\frac{M^2}{9}(J+\frac12)^2} \nn \\
& & \hspace{3mm} \times \begin{bmatrix} 0        &        &        &        &         &         &        &       \\
                         &   0    &        &        &         &         &        &        \\
                         &         & \frac{M}{3}(J+\frac12) &   &  &         &        &  q_2   \\
                         &       &      & \frac{M}{3}(J+\frac12) &      &   &        &  q_1   \\
                         &        &        &         &     0  &        &        &         \\
                         &        &        &        &         &  0    &         &         \\
                        &        &         &         &        &         &  0    &         \\
                        &        &  -q_2   & -q_1    &       &          &       & \frac{M}{3}(J+\frac12) \end{bmatrix} 
\nn 
\eea
\bea
& & +\frac{1}{q^2+\frac{M^2}{9}(J+\frac12)(J+\frac56)} 
\begin{bmatrix}   0     &         &        &         &          &        &          &         \\
                        &   0     &        &         &          &        &           &        \\
                        &          &  0    &         &          &        &  q_1      &        \\
                        &          &        &   0    &           &       &  -q_2     &         \\
                        &          &        &          &    0   &        &           &         \\
                        &         &        &         &           &   0   &           &         \\
                        &         & -q_1   & q_2     &          &        & \frac{M}{3}(J+\frac12) &    \\
                        &         &        &         &          &        &           &    0    \end{bmatrix}  
\nn \\
& & +\frac{1}{q^2 + \frac{M^2}{9}(J+\frac56)(J+\frac76)} \nn \\
& & \hspace{7mm}\times 
\begin{bmatrix} \frac{M}{3}(J+\frac56) &          &        &      &   q_1    &    -q_2      &    &    \\
                              & \frac{M}{3}(J+\frac56) &   &     &   q_2    &   q_1       &     &   \\
                                     &               &   0  &      &         &             &     &    \\
                                     &              &        & 0   &         &            &      &    \\
                          -q_1     &   -q_2     &        &      & \frac{M}{3}(J+\frac76) &    &   &   \\
                            q_2    &  -q_1      &        &      &     & \frac{M}{3}(J+\frac76) &  &   \\
                                   &             &         &    &               &          &  0   &    \\
                                   &             &       &       &          &          &        &  0  \end{bmatrix}
\nn \\
& & + \frac{\frac{M}{9}}{[q^2+\frac{M^2}{9}(J+\frac12)^2][q^2+\frac{M^2}{9}(J+\frac12)(J+\frac56)]} 
\begin{bmatrix}  0  &   &        &           &     &      &    &     \\
                    & 0 &        &            &    &      &    &     \\
                    &   & q_1^2 &  -q_1q_2  &     &       &     &     \\
                    &    & -q_1q_2 & q_2^2  &     &      &      &     \\
                    &    &       &         &  0   &      &      &     \\
                    &    &       &         &     &  0    &     &      \\
                    &    &        &        &     &      &  0  &       \\
                    &    &        &        &     &      &     &   0    \end{bmatrix}.  \nn \\
 & & 
\eea             

\section{One-point functions at one-loop}
\label{sec:1pt}
Due to the $Q_\pm$ supersymmetries, gauge invariant one-point functions should not be induced
radiatively for any $n=2j+1$ and $M$. 
As a warm-up exercise, let us check this at the one-loop level by computing tadpole diagrams. 
We also see that the one-point functions do not contribute 
to the one-loop effective action in the Moyal limit. 
In this section, $\bar{p}$ and $(\bar{J},\,\bar{m})$ denote external momentum and angular momentum, respectively. 

\subsection{One-point functions of $x_i$ and $a_\mu$}
The one-loop contribution to $\vev{\frac{1}{k}\tr_k\,x_{i, \,\bar{J}\, \bar{m}}(\bar{p})}$ ($i=3,4,5,6$) comes from 
the Wick contractions among $\frac{1}{k}\tr_k\,x_{i, \,\bar{J}\, \bar{m}}(\bar{p})$ and $-S_{3, F}^{\cF}$. 
We obtain 
\begin{align}
\vev{\frac{1}{k}\tr_k\,x_{i, \,\bar{J}\, \bar{m}}(\bar{p})} 
=&  \frac{g_{2d}}{\sqrt{2n}} \,k\,\delta^2(\bar{p})
\,(-1)^{-\bar{m}}\int^{\Lambda_p} d^2p \sum_{\kappa =\pm 1} 
\sum_{J=\frac12\delta_{\kappa,\,-1}}^{2j-\frac12\delta_{\kappa, \,-1}}
\sum_{m=-U}^U
\nn \\
 & \times \hat{\cF}^{J\,-m\,(jj)\,\kappa}_{J\, -m\,(jj) \, \kappa\,\bar{J}\,-\bar{m}\,(jj)}
\frac{1}{\frac{M^2}{9}(\bar{J}+\frac13)(\bar{J}+\frac23)}\,\tr_8\left(\hat{\gamma}_i\hat{D}_{J,\,\kappa}(p)^{-1}\right)
\end{align}
at the one-loop order. 
We introduced a cutoff $\Lambda_p$ for the loop momentum integration. ``$\tr_8$'' stands for trace over spinor indices.  
It is easy to see that $\tr_8\left(\hat{\gamma}_i\hat{D}_{J,\,\kappa}(p)^{-1}\right)$ vanishes for each $i$ and $\kappa$. Hence, 
\be
\vev{\frac{1}{k}\tr_k\,x_{i, \,\bar{J}\, \bar{m}}(\bar{p})} =0 \qquad \mbox{at the one-loop order}.
\ee  

The one-loop contribution to 
$\vev{\frac{1}{k}\tr_k\,x_{7, \,\bar{J}\, \bar{m}}(\bar{p})}$ can be written as 
\be
\vev{\frac{1}{k}\tr_k\,x_{7, \,\bar{J}\, \bar{m}}(\bar{p})} 
= \vev{\frac{1}{k}\tr_k\,x_{7, \,\bar{J}\, \bar{m}}(\bar{p})\,
(-S_{3, B}^{C}-S_{3, B}^{\cD} -S_{3,F}^{\cF})}_{\rm 1-loop},
\ee
where the subscript ``1-loop'' means the Wick contractions generating one-loop diagrams. 
We see that each of the three contributions arising from the contractions with  
$-S_{3, B}^{C}$, 
$-S_{3, B}^{\cD}$ and 
$-S_{3,F}^{\cF}$
vanishes separately, because the integrands are odd functions of loop momenta 
or vanish by themselves due to 
\be
p_\mu\,\left(\tilde{\Delta}_J(p)^{-1}\right)_{\mu\,3} = 
p_\mu\,\left(\tilde{\Delta}_J(p)^{-1}\right)_{3 \,\mu} = 0, \qquad \tr_8\left(\hat{\gamma}_7\hat{D}_{J,\,\kappa}(p)^{-1}\right)=0, 
\ee
which leads to 
\be
\vev{\frac{1}{k}\tr_k\,x_{7, \,\bar{J}\, \bar{m}}(\bar{p})} =0 \qquad \mbox{at the one-loop order}.
\ee  

$\vev{\frac{1}{k}\tr_k\,a_{\mu, \,\bar{J}\, \bar{m}}(\bar{p})}$ is similar to the situation for $x_7$: 
\be
\vev{\frac{1}{k}\tr_k\,a_{\mu, \,\bar{J}\, \bar{m}}(\bar{p})} 
= \vev{\frac{1}{k}\tr_k\,a_{\mu, \,\bar{J}\, \bar{m}}(\bar{p})\,
(-S_{3, B}^{C}-S_{3, B}^{\cD} -S_{3,F}^{\cF})}_{\rm 1-loop}, 
\ee
and all the loop integrals vanish for the same reason as in $x_7$;
\be
\vev{\frac{1}{k}\tr_k\,a_{\mu, \,\bar{J}\, \bar{m}}(\bar{p})} =0 \qquad \mbox{at the one-loop order}.
\ee  

\subsection{One-point function of $y$}
The one-point function of $y$ can be written as 
\be
\vev{\frac{1}{k}\tr_k\,y_{\bar{J}\, \bar{m}, \,\bar{\rho}}(\bar{p})} 
= \vev{\frac{1}{k}\tr_k\,y_{\bar{J}\, \bar{m}, \,\bar{\rho}}(\bar{p})\,
(-S_{3, B}^{\cE}-S_{3, B}^{\cD} -S_{3,F}^{\cG})}_{\rm 1-loop}.
\ee
The contractions with $-S_{3, B}^{\cE}$ and $-S_{3, B}^{\cD}$ lead to diagrams with loops of 
bosons or ghosts, and the contractions with $-S_{3,F}^{\cG}$ to loop diagrams of fermions. 
This time each diagram does not vanish separately. 
Let us compute these three contributions explicitly.

\paragraph{First contribution} 
The first contribution is tadpoles of $y$-loops: 
\bea
& & \vev{\frac{1}{k}\tr_k\,y_{\bar{J}\, \bar{m}, \,\bar{\rho}}(\bar{p})\,(-S_{3, B}^{\cE})}_{\rm 1-loop} \nn \\
& & \hspace{3mm}= -\frac{g_{2d}}{\sqrt{2n}}\,k\,i\frac{M}{3}\,\delta^2(\bar{p})\,(-1)^{-\bar{m}}\tilde{\Delta}_{(\bar{\rho},\bar{J})}(\bar{p}) 
\sum_{\rho'=-1}^1\sum_{J'=\delta_{\rho',\,0}}^{2j-\delta_{\rho',\,-1}}
\left[\bar{\rho}(\bar{J}+1) + 2\rho'(J'+1)\right] \nn \\
& & \hspace{7mm} \times \left(\sum_{m'=-Q'}^{Q'} (-1)^{-m'}
\hat{\cE}_{\bar{J}\,-\bar{m}\,(jj)\,\bar{\rho}\,J'\,m'\,(jj) \rho'\,J'\,-m'\,(jj)\,\rho'} \right)
\int^{\Lambda_p}d^2p \,\tilde{\Delta}_{(\rho',J')}(p) . 
\label{y_E}
\eea 

The sum of $\hat{\cE}$ is calculated in appendix~\ref{app:sum_E}. Plugging the result (\ref{sum_E_3}) 
into (\ref{y_E}) leads to 
\bea
& & \hspace{-7mm}\vev{\frac{1}{k}\tr_k\,y_{\bar{J}\, \bar{m}, \,\bar{\rho}}(\bar{p})\,(-S_{3, B}^{\cE})}_{\rm 1-loop}
=\frac{g_{2d}}{\sqrt{2n}}\,i\frac{3k}{2M}\,\frac{1}{\sqrt{j(j+1)}}\,\delta^2(\bar{p})\,
\delta_{\bar{\rho},\,-1}\,\delta_{\bar{J}\, 0}\,\delta_{\bar{m}\,0} \nn \\
& & \hspace{-4mm}\times \sum_{J'=1}^{2j}\int^{\Lambda_p}d^2p 
\left\{\frac{2J'+1}{p^2 +\frac{M^2}{9}J'(J'+1)} 
+\frac{J'(2J'+1)(2J'+3)}{p^2+\frac{M^2}{9}(J'+1)^2} 
+\frac{(J'+1)(2J'-1)(2J'+1)}{p^2+\frac{M^2}{9}J'^2} \right\}. \nn \\
& & 
\label{y_E_3}
\eea 

\paragraph{Second contribution} 
The second contribution is tadpoles of boson ($a_\mu,\,x_i)$ loops and ghost loops: 
\bea
& & \vev{\frac{1}{k}\tr_k\,y_{\bar{J}\, \bar{m}, \,\bar{\rho}}(\bar{p})\,(-S_{3, B}^{\cD})}_{\rm 1-loop}
= \frac{g_{2d}}{\sqrt{2n}}\,\frac{kM}{3}\,\delta^2(\bar{p})\,(-1)^{-\bar{m}}\tilde{\Delta}_{(\bar{\rho}, \bar{J})}(\bar{p})
\nn \\
& & \hspace{4mm}\times \sum_{J=1}^{2j} \sqrt{J(J+1)}\left\{\sum_{m=-J}^J
\left(\hat{\cD}^{J\,m\,(jj)}_{\bar{J}\,-\bar{m}\,(jj)\,\bar{\rho}\,J\,m\,(jj)\,0}
-\hat{\cD}^{J\,m\,(jj)}_{J\,m\,(jj)\,0\, \bar{J}\,-\bar{m}\,(jj)\,\bar{\rho}}\right)\right\} \nn \\
& & \hspace{7mm} \times \int^{\Lambda_p}d^2p \left\{\left(\tilde{\Delta}_J(p)^{-1}\right)_{KK} 
+ \frac{4}{p^2+\frac{M^2}{9}(J+\frac13)(J+\frac23)} +\frac{-1}{p^2+\frac{M^2}{9}J(J+1)}\right\}.\nn \\
& & \label{y_D}
\eea
{}From the result of the sum of $\hat{\cD}$, (\ref{sum_D_3}), we find 
\bea
& & \hspace{-7mm}\vev{\frac{1}{k}\tr_k\,y_{\bar{J}\, \bar{m}, \,\bar{\rho}}(\bar{p})\,(-S_{3, B}^{\cD})}_{\rm 1-loop}
\nn \\
& & \hspace{-3mm} 
= \frac{g_{2d}}{\sqrt{2n}}\,i\frac{3k}{M}\,\frac{1}{\sqrt{j(j+1)}} \,\delta^2(\bar{p})\,
\delta_{\bar{\rho},\,-1}\,\delta_{\bar{J}\,0}\,\delta_{\bar{m}\,0}  
\,\sum_{J=1}^{2j} J(J+1)(2J+1) \nn \\
& & \hspace{3mm} \times \int^{\Lambda_p}d^2p \left\{
\frac{2}{p^2+\frac{M^2}{9}J(J+1)} + \frac{4}{p^2+\frac{M^2}{9}(J+\frac13)(J+\frac23)} \right. \nn \\
& & \hspace{24mm} + \frac{-\frac{2M^2}{81}}{[p^2+\frac{M^2}{9}J(J+\frac13)][p^2+\frac{M^2}{9}(J+1)(J+\frac23)]} 
\nn \\
& & \left. \hspace{24mm} 
+ \frac{-\frac{8M^2}{81}p^2}{[p^2+\frac{M^2}{9}J(J+1)][p^2+\frac{M^2}{9}J(J+\frac13)][p^2+\frac{M^2}{9}(J+1)(J+\frac23)]} \right\}.
\nn \\
& & 
\label{y_D_2}
\eea

\paragraph{Third contribution} 
The third contribution is tadpoles of fermion loops: 
\bea
& & \vev{\frac{1}{k}\tr_k\,y_{\bar{J}\, \bar{m}, \,\bar{\rho}}(\bar{p})\,(-S_{3, F}^{\cG})}_{\rm 1-loop}
 = \frac{g_{2d}}{\sqrt{2n}}\,k\,\delta^2(\bar{p})\,(-1)^{-\bar{m}+1}\,\tilde{\Delta}_{(\bar{\rho},\,\bar{J})}(\bar{p})
\nn \\
& & \hspace{4mm} \times \sum_{\kappa=\pm 1}\sum_{J=\frac12\delta_{\kappa,\,-1}}^{2j-\frac12\delta_{\kappa,\,-1}}
\left(\sum_{m=-U}^U\hat{\cG}^{J\,-m \,(jj) \,\kappa}_{J \,-m\,(jj)\,\kappa\,\bar{J}\,-\bar{m}\,(jj)\,\bar{\rho}}
\right) \int^{\Lambda_p}d^2p\,\tr_8\left(\hat{D}_{J,\,\kappa}(p)^{-1}\right) .
\label{y_G}
\eea
Here, the spinor trace reads 
\bea
& & \hspace{-7mm}\tr_8\left(\hat{D}_{J,\,\kappa=1}(p)^{-1}\right) =  
\frac{-M(J+1)}{p^2+\frac{M^2}{9}(J+1)^2} + \frac{-\frac{M}{3}(J+1)}{p^2+\frac{M^2}{9}(J+1)(J+\frac23)} \nn \\
& & + \frac{-\frac{M}{3}(4J+2)}{p^2+\frac{M^2}{9}(J+\frac13)(J+\frac23)} 
+ \frac{\frac{M}{9}p^2}{[p^2+\frac{M^2}{9}(J+1)^2][p^2+\frac{M^2}{9}(J+1)(J+\frac23)]} 
 \label{tr8_D1} 
\eea
and 
\bea
& & \hspace{-7mm}\tr_8\left(\hat{D}_{J,\,\kappa=-1}(p)^{-1}\right) =  
\frac{M(J+\frac12)}{p^2+\frac{M^2}{9}(J+\frac12)^2} 
+ \frac{\frac{M}{3}(J+\frac12)}{p^2+\frac{M^2}{9}(J+\frac12)(J+\frac56)} \nn \\
& & + \frac{\frac{M}{3}(4J+4)}{p^2+\frac{M^2}{9}(J+\frac56)(J+\frac76)} 
+ \frac{\frac{M}{9}p^2}{[p^2+\frac{M^2}{9}(J+\frac12)^2][p^2+\frac{M^2}{9}(J+\frac12)(J+\frac56)]}. 
 \label{tr8_D-1}
\eea
The sum of $\hat{\cG}$ is computed in (\ref{sum_G_3}). Plugging these into (\ref{y_G}), we obtain 
\bea
& & \hspace{-7mm}\vev{\frac{1}{k}\tr_k\,y_{\bar{J}\, \bar{m}, \,\bar{\rho}}(\bar{p})\,(-S_{3, F}^{\cG})}_{\rm 1-loop}
 =\frac{g_{2d}}{\sqrt{2n}}\,i\frac{9k}{M}\,\frac{1}{\sqrt{j(j+1)}}\,\delta^2(\bar{p})\,
\delta_{\bar{\rho},\,-1}\,\delta_{\bar{J}\,0}\,\delta_{\bar{m}\,0} \sum_{J=1}^{2j} J(J+1) \nn \\
& & \hspace{-3mm}\times \int^{\Lambda_p}d^2p\,\left\{
\frac{-(J+1)}{p^2+\frac{M^2}{9}(J+1)^2} + \frac{-\frac13(J+1)}{p^2+\frac{M^2}{9}(J+1)(J+\frac23)} \right. \nn \\
& & \hspace{17mm}+\frac{-J}{p^2+\frac{M^2}{9}J^2} + \frac{-\frac13 J}{p^2+\frac{M^2}{9}J(J+\frac13)} 
+\frac{-\frac83 (J+\frac12)}{p^2+\frac{M^2}{9}(J+\frac13)(J+\frac23)} \nn \\
& & \hspace{17mm}\left.+\frac{\frac19 p^2}{[p^2+\frac{M^2}{9}(J+1)^2][p^2+\frac{M^2}{9}(J+1)(J+\frac23)]}
+\frac{-\frac19 p^2}{[p^2+\frac{M^2}{9}J^2][p^2+\frac{M^2}{9}J(J+\frac13)]}\right\}. \nn \\
& & 
\label{y_G_3}
\eea

\paragraph{Total contribution}
Gathering the three contributions (\ref{y_E_3}), (\ref{y_D_2}), and (\ref{y_G_3}), the one-point function becomes 
\bea
& & \hspace{-3mm}\vev{\frac{1}{k}\tr_k\,y_{\bar{J}\, \bar{m}, \,\bar{\rho}}(\bar{p})} 
= \frac{g_{2d}}{\sqrt{2n}}\,i\frac{3k}{2M}\,\frac{1}{\sqrt{j(j+1)}}\,\delta^2(\bar{p})\,
\delta_{\bar{\rho},\,-1}\,\delta_{\bar{J}\,0}\,\delta_{\bar{m}\,0}\nn \\
& & \times \sum_{J=1}^{2j}\int^{\Lambda_p} d^2p\,\left\{
\frac{2J+1}{p^2+\frac{M^2}{9}J(J+1)} +\frac{-(J+1)}{p^2+\frac{M^2}{9}J^2} 
+\frac{-J}{p^2+\frac{M^2}{9}(J+1)^2}\right\}.
\label{y_1loop_2}
\eea  
Here, we see that the $\cO(1/p^2)$ and $\cO(1/p^4)$ terms of the large-$|p|$ 
expansion of the integrand vanish and the integral converges 
owing to the $Q_\pm$ supersymmetries. 
In fact, after the $p$-integrals and the summation of $J$, we end up with  
\be
\vev{\frac{1}{k}\tr_k\,y_{\bar{J}\, \bar{m}, \,\bar{\rho}}(\bar{p})}
 = \frac{g_{2d}}{\sqrt{2n}}\,i\frac{3k}{M}\,\frac{-\pi\ln n}{\sqrt{n^2-1}}\,\delta^2(\bar{p})\,
\delta_{\bar{\rho},\,-1}\,\delta_{\bar{J}\,0}\,\delta_{\bar{m}\,0}.
\label{y_1loop_3}    
\ee
This indicates that $\tr_k\,y_{0\, 0, \,\rho=-1}$ is generated by a one-loop effect. 
However, we should note that 
$\tr_k\,y_{0\, 0, \,\rho=-1}$ is not gauge invariant, as discussed in appendix~\ref{app:gauge_y}. 

Let us consider the gauge invariant combination from (\ref{gauge_Y}): 
\be
\Tr\,\vec{\tilde{Y}}(p) =  \sum_{\rho=-1}^1\sum_{J=\delta_{\rho,\,0}}^{2j-\delta_{\rho,\,-1}}\sum_{m=-Q}^Q
\left(\tr_n\,\vec{\hat{Y}}^\rho_{J\,m\,(jj)}\right)\,\left(\tr_k\,y_{J\,m,\,\rho}(p)\right). 
\ee
Its expectation value at the one-loop level becomes 
\be
\vev{\Tr\,\vec{\tilde{Y}}(p) } = \left(\tr_n\,\vec{\hat{Y}}^{\rho=-1}_{0\,0\,(jj)}\right)\,
\vev{\tr_k\,y_{J\,m,\,\rho}(p)}, 
\ee
since $\vev{\tr_k\,y_{J\,m,\,\rho}(p)}\propto  \delta_{\rho,\,-1}\,\delta_{J\,0}\,\delta_{m\,0}$ 
from (\ref{y_1loop_3}). 
It is shown in appendix~\ref{app:tr_vecY} that 
\be
\tr_n\,\vec{\hat{Y}}^{\rho=-1}_{0\,0\,(jj)} = 0.
\label{tr_vecY}
\ee
Thus, we conclude that the expectation value of the gauge invariant part $\vev{\Tr\,\vec{\tilde{Y}}(p) }$ 
vanishes at the one-loop level. 

\subsection{Summary of the one-point functions}
The results obtained in this section give an explicit check at the one-loop order for the statement 
that any gauge invariant one-point operators 
are not radiatively induced for arbitrary $n$ and $M$. 
   
Equation (\ref{y_1loop_3}) multiplied by $M^2/9$, which is the one-point function with the external line truncated, is seen to vanish in the 
Moyal limit (Step 1 in the successive limits). 
Therefore, the one-point functions give no contribution in the one-loop effective action in the successive limits.

\section{Two-point functions of scalar fields at one-loop}
\label{sec:2pt}

We express the two-point functions of scalar fields $X_i$ as 
\begin{align}
-\frac{g_{2d}^2}{2n}\left(\frac{3}{M}\right)^2\int \frac{d^2p}{(2\pi)^2}\sum_{(J,m)}\sum_{(\bar{J},\bar{m})}
(-1)^m \Bigl\{
& k\,\tr_k\left(x_{i,\,J\,m}(p)x_{\bar{i},\,\bar{J}\,\bar{m}}(-p)\right) \cA_{i,\bar{i}}(p;\,J\,m;\,\bar{J}\,\bar{m}) \nn \\
 &+ \tr_k\left(x_{i,\,J\,m}(p)\right)\tr_k\left(x_{\bar{i},\,\bar{J}\,\bar{m}}(-p)\right) 
 {\cB}_{i,\bar{i}}(p;\,J\,m;\,\bar{J}\,\bar{m})
 \Bigr\}. 
\label{2pt func}
\end{align}
Here, $\cA_{i,\bar{i}}(p;\,J\,m;\,\bar{J}\,\bar{m})$ and $\cB_{i,\bar{i}}(p;\,J\,m;\,\bar{J}\,\bar{m})$
are contributions from planar and non-planar diagrams, respectively, 
whose external legs are removed. 

\subsection{List of one-loop graphs}
\label{sec:list_2pt_1-loop}
Let us write each of $\cA_{i,\bar{i}}$ and $\cB_{i,\bar{i}}$ as a summation of 
six contributions:
\begin{align}
 \cA_{i,\bar{i}}&=\cA_{i,\bar{i}}^{4CC}+\cA_{i,\bar{i}}^{4C\cD}+\cA_{i,\bar{i}}^{3\cD\cD}+\cA_{i,\bar{i}}^{3\cF\cF}
 +\cA_{i,\bar{i}}^{3CC(1)}+\cA_{i,\bar{i}}^{3CC(2)}, \nn \\
 \cB_{i,\bar{i}}&=\cB_{i,\bar{i}}^{4CC}+\cB_{i,\bar{i}}^{4C\cD}
 +\cB_{i,\bar{i}}^{3\cD\cD}+\cB_{i,\bar{i}}^{3\cF\cF}
 +\cB_{i,\bar{i}}^{3CC(1)}+\cB_{i,\bar{i}}^{3CC(2)}.
 \label{separation}
\end{align}
The superscript $4XY$ means contributions from diagrams containing a single 4-point vertex that has 
the vertex coefficients $\hat{X}$ and $\hat{Y}$, while $3XY$ stands for contributions from diagrams consisting of 
two 3-point vertices one of which includes the vertex coefficient $\hat{X}$ and the other has $\hat{Y}$ 
($X, Y=C,\,\cD,\,\cE,\,\cF,\,\cG$).
In (\ref{separation}), 
we have further divided each of $\cA_{i,\bar{i}}^{3CC}$ and $\cB_{i,\bar{i}}^{3CC}$
into two parts with the symbols ``(1)'' and ``(2)'', 
in which the latter reflects that 
the interaction terms $S_{3,\,B}^C$ including 
$x_{3,\,J\,m}(p)$, $x_{4,\,J\,m}(p)$, and $x_{7,\,J\,m}(p)$  
yield extra contributions to 
$\cA_{3,3}^{3CC}$, $\cA_{4,4}^{3CC}$, $\cA_{3,4}^{3CC}$,
$\cA_{7,7}^{3CC}$, 
$\cB_{3,3}^{3CC}$, $\cB_{4,4}^{3CC}$, $\cB_{3,4}^{3CC}$, 
and $\cB_{7,7}^{3CC}$.  
In addition, due to the propagator connecting $x_7$ and $a_\mu$ (\ref{propa_ax7}), 
the expressions of $\cA_{i,\bar{i}}$ and $\cB_{i,\bar{i}}$ are different depending on whether the external line 
includes $x_{7,Jm}(p)$ or not.
So we separately treat $\cA_{i,\bar{i}}$, $\cB_{i,\bar{i}}$ with 
$(i,\bar{i})\in \{3,\cdots,6\}$ and 
$\cA_{7,7}$, $\cB_{7,7}$. 
Note that $\cA_{i, 7}$ and $\cB_{i,7}$ are identically zero. 

In the following, we explicitly present all the contributions that do not trivially vanish. 
In order to simplify notations, the following symbols are employed: 
\be
\sum_{(J,m)} \equiv \sum_{J=0}^{2j} \sum_{m=-J}^{J}, \qquad
\sum_{(\rho,J,m)} \equiv \sum_{\rho=-1}^{1} 
\sum_{J=\delta_{\rho,0}}^{2j-\delta_{\rho,-1}}
\sum_{m=-Q}^Q, \qquad
\sum_{[\kappa,J,m]} \equiv \sum_{\kappa=\pm 1}
\sum_{J=\frac{1}{2}\delta_{\kappa,-1}}^{2j-\frac{1}{2}\delta_{\kappa,-1}}
\sum_{m=-U}^U 
\label{symbols_sum}
\ee
with $Q=J+\delta_{\rho,1}$ and $U=J+\frac{1}{2}\delta_{\kappa,1}$. 
In expressing index structures of planar and non-planar contributions, the relations (\ref{Chat_23_32}), (\ref{cD_12_21}), and 
(\ref{cF_12_21}) are used.  

\subsubsection{$\cA_{i,\bar{i}}$ and $\cB_{i,\bar{i}}$ with $i,\bar{i}=3,\cdots,6$}
The planar contributions are~\footnote{Here, $i$ and $\bar{i}$ run from $3$ to $6$, excluding $7$. 
In what follows, the sum over $i$ is not assumed in $\cA_{i,i}$ or $\cB_{i,i}$.}  
\begin{align}
\label{Pii4CC}
\cA_{i,i}^{4CC}=& (-1)^m \left( \frac{M}{3} \right)^2 
\sum_{(J',m')}\sum_{(J'',m'')}(-1)^{m''-m'}
\hat{C}^{J''\,m''\,(jj)}_{J\,m\,(jj)\,\bar{J}\,\bar{m}\,(jj)} 
\hat{C}^{J''\,-m''\,(jj)}_{J'\,m'\,(jj)\,J'\,-m'\,(jj)} \nn \\
& \hspace{7mm} \times\int \frac{d^2q}{(2\pi)^2}
\left\{\left(\tilde{\Delta}_{J'}(q)^{-1}\right)_{KK} 
+\frac{3}{q^2+\frac{M^2}{9}(J'+\frac13)(J'+\frac23)}\right\},  \\
\label{Pii4CD}
\cA_{i,i}^{4C\cD}=& (-1)^m \left( \frac{M}{3} \right)^2 
\sum_{(\rho', J',m')}\sum_{(J'',m'')}
(-1)^{m''-m'+1} \hat{C}^{J''\,m''\,(jj)}_{J\,m\,(jj)\,\bar{J}\,\bar{m}\,(jj)}\,
\hat{\cD}^{J''\,-m''\,(jj)}_{J'\,m'\,(jj)\,\rho'\,J'\,-m'\,(jj)\,\rho'} \nn \\
& \hspace{4mm} \times\int\frac{d^2q}{(2\pi)^2}\,
\tilde{\Delta}_{(\rho',J')}(q)^{-1},  \\
\label{Pii3DD}
 \cA_{i,i}^{3\cD\cD}=& (-1)^{m} \left( \frac{M}{3} \right)^2 
 \sum_{(J',m')}\sum_{(\rho'',J'',m'')} (-1)^{-m'-m''} \nn \\
& \hspace{-14mm}\times\left((-1)^{m'}\sqrt{J(J+1)}\hat{\cD}^{J'\,-m'\,(jj)}_{J''\,m''\,(jj)\,\rho''\,J\,m\,(jj)\,0}
-(-1)^m\sqrt{J'(J'+1)}\hat{\cD}^{J\,-m\,(jj)}_{J'\,m'\,(jj)\,0\,J''\,m''\,(jj)\,\rho''}\right) \nn \\
& \hspace{-14mm}\times\left((-1)^{\bar{m}}\sqrt{J'(J'+1)}\hat{\cD}^{\bar{J}\,-\bar{m}\,(jj)}_{J''\,-m''\,(jj)\,\rho''\,J'\,-m'\,(jj)\,0} 
-(-1)^{-m'}\sqrt{\bar{J}(\bar{J}+1)}\hat{\cD}^{J'\,m'\,(jj)}_{\bar{J}\,\bar{m}\,(jj)\,0\,J''\,-m''\,(jj)\,\rho''}\right)
\nn \\
& \hspace{4mm}\times \frac{M^2}{9}\int\frac{d^2q}{(2\pi)^2}\,
\frac{1}{q^2+\frac{M^2}{9}(J'+\frac13)(J'+\frac23)}\,
\tilde{\Delta}_{(\rho'',J'')}(p+q)^{-1}, \\
\label{Pii3FF}
\cA_{i,\bar{i}}^{3\cF\cF}=& (-1)^{m} \left( \frac{M}{3} \right)^2
\sum_{[\kappa',J',m']}\sum_{[\kappa'',J'',m'']} 
\hat{\cF}^{J'\,m'\,(jj)\,\kappa'}_{J''\,m''\,(jj)\,\kappa''\,\bar{J}\,\bar{m}\,(jj)}\,
\hat{\cF}^{J''\,m''\,(jj)\,\kappa''}_{J'\,m'\,(jj)\,\kappa'\,J\,m\,(jj)} \nn \\
& \hspace{4mm} \times \frac{1}{2}\int \frac{d^2q}{(2\pi)^2} \,
\tr_8\left(\hat{\gamma}_i \hat{D}_{J',\kappa'}(q)^{-1}\hat{\gamma}_{\bar{i}} 
\hat{D}_{J'',\kappa''}(p+q)^{-1}\right), \\
\label{Pii3CC}
\cA_{i,i}^{3CC(1)}=& (-1)^{m+1} \left( \frac{M}{3} \right)^2 \sum_{(J',m')}\sum_{(J'',m'')}
(-1)^{m'-m''}
 \hat{C}_{J \,m \,(jj)\,J''\,m''\, (jj)}^{J' \,m'\,(jj)}
\hat{C}_{J''\, -m'' \,(jj) \bar{J} \, \bar{m}\, (jj)}^{J'\, -m'\,(jj)} \nn \\
&\hspace{-10mm} \times \int \frac{d^2q}{(2\pi)^2} 
(q_\mu+2p_\mu)(q_\nu+2p_\nu) \left( \tDelta_{J'}(q)^{-1} \right)_{\mu\nu}
\frac{1}{(p+q)^2+\frac{M^2}{9}(J''+\frac13)(J''+\frac23)}, \\
\label{P333CC}
\cA_{3,3}^{3CC(2)}=&\cA_{4,4}^{3CC(2)}=
(-1)^{m}\left( \frac{M}{3} \right)^2  
\sum_{(J',m')}\sum_{(J'',m'')}(-1)^{m'-m''} 
\hat{C}^{J'\,m'\,(jj)}_{J\,m\,(jj)\,J''\,m''\,(jj)}\,
\hat{C}^{J'\,-m'\,(jj)}_{J''\,-m''\,(jj)\,\bar{J}\,\bar{m}\,(jj)} 
\nn \\
& \hspace{7mm} 
\times \left(\frac{2M}{9}\right)^2 
\int\frac{d^2q}{(2\pi)^2} 
 \left(\tilde{\Delta}_{J'}(q)^{-1}\right)_{33} 
\frac{1}{(p+q)^2+\frac{M^2}{9}(J''+\frac13)(J''+\frac23)}, \\
\cA_{3,4}^{3CC(2)}=&-\cA_{4,3}^{3CC(2)}
= (-1)^{m+1}\left( \frac{M}{3} \right)^2  \sum_{(J',m')}\sum_{(J'',m'')} (-1)^{m'-m''} \nn \\
& \hspace{69mm} \times 
\hat{C}^{J'\,m'\,(jj)}_{J\,m\,(jj)\,J''\,m''\,(jj)}\,\hat{C}^{J'\,-m'\,(jj)}_{J''\,-m''\,(jj)\,\bar{J}\,\bar{m}\,(jj)}
\nn \\
& \hspace{-4mm}\times \frac{4M}{9}\int\frac{d^2q}{(2\pi)^2}  (q_\mu+2p_\mu) \left(\tilde{\Delta}_{J'}(q)^{-1}\right)_{\mu 3} 
\frac{1}{(p+q)^2+\frac{M^2}{9}(J''+\frac13)(J''+\frac23)}. 
\label{P343CC}
\end{align}
Corresponding to the second expression of $S_4^{C\cD}$ (\ref{S4_CD2}), $\cA_{i,i}^{4C\cD}$ can also be 
written as 
\begin{align}
\label{Pii4CD_2}
\cA_{i,i}^{4C\cD} =& (-1)^{\bar{m}}\left( \frac{M}{3} \right)^2 \sum_{(\rho',J',m')}\sum_{(\rho'',J'',m'')} 
(-1)^{m''-m'}\nn \\
& \hspace{4mm}\times\hat{\cD}^{J\,-m\,(jj)}_{J''\,m''\,(jj)\,\rho''\,J'\,-m'\,(jj)\,\rho'}\,
\hat{\cD}^{\bar{J}\,-\bar{m}\,(jj)}_{J'\,m'\,(jj)\,\rho'\,J''\,-m''\,(jj)\,\rho''} 
 \int\frac{d^2q}{(2\pi)^2}\,\tilde{\Delta}_{(\rho',J')}(q)^{-1}. 
\end{align}
We can see that off-diagonal parts of $\cA_{i,\bar{i}}^{3\cF\cF}$ with $(i,\bar{i})=(3,4)$ and $(5,6)$ are identically zero, and 
that those with $(i,\bar{i})=(3,5)$, $(3,6)$, $(4,5)$ and $(4,6)$ are nonzero but become irrelevant in 
the Moyal limit, for instance, by using Theorem 1 and 2 in section~\ref{sec:strategy}.

The non-planar contributions are 
\begin{align}
\label{NPii4CC}
\cB_{i,i}^{4CC}
=& \left( \frac{M}{3} \right)^2 \sum_{(J',m')}\sum_{(J'',m'')} (-1)^{J'+J''-J+1}\,(-1)^{m''-m'} 
\hat{C}^{J''\,m''\,(jj)}_{J'\,m'\,(jj)\,J\,m\,(jj)} \hat{C}^{J'\,m'\,(jj)}_{J''\,m''\,(jj)\,\bar{J}\,\bar{m}\,(jj)}\nn \\
& \hspace{7mm}\times\int\frac{d^2q}{(2\pi)^2}
\left\{\left(\tilde{\Delta}_{J'}(q)^{-1}\right)_{KK} +\frac{3}{q^2+\frac{M^2}{9}(J'+\frac13)(J'+\frac23)}
\right\}, \\
\label{NPii4DD}
 \cB_{i,i}^{4\cD\cD}=
 & (-1)^{\bar{m}} \left( \frac{M}{3} \right)^2 \sum_{(\rho',J',m')}\sum_{(\rho'',J'',m'')} (-1)^{\tilde{Q}'+\tilde{Q}''-J+1}\,(-1)^{m''-m'} \nn \\
 & 
\times \hat{\cD}^{J\,-m\,(jj)}_{J''\,m''\,(jj)\,\rho''\,J'\,-m'\,(jj)\,\rho'}\, 
\hat{\cD}^{\bar{J}\,-\bar{m}\,(jj)}_{J'\,-m'\,(jj)\,\rho'\,J''\,m''\,(jj)\,\rho''} 
\int\frac{d^2q}{(2\pi)^2}\,\tilde{\Delta}_{(\rho',J')}(q)^{-1}, \\
\label{NPii3DD}
 \cB_{i,i}^{3\cD\cD} =
& (-1)^{m}  \left( \frac{M}{3} \right)^2 \sum_{(J',m')}\sum_{(\rho'',J'',m'')} (-1)^{J'+\tilde{Q}''-J+1}\,(-1)^{-m'-m''} \nn \\
& \hspace{-14mm}\times\left((-1)^{m'}\sqrt{J(J+1)}\hat{\cD}^{J'\,-m'\,(jj)}_{J''\,m''\,(jj)\,\rho''\,J\,m\,(jj)\,0} 
-(-1)^m\sqrt{J'(J'+1)}\hat{\cD}^{J\,-m\,(jj)}_{J'\,m'\,(jj)\,0\,J''\,m''\,(jj)\,\rho''}\right) \nn \\
& \hspace{-14mm}\times\left((-1)^{\bar{m}}\sqrt{J'(J '+1)}\hat{\cD}^{\bar{J}\,-\bar{m}\,(jj)}_{J''\,-m''\,(jj)\,\rho''\,J'\,-m'\,(jj)\,0} 
-(-1)^{m'}\sqrt{\bar{J}(\bar{J}+1)}\hat{\cD}^{J'\,m'\,(jj)}_{\bar{J}\,\bar{m}\,(jj)\,0\,J''\,-m''\,(jj)\,\rho''}\right)\nn \\
& \hspace{4mm}\times \frac{M^2}{9}\int\frac{d^2q}{(2\pi)^2}\,\frac{1}{q^2+\frac{M^2}{9}(J'+\frac13)(J'+\frac23)}\,
\tilde{\Delta}_{(\rho'',J'')}(p+q)^{-1}, \\
\label{NPii3FF}
\cB_{i,\bar{i}}^{3\cF\cF}=
& (-1)^{m} \left( \frac{M}{3} \right)^2 \sum_{[\kappa',J',m']}\sum_{[\kappa'',J'',m'']} (-1)^{\tilde{U}'+\tilde{U}''-J+1}\,
\hat{\cF}^{J''\,m''\,(jj)\,\kappa''}_{J'\,m'\,(jj)\,\kappa'\,J\,m\,(jj)}\,
\hat{\cF}^{J'\,m'\,(jj)\,\kappa'}_{J''\,m''\,(jj)\,\kappa''\,\bar{J}\,\bar{m}\,(jj)} \nn \\
& \hspace{34mm}\times \frac{1}{2}\int\frac{d^2q}{(2\pi)^2} \,
\tr_8\left(\hat{\gamma}_i \hat{D}_{J',\kappa'}(q)^{-1}\hat{\gamma}_{\bar{i}} \hat{D}_{J'',\kappa''}(p+q)^{-1}\right), \\
\label{NPii3CC}
\cB_{i,i}^{3CC(1)} =
& (-1)^{m+1} \left( \frac{M}{3} \right)^2 \sum_{(J',m')}\sum_{(J'',m'')}(-1)^{J'+J''-J+1}\,(-1)^{m'-m''} \nn \\
& \hspace{47mm} \times\hat{C}^{J'\,m'\,(jj)}_{J\,m\,(jj)\,J''\,m''\,(jj)}\,\hat{C}^{J'\,-m'\,(jj)}_{J''\,-m''\,(jj)\,\bar{J}\,\bar{m}\,(jj)} 
\nn \\
& \hspace{-14mm}\times \int\frac{d^2q}{(2\pi)^2} (q_\mu+2p_\mu)(q_\nu+2p_\nu) \left(\tilde{\Delta}_{J'}(q)^{-1}\right)_{\mu\nu} 
\,\frac{1}{(p+q)^2+\frac{M^2}{9}(J''+\frac13)(J''+\frac23)}, \\
\label{NP333CC1}
\cB_{3,3}^{3CC(2)}=&\cB_{4,4}^{3CC(2)}
= (-1)^{m} \left( \frac{M}{3} \right)^2 \sum_{(J',m')}\sum_{(J'',m'')}(-1)^{J'+J''-J}\,(-1)^{m'-m''} \nn \\
& \hspace{65mm} \times \hat{C}^{J'\,m'\,(jj)}_{J\,m\,(jj)\,J''\,m''\,(jj)}\,\hat{C}^{J'\,-m'\,(jj)}_{J''\,-m''\,(jj)\,\bar{J}\,\bar{m}\,(jj)}
 \nn \\
& \hspace{7mm} \times \left(\frac{2M}{9}\right)^2 
\int\frac{d^2q}{(2\pi)^2}  \left(\tilde{\Delta}_{J'}(q)^{-1}\right)_{33}  \frac{1}{(p+q)^2+\frac{M^2}{9}(J''+\frac13)(J''+\frac23)}, \\
\label{NP343CC2}
\cB_{3,4}^{3CC(2)}=&-\cB_{4,3}^{3CC(2)}
= (-1)^{m+1} \left( \frac{M}{3} \right)^2 \sum_{(J',m')}\sum_{(J'',m'')} (-1)^{J'+J''-J+1}\,(-1)^{m'-m''} \nn \\
& \hspace{65mm} \times 
\hat{C}^{J'\,m'\,(jj)}_{J\,m\,(jj)\,J''\,m''\,(jj)}\,\hat{C}^{J'\,-m'\,(jj)}_{J''\,-m''\,(jj)\,\bar{J}\,\bar{m}\,(jj)} \nn \\
& \hspace{-4mm}\times \frac{4M}{9}\int\frac{d^2q}{(2\pi)^2}  (q_\mu+2p_\mu) \left(\tilde{\Delta}_{J'}(q)^{-1}\right)_{\mu 3} 
\frac{1}{(p+q)^2+\frac{M^2}{9}(J''+\frac13)(J''+\frac23)}.
\end{align}
Similarly to the planar case, the off-diagonal parts of $\cB_{i,\bar{i}}^{3\cF\cF}$ with $(i,\bar{i})=(3,4)$ and $(5,6)$ are 
exactly zero, and 
those with $(i,\bar{i})=(3,5)$, $(3,6)$, $(4,5)$, and $(4,6)$ turn out to disappear in 
the Moyal limit, for instance, from Theorem 3 in section~\ref{sec:strategy}.

\subsubsection{$\cA_{7,7}$ and $\cB_{7,7}$}

The planar contributions are 
\begin{align}
\label{P774CC}
\cA_{7,7}^{4CC}=
&(-1)^m \left(\frac{M}{3}\right)^2 \sum_{(J',m')}\sum_{(J'',m'')}(-1)^{m''-m'}
\hat{C}^{J''\,m''\,(jj)}_{J\,m\,(jj)\,\bar{J}\,\bar{m}\,(jj)} 
\hat{C}^{J''\,-m''\,(jj)}_{J'\,m'\,(jj)\,J'\,-m'\,(jj)}  \nn \\
& \hspace{24mm} \times\int \frac{d^2q}{(2\pi)^2}\left\{
\left(\tilde{\Delta}_{J'}(q)^{-1}\right)_{\mu\mu} 
+\frac{4}{q^2+\frac{M^2}{9}(J'+\frac13)(J'+\frac23)}\right\} \\
\label{P774CD}
\cA_{7,7}^{4C\cD}=& (-1)^m \left( \frac{M}{3} \right)^2 
\sum_{(\rho', J',m')}\sum_{(J'',m'')}
(-1)^{m''-m'+1} \hat{C}^{J''\,m''\,(jj)}_{J\,m\,(jj)\,\bar{J}\,\bar{m}\,(jj)}\,
\hat{\cD}^{J''\,-m''\,(jj)}_{J'\,m'\,(jj)\,\rho'\,J'\,-m'\,(jj)\,\rho'} \nn \\
& \hspace{24mm} \times\int\frac{d^2q}{(2\pi)^2}\,
\tilde{\Delta}_{(\rho',J')}(q)^{-1},  \\
\label{P773DD}
\cA_{7,7}^{3\cD\cD}=& 
(-1)^{m}\left( \frac{M}{3} \right)^2 
\sum_{(J',m')}\sum_{(\rho'',J'',m'')} (-1)^{-m'-m''} \nn \\
& \hspace{-16mm}\times \left((-1)^{m'}\sqrt{J(J+1)}\hat{\cD}^{J'\,-m'\,(jj)}_{J''\,m''\,(jj)\,\rho''\,J\,m\,(jj)\,0}
-(-1)^m\sqrt{J'(J'+1)}\hat{\cD}^{J\,-m\,(jj)}_{J'\,m'\,(jj)\,0\,J''\,m''\,(jj)\,\rho''}\right) \nn \\
& \hspace{-16mm}\times\left((-1)^{\bar{m}}\sqrt{J'(J'+1)}\hat{\cD}^{\bar{J}\,-\bar{m}\,(jj)}_{J''\,-m''\,(jj)\,\rho''\,J'\,-m'\,(jj)\,0} 
-(-1)^{-m'}\sqrt{\bar{J}(\bar{J}+1)}\hat{\cD}^{J'\,m'\,(jj)}_{\bar{J}\,\bar{m}\,(jj)\,0\,J''\,-m''\,(jj)\,\rho''}\right)
\nn \\
& \hspace{4mm}\times \frac{M^2}{9}\int\frac{d^2q}{(2\pi)^2}\left(\tilde{\Delta}_{J'}(q)^{-1}\right)_{33} 
\tilde{\Delta}_{(\rho'',J'')}(p+q)^{-1}, \\
\label{P773FF}
\cA_{7,7}^{3\cF\cF}=& (-1)^{m} \left( \frac{M}{3} \right)^2
\sum_{[\kappa',J',m']}\sum_{[\kappa'',J'',m'']} 
\hat{\cF}^{J'\,m'\,(jj)\,\kappa'}_{J''\,m''\,(jj)\,\kappa''\,\bar{J}\,\bar{m}\,(jj)}\,
\hat{\cF}^{J''\,m''\,(jj)\,\kappa''}_{J'\,m'\,(jj)\,\kappa'\,J\,m\,(jj)} \nn \\
& \hspace{4mm} \times \frac{1}{2}\int \frac{d^2q}{(2\pi)^2} \,
\tr_8\left(\hat{\gamma}_7 \hat{D}_{J',\kappa'}(q)^{-1}\hat{\gamma}_{7} 
\hat{D}_{J'',\kappa''}(p+q)^{-1}\right), \\
\label{P773CC1}
\cA_{7,7}^{3CC(1)}
=& (-1)^{m+1}\left( \frac{M}{3} \right)^2 \sum_{(J',m')}\sum_{(J'',m'')}\Biggl[
\hat{C}^{J' \,m'\,(jj)}_{J''\,m''\,(jj)\,J\,m\,(jj)}\hat{C}^{J''\,m''\,(jj)}_{J'\,m'\,(jj)\,\bar{J}\,\bar{m}\,(jj)} \nn \\
& \hspace{14mm} \times \int\frac{d^2q}{(2\pi)^2}\,(q_\mu+2p_\mu)(q_\nu-p_\nu)  
\left(\tilde{\Delta}_{J'}(q)^{-1}\right)_{\mu 3}\left(\tilde{\Delta}_{J''}(-p-q)^{-1}\right)_{3\nu} \nn \\
& \hspace{7mm} + (-1)^{m'-m''}
\hat{C}^{J'\,m'\,(jj)}_{J\,m\,(jj)\,J''\,m''\,(jj)}\hat{C}^{J'\,-m'\,(jj)}_{J''\,-m''\,(jj)\,\bar{J}\,\bar{m}\,(jj)} \nn \\
& \hspace{14mm}  \times \int \frac{d^2q}{(2\pi)^2}\,(q_\mu+2p_\mu)(q_\nu+2p_\nu) 
\left(\tilde{\Delta}_{J'}(q)^{-1}\right)_{\mu \nu}\left(\tilde{\Delta}_{J''}(-p-q)^{-1}\right)_{33} 
\Biggr] \nn\\
&\hspace{-4mm} +(-1)^{\bar{m}+1}\left( \frac{M}{3} \right)^2 
\sum_{(J',m')}\sum_{(J'',m'')}(-1)^{-m'-m''} \hat{C}^{J\,-m\,(jj)}_{J'\,m'\,(jj)\,J''\,m''\,(jj)}\,\hat{C}^{\bar{J}\,-\bar{m}\,(jj)}_{J''\,-m''\,(jj)\,J'\,-m'\,(jj)} \nn \\
& \hspace{14mm}\times \left(\frac{2M}{9}\right)^2\left\{\frac{1}{2}\int\frac{d^2q}{(2\pi)^2}
 \left(\left(\tilde{\Delta}_{J'}(q)^{-1}\right)_{12} \left(\tilde{\Delta}_{J''}(-p-q)^{-1}\right)_{21} 
\right.\right. \nn \\
& \hspace{56mm} \left.+\left(\tilde{\Delta}_{J'}(q)^{-1}\right)_{21} \left(\tilde{\Delta}_{J''}(-p-q)^{-1}\right)_{12}\right) \nn \\
& \hspace{39mm}-\int\frac{d^2q}{(2\pi)^2}\,
\frac{1}{q^2+\frac{M^2}{9}(J'+\frac13)(J'+\frac23)} \nn \\
& \hspace{56mm} \left. \times\frac{1}{(p+q)^2+\frac{M^2}{9}(J''+\frac13)(J''+\frac23)}\right\}, \\
\label{P773CC2}
\cA_{7,7}^{3CC(2)}
=& \left( \frac{M}{3} \right)^2 \sum_{(J',m')}\sum_{(J'',m'')}(-1)^{-m'} 
\hat{C}^{J\,-m\,(jj)}_{J''\,m''\,(jj)\,J'\,m'\,(jj)}\,\hat{C}^{J''\,m''\,(jj)}_{\bar{J}\,\bar{m}\,(jj)\,J'\,-m'\,(jj)}  \nn \\
& \hspace{14mm} \times \frac{4M}{9}\int\frac{d^2q}{(2\pi)^2} (q_\mu-p_\mu) \left(
\left(\tilde{\Delta}_{J'}(q)^{-1}\right)_{13}\left(\tilde{\Delta}_{J''}(-p-q)^{-1}\right)_{2\mu} \right. \nn \\
& \hspace{63mm}\left.-  \left(\tilde{\Delta}_{J'}(q)^{-1}\right)_{23}\left(\tilde{\Delta}_{J''}(-p-q)^{-1}\right)_{1\mu} \right) \nn \\
& + (-1)^{\bar{m}}\left( \frac{M}{3} \right)^2 \sum_{(J',m')}\sum_{(J'',m'')}(-1)^{-m'-m''} 
\hat{C}^{J\,-m\,(jj)}_{J'\,m'\,(jj)\,J''\,m''\,(jj)}\,\hat{C}^{\bar{J}\,-\bar{m}\,(jj)}_{J''\,-m''\,(jj)\,J'\,-m'\,(jj)}  \nn \\
& \hspace{14mm}\times \left(\frac{2M}{9}\right)^2\int\frac{d^2q}{(2\pi)^2} 
\left(\tilde{\Delta}_{J'}(q)^{-1}\right)_{11}\left(\tilde{\Delta}_{J''}(-p-q)^{-1}\right)_{22}. 
\end{align}
By using (\ref{S4_CD2}), $\cA_{7,7}^{4C\cD}$ can also be 
expressed  as 
\begin{align}
\label{P774CD_2}
\cA_{7,7}^{4C\cD} =& (-1)^{\bar{m}}\left( \frac{M}{3} \right)^2 \sum_{(\rho',J',m')}\sum_{(\rho'',J'',m'')} (-1)^{m''-m'}\nn \\
& \hspace{4mm}\times\hat{\cD}^{J\,-m\,(jj)}_{J''\,m''\,(jj)\,\rho''\,J'\,-m'\,(jj)\,\rho'}\,
\hat{\cD}^{\bar{J}\,-\bar{m}\,(jj)}_{J'\,m'\,(jj)\,\rho'\,J''\,-m''\,(jj)\,\rho''} 
 \int\frac{d^2q}{(2\pi)^2}\,\tilde{\Delta}_{(\rho',J')}(q)^{-1}. 
\end{align}
The non-planar contributions are 
\begin{align}
\label{NP774CC}
\cB_{7,7}^{4CC}=& 
(-1)^m \left(\frac{M}{3}\right)^2 
\sum_{(J',m')}\sum_{(J'',m'')}(-1)^{m''-m'+1}
\hat{C}^{J''\,m''\,(jj)}_{J'\,m'\,(jj)\,J\,m\,(jj)} 
\hat{C}^{J''\,-m''\,(jj)}_{J'\,-m'\,(jj)\,\bar{J}\,\bar{m}\,(jj)} \nn \\
& \hspace{24mm} \times\int \frac{d^2q}{(2\pi)^2}\left\{
\left(\tilde{\Delta}_{J'}(q)^{-1}\right)_{\mu\mu} 
+\frac{4}{q^2+\frac{M^2}{9}(J'+\frac13)(J'+\frac23)}\right\}, \\
\label{NP774DD}
\cB_{7,7}^{4\cD\cD}
=& (-1)^{\bar{m}} \left(\frac{M}{3}\right)^2 \sum_{(\rho',J',m')}\sum_{(\rho'',J'',m'')}(-1)^{\tilde{Q}'+\tilde{Q}''-J+1}\, (-1)^{m''-m'}\nn \\
& \times\hat{\cD}^{J\,-m\,(jj)}_{J''\,m''\,(jj)\,\rho''\,J'\,-m'\,(jj)\,\rho'}\, 
\hat{\cD}^{\bar{J}\,-\bar{m}\,(jj)}_{J'\,-m'\,(jj)\,\rho'\,J''\,m''\,(jj)\,\rho''} 
\int \frac{d^2q}{(2\pi)^2}\,\tilde{\Delta}_{(\rho',J')}(q)^{-1}.  \\
\label{NP773DD}
\cB_{7,7}^{3\cD\cD}
=& (-1)^m \left(\frac{M}{3}\right)^2 \sum_{(J',m')}\sum_{(\rho'',J'',m'')} (-1)^{J'+\tilde{Q}''-J+1}\,(-1)^{-m'-m''+1} \nn \\
& \hspace{-14mm}\times\left((-1)^{m'}\sqrt{J(J+1)}\hat{\cD}^{J'\,-m'\,(jj)}_{J''\,m''\,(jj)\,\rho''\,J\,m\,(jj)\,0}
-(-1)^m\sqrt{J'(J'+1)}\hat{\cD}^{J\,-m\,(jj)}_{J'\,m'\,(jj)\,0\,J''\,m''\,(jj)\,\rho''}\right) \nn \\
& \hspace{-14mm}\times\left((-1)^{\bar{m}}\sqrt{J'(J'+1)}\hat{\cD}^{\bar{J}\,-\bar{m}\,(jj)}_{J''\,-m''\,(jj)\,\rho''\, J'\,-m'\,(jj)\,0} 
-(-1)^{m'}\sqrt{\bar{J}(\bar{J}+1)}\hat{\cD}^{J'\,m'\,(jj)}_{\bar{J}\,\bar{m}\,(jj)\,0\,J''\,-m''\,(jj)\,\rho''}\right)
\nn \\
& \hspace{4mm}\times \frac{M^2}{9}\int\frac{d^2q}{(2\pi)^2}\left(\tilde{\Delta}_{J'}(q)^{-1}\right)_{33} 
\tilde{\Delta}_{(\rho'',J'')}(p+q)^{-1}, \\
\label{NP773FF}
\cB_{7,7}^{3\cF\cF}
=& (-1)^{m}\left(\frac{M}{3}\right)^2 \sum_{[\kappa',J',m']}\sum_{[\kappa'',J'',m'']} (-1)^{\tilde{U}'+\tilde{U}''-J+1}\,
\hat{\cF}^{J''\,m''\,(jj)\,\kappa''}_{J'\,m'\,(jj)\,\kappa'\,J\,m\,(jj)}\,
\hat{\cF}^{J'\,m'\,(jj)\,\kappa'}_{J''\,m''\,(jj)\,\kappa''\,\bar{J}\,\bar{m}\,(jj)} \nn \\
& \hspace{24mm}\times \frac{1}{2}\int \frac{d^2q}{(2\pi)^2} \,
\tr_8\left(\hat{\gamma}_7 \hat{D}_{J',\kappa'}(q)^{-1}\hat{\gamma}_7 \hat{D}_{J'',\kappa''}(p+q)^{-1}\right), \\
\label{NP773CC1}
\cB_{7,7}^{3CC(1)}
=& (-1)^{m+1}\left(\frac{M}{3}\right)^2 \sum_{(J',m')}\sum_{(J'',m'')}(-1)^{J'+J''-J+1}\Biggl[
\hat{C}^{J' \,m'\,(jj)}_{J''\,m''\,(jj)\,J\,m\,(jj)}\,\hat{C}^{J''\,m''\,(jj)}_{J'\,m'\,(jj)\,\bar{J}\,\bar{m}\,(jj)}  \nn \\
& \hspace{14mm}\times \int\frac{d^2q}{(2\pi)^2}\,(q_\mu+2p_\mu)(q_\nu-p_\nu)
\left(\tilde{\Delta}_{J'}(q)^{-1}\right)_{\mu 3}\left(\tilde{\Delta}_{J''}(-p-q)^{-1}\right)_{3\nu}  \nn\\
& \hspace{7mm} + (-1)^{m'-m''}\hat{C}^{J'\,m'\,(jj)}_{J\,m\,(jj)\,J''\,m''\,(jj)}\,
\hat{C}^{J'\,-m'\,(jj)}_{J''\,-m''\,(jj)\,\bar{J}\,\bar{m}\,(jj)} \nn \\
& \hspace{14mm} \times \int \frac{d^2q}{(2\pi)^2}\,(q_\mu+2p_\mu)(q_\nu+2p_\nu)
\left(\tilde{\Delta}_{J'}(q)^{-1}\right)_{\mu\nu}\left(\tilde{\Delta}_{J''}(-p-q)^{-1}\right)_{33}\Biggr] \nn \\
& \hspace{-4mm}+ (-1)^{\bar{m}+1}\left(\frac{M}{3}\right)^2 \sum_{(J',m')}\sum_{(J'',m'')}(-1)^{-m'-m''} 
\hat{C}^{J\,-m\,(jj)}_{J'\,m'\,(jj)\,J''\,m''\,(jj)}\,\hat{C}^{\bar{J}\,-\bar{m}\,(jj)}_{J'\,-m'\,(jj)\,J''\,-m''\,(jj)}  \nn \\
& \hspace{14mm}\times \left(\frac{2M}{9}\right)^2\left\{\frac{1}{2}\int\frac{d^2q}{(2\pi)^2}
 \left(\left(\tilde{\Delta}_{J'}(q)^{-1}\right)_{11} 
 \left(\tilde{\Delta}_{J''}(-p-q)^{-1}\right)_{22} \right.\right. \nn \\
& \hspace{56mm} \left.+\left(\tilde{\Delta}_{J'}(q)^{-1}\right)_{22} 
\left(\tilde{\Delta}_{J''}(-p-q)^{-1}\right)_{11}\right) \nn \\
& \hspace{39mm}+\int\frac{d^2q}{(2\pi)^2}\,\frac{1}{q^2+\frac{M^2}{9}(J'+\frac13)(J'+\frac23)}\nn \\
& \hspace{56mm} \left. \times\frac{1}{(p+q)^2+\frac{M^2}{9}(J''+\frac13)(J''+\frac23)}\right\}, 
\\
\label{NP773CC2}
\cB_{7,7}^{3CC(2)}
=&  \left(\frac{M}{3}\right)^2 \sum_{(J',m')}\sum_{(J'',m'')}(-1)^{J'+J''-J+1}\,(-1)^{-m'} 
\hat{C}^{J\,-m\,(jj)}_{J''\,m''\,(jj)\,J'\,m'\,(jj)}\,\hat{C}^{J''\,m''\,(jj)}_{\bar{J}\,\bar{m}\,(jj)\,J'\,-m'\,(jj)}  \nn \\
& \hspace{14mm} \times \frac{4M}{9}\int\frac{d^2q}{(2\pi)^2} (q_\mu-p_\mu) \left(
 \left(\tilde{\Delta}_{J'}(q)^{-1}\right)_{13}\left(\tilde{\Delta}_{J''}(-p-q)^{-1}\right)_{2\mu} \right. \nn \\
& \hspace{60mm}\left.-  \left(\tilde{\Delta}_{J'}(q)^{-1}\right)_{23}\left(\tilde{\Delta}_{J''}(-p-q)^{-1}\right)_{1\mu} \right) \nn \\
& + (-1)^{\bar{m}} \left(\frac{M}{3}\right)^2 \sum_{(J',m')}\sum_{(J'',m'')}(-1)^{J'+J''-J}\,(-1)^{-m'-m''} \nn \\
& \hspace{50mm} \times\hat{C}^{J\,-m\,(jj)}_{J'\,m'\,(jj)\,J''\,m''\,(jj)}\,
\hat{C}^{\bar{J}\,-\bar{m}\,(jj)}_{J''\,-m''\,(jj)\,J'\,-m'\,(jj)}  \nn \\
& \hspace{14mm}\times \left(\frac{2M}{9}\right)^2\int\frac{d^2q}{(2\pi)^2}  
\left(\tilde{\Delta}_{J'}(q)^{-1}\right)_{12}\left(\tilde{\Delta}_{J''}(-p-q)^{-1}\right)_{21}. 
\end{align}

\subsection{Strategy to evaluate the one-loop diagrams}
\label{sec:strategy}
In the following, we compute the one-loop diagrams listed 
in (\ref{Pii4CC})--(\ref{NP773CC2}) in the Moyal limit of Step 1:
\begin{equation}
 M\to0, \quad n=2j+1 \to \infty, \quad \frac{1}{n(M/3)^2} \equiv \frac{\Theta}{2}:{\rm fixed}. 
 \label{continuum limit}
\end{equation}
As in the computation of the one-point functions, 
UV divergences from the momentum integrations are regularized by the UV cutoff $\Lambda_p$~\footnote{
As we will see, $\Lambda_p$-dependent terms eventually cancel each other and thus the gauge symmetry is not spoiled due to the cutoff. 
Alternatively, we can see at least at the one-loop level 
that the cutoff regularization corresponds to the dimensional reduction regularization ($d=2-2\epsilon$) 
which respects the gauge symmetry, by 
$
\ln\Lambda_p^2\Leftrightarrow \frac{1}{\epsilon}-\gamma+\ln(4\pi) 
$ 
with $\gamma$ being the Euler constant.}. 
Even after setting the cutoff, we often encounter divergences in the sums over $J'$ and $J''$ from the region of $J'=0$ or $J''=0$. 
Since they are angular momenta in the fuzzy $S^2$, they can be regarded as ``IR'' divergences on the $S^2$. 
We consider the case with the external angular momentum $J$ nonzero such that the Moyal limit of the momentum 
$u\equiv \frac{M}{3}\cdot J$ is kept finite as $M \to 0$. 
The external momentum $|p|$ in the original $\R^2$ is assumed to be of the same order as $u$. 
{}From the triangular inequality 
\begin{equation}
|J'-J''|\leq J \leq J'+J'', 
\end{equation}
the IR divergences may arise when one of $J'$ and $J''$ (say $J''$) is equal to zero. 
We first remove the region of $J''=0$ from the summation in that case, and call the remaining part the ``UV part''. 
The $J''=0$ part is treated separately by introducing the IR cutoff $\delta$, 
which is removed after taking the Moyal limit (\ref{continuum limit}). 
By using the expression of the vertex coefficients (\ref{VC_FSH}), (\ref{VC_D}), and (\ref{VC_F}), 
we can carry out the sum over variables other than $J'$ and $J''$ in 
the UV part of the planar contributions (\ref{Pii4CC})--(\ref{Pii4CD_2}) and (\ref{P774CC})--(\ref{P774CD_2}). 
The result of each contribution is expressed as a sum of the following building block: 
\begin{align}
\cA^{\rm UV}  \equiv \left(\frac{M}{3}\right)^2 \sum_{J'=0}^{2j} \sum_{J''=1}^{2j} 
& n \,f(J',J'';J)\, 
\Wsj^2\nn \\ 
&\times \int^{\tilde{\Lambda}_p} \frac{d^2\tq}{(2\pi)^2}
\frac{g(\tp,\tq)}{M_{a,b}\left(\tP_i(J') ; \tQ_k(J'') ; \tq,\tp
 \right)}.
\label{general A}
\end{align}
Similarly, for the non-planar contribution in (\ref{NPii4CC})--(\ref{NP343CC2}) and (\ref{NP774CC})--(\ref{NP773CC2}), 
the corresponding building block takes the form~\footnote{Note that $\cA_{i,i}^{4C\cD}$ corresponds to $\cB_{i,i}^{4\cD\cD}$ 
from the expressions (\ref{Pii4CD_2}) and (\ref{NPii4DD}). This is also similar for $\cA_{7,7}^{4C\cD}$ and $\cB_{7,7}^{4\cD\cD}$.} 
\begin{align}
\cB ^{\rm UV}  \equiv -\left(\frac{M}{3}\right)^2 \sum_{J'=0}^{2j} \sum_{J''=1}^{2j} 
& n (-1)^{J'+J''+J} f(J',J'';J)\, 
\Wsj^2 \nn \\
&\times \int^{\tilde{\Lambda}_p} \frac{d^2\tq}{(2\pi)^2}
\frac{g(\tp,\tq)}{M_{a,b}\left(\tP_i(J') ; \tQ_k(J'') ; \tq,\tp
\right)}. 
\label{general B}
\end{align}
We have rescaled the external and internal momenta $p_\mu$ and $q_\mu$ as
$p_\mu=\left(\frac{M}{3}\right)\tp_\mu$ and 
$q_\mu=\left(\frac{M}{3}\right)\tq_\mu$. 
Correspondingly, we have rescaled the UV cutoff as $\Lambda_p=\left(\frac{M}{3}\right) \tilde\Lambda_p$ 
and the momentum integral $\int^{\tilde{\Lambda}_p}$ denotes the integration 
with the rescaled UV cutoff. 
$f(J',J'';J)$ is a function of 
the form
\begin{equation}
f(J',J'';J) = C \ J^{N_J} (J'-J'')^{N_\Delta}(J')^{N_1}(J'')^{N_2},
\label{form_f}
\end{equation}
where $C$ is an $\cO(1)$ constant, and $N_J$, $N_\Delta$, $N_1$ and $N_2$ are 
integers.  
$g(\tp,\tq)$ is a homogeneous polynomial of $\tp$ and $\tq$ consisting of 
$\tq^2$, $\tp^2$, $\tp\cdot\tq$, $(\tp\times\tq)^2$.  
$P_i(J)=(\frac{M}{3})^2\tP_i(J)$ $(i=1,\cdots,a)$ and $Q_k(J)=(\frac{M}{3})^2\tQ_k(J)$ $(k=1,\cdots,b)$ 
are monic quadratic polynomials of $J$ with 
$0\le J \le 2j$. 
The symbol $M_{a,b}(\tP_i(J') ; \tQ_k(J'') ; \tq,\tp)$ is defined as
\begin{align}
 M_{a,b}(\tP_i(J') ; \tQ_k(J'') ; \tq,\tp)
 &\equiv\prod_{i=1}^a \left(\tq^2+\tP_i(J')\right)\cdot 
 \prod_{k=1}^b \left((\tq+\tp)^2+\tQ_k(J'')\right)
 \nn \\
& \hspace{-8mm}= \left(\frac{M}{3}\right)^{-2(a+b)}\prod_{i=1}^a  \left(q^2+P_i(J')\right)\cdot 
 \prod_{k=1}^b\left((q+p)^2+Q_k(J'')\right).  
\end{align}
For the case of $b=0$, we have 
\be
M_{a,0}(\tP_i(J');\tq)=\prod_{i=1}^a\left(\tq^2+\tP_i(J')\right)
=\left(\frac{M}{3}\right)^{-2a}\prod_{i=1}^a\left(q^2+P_i(J')\right).
\label{Ma0}
\ee 
In the calculation, the function
\begin{align}
&\hat{L}(\tP(J'),\tQ(J'');\tp) \equiv 
\int \frac{d^2 \tq}{(2\pi)^2} 
\frac{4\pi}{M_{1,1}\left( \tP(J'),\tQ(J'');\tq,\tp \right)} \nn \\
&=
\frac{1}{\sqrt{(\tp^2)^2+2(\tP(J')+\tQ(J''))\tp^2+(\tP(J')-\tQ(J''))^2}} \nn \\
&\hspace{7mm}\times \ln\left(
\frac{\tp^2+\tP(J')+\tQ(J'')+\sqrt{(\tp^2)^2+2(\tP(J')+\tQ(J''))\tp^2+(\tP(J')-\tQ(J''))^2}}
{\tp^2+\tP(J')+\tQ(J'')-\sqrt{(\tp^2)^2+2(\tP(J')+\tQ(J''))\tp^2+(\tP(J')-\tQ(J''))^2}}
\right)
\label{hatL}
\end{align}
is used, and the polynomials
\begin{align}
 \tA(J) &\equiv (J+\frac{1}{3})(J+\frac{2}{3}),\quad 
 \tB(J) \equiv J(J+\frac{1}{3}), \quad 
 \tC(J) \equiv (J+1)(J+\frac{2}{3}), \nn \\
 \tD(J) &\equiv J(J+1), \quad
 \tE(J) \equiv (J+1)^2, \quad
 \tF(J) \equiv J^2 
\label{tPtQ_actual} 
\end{align}
appear as 
$\tP_i(J)$ and $\tQ_k(J)$. 
For $\cA_{7,7}^{3CC(1)}, \,\cA_{7,7}^{3CC(2)}$ and $\cB_{7,7}^{3CC(1)}, \,\cB_{7,7}^{3CC(2)}$, 
the expressions (\ref{general A}) and (\ref{general B}) hold respectively up to additive terms 
that are irrelevant in the limit (\ref{continuum limit}). 
It should be noted that the summand of (\ref{general B}) is different from that of (\ref{general A}) 
just by the sign factor $-(-1)^{J'+J''+J}$. 

\begin{figure}
\begin{center}
\includegraphics[height=7cm, width=7cm, clip]{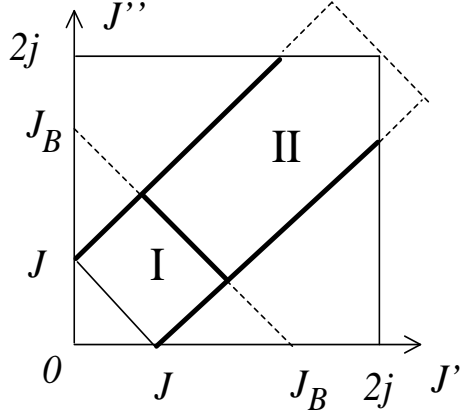}
\end{center}
\caption{\small Regions I and II are surrounded by solid lines. Note that the two triangular regions at the top right  
give no contribution since the $6j$ symbol is null there.}
\label{fig:regionI_II}
\end{figure}

Note that the $6j$ symbol $\Wsj$ vanishes outside the region of 
$J\le J'+J'' \le 4j$ and $|J'-J''| \le J$. 
In order to estimate the expressions (\ref{general A}) and (\ref{general B})
in the limit (\ref{continuum limit}), 
we separate the range of the summation of angular momenta $J'$ and $J''$, 
 \begin{equation}
  0\le J' \le 2j, \quad 1 \le J'' \le 2j, \quad -J \le J'-J'' \le J, 
\label{total_region}  
\end{equation}
into two regions satisfying
\begin{equation}
\begin{cases}
{\rm Region\ I\,:} \qquad & J \le J'+J'' < J_B \\
{\rm Region\ II\,:} \qquad & J_B \le J'+J'' \le 4j
\end{cases}  \qquad 
\left(J_B=\cO(M^{-2\alpha}), \quad \frac{1}{2}<\alpha<1\right). 
\label{regions}
\end{equation}
The regions are depicted in Fig.~\ref{fig:regionI_II}. 
In Region I, we can evaluate the summations of $J'$ and $J''$ by integrations 
in the Moyal limit, but this it is not justified in Region II. 
%
Since $J$ is typically of the order of $\cO(M^{-1})$, $J\ll J_B \ll j$ in the limit $M\to 0$. 
Correspondingly, we divide $\cA^{\rm UV}$ into two parts: 
\begin{align}
 \cA^{\rm UV}_{I}\equiv &\left(\frac{M}{3}\right)^2 \sum_{J',J''\in \text{Region I}} 
n \,f(J',J'',J)\, 
\Wsj^2 \nn \\
& \hspace{33mm}\times \int^{\tilde{\Lambda}_p} \frac{d^2\tq}{(2\pi)^2}
\frac{g(\tp,\tq)}{M_{a,b}\left(\tP_i(J') ; \tQ_k(J'') ; \tq,\tp
 \right)},
\label{cAUV_I}  \\
 \cA^{\rm UV}_{II} \equiv &\left(\frac{M}{3}\right)^2 \sum_{J',J''\in \text{Region II}} 
n \,f(J',J'',J)\, 
\Wsj^2 \nn \\
& \hspace{33mm}\times \int^{\tilde{\Lambda}_p} \frac{d^2\tq}{(2\pi)^2}
\frac{g(\tp,\tq)}{M_{a,b}\left(\tP_i(J') ; \tQ_k(J'') ; \tq,\tp
 \right)};
 \label{cAUV_II}
\end{align}
we do the same for $\cB^{\rm UV}$.

Here 
the amount of computation is considerably reduced 
by looking at the asymptotic behavior of the integrands. 
We assume that the homogeneous polynomial $g(\tp,\tq)$ behaves asymptotically as 
\begin{equation}
g(\tp,\tq) \sim |\tp|^{n_p}|\tq|^{n_q} \left(1 + \cO(|\tp||\tq|^{-1})\right)
\label{asympt g}
\end{equation}
for $|\tilde{q}|\gg 1$. 
Then we can claim the following theorem with respect to the summation in 
Region II: 
\noindent 
\begin{description}
\item\underline{\bf Theorem 1 (Region II)}  \\
{\em 
Let us define 
\begin{equation}
 w\equiv N_1+N_2 -2(a+b) + n_q + 2. 
\end{equation}
Then $\cA^{\rm UV}_{II}$ vanishes in the Moyal limit (\ref{continuum limit}) 
if one of the following conditions is satisfied; 
\begin{enumerate}
\item{$w>0$ and $W^+ \equiv -N_J-N_\Delta-n_p +2-2 w>0$}
\item{$w\leq 0$ and $W^- \equiv -N_J-N_\Delta-n_p+2 -w >0$}.
\end{enumerate}
}
\end{description}

In Region I, on the other hand, 
the summation over $J'$ and $J''$ can be 
approximated by an integration over the variables 
$u'\equiv \frac{M}{3}\cdot J'$ and $u'' \equiv \frac{M}{3}\cdot J''$ when $M$ is sufficiently small. 
Under the assumption that $\cA_{II}^{\rm UV}$ vanishes in the Moyal limit, the IR property of the function~\footnote{
The argument $q-tp$ arises by introducing the Feynman parameter $t$.} 
$g(p,q-tp)$ has a key role, as discussed in Appendix~\ref{app:proof R1}. 
In the expansion 
\be
g(p, q-tp) = \sum_{\ell\geq 0}\alpha_\ell(t)q^{2\ell}, 
\ee 
the coefficient $\alpha_\ell(t)$, which is a polynomial of $t$, is assumed to behave as 
\begin{align}
 \alpha_\ell(t)&= 
\begin{cases}
	t^{a_\ell^{(0)}} +\cO\left(t^{a_\ell^{(0)}+1}\right) & (t\sim 0) \\
	(1-t)^{a_\ell^{(1)}}+\cO\left((1-t)^{a_\ell^{(1)}+1}\right) & (t\sim 1)
\end{cases}  
\label{alpha_l(t)}
\end{align}
up to multiplicative $\cO(1)$ factors.  
Then we can prove the theorem with respect to the summation over Region I: 
\begin{description}
\item \underline{\bf Theorem 2 (in Region I)} \\
{\em 
Suppose that $\cA^{\rm UV}_{II}$ vanishes in the Moyal limit by satisfying 
the condition of Theorem 1.
In the case of $a,\,b\geq 1$, $\cA^{\rm UV}_{I}$ vanishes in the same limit 
if both of 
\begin{equation}
	D_0 \equiv W^- + N_1-2a+3 + 2\min_\ell ( \ell + a^{(0)}_\ell ) > 0 
\end{equation}
and 
\begin{equation}
	D_1 \equiv W^- + N_2-2b+3 + 2\min_\ell ( \ell + a^{(1)}_\ell ) > 0 
\end{equation}
are satisfied. When one of $a$ and $b$ is zero (say $b=0$), the condition for $\cA^{\rm UV}_{I}\to 0$ is given by 
\be
D_0^{(b=0)}\equiv W^-+N_1-2a+2\ell_0+3>0,  
\ee
where $\ell_0$ is the smallest $\ell$ such that $\alpha_\ell(0)\neq 0$. 
}
\end{description}
Proofs of these theorems are given in Appendix \ref{app:proof}. 

As we mentioned above, when the planar contribution has the form (\ref{general A}), 
the corresponding non-planar contribution takes the form (\ref{general B}). 
In Region I, 
where $J,J',J'' \ll j$ is satisfied,  
the Wigner $6j$ symbol can be approximated as (\ref{6j_RI}). 
Since it is negligible for $J'+J''+J$ odd, 
the phase factor $(-1)^{J'+J''+J}$ in (\ref{general B}) is irrelevant 
in Region I. 
In addition, looking at the proof in Appendix \ref{app:proof R2},  
we can evaluate the non-planar contribution in exactly the same way 
as the planar contribution because the difference between them 
is only the factor $(-1)^{J'+J''+J}$. 
Namely, $|\cB^{\rm UV}_{II}|$ can also be bounded from the above by the r.h.s. of (\ref{AII_bound}). 
Therefore, we can claim 
\begin{description}
\item \underline{\bf Theorem 3} \\
{\em 
When $\cA^{\rm UV}_{I}$ and $\cA^{\rm UV}_{II}$ vanish in the Moyal limit, 
the corresponding $\cB^{\rm UV}_{I}$ and $\cB^{\rm UV}_{II}$ also vanish in the same limit. 
}
\end{description}
These three theorems are useful in evaluating (\ref{Pii4CC})--(\ref{NP773CC2}).

\subsection{Contributions from one-loop diagrams}

We explicitly calculate the diagrams listed in section~\ref{sec:list_2pt_1-loop}. 
Although there are a number of diagrams, applying the theorems avoids carrying out 
brute-force computation for all of them. 
We first look at the diagrams that concern to the two-point function 
$\langle x_{i}x_{i} \rangle$ $(i=3,\ldots,6)$, where the computations of (\ref{Pii4CC}), (\ref{NPii4CC}),
 (\ref{Pii3CC}), and (\ref{NPii3CC}) are presented 
as typical examples. 
For the other diagrams, we simply show the results. 
After that, it is shown that the diagrams for $\langle x_{7}x_{7} \rangle$
give identical results with those for $\langle x_{i}x_{i} \rangle$ $(i=3,\ldots,6)$
in the Moyal limit. 

\subsubsection{Diagrams from 4-point vertices: 
${\cal A}^{4CC}_{i,i}$ and ${\cal B}^{4CC}_{i,i}$}
Let us first look at the planar contribution (\ref{Pii4CC}). 
By using 
\bea
& & \sum_{m'=-J'}^{J'}(-1)^{-m'}C^{J'\,-m''}_{J'\,m'\,J'\,-m'}=(-1)^{-J}\sqrt{2J'+1}\,\delta_{J''\,0}\delta_{m''\,0}, \\
& & \left\{\begin{matrix} J_1&J_2&0\\ j&j&j \end{matrix}\right\}
 = (-1)^{J_1+2j}\frac{\delta_{J_1\,J_2}}{\sqrt{n(2J_1+1)}}, 
 \label{6j_formula2}
\eea
the UV part can be written as 
\begin{align}
{\cal A}^{4CC,{\rm UV}}_{i,i} 
=&  \delta_{J\,\bar{J}}\,\delta_{m\,-\bar{m}}\left( \frac{M}{3} \right)^2 \sum_{J'=0}^{2j} \sum_{J''=1}^{2j} 
n (2J'+1)(2J''+1) 
\left\{ \begin{matrix} J' & J'' & J \\ j & j & j  \end{matrix} \right\}^2 \nn \\
& \hspace{-16mm}\times \int^{\tLambda_p} \frac{d^2\tq}{(2\pi)^2} \Biggl\{
\frac{2}{\tq^2 + \tD(J'')} + \frac{4}{\tq^2+\tA(J'')}  \nn \\
& \hspace{1mm}-\frac{4}{9} \frac{\tq^2}{M_{3,0}(\tB(J''),\tC(J''),\tD(J'');\tq)}
-\frac{4}{9} \frac{\tq^2}{M_{3,0}(\tA(J''),\tB(J''),\tC(J'');\tq)}
\Biggr\} .
\label{Pii4CCUVpre}
\end{align}
For example, $M_{3,0}(\tB(J''),\tC(J''),\tD(J'');\tq)$ stands for (\ref{Ma0}) with $a=3$, $\tP_1=\tB$, $\tP_2=\tC$, and $\tP_3=\tD$. 
As a result of straightforward calculations together with 
the identity 
\begin{equation}
 \sum_{J'=0}^{2j}n(2J'+1)
 \left\{\begin{matrix}J'&J''&J\\ j&j&j
 \end{matrix}\right\}^2 = 1 , 
 \label{6j_formula1}
\end{equation} 
it turns out that the last two terms do not contribute in the Moyal limit, and we have 
\begin{align}
{\cal A}_{i,i}^{4CC,{\rm UV}} =& \delta_{J\,\bar{J}}\,\delta_{m\,-\bar{m}}\,\frac{1}{4\pi} \left( \frac{M}{3} \right)^2 
\sum_{J''=1}^{2j}
(2J''+1)\Bigl(
6\ln\tLambda_p^2
-2 \ln \tD(J'') - 4 \ln \tA(J'') 
\Bigr) + \cdots.  
\label{Pii4CC-UV}
\end{align}
Here and in what follows, the ellipsis ($\cdots$) 
expresses terms that vanish in the Moyal limit~\footnote{
The last two terms $(a=3, b=0)$ have 
$(N_1,N_2)=(1,1), (1,0), (0,1), (0,0)$, $N_J=N_\Delta = 0$, $(n_p, n_q)=(0,2)$, $\ell_0=1$, 
which leads to $D_0^{(b=0)}\geq 2>0$. 
Thus, Theorems 1 and 2 also tell that the terms are negligible in the Moyal limit.}. 

For the IR part of (\ref{Pii4CC}) (contribution from $J''=0$), 
introducing the IR cutoff $J''=\delta$ ($0<\delta \ll 1$) regularizes it as 
\begin{align}
{\cal A}^{4CC,{\rm IR}}_{i,i} 
=&  \delta_{J\,\bar{J}}\,\delta_{m\,-\bar{m}}
\left( \frac{M}{3} \right)^2 \sum_{J'=0}^{2j} 
n (2J'+1)
\left\{ \begin{matrix} J & J' & \delta \\ j & j & j  \end{matrix} \right\}^2 \nn \\
  &\hspace{-14mm}\times \int^{\tLambda_p} \frac{d^2\tq}{(2\pi)^2} \Biggl\{
\frac{2}{\tq^2 + \tD(\delta)} + \frac{4}{\tq^2+\tA(\delta)}  \nn \\
&\hspace{12mm} -\frac{4}{9} \frac{\tq^2}{M_{3,0}(\tB(\delta),\tC(\delta),\tD(\delta);\tq)}
-\frac{4}{9} \frac{\tq^2}{M_{3,0}(\tA(\delta),\tB(\delta),\tC(\delta);\tq)}
\Biggr\}  \nn \\
=& 
\delta_{J\,\bar{J}}\,\delta_{m\,-\bar{m}}\,\frac{1}{4\pi} \left( \frac{M}{3} \right)^2
 \Bigl\{
6h_j^{({\rm p})}(\delta)\ln\tLambda_p^2 -\frac{4}{3} \ln \delta 
\Bigr\}  + \cdots
\label{Pii4CC-IR}
\end{align}
with
\be
h_j^{({\rm p})}(\delta)\equiv 
(1+2\delta)\sum_{J'=0}^{2j}n(2J'+1)\left\{\begin{matrix} J & J' & \delta \\ j & j & j  \end{matrix} \right\}^2. 
\ee
We have assumed 
\be
\left\{\begin{matrix} J & J' & \delta \\ j & j & j  \end{matrix} \right\}
= \left\{\begin{matrix} J & J' & 0 \\ j & j & j  \end{matrix} \right\} + o(\delta^0),
\ee
and used (\ref{6j_formula2}).  

We next evaluate the non-planar contribution (\ref{NPii4CC}). 
Theorem 3 leads to its UV part as
\begin{align}
{\cal B}^{4CC,{\rm UV}}_{i,i} 
=&  -\delta_{J\,\bar{J}}\,\delta_{m\,-\bar{m}}\,\frac{1}{4\pi} \left( \frac{M}{3} \right)^2 \sum_{J'=0}^{2j} \sum_{J''=1}^{2j} 
(-1)^{J'+J''+J} n (2J'+1)(2J''+1) 
\left\{ \begin{matrix} J' & J'' & J \\ j & j & j  \end{matrix} \right\}^2 \nn \\
&\times \Bigl(
6\ln\tLambda_p^2-2\ln\tD(J'')-4\ln\tA(J'') 
\Bigr)+\cdots.
\label{NPii4CCUV}
\end{align}
Similarly to (\ref{Pii4CC-IR}), the IR part of 
(\ref{NPii4CC}) becomes 
\begin{align}
{\cal B}^{4CC,{\rm IR}}_{i,i} 
=&  
-\delta_{J\,\bar{J}}\,\delta_{m\,-\bar{m}}\left( \frac{M}{3} \right)^2 \sum_{J'=0}^{2j} 
(-1)^{J+J'+\delta}n (2J'+1) (1+2\delta)
\left\{ \begin{matrix} J & J' & \delta \\ j & j & j  \end{matrix} \right\}^2 \nn \\
  &\hspace{-14mm}\times \int^{\tLambda_p} \frac{d^2\tq}{(2\pi)^2} \Biggl\{
\frac{2}{\tq^2 + \tD(\delta)} + \frac{4}{\tq^2+\tA(\delta)}  \nn \\
&\hspace{12mm} -\frac{4}{9} \frac{\tq^2}{M_{3,0}(\tB(\delta),\tC(\delta),\tD(\delta);\tq,\tp)}
-\frac{4}{9} \frac{\tq^2}{M_{3,0}(\tA(\delta),\tB(\delta),\tC(\delta);\tq,\tp)}
\Biggr\}  \nn \\
=& 
-\delta_{J\,\bar{J}}\,\delta_{m\,-\bar{m}}\,\frac{1}{4\pi}\left( \frac{M}{3} \right)^2
 \Bigl\{
6h_j^{({\rm np})}(\delta)\ln\tilde{\Lambda}_p^2 -\frac{4}{3} \ln \delta
\Bigr\} + \cdots
\label{NPii4CC-IR}
\end{align}
with
\be
h_j^{({\rm np})}(\delta)\equiv 
(1+2\delta)\sum_{J'=0}^{2j}(-1)^{J+J'+\delta}n(2J'+1)\left\{\begin{matrix} J & J' & \delta \\ j & j & j  \end{matrix} \right\}^2. 
\ee

\subsubsection{Diagrams from 3-point vertices: 
${\cal A}^{3CC(1)}_{i,i}$ and ${\cal B}^{3CC(1)}_{i,i}$}

The UV part of the planar contribution (\ref{Pii3CC}) can be expressed as 
\begin{align}
&\cA_{i,i}^{3CC(1),{\rm UV}}
= \delta_{J\,\bar{J}}\,\delta_{m\,-\bar{m}}\left( \frac{M}{3} \right)^2
\sum_{J'=0}^{2j} \sum_{J''=1}^{2j} n(2J'+1)(2J''+1)
\left\{ \begin{matrix} J' & J'' & J \\ j & j & j  \end{matrix} \right\}^2 \nn \\
 &\times \int^{\tLambda_p} \frac{d^2\tq}{(2\pi)^2} 
\Biggl\{
-\frac{ (\tq-\tp)^2 }{ M_{1,1}(\tA(J');\tD(J'');\tq,\tp)} 
+\frac{16}{9}  \frac{ (\tq \times \tp)^2 }{ M_{1,3}(\tA(J');\tB(J''),\tC(J''),\tD(J'');\tq,\tp) }
\Biggr\}  .
\label{Pii3CC_UVpre1}
\end{align}
It is easy to see that the second term in the integrand does not contribute in the Moyal limit 
due to Theorems 1 and 2. 
For the first term, we rewrite the numerator as
\begin{equation}
(\tq-\tp)^2 = - \left( (\tq+\tp)^2 + \tD(J'') \right) 
+ 2 \left( \tq^2 + \tA(J') \right)
+2\tp^2 + \tD(J'') -2 \tA(J'), 
\end{equation}
and apply the theorems, leading to 
\begin{align}
\cA_{i,i}^{3CC(1), {\rm UV}} =&\delta_{J\,\bar{J}}\,\delta_{m\,-\bar{m}}
 \left( \frac{M}{3} \right)^2 \sum_{J'=0}^{2j} \sum_{J''=1}^{2j}
 n 
 \left\{ \begin{matrix} J' & J'' & J \\ j & j & j  \end{matrix} \right\}^2 
\nn \\
 &\hspace{3mm}\times\int^{\tLambda_p} \frac{d^2\tq}{(2\pi)^2} \left[(2J'+1)(2J''+1)\biggl\{ \frac{1}{\tq^2+\tA(J')} 
   - \frac{2}{\tq^2+\tD(J'')}\biggr\} \right. \nn \\
 &\left.\hspace{29mm}  +\frac{4J'J''\left( -2\tp^2 + 2J'^2 - J''^2 \right) }
{M_{1,1}(\tA(J'),\tD(J'');\tq,\tp)}
   \right] +\cdots \nn \\
=&\delta_{J\,\bar{J}}\,\delta_{m\,-\bar{m}}\,\frac{1}{4\pi} \left( \frac{M}{3} \right)^2 \Biggl[
\sum_{J''=1}^{2j} (2J''+1) \left( -\ln\tLambda_p^2 -\ln\tA(J'') + 2 \ln \tD(J'') \right) \nn \\
&\hspace{3mm} +\sum_{J'=0}^{2j} \sum_{J''=1}^{2j}n\Wsj^2 
4J'J''\left( -2\tp^2 + 2J'^2 - J''^2 \right) \hL(\tA(J'),\tD(J'');\tp) \Biggr]
\nn \\
&  +\cdots, 
 \label{Pii3CC_UV}
 \end{align}
where 
the function $\hL(\tA,\tD;\tp)$ is given by (\ref{hatL}). 

The IR part of (\ref{Pii3CC}) from 
$J''=0$ is regularized by the IR cutoff $\delta$ and becomes  
\begin{align}
\cA_{i,i}^{3CC(1),{\rm IR}}
=&  \delta_{J\,\bar{J}}\,\delta_{m\,-\bar{m}}\left( \frac{M}{3} \right)^2
\sum_{J'=0}^{2j} n(2J'+1)(1+2\delta)
 \left\{ \begin{matrix} J & J' & \delta \\ j & j & j  \end{matrix} \right\}^2 \nn \\
&\hspace{-12mm}\times \int^{\tLambda_p} \frac{d^2\tq}{(2\pi)^2} 
\Biggl\{ 
-\frac{ (\tq-\tp)^2 }{ M_{1,1}(\tA(J');\tD(\delta);\tq,\tp)} 
+\frac{16}{9}  \frac{ (\tq \times \tp)^2 }
{ M_{1,3}(\tA(J');\tB(\delta),\tC(\delta),\tD(\delta);\tq,\tp) }
\Biggr\}  \nn \\
=& \delta_{J\,\bar{J}}\,\delta_{m\,-\bar{m}}\,\frac{1}{4\pi} \left( \frac{M}{3} \right)^2
\left\{ -h_j^{({\rm p})}(\delta) \ln \tLambda_p^2
+\frac{4}{3}\left(\frac{\tp^2-\tA(J)}{\tp^2+\tA(J)} + 1 \right)
  \ln \delta \right\} +\cdots.
\label{Pii3CC-IR}  
\end{align}

For the non-planar contribution (\ref{NPii3CC}), 
Theorem 3 gives its UV part as 
\begin{align}
\cB_{i,i}^{3CC(1),{\rm UV}} =&
 -\delta_{J\,\bar{J}}\,\delta_{m\,-\bar{m}}\left( \frac{M}{3} \right)^2 \sum_{J'=0}^{2j} \sum_{J''=1}^{2j}
 (-1)^{J'+J''+J} n 
 \left\{ \begin{matrix} J' & J'' & J \\ j & j & j  \end{matrix} \right\}^2 \nn \\
&\hspace{15mm} \times 
 \int^{\tLambda_p} \frac{d^2\tq}{(2\pi)^2} \left[(2J'+1)(2J''+1)\biggl\{ \frac{1}{\tq^2+\tA(J')} 
   - \frac{2}{\tq^2+\tD(J'')} 
   \biggr\} \right.\nn \\
 &\left.\hspace{42mm}+ 
\frac{4J'J''\left( -2\tp^2 + 2J'^2 - J''^2 \right) }
{M_{1,1}(\tA(J'),\tD(J'');\tq,\tp)}\right]+\cdots \nn \\
=&-\delta_{J\,\bar{J}}\,\delta_{m\,-\bar{m}}\,\frac{1}{4\pi} \left( \frac{M}{3} \right)^2 
\sum_{J'=0}^{2j}\sum_{J''=1}^{2j}(-1)^{J'+J''+J} n\Wsj^2 \nn \\
& \hspace{15mm} \times \left[(2J'+1)(2J''+1)\left\{ -\ln\tLambda_p^2 -\ln\tA(J'') + 2 \ln \tD(J'') \right\} \right.\nn \\
&\left.\hspace{21mm} +4J'J''\left( -2\tp^2 + 2J'^2 - J''^2 \right) \hL(\tA(J'),\tD(J'');\tp) \right]
+\cdots. 
\label{NPii3CC_UVpre1}
\end{align}

We have the IR part of (\ref{NPii3CC}):
\begin{align}
\cB_{i,i}^{3CC(1),{\rm IR}}
=&  -\delta_{J\,\bar{J}}\,\delta_{m\,-\bar{m}}\left( \frac{M}{3} \right)^2
\sum_{J'=0}^{2j} (-1)^{J+J'+\delta} n(2J'+1)(1+2\delta)
 \left\{ \begin{matrix} J & J' & \delta \\ j & j & j  \end{matrix} \right\}^2 \nn \\
&\hspace{-12mm}
\times \int^{\tLambda_p} \frac{d^2\tq}{(2\pi)^2} 
\Biggl\{ 
-\frac{ (\tq-\tp)^2 }{ M_{1,1}(\tA(J');\tD(\delta);\tq,\tp)} 
+\frac{16}{9}  \frac{ (\tq \times \tp)^2 }
{ M_{1,3}(\tA(J');\tB(\delta),\tC(\delta),\tD(\delta);\tq,\tp) }
\Biggr\}  \nn \\
=& -\delta_{J\,\bar{J}}\,\delta_{m\,-\bar{m}}\,\frac{1}{4\pi} \left( \frac{M}{3} \right)^2
\left\{ -h_j^{({\rm np})}(\delta)\ln \tLambda_p^2
 +\frac{4}{3}\left(\frac{\tp^2-\tA(J)}{\tp^2+\tA(J)}+1 \right) \ln \delta \right\} +\cdots.
\label{NPii3CC-IR}
\end{align}

\subsubsection{Other diagrams for $\langle x_ix_i\rangle$ ($i = 3,...,6$)}
Here we present the results of other one-loop diagrams for $\langle x_ix_i\rangle$ ($i = 3,...,6$), 
except for $\cA^{3CC(2)}$ and $\cB^{2CC(2)}$. 
The results of the UV parts of the planar diagrams are expressed as 

\begin{align}
\label{Pii4CD-UV}
\cA_{i,i}^{4C\cD,{\rm UV}}=& 
\delta_{J\,\bar{J}}\,\delta_{m\,-\bar{m}}\,\frac{1}{4\pi} \left( \frac{M}{3} \right)^2 \sum_{J''=1}^{2j}
\Bigl\{
3(2J''+1) \ln \tLambda_p^2 
-(2J''+1) \ln \tD(J'')
\nn \\
&\hspace{44mm}  
-(2J''+3) \ln \tE(J'') -(2J''-1) \ln \tF(J'') 
\Bigr\} \nn \\
&+ \cdots,   \\
\label{Pii3DD-UV}
\cA_{i,i}^{3\cD\cD,{\rm UV}}=&
\delta_{J\,\bar{J}}\,\delta_{m\,-\bar{m}}\,\frac{1}{4\pi} \left( \frac{M}{3} \right)^2 \sum_{J'=0}^{2j}\sum_{J''=1}^{2j}
n\left\{ \begin{matrix} J' & J'' & J \\ j & j & j  \end{matrix} \right\}^2 \left(-\frac{2J'}{J''}\right) \nn \\
&\hspace{-12mm}\times\Bigl[2\left( J'^2 - J^2\right)^2\hL(\tA(J'),\tD(J'');\tp) \nn \\
&\hspace{-7mm}+\left((J'+J'')^2-J^2\right) \left( J^2 - (J'-J'')^2 \right)\left\{\hL(\tA(J'),\tE(J'');\tp)+\hL(\tA(J'),\tF(J'');\tp)\right\}\Bigr] 
\nn \\
&\hspace{-14mm} + \cdots, \\
\label{Pii3FF-UV}
\cA_{i,i}^{3\cF\cF,{\rm UV}}=&  
\delta_{J\,\bar{J}}\,\delta_{m\,-\bar{m}}\,\frac{1}{4\pi} \left( \frac{M}{3} \right)^2 
\left[\sum_{J''=1}^{2j}\left\{-8(2J''+1)\ln\tLambda_p^2+4(2J''+1)\ln \tA(J'') \right. \right.\nn \\
& \hspace{23mm}\left.+2(J''+1)\left(\ln\tC(J'') +\ln\tD(J'')\right) +2J''\left(\ln\tB(J'')+\ln\tF(J'')\right)\right\}\nn \\
&\quad+\sum_{J'=0}^{2j} \sum_{J''=1}^{2j} n \Wsj^2 4J'J'' \left( \tp^2 + J^2 \right) 
\Bigl\{\hL(\tA(J'),\tB(J'');\tp) \nn \\
&\hspace{24mm}\left.+\hL(\tA(J'),\tC(J'');\tp) +\hL(\tA(J'),\tE(J'');\tp) +\hL(\tA(J'),\tF(J'');\tp)\Bigr\}\right] \nn \\
&+ \cdots . 
\end{align}

The IR parts of the planar contributions are 
\begin{align}
\label{Pii4CD-IR}
\cA_{i,i}^{4C\cD,{\rm IR}}=& 
\delta_{J\,\bar{J}}\,\delta_{m\,-\bar{m}}\,\frac{1}{4\pi} \left( \frac{M}{3} \right)^2
 \Bigl(
3 h^{({\rm p})}_j(\delta) \ln \tLambda_p^2 
\Bigr) +\cdots, \\
\label{Pii3DD-IR}
 \cA_{i,i}^{3\cD\cD,{\rm IR}}=& 
\cdots , \\
\label{Pii3FF-IR}
\cA_{i,i}^{3\cF\cF,{\rm IR}}=&  
\delta_{J\,\bar{J}}\,\delta_{m\,-\bar{m}}\,\frac{1}{4\pi} \left( \frac{M}{3} \right)^2
\left( -8 h^{({\rm p})}_j(\delta)\ln\tLambda_p^2 
\right)  
+\cdots .
\end{align}

For the non-planar diagrams, 
the results of the UV parts are obtained as 
\begin{align}
\label{NPii4CD-UV}
\cB_{i,i}^{4C\cD,{\rm UV}}=& 
-\delta_{J\,\bar{J}}\,\delta_{m\,-\bar{m}}\,\frac{1}{4\pi} \left( \frac{M}{3} \right)^2 \sum_{J'=0}^{2j} \sum_{J''=1}^{2j} 
(-1)^{J'+J''+J} \,n(2J'+1) \Wsj^2  \nn \\
&\hspace{38mm}\times \Bigl\{
3(2J''+1) \ln \tLambda_p^2 
-(2J''+1) \ln \tD(J'') 
 \nn \\
&\hspace{44mm} 
-(2J''+3) \ln \tE(J'')-(2J''-1) \ln \tF(J'')
\Bigr\}\nn \\
& + \cdots, \\
\label{NPii3DD-UV}
 \cB_{i,i}^{3\cD\cD,{\rm UV}}=& 
-\delta_{J\,\bar{J}}\,\delta_{m\,-\bar{m}}\,\frac{1}{4\pi} \left( \frac{M}{3} \right)^2 \sum_{J'=0}^{2j}\sum_{J''=1}^{2j}
(-1)^{J'+J''+J} \,n\left\{ \begin{matrix} J' & J'' & J \\ j & j & j  \end{matrix} \right\}^2 \left(-\frac{2J'}{J''}\right) \nn \\
&\hspace{-12mm}\times \Bigl[2\left( J'^2 - J^2 \right)^2\hL(\tA(J'),\tD(J'');\tp)\nn \\
&\hspace{-7mm}+\left((J'+J'')^2-J^2\right) \left( J^2 - (J'-J'')^2 \right)
\left\{\hL(\tA(J'),\tE(J'');\tp)+\hL(\tA(J'),\tF(J'');\tp)\right\}
\Bigr]\nn \\
&\hspace{-14mm}+ \cdots, \\
\label{NPii3FF-UV}
\cB_{i,i}^{3\cF\cF,{\rm UV}}=&  
-\delta_{J\,\bar{J}}\,\delta_{m\,-\bar{m}}\,\frac{1}{4\pi} \left( \frac{M}{3} \right)^2 \sum_{J'=0}^{2j} \sum_{J''=1}^{2j} 
(-1)^{J'+J''+J} \,n \left\{ \begin{matrix} J' & J'' & J \\ j & j & j  \end{matrix} \right\}^2 \nn \\
&\hspace{3mm}\times \Bigl[
-8 (2J'+1)(2J''+1) \ln\tLambda_p^2  +4(2J'+1)(2J''+1)\ln\tA(J'')\nn \\
&\hspace{9mm} + 2(2J'+1)(J''+1) \left( \ln \tC(J'') + \ln \tE(J'')\right)\nn \\
&\hspace{9mm}+ 2(2J'+1)J'' \left( \ln \tB(J'') + \ln \tF(J'')\right) \nn \\
&\hspace{9mm} +4J'J'' \left( \tp^2 + J^2 \right)\left\{\hL(\tA(J'),\tB(J'');\tp)+\hL(\tA(J'),\tC(J'');\tp) \right. \nn \\
&\hspace{40mm}\left.\ +\hL(\tA(J'),\tE(J'');\tp)+\hL(\tA(J'),\tF(J'');\tp) \right\}
\Bigr]\nn \\
& + \cdots. 
%
\end{align}

The IR parts are given by 
\begin{align}
\label{NPii4CD-IR}
\cB_{i,i}^{4C\cD,{\rm IR}}=& 
-\delta_{J\,\bar{J}}\,\delta_{m\,-\bar{m}}\,\frac{1}{4\pi} \left( \frac{M}{3} \right)^2
 \Bigl(
3 h^{({\rm np})}_j(\delta) \ln \tLambda_p^2   
\Bigr)
+\cdots, \\
\label{NPii3DD-IR}
 \cB_{i,i}^{3\cD\cD,{\rm IR}}=& 
\cdots , \\
\label{NPii3FF-IR}
\cB_{i,i}^{3\cF\cF,{\rm IR}}=&  
-\delta_{J\,\bar{J}}\,\delta_{m\,-\bar{m}}\,\frac{1}{4\pi} \left( \frac{M}{3} \right)^2
\Bigl( -8 h^{({\rm np})}_j(\delta) \ln\tLambda_p^2 
\Bigr)+\cdots . 
%
\end{align}

\subsubsection{Diagrams for $\langle x_7 x_7 \rangle$ and extra diagrams for 
$\langle x_{3,4} x_{3,4} \rangle$}

We evaluate the diagrams (\ref{P774CC})--(\ref{P773CC2}) 
and (\ref{NP774CC})--(\ref{NP773CC2}) by taking the differences between 
(\ref{Pii4CC})--(\ref{Pii3CC}) 
and 
(\ref{NPii4CC})--(\ref{NPii3CC})
with help of the theorems. 

The difference between (\ref{P774CC}) and (\ref{Pii4CC}) reads 
\begin{align}
 \Delta{\cal A}^{4CC}\equiv& {\cal A}_{7,7}^{4CC}-{\cal A}_{i,i}^{4CC} \nn \\
 =& \delta_{J\,\bar{J}}\,\delta_{m\,-\bar{m}} \left(\frac{M}{3}\right)^2\sum_{J',J''=0}^{2j} n(2J'+1)(2J''+1) \Wsj^2  \nn \\
&\times  \int^{\tilde{\Lambda}_p} \frac{d^2\tq}{(2\pi)^2} 
\frac{\frac{4}{9} \tq^2}
{M_{3,0}(\tA(J''),\tB(J''),\tC(J'');\tq)} ,
\label{delta 4CC}
\end{align}
whose UV part vanishes in the Moyal limit from the same reason why the last term in (\ref{Pii4CCUVpre}) vanishes. 
{}From Theorem 3, the UV part of the non-planar counterpart does not contribute. 
It is easy to see that the IR parts of (\ref{delta 4CC}) and the corresponding 
expression for the non-planar diagrams 
(the contribution from $J''=0$) both vanish in the limit. 
Therefore, (\ref{P774CC}) and (\ref{NP774CC}) coincide to 
(\ref{Pii4CC}) and (\ref{NPii4CC}) in the Moyal limit, respectively.

Let us next evaluate the difference between (\ref{P773DD}) and (\ref{Pii3DD}): 
\begin{align}
&\Delta\cA^{3\cD\cD} \equiv \cA^{3\cD\cD}_{7,7} - \cA^{3\cD\cD}_{i,i} \nn \\
&=\delta_{J\,\bar{J}}\,\delta_{m\,-\bar{m}} \left(\frac{M}{3}\right)^2\left[
\sum_{J'=0}^{2j}\sum_{J''=1}^{2j}
\frac{(2J'+1)(2J''+1)}{J''(J''+1)}\left(J(J+1)-J'(J'+1)\right)^2 \right.
\nn \\
&\hspace{38mm} \times n\Wsj^2 \int \frac{d^2\tq}{(2\pi)^2}
\frac{\frac{4}{9}\tq^2}{M_{3,1}\left( \tA(J'),\tB(J'),\tC(J');\tD(J'');\tq, \tp \right)} \nn \\
& \quad+\sum_{J'=0}^{2j}\sum_{J''=0}^{2j}
\frac{2J'+1}{J''+1}(J'+J''+J+2)(J'-J''+J)(-J'+J''+J+1)(J'+J''-J+1) 
 \nn \\
&\hspace{23mm}\times n\Wsj^2\int \frac{d^2\tq}{(2\pi)^2} 
\frac{\frac{4}{9}\tq^2}{M_{3,1}\left( \tA(J'),\tB(J'),\tC(J');\tE(J'');\tq, \tp  \right)} \nn \\
&\quad+\sum_{J'=0}^{2j}\sum_{J''=1}^{2j}
\frac{2J'+1}{J''}(J'+J''+J+1)(J'-J''+J+1)(-J'+J''+J)(J'+J''-J)
\nn \\
&\left.\hspace{22mm} \times n\Wsj^2 \int \frac{d^2\tq}{(2\pi)^2} 
\frac{\frac{4}{9}\tq^2}{M_{3,1}\left( \tA(J'),\tB(J'),\tC(J');\tF(J'');\tq, \tp  \right)} \right].
\label{delta 3DD}
\end{align}
We read off the parameters in the theorems for each of the terms that satisfy $w\le 0$, 
$W^- \ge 2$, $D_0 \ge 1$, and $D_1 \ge 2$, meaning that the UV parts of 
$\cA^{3\cD\cD}_{7,7}$ and $\cB^{3\cD\cD}_{7,7}$ coincide 
with those of $\cA^{3\cD\cD}_{i,i}$ and $\cB^{3\cD\cD}_{i,i}$ in the Moyal limit, 
respectively. 
In addition, the IR parts of (\ref{delta 3DD}) 
and the corresponding non-planar contributions both behave as 
$(M/3)^2\times \cO(\delta^0)$, which is irrelevant. 
Therefore, we can say that $\cA^{3\cD\cD}_{7,7}$ and $\cB^{3\cD\cD}_{7,7}$ has the same contribution 
as $\cA^{3\cD\cD}_{i,i}$ and $\cB^{3\cD\cD}_{i,i}$ in the 
limit, 
respectively. 

By repeating the same manipulation,  
the quantities (\ref{P774CC})--(\ref{P773CC2}) 
and (\ref{NP774CC})--(\ref{NP773CC2})
coincide with 
(\ref{Pii4CC})--(\ref{Pii3CC}) 
and 
(\ref{NPii4CC})--(\ref{NPii3CC}), respectively. 
Furthermore, we can show that
the residual diagrams (\ref{P333CC}), (\ref{P343CC}), 
(\ref{NP333CC1}) and (\ref{NP343CC2}) for $\langle x_{3,4} x_{3,4} \rangle$ 
vanish 
in the same way.

Combining the results obtained above, 
we see that the amplitudes $\cA_{i, \bar{i}}(p;Jm;\bar{J}\bar{m})$ 
and $\cB_{i,\bar{i}}(p;Jm;\bar{J}\bar{m})$ $(i,\bar{i}=1,\cdots,7)$ in (\ref{2pt func}) 
vanish for $i\ne \bar{i}$ and those for $i = \bar{i}$ become independent of 
the value of $i$ in the Moyal limit. 

\section{Amplitudes and effective action}
\label{sec:decpt limit}
In this section, we sum up the contributions from the various diagrams obtained in the previous section, and evaluate 
the summations with respect to $J'$ and $J''$ to obtain the one-loop effective action 
for the scalar kinetic terms in the successive limits (the Moyal limit and the commutative limit).
 
\subsection{Planar diagrams}
The UV part of the planar contributions is given by the summation of 
(\ref{Pii4CC-UV}), (\ref{Pii3CC_UV}), (\ref{Pii4CD-UV})--(\ref{Pii3FF-UV}): 
\begin{align}
{\cal A}_{i,i}^{\rm UV} 
\equiv {\cal A}_{i,i}^{\rm UV,\, single} +  {\cal A}_{i,i}^{\rm UV,\, double} + \cdots
\end{align}
with 
\begin{align}
\label{Ptotal_UV_single}
 {\cal A}_{i,i}^{\rm UV,\, single} 
 \equiv &\, \delta_{J\,\bar{J}}\,\delta_{m\,-\bar{m}}\,\frac{1}{4\pi}\left( \frac{M}{3} \right)^2 \sum_{J'=1}^{2j} \Bigl\{
-\ln (J'+ 1/3) + \ln (J'+2/3) -\ln(J'+1) + \ln J'
\Bigr\}, \\
 {\cal A}_{i,i}^{\rm UV,\, double} \equiv& 
 \delta_{J\,\bar{J}}\,\delta_{m\,-\bar{m}}\,\frac{1}{4\pi}\left( \frac{M}{3} \right)^2 \sum_{J'=0}^{2j} \sum_{J''=1}^{2j} 
n \left\{ \begin{matrix} J' & J'' & J \\ j & j & j  \end{matrix} \right\}^2
 \nn \\
& \times\Biggl[ 4 J' J'' \left(\tp^2+J^2\right)\left\{\hL(\tA(J'),\tB(J'');\tp) +\hL(\tA(J'),\tC(J'');\tp)\right\} \nn \\
&\hspace{6mm} + \left\{4J'J'' \tp^2 +\frac{2J'}{J''} \left[ \left(J'^2-J''^2\right)^2 - 2J^2 J'^2 + J^4 \right]\right\}\nn \\
&\hspace{9mm} \times\left\{-2\hL(\tA(J'),\tD(J'');\tp)+\hL(\tA(J'),\tE(J'');\tp)+\hL(\tA(J'),\tF(J'');\tp)\right\}  
\Biggr], 
\label{Ptotal_UV}
\end{align}
and the IR part is given by the summation of 
(\ref{Pii4CC-IR}), (\ref{Pii3CC-IR}), (\ref{Pii4CD-IR})--(\ref{Pii3FF-IR}): 
\begin{align}
\cA_{i,i}^{\rm IR}
\equiv&\, \delta_{J\,\bar{J}}\,\delta_{m\,-\bar{m}}\,\frac{1}{4\pi} \left(\frac{M}{3}\right)^2 
\frac{\tp^2-\tA(J)}{\tp^2+\tA(J)}
\ln \delta + \cdots. 
\label{Ptotal_IR}
\end{align}
Note that the UV divergences containing $\ln \tLambda_p$ are canceled in the sums and do not appear 
in either (\ref{Ptotal_UV}) or (\ref{Ptotal_IR}), 
which is expected from the supersymmetry and supports the softness of the mass $M$. 
We easily see that
(\ref{Ptotal_UV_single}) and (\ref{Ptotal_IR}) 
vanish in the Moyal limit (\ref{continuum limit}) 
(followed by $\delta \to 0$ for the IR part) : 
\begin{equation}
\cA_{i,i}^{\rm UV,\,single} \to 0, \quad 
\cA_{i,i}^{\rm IR} \to 0,  
\label{PIR}
\end{equation}
because 
(\ref{Ptotal_UV_single}) behaves as $\cO\left(\frac{1}{n}\,\ln n\right) \to 0$ as $n=2j+1 \to \infty$.
Therefore only (\ref{Ptotal_UV}) has a sensible contribution in the Moyal limit. 
In the following, we separately evaluate the contributions of (\ref{Ptotal_UV}) 
from Region I and Region II: 
\begin{align}
{\cal A}_{i,i}^{\rm UV,\,double} = 
{\cal A}_{i,i}^{\rm UV,\,double,\,I} + 
{\cal A}_{i,i}^{\rm UV,\,double,\,II}. 
\end{align}

\subsubsection{${\cal A}_{i,i}^{\rm UV,\,double,\,I}$}
\label{subsubsec:planar RI}

In terms of the rescaled variables 
\begin{equation}
 u=\frac{M}{3}\,J, \qquad 
 u'=\frac{M}{3}\,J', \qquad 
 u''=\frac{M}{3}\,J'',  
\end{equation}
the $6j$ symbol can be approximated as (\ref{WRI}).
Also,
\begin{align}
\tA(J') \simeq  \left( \frac{M}{3} \right)^2 (u')^2, \quad 
\tB(J'') \simeq  \tC(J'') \simeq \tD(J'') \simeq \tE(J'') \simeq \tF(J'') 
\simeq  \left( \frac{M}{3} \right)^2 (u'')^2, 
\end{align}
and all the $\hL$ functions appearing in (\ref{Ptotal_UV}) have the same leading-order behavior: 
\begin{align}
\hL(\tP(J'),\tQ(J'');\tp^2) & \simeq  \left( \frac{M}{3} \right)^2
\frac{1}{\sqrt{\left[p^2 + (u'+u'')^2 \right]  \left[p^2 + (u'-u'')^2 \right]}} \nn \\
&\times \ln\left(
\frac{ p^2 + (u')^2 + (u'')^2 + \sqrt{\left[p^2 + (u'+u'')^2 \right]  \left[p^2 + (u'-u'')^2 \right]} }
{ p^2 + (u')^2 + (u'')^2 - \sqrt{\left[p^2 + (u'+u'')^2 \right]  \left[p^2 + (u'-u'')^2 \right]} }
\right). 
\end{align}
Recall that the external momentum $|p|$ is assumed to be the same order as $u$. 
In this region, 
the summation can be approximated by the integral 
\begin{equation}
 \left( \frac{M}{3} \right)^2 \sum_{J',J''} \simeq \frac{1}{2}   
\int \int _{u\le u'+u''\le u_B,\, -u \le u'-u'' \le u}
du' du'', 
\end{equation}
where \footnote{Recall that $1/2<\alpha<1$. } 
\begin{equation}
u_B=\frac{M}{3} J_B =\cO\left((1/M)^{2\alpha-1}\right)\gg 1 
\end{equation}
and the prefactor $\frac{1}{2}$ reflects  
the fact that only the cases of $J'+J''+J$ being even contribute to the summation. 

Then the double sum part in Region I can be expressed as 
\begin{align}
{\cal A}_{i,i}^{\rm UV,\, double,\,I} \simeq &\delta_{J\,\bar{J}}\,\delta_{m\,-\bar{m}}\,
\frac{2}{\pi^2} (p^2+u^2)
\int \int _{u\le u'+u''\le u_B, \,-u \le u'-u'' \le u}
du' du'' \nn \\
&\times \frac{u'u''}{\sqrt{
\left[p^2 + (u'+u'')^2 \right]  \left[p^2 + (u'-u'')^2 \right]
\left[ (u'+u'')^2 - u^2 \right]  \left[  u^2 - (u'-u'')^2  \right]
}} \nn \\
&\times \ln\left(
\frac{ p^2 + (u')^2 + (u'')^2 + \sqrt{\left[p^2 + (u'+u'')^2 \right]  \left[p^2 + (u'-u'')^2 \right]} }
{ p^2 + (u')^2 + (u'')^2 - \sqrt{\left[p^2 + (u'+u'')^2 \right]  \left[p^2 + (u'-u'')^2 \right]} }
\right) \nn \\
=&\delta_{J\,\bar{J}}\,\delta_{m\,-\bar{m}}\, \frac{1}{\pi^2} (p^2+u^2) \int_1^{\tLambda} dz \int_a^1 dw \,f(z,w), 
\label{pre-region1}
\end{align}
where the integration variables have been changed from $(u',u'')$ to $(z,w)$ by 
\begin{equation}
u'+ u''=\sqrt{ (p^2+u^2) z^2 - p^2 }, \qquad 
|u'- u''|=\sqrt{ (p^2+u^2) w^2 - p^2 },
\end{equation}
and we have defined 
\bea
& & \tLambda\equiv \sqrt{ \frac{p^2+u_B^2}{p^2+u^2} }\simeq \frac{u_B}{\sqrt{p^2+u^2}}, \qquad 
a\equiv \sqrt{ \frac{p^2}{p^2+u^2} }, \\
& & f(z,w)\equiv \frac{z^2-w^2}{\sqrt{(z^2-1)(z^2-a^2)(1-w^2)(w^2-a^2)}} 
\ln\left( \frac{z+w}{z-w} \right). 
\eea

The leading contribution of the integration (\ref{pre-region1}) 
comes from the region $z\sim \infty$ where the integrand behaves as
\begin{equation}
f(z,w) \sim \frac{2w}{\sqrt{(1-w^2)(w^2-a^2)}} \,\frac{1}{z} \equiv f_0(z,w), 
\end{equation}
which gives a singular behavior upon the integration: 
\begin{align}
\int_1^{\tLambda} dz \int_a^1 dw\, f_0(z,w) 
= \int_1^{\tLambda} dz \,\frac{\pi}{z} = \pi\ln\tLambda  
\simeq  \pi \ln \left( \frac{u_B}{ \sqrt{ p^2 + u^2 }} \right). 
\end{align}
For the quantity subtracted by the singular part, 
\begin{equation}
I(a) \equiv \lim_{\tLambda\to \infty}\int_1^{\tLambda} dz \int_a^1 dw \left[ f(z,w) - f_0(z,w) \right], 
\end{equation}
we have analytically computed both $I(0)$ and $\lim_{a\to 1-0}I(a)$ to provide the identical result $\pi \left( - \ln 2 + 1 \right)$. 
Furthermore, numerical computations for general $0<a<1$ (from, e.g., Mathematica) 
strongly suggest that $I(a)$ is indeed a constant independent of $a$: 
\begin{equation}
I(a) = \pi \left( - \ln 2 + 1 \right) \qquad \mbox{for}\qquad 0\leq a<1. 
\label{I(a)}
\end{equation}
We proceed assuming that this is correct.

Combining the above results, we eventually have 
\begin{align}
{\cal A}_{i,i}^{\rm UV,\,double,\,I} &\simeq 
\delta_{J\,\bar{J}}\,\delta_{m\,-\bar{m}}\,\frac{1}{\pi} \left( p^2+u^2 \right) 
\left(
\ln u_B - \frac{1}{2} \ln\left(p^2+u^2\right) - \ln 2 + 1 
\right)
\label{PUV-RI}
\end{align}
in the Moyal limit. 

\subsubsection{${\cal A}_{i,i}^{\rm UV,\,double,\,II}$}
\label{subsubsec:planar RII}

In Region II, $J'$ and $J''$ satisfy 
\begin{equation}
 J',\, J'' \ge \frac{J_B - J}{2}, \quad 
 J\ge |\Delta |. \qquad (\Delta \equiv J'-J'')
\end{equation}
Recalling that $J_B =\cO\left(j^\alpha\right)$ and $J=\cO\left(j^{1/2}\right)$, we see  
\begin{equation}
j, J', J'' \gg J, |\Delta| \qquad \mbox{for}\qquad  j \gg 1. 
\end{equation}
In this region, 
the summation of $J'$ and $J''$ cannot be evaluated by integrals 
as we have done in Region I. 
As seen in appendix~\ref{app:proof R2}, 
the $6j$ symbol $\Wsj$ in this region 
can be well approximated by using Edmonds' formula (\ref{Edmonds}), 
which leads to  
\begin{equation}
\begin{split}
  \Wsj^2 & \simeq \frac{1}{n} \frac{1}{2J''+1}\,\left( d_{0\Delta}^J(\beta) \right)^2 , \\
\cos\beta  = \frac{1}{2} \sqrt{ \frac{J''(J''+1)}{j(j+1)}} 
&= \frac{J_+}{4j}\left[1+\cO\left(\frac{\Delta}{J_+}, \frac{1}{J''}\right)\right] 
\quad \mbox{for} \quad 0<\beta<\frac{\pi}{2}, 
\label{cosbeta2}  
\end{split}
\end{equation}
where $J_+\equiv J'+J''$ and 
$d_{M\,M'}^J(\beta)$ is a real function related to the Wigner $D$-function 
(see (\ref{d-function})). 
Since $p^2$ is assumed to be of the same order as $u^2=\cO(M^0)$, 
we can also see that all the functions 
$\hL$ appearing in $\cA_{i,i}^{\rm UV,\,double,\,II}$ commonly behave as 
\be
\hL(\tP(J'),\tQ(J'');\tp^2)\simeq \frac{4}{J_+^2} \left( 1 + {\cal O}\left(\frac{\Delta}{J_+}\right) \right).   
\ee
Together with the above approximations, 
it turns out that (\ref{Ptotal_UV}) in Region II takes a simple form: 
\be
\cA_{i,i}^{\rm UV,\,double,\,II}\simeq \delta_{J\,\bar{J}}\,\delta_{m\,-\bar{m}}\,\frac{2}{\pi}(p^2+u^2+\cO(M))
\sum_{\Delta=-J}^J\sum_{J_B\leq J_+\leq 4j}^{\rm (2-step)}\frac{\left( d_{0\Delta}^J(\beta) \right)^2}{J_+}\,
\left( 1 + {\cal O}\left(\frac{\Delta}{J_+}\right) \right).
\label{UV region II pre}
\ee
Note that no $\Delta$-dependence remains in the leading term except for the $d$-function.

The variable $J_+$ in the sum runs by two steps as signified, 
because $J_+$ must take even (odd) integers for a fixed $\Delta$ being even (odd). 
Thus the summation separates into those over 
even integers for both $\Delta$ and $J_+$ 
and 
odd integers for both $\Delta$ and $J_+$.  
Let us consider replacing the latter summation over odd $\Delta$ and odd $J_+$ 
with the summation over odd $\Delta$ and even $J_+$ 
by increasing or reducing the value of $J_+$ by one. 
Since the error induced by this replacement 
is of the order $\cO\left(J_+^{-2}\right)$, we can rewrite the summation 
in (\ref{UV region II pre}) as 
\be
\sum_{\Delta=-J}^J\sum_{J_B\leq J_+\leq 4j}^{\rm (2-step)}\frac{\left( d_{0\Delta}^J(\beta) \right)^2}{J_+}\,
\left( 1 + {\cal O}\left(\frac{\Delta}{J_+}\right) \right) 
= \sum_{J_B\leq J_+\leq 4j}^{(J_+:\,{\rm even})}\frac{1}{J_+}
\sum_{\Delta=-J} ^J\left(d^J_{0\Delta}(\beta)\right)^2 +\cR(J), 
\ee
where 
\be
\cR(J)\equiv \sum_{J_B\leq J_+\leq 4j}^{(J_+:\,{\rm even})}\cO\left(J_+^{-2}\right) 
\sum_{-J\leq \Delta \leq J}^{(\Delta:\,{\rm odd})} \left(d^J_{0\Delta}(\beta)\right)^2.
\label{cR}
\ee
By using the identity $\sum_{\Delta=-J}^J\left(d^J_{0\Delta}(\beta)\right)^2=1$ (see (\ref{d2formula1})), 
we see that $\cR(J)$ does not contribute in the Moyal limit as 
\bea
|\cR(J)| 
& \leq & \left|\sum_{J_B\leq J_+\leq 4j}^{(J_+:\,{\rm even})}\cO\left(J_+^{-2}\right) \right| 
\leq  (\mbox{const.}) \left(\frac{1}{J_B}-\frac{1}{4j}\right) \to 0.
\eea
Thus we end up with a simple sum to be evaluated as
\be
\sum_{J_B\leq J_+\leq 4j}^{(J_+:\,{\rm even})}\frac{1}{J_+} = 
\frac12\int^1_{\frac{J_B}{4j}}\frac{dX}{X}= \frac12\ln\frac{4j}{J_B} = \frac12\ln \frac{2\Lambda_j}{u_B} 
\label{planar_f}
\ee
with $\Lambda_j\equiv\frac{M}{3}\cdot 2j$, and the amplitude becomes 
\be
\cA_{i,i}^{\rm UV,\,double,\,II}\simeq \delta_{J\,\bar{J}}\,\delta_{m\,-\bar{m}}\,\frac{1}{\pi}(p^2+u^2)\ln \frac{2\Lambda_j}{u_B}. 
\label{PUV-RII}
\ee

\subsubsection{Total contribution from the planar diagrams in the Moyal limit}

Combining the results (\ref{PIR}), (\ref{PUV-RI}), and (\ref{PUV-RII}), 
we obtain the Moyal limit of the total contribution from the planar diagrams: 
\begin{align}
{\cal A}_{i,i} &= \cA_{i,i}^{\rm IR}+ \cA_{i,i}^{\rm UV} ={\cal A}_{i,i}^{\rm UV,\,double,\,I} 
+{\cal A}_{i,i}^{\rm UV,\,double,\,II} +\cdots
\nn \\
&\simeq \delta_{J\,\bar{J}}\,\delta_{m\,-\bar{m}}\,\frac{1}{\pi} (p^2+u^2)\left(\ln\Lambda_j-\frac12 \ln(p^2+u^2) + 1 \right). 
\label{Pfinal}
\end{align}
The dependence on $u_B$ cancels between the contributions from Region I and Region II as it should. 
The amplitude depends on the external momenta in the 2D plane $p$ and in the decompactified fuzzy sphere 
$u$ only through the combination $p^2+u^2$, which suggests the restoration of 4D rotational symmetry 
from $\R^2\times (\mbox{fuzzy }S^2)$ in the Moyal limit. 

\subsection{Non-planar diagrams}

In the non-planar diagrams, the UV contributions 
(\ref{NPii4CCUV}), (\ref{NPii3CC_UVpre1}), (\ref{NPii4CD-UV})--(\ref{NPii3FF-UV}) are summed as 
\begin{align}
{\cal B}_{i,i}^{\rm UV}  \equiv {\cal B}_{i,i}^{\rm UV,\, single}  
+ {\cal B}_{i,i}^{\rm UV,\, double} + \cdots  
\end{align}
with
\begin{align}
\label{NPtotal_UV_single}
{\cal B}_{i,i}^{\rm UV,\, single} 
\equiv & -\delta_{J\,\bar{J}}\,\delta_{m\,-\bar{m}}\,\frac{1}{4\pi}\left( \frac{M}{3} \right)^2 \sum_{J'=0}^{2j} \sum_{J''=1}^{2j}
(-1)^{J'+J''+J} n (2J'+1) \Wsj^2  \nn \\
& \qquad \times \Bigl\{
-\ln (J''+ 1/3) + \ln (J''+2/3) -\ln(J''+1) + \ln J''
\Bigr\},  \\
{\cal B}_{i,i}^{\rm UV,\, double} \equiv &- \delta_{J\,\bar{J}}\,\delta_{m\,-\bar{m}}\,\frac{1}{4\pi}\left( \frac{M}{3} \right)^2 \sum_{J'=0}^{2j} \sum_{J''=1}^{2j} 
(-1)^{J'+J''+J}\,
n \left\{ \begin{matrix} J' & J'' & J \\ j & j & j  \end{matrix} \right\}^2  \nn \\
&\hspace{3mm} \times\Biggl[ 4 J' J'' \left(\tp^2+J^2\right)\left\{\hL(\tA(J'),\tB(J'');\tp) +\hL(\tA(J'),\tC(J'');\tp)\right\} \nn \\
&\hspace{9mm} + \left\{4J'J'' \tp^2 +\frac{2J'}{J''} \left[ \left(J'^2-J''^2\right)^2 - 2J^2 J'^2 + J^4 \right]\right\}\nn \\
&\hspace{12mm} \times\left\{-2\hL(\tA(J'),\tD(J'');\tp)+\hL(\tA(J'),\tE(J'');\tp)+\hL(\tA(J'),\tF(J'');\tp)\right\}  
\Biggr], 
\label{NPtotal_UV}
\end{align}
and the IR contribution is given by the summation of 
(\ref{NPii4CC-IR}), (\ref{NPii3CC-IR}), (\ref{NPii4CD-IR})--(\ref{NPii3FF-IR}): 
\begin{align}
\cB_{i,i}^{\rm IR}
\equiv& -\delta_{J\,\bar{J}}\,\delta_{m\,-\bar{m}}\,\frac{1}{4\pi} \left(\frac{M}{3}\right)^2 
\frac{\tp^2-\tA(J)}{\tp^2+\tA(J)}
 \ln \delta + \cdots. 
\label{NPtotal_IR}
\end{align}
We see that cancellation of UV singular terms of $\ln\tLambda_p$ also occurs in the non-planar contributions.  

Similarly to the planar case, (\ref{NPtotal_UV_single}) and
(\ref{NPtotal_IR}) vanish in the Moyal limit, 
\begin{equation}
{\cal B}_{i,i}^{\rm UV,\, single} \to 0, \quad 
{\cal B}_{i,i}^{\rm IR} \to 0, 
\label{NPIR}
\end{equation}
and the remaining terms (\ref{NPtotal_UV}) are computed separately 
in Region I and Region II: 
\begin{align}
{\cal B}_{i,i}^{\rm UV,\,double} = 
{\cal B}_{i,i}^{\rm UV,\,double,\,I} + 
{\cal B}_{i,i}^{\rm UV,\,double,\,II}. 
\end{align}
We can evaluate ${\cal B}_{i,i}^{\rm UV,\,double,\,I}$ 
and ${\cal B}_{i,i}^{\rm UV,\,double,\,II}$ by using almost the same 
manipulations to derive (\ref{Pfinal}). 
The only difference from the planar counterparts 
$\cA_{i,i}^{\rm UV,\,double,\,I}$ and $\cA_{i,i}^{\rm UV,\,double,\,II}$ 
is the existence of the sign factor $-(-1)^{J'+J''+J}$ in the summation. 

In Region I, 
the $6j$ symbol is negligible unless $(-1)^{J'+J''+J}=1$ 
from (\ref{6j in RI}).
Hence we can repeat the same calculation as in the planar case (section \ref{subsubsec:planar RI}) 
to evaluate ${\cal B}_{i,i}^{\rm UV,\,double,\,I}$, 
which leads to 
\begin{equation}
{\cal B}_{i,i}^{\rm UV, \,double,\,I} \simeq
-\delta_{J\,\bar{J}}\,\delta_{m\,-\bar{m}}\,\frac{1}{\pi} \left( p^2+u^2 \right) 
\left(
\ln u_B - \frac{1}{2} \ln\left(p^2+u^2\right) - \ln 2 + 1 
\right). 
\label{NPUV-R1}
\end{equation}

In Region II, the expression reduces to 
\be
{\cal B}_{i,i}^{\rm UV, \,double,\,II} \simeq -\delta_{J\,\bar{J}}\,\delta_{m\,-\bar{m}}\,(-1)^J\frac{2}{\pi}(p^2+u^2+\cO(M))
\sum_{J_B\leq J_+\leq 4j}^{(J_+:\,{\rm even})}\frac{1}{J_+}
\sum_{\Delta =-J}^J (-1)^\Delta\left(d^J_{0\Delta}(\beta)\right)^2. 
\label{cB-RII}
\ee
Combining (\ref{formula-d1}) and the formula~\footnote{See, e.g., eq. (7) in Chapter 4.7.2 of \cite{QTAM}.} 
\be
\sum_{M''=-J}^J d_{MM''}^J(\beta_1)\,d^J_{M''M'}(\beta_2)=d^J_{MM'}(\beta_1+\beta_2) 
\qquad
\mbox{for}\qquad \beta_1+\beta_2\leq \pi,
\label{formula-d3}
\ee
we obtain 
\be
\sum_{\Delta=-J}^J(-1)^\Delta\left(d^J_{0\Delta}(\beta)\right)^2=d^J_{00}(2\beta). 
\label{d2formula2}
\ee
Equation (\ref{d-function}) gives the explicit form of the function 
$d^J_{00}(2\beta)$ as 
\begin{align}
d^J_{00}(2\beta) & =  \frac{(-1)^J}{2^J \,J!}\frac{d^J}{dY^J}\left\{(1-Y)^J(1+Y)^J\right\} \nn \\
& =  \frac{(-1)^J}{2^J}\sum_{r=0}^J(-1)^r\begin{pmatrix}J \\ r\end{pmatrix}^2\, (1+Y)^r(1-Y)^{J-r} \nn \\
&= (-1)^J\left[1+\sum_{r=1}^J(-1)^r\begin{pmatrix}J \\ r\end{pmatrix}X^{2r} 
+\sum_{r=1}^J(-1)^r\begin{pmatrix}J \\ r\end{pmatrix}^2\, X^{2r}(1-X^2)^{J-r}\right], 
\end{align}
where  $Y\equiv \cos(2\beta)$ and $X\equiv \cos\beta$. 
On the other hand, the summation $\sum_{J_B\leq J_+\leq 4j}^{(J_+:\,{\rm even})}\frac{d^J_{00}(2\beta)}{J_+}$ can be converted into an integral as 
\be
\sum_{J_B\leq J_+\leq 4j}^{(J_+:\,{\rm even})}\frac{d^J_{00}(2\beta)}{J_+}
\to \frac12\int^1_{\frac{J_B}{4j}}\frac{dX}{X}\,d^J_{00}(2\beta). 
\ee
Then we can evaluate the summation in (\ref{cB-RII}) as 
\be
-(-1)^J\sum_{J_B\leq J_+\leq 4j}^{(J_+:\,{\rm even})}\frac{d^J_{00}(2\beta)}{J_+} 
 =  -\frac12\ln \frac{2\Lambda_j}{u_B} +\frac12\,H_J,  
\label{cBUV-RII}
\ee
where $H_J$ denotes the harmonic number 
\be
H_J\equiv \sum_{n=1}^J\frac{1}{n}=\gamma+\psi(J+1)=-\sum_{r=1}^J\frac{(-1)^r}{r}\begin{pmatrix}J \\ r\end{pmatrix} 
\label{H_J}
\ee
and $\gamma$ is the Euler constant.  
In particular, $H_J$ is evaluated as 
\be
H_J\simeq \ln\left(\frac{3}{M}u\right)+\gamma +\cO(M) \label{H_J_asymp}
\ee
for small $M$. 
Plugging these results into (\ref{cB-RII}), we obtain the non-planar contribution in Region II: 
\be
{\cal B}_{i,i}^{\rm UV, \,double,\,II} \simeq -\delta_{J\,\bar{J}}\,\delta_{m\,-\bar{m}}\,\frac{1}{\pi}(p^2+u^2)
\left(\ln \frac{2\Lambda_j}{u_B}-\ln\left(\frac{3}{M}u\right)-\gamma \right).
\label{NPUV-R2}
\ee

Combining (\ref{NPIR}), (\ref{NPUV-R1}) and (\ref{NPUV-R2}), 
the Moyal limit of the total contribution from the non-planar diagrams becomes
\begin{align}
{\cal B}_{i,i} & = \cB_{i,i}^{\rm IR}+\cB_{i,i}^{\rm UV} ={\cal B}_{i,i}^{\rm UV,\,double,\,I} + 
{\cal B}_{i,i}^{\rm UV,\,double,\,II} +\cdots \nn \\
& \simeq -\delta_{J\,\bar{J}}\,\delta_{m\,-\bar{m}}\,\frac{1}{\pi} (p^2+u^2) \left(\ln \Lambda_j -\frac12\ln (p^2+u^2) + 1 -\ln\left(\frac{3}{M}u\right)-\gamma
\right). 
\label{NPfinal}
\end{align}
The last two terms $-\ln\left(\frac{3}{M}u\right)-\gamma$ in the non-planar amplitude have no counterpart in the planar 
contribution (\ref{Pfinal}). They arise from the asymptotic behavior of the harmonic number (\ref{H_J_asymp}) that 
has been recognized as a ``noncommutative anomaly'' in scalar field theory on fuzzy $S^2$~\cite{Chu:2001xi}. 
Although this anomaly is finite in the theory on the fuzzy $S^2$ 
since the external angular momentum $J$ is finite,  
it becomes singular in the Moyal limit; i.e., the large radius limit of 
the fuzzy sphere to the Moyal plane $\R^2_\Theta$. 
Actually, in (\ref{NPfinal}), $3/M$ is nothing 
but the radius of the fuzzy $S^2$ which diverges 
in sending $M\to 0$ with fixed $u$. 
Note that the terms are expressed as 
$-\ln\left(\sqrt{\Theta}\,u\right)-\frac12 \ln\frac{n}{2}-\gamma$, 
in which the first term signifies the UV/IR mixing 
phenomenon~\cite{Minwalla:1999px}. 
Due to their $u$-dependence, the noncommutative anomaly in the non-planar amplitude prevents restoration of 
the 4D rotational symmetry, 
which makes a contrast to the planar case.

\subsection{Moyal limit of the modes}

In order to obtain the one-loop effective action 
in the Moyal limit, we have to know the concrete form of mapping from 
the modes $x_{i,\,J\,m}(p)$ in the expansion by the fuzzy spherical harmonics (\ref{expand scalar}) 
to the modes $x_i(p,p')$ expanded by plane waves on the Moyal plane, 
\begin{equation}
X_i(x) = \int \frac{d^2p}{(2\pi)^2} e^{ip\cdot x} 
\int \frac{d^2 p'}{(2\pi)^2} e^{ip'\cdot \hat{x}} \otimes x_i(p,p'),
\end{equation}
where $\hat{x}=(\hat\xi, \hat\eta)$ 
are the coordinates of the Moyal plane ($\R^2_\Theta$) satisfying 
\begin{equation}
  [ \hat\xi, \hat\eta ] = i\Theta, 
\end{equation}
and $p'\cdot\hat{x}=p_1'\hat{\xi}+p_2'\hat{\eta}$. 
When $n(=2j+1)$ is large, $x_{i,\,J\,m}(p)$ can be expressed by $x_i(p,p')$: 
\begin{equation}
x_{i,\,J\,m}(p)=\int \frac{d^2p'}{(2\pi)^2} 
\frac{1}{n}\tr_n\left( \hat{Y}^{(jj)\dagger}_{Jm} e^{ip'\cdot \hat{x}}\right)
x_i(p,p'). 
\label{xp_to_xJ}
\end{equation}

Since we are eventually interested in the commutative limit $\Theta\to 0$, 
let us first consider the Moyal limit with $\Theta$ being small and $j\gg 1$. 
According to (\ref{FSH_Jsmall}) and (\ref{FSH_small_Theta}) in appendix~\ref{app:FSH flat limit}, 
the fuzzy spherical harmonics $\hat{Y}^{(jj)}_{J\,m}$ can be approximated as
\begin{align}
\hat{Y}_{J\,m}^{(jj)} \simeq 
\begin{cases}
\displaystyle
\delta_{m\,0} \sqrt{2J+1} \id_n  & {\rm for}\ J\le J_\ve , \\
\displaystyle
2\pi(-i)^m  \sqrt{2J}
\int \frac{d^2 p'}{(2\pi)^2} 
\frac{e^{im\varphi_{p'}} }{|p'|} \delta(|p'|-u) 
e^{-ip'\cdot \hat{x}} & {\rm for}\ J>J_\ve
\end{cases} 
\label{FSH approx}
\end{align}
with $J_\ve$ being an integer of the order of $\cO(M^{-\ve})$ $(0<\ve\ll 1)$. 
$\varphi_{p'}$ is a phase of the complex combination of the momentum: 
$p'_1+ip'_2=|p'|e^{i\varphi_{p'}}$. 
Plugging (\ref{FSH approx}) into (\ref{xp_to_xJ}) leads to 
\begin{align}
x_{i,\,J\,m}(p)=
\begin{cases}
\displaystyle
\vspace{1mm}
\frac{1}{(2\pi\Theta) n} \sqrt{2J+1}\, \delta_{m\,0}\, x_i(p,0) 
&{\rm for}\ J\le J_\ve, \\
\displaystyle
\frac{i^{m} \sqrt{2J} }{(2\pi \Theta) n}\, 
\int^{2\pi}_0 \frac{d \varphi_{p'}}{2\pi} e^{im\varphi_{p'}} x_i(p,ue^{i\varphi_{p'}})
&{\rm for}\ J > J_\ve ,
\end{cases}
\label{moyal modes}
\end{align}
where 
we have used
\begin{equation}
\tr_n\left(e^{ip\cdot \hat{x}} e^{iq\cdot \hat{x}}\right) = \frac{2\pi}{\Theta}
\delta^2(p+q) \quad \mbox{for}\quad n\sim\infty, 
\end{equation}
and the second argument of $x_i(p,ue^{i\varphi_{p'}})$ specifies the momentum in $\R_\Theta^2$ in the form of 
the complex combination.
In addition, we divide the summation over $J$ into two parts: 
\begin{equation}
\sum_{J=0}^{2j} = \sum_{J=0}^{J_\ve} + \sum_{J=J_\ve+1}^{2j}, 
\end{equation}
and transcribe the latter summation into the integral as
\begin{equation}
 \sum_{J=J_\ve+1}^{2j} \sum_{m=-J}^J \to \frac{3}{M} \int_0^\infty du \sum_{m\in \Z}. 
\end{equation}

\subsection{Scalar kinetic terms in the one-loop effective action}
\label{sec:Seff_scalar}
Let us first consider rewriting the tree-level kinetic terms of the scalar fields in terms of the modes $x_i(p,p')$: 
\begin{align}
S_{2,\,B,\,x_i}^{\rm tree}&=\frac{n}{g_{2d}^2} \int \frac{d^2 p}{(2\pi)^2} 
\sum_{J=0}^{2j} \sum_{m=-J}^{J} (-1)^m
\left( p^2 + \frac{2M^2}{81}+ \frac{M^2}{9}J(J+1) \right)\nn \\
&\hspace{42mm}\times \tr_k \left[x_{i,\,J\,-m}(-p) \,x_{i,\,J\,m}(p)\right].  
\label{S2Bx}
\end{align}
In the Moyal limit, 
we replace the modes with (\ref{moyal modes}) and take the limit 
of $M\to 0$ and $n\to \infty$ with fixing $\Theta=\frac{18}{M^2n}$. 
It turns out that the contribution from $0\le J \le J_\ve$ disappears,   
and the result reads 
\begin{align}
S_{2,\,B,\,x_i}^{\rm tree} &\to 
\frac{1}{g_{4d}^2} \int \frac{d^4 \bfp}{(2\pi)^4}  \,
 \tr_k\left(  {x}_i(-\bfp) {x}_i(\bfp)  \right) \,\bfp^2
 \label{tree S}
\end{align}
with the 4D coupling $g_{4d}^2\equiv 2\pi \Theta g_{2d}^2$ and four-momentum $\bfp\equiv(p,p')$. 
We have also used $x_i(\bfp)=x_i(p,p')$, $\bfp^2=p^2+u^2$ and 
\begin{equation}
\int \frac{d^4 {\mathbf p}}{(2\pi)^4} =
\int \frac{d^2 {p}}{(2\pi)^2} \int_0^\infty du \,u \int^{2\pi}_0 \frac{d \varphi_p'}{(2\pi)^2}. 
\end{equation}
This reproduces the tree-level kinetic terms of 4D scalar fields.

We repeat the same procedure for 
the one-loop part of the effective action for the operators 
$\tr_k(x_{i,\,J\,m}(-p)x_{i,\,J\,-m}(p))$   
and 
$\tr_k(x_{i,\,J\,m}(-p))\,\tr_k(x_{i,\,J\,-m}(p))$ 
which is nothing but the negative of 
the scalar two-point function (\ref{2pt func}) 
with (\ref{Pfinal}) and (\ref{NPfinal}). 
Since all the fields were rescaled as (\ref{field_rescale})
in the perturbative calculation, we rescale them back to the original expression. 
Then, the result in the Moyal limit becomes
\begin{align}
S_{2,\,B,\,x_i}^{\rm 1-loop} 
\to &
\int\frac{d^4\bfp}{(2\pi)^4}\,\tr_k\left[x_i(-\bfp)x_i(\bfp)\right]\,\bfp^2\times \frac{k}{4\pi^2}
\left\{\ln \Lambda_j -\frac12\ln(\bfp^2)+1\right\} \nn \\
&  +\int\frac{d^4\bfp}{(2\pi)^4}\,\tr_k\left(x_i(-\bfp)\right)\tr_k\left(x_i(\bfp)\right)\,\bfp^2\nn \\
&  \hspace{21mm}\times 
\frac{-1}{4\pi^2}\left\{\ln \Lambda_j -\frac12\ln(\bfp^2)+1-\ln\left(\frac{3}{M}u\right)-\gamma\right\}.
\label{1-loop Seff}
\end{align}

We decompose the modes to the $SU(k)$ part and the overall $U(1)$ part: 
\be
x_i(\bfp)=x^{SU(k)}_i(\bfp)+x^{U(1)}_i(\bfp) 
\quad \mbox{with} \quad \tr_k\left( x^{SU(k)}_i(\bfp)\right)=0, \quad x^{U(1)}_i(\bfp) \propto \id_k, 
\ee
and express the effective action up to the one-loop order ((\ref{tree S}) $+$ ({\ref{1-loop Seff})): 
\bea
\Gamma_{2,\,B,\,x_i}^{\rm 1-loop}& \equiv & S_{2,\,B,\,x_i}^{\rm tree} + S_{2,\,B,\,x_i}^{\rm 1-loop} \nn \\
& = & \frac{1}{g_{4d}^2}\int\frac{d^4\bfp}{(2\pi)^4}\,\frac{1}{k}\,\tr_k\left(x^{U(1)}_i(-\bfp)\right)
\tr_k\left(x^{U(1)}_i(\bfp)\right)\,\bfp^2\nn \\
& & \hspace{28mm}\times \left[1+\frac{g_{4d}^2k}{4\pi^2}\left\{\ln\left(\frac{3}{M}u\right)+\gamma\right\}\right] \nn \\
& & +\frac{1}{g_{4d}^2}\int\frac{d^4\bfp}{(2\pi)^4}\,
\tr_k\left[x^{SU(k)}_i(-\bfp)x^{SU(k)}_i(\bfp)\right] \,\bfp^2 \nn \\
& & \hspace{28mm}\times \left[1+\frac{g_{4d}^2k}{4\pi^2}\left\{\ln \Lambda_j -\frac12\ln(\bfp^2)+1\right\}\right].
\label{Seff}
\eea

The wave function renormalization: 
\bea
x^{SU(k)\,(R)}_i(\bfp) & \equiv & \left(1+\frac{g_{4d}^2k}{4\pi^2}\ln\frac{\Lambda_j}{\mu_R}\right)^{1/2} \,
x^{SU(k)}_i(\bfp), \\
x^{U(1)\,(R)}_i(\bfp) & \equiv & \left(1+\frac{g_{4d}^2k}{4\pi^2}\ln\left(\frac{3}{M}\,e^\gamma\mu_R\right)\right)^{1/2} \,
x^{U(1)}_i(\bfp)
\label{wf-ren_U(1)}
\eea
($\mu_R$ is the renormalization point) absorbs all the divergences arising in the Moyal limit and 
recasts the effective action as 
\bea
\Gamma_{2,\,B,\,x_i}^{\rm 1-loop}& = & 
\frac{1}{g_{4d}^2}\int\frac{d^4\bfp}{(2\pi)^4}\,\frac{1}{k}\,\tr_k\left(x^{U(1)\,(R)}_i(-\bfp)\right)
\tr_k\left(x^{U(1)\,(R)}_i(\bfp)\right)\,\bfp^2 + \Delta \Gamma\nn \\
& & +\frac{1}{g_{4d}^2}\int\frac{d^4\bfp}{(2\pi)^4}\,
\tr_k\left[x^{SU(k)\,(R)}_i(-\bfp)\,x^{SU(k)\,(R)}_i(\bfp)\right] \,\bfp^2 \nn \\
& & \hspace{28mm}\times \left[1+\frac{g_{4d}^2k}{4\pi^2}\left\{-\frac12\ln\frac{\bfp^2}{\mu_R^2}+1\right\} 
+\cO(g_{4d}^4)\right],
\label{Seff2}
\eea
where
\be
\Delta \Gamma\equiv\int\frac{d^4\bfp}{(2\pi)^4}\, \tr_k\left(x^{U(1)\,(R)}_i(-\bfp)\right)
\tr_k\left(x^{U(1)\,(R)}_i(\bfp)\right)\,\frac{\bfp^2}{4\pi^2}\ln\frac{u}{\mu_R}+\cO(g_{4d}^2).
\label{Seff_Delta}
\ee
At this stage, the limit of Step 2 (commutative limit $\Theta\to 0$) can be trivially taken to give the final result 
(\ref{Seff2}) and (\ref{Seff_Delta}). 
The $U(1)$ part is not $SO(4)$ invariant due to the noncommutative anomaly in $\Delta \Gamma$ 
(recall that $u$ is the momentum in the plane obtained from the fuzzy $S^2$), 
while the anomaly is harmless 
in the $SU(k)$ part at the one-loop level. 
Beyond the one-loop order, however,
such $SO(4)$ breaking could also affect the $SU(k)$ sector in the kinetic terms. 
Here, the $U(1)$ part does not couple with the $SU(k)$ part in the quadratic terms of the effective action, which 
is the case for any quadratic term to all the orders for a group theoretical reason. 
We also expect that the interaction terms receive no radiative corrections except those absorbed by the wave function renormalization 
which is the same as the situation in the ordinary $\cN=4$ SYM on $\R^4$~\cite{Mandelstam:1982cb,Brink:1982wv,Howe:1983sr}. 
For $n$-point amplitudes with $n\geq 3$, the UV divergence will be at most logarithmic from the power counting and parity invariance 
in the Moyal limit. 
The leading divergence would be canceled by supersymmetry as seen in the two-point amplitudes, and 
the result would be UV finite. In such UV-finite amplitudes, there is no obstruction in the commutative limit 
on the convergence to the corresponding results in the ordinary $\cN=4$ SYM. 
%
As in the ordinary $\cN=4$ SYM~\cite{Mandelstam:1982cb}, 
radiative corrections to the quadratic terms in the effective action would be gauge-dependent,  
and thus the noncommutative anomaly that appears accompanied by the wave function renormalization 
would be a gauge artifact not affecting gauge invariant observables.  
This is supported by the analysis of the 4D $\cN=4$ SYM theory in 
$\R^2\times \R^2_\Theta$ in the light-cone gauge~\cite{Hanada:2014ima}, 
which shows that the limit $\Theta\to 0$ is continuous to the ordinary theory defined on $\R^4$ 
to all the orders in perturbation theory; namely, the noncommutative anomaly does not appear.

\section{Conclusion and discussion}
\label{sec:conclusion}
Starting with the mass deformation of 2D $\cN=(8,8)$ $U(N)$ SYM ,which preserves two supercharges, 
we have obtained 4D $\cN=4$ $U(k)$ SYM on $\R^2\times$ (fuzzy $S^2$) around the fuzzy sphere 
classical solution of the 2D theory. 
The radius of the fuzzy $S^2$ is proportional to the inverse of the mass $M$
and 
the noncommutativity $\Theta$ is proportional to the inverse of the product of $n=N/k$ 
and the mass squared $M^2$. 
It is clear at the classical level that the two successive 
procedures, (1) decompactify the fuzzy $S^2$ to a noncommutative plane and 
(2) turn off the noncommutativity of the plane, derive the ordinary $\cN=4$ SYM on $\R^4$. 
As a nontrivial check at the quantum level, we have computed the one-loop effective action with respect to 
the kinetic terms for scalar fields $X_i$ ($i=3,\cdots, 7$), 
where the gauge is fixed to a Feynman-type gauge. 
The IR singularities turn out to be harmless in the above limits 
by introducing the IR cutoff $\delta$ at 
the intermediate step in the computation. 

For the $SU(k)$ sector in the gauge group $U(k)$ of the 4D theory, 
which contains only the contribution of planar diagrams,
the result coincides with the ordinary 4D SYM on $\R^4$ after the wave function renormalization. 
In particular, the $SO(4)$ rotational symmetry in $\R^4$ is not ruined by the quantum correction. 
On the other hand, the overall $U(1)$ sector including the contribution of non-planar diagrams 
has shown a ``noncommutative anomaly'', which has no counterpart in the ordinary SYM. 
Due to this anomaly, the $SO(4)$ symmetry does not appear to be restored. 
Also, such an anomaly may affect the $SU(k)$ sector beyond the one-loop order. 
However, we expect that it arises only accompanied by the wave function renormalization. 
Since the wave function renormalization is gauge dependent as in the ordinary 4D $\cN=4$ SYM, 
the anomaly is expected to be also a gauge artifact 
and will not arise in computing gauge invariant observables 
as far as the gauge symmetry is respected. 
Of course, it is desirable to calculate other kinetic terms 
and interaction terms of the one-loop effective action as well as 
higher-loop corrections in order to make the expectation firmer. 
Due to the technical complexity, we will leave this for future work. 
It will be important to confirm the harmlessness of the noncommutative anomaly 
by numerical simulations. 
    
Rigorously speaking, the gauge invariant observables should be invariant under the gauge transformation of the theory on 
$\R^2\times (\mbox{fuzzy }S^2)$ before taking the Moyal limit. 
They should be nonlocal in the fuzzy $S^2$ directions and have the angular momentum $J=0$. 
Interestingly, however, for field variables in the effective action with $J\ll j$, 
we can consider local observables with nonzero $J$ in the following reason. 
The scalar field $X_i(x)$, whose mode expansion is given by (\ref{expand scalar}), 
transforms under the gauge transformation with 
the parameter $\Omega(x)=\int\frac{d^2p}{(2\pi)^2}\,\tilde{\Omega}(p)$ and (\ref{gauge_omega}) as 
\begin{align}
\delta X_i(x) &= i[\Omega(x), X_i(x)] \nn \\
&= i\int\frac{d^2p_1}{(2\pi)^2}\int\frac{d^2p_2}{(2\pi)^2}\,e^{i(p_1+p_2)\cdot x}\sum_{(J,m)}\sum_{(J_1,m_1)}\sum_{(J_2,m_2)}
\hat{C}^{J\,m\,(jj)}_{J_1\,m_1\,(jj)\,J_2\,m_2\,(jj)} \nn \\
& \hspace{7mm}\times\hat{Y}^{(jj)}_{J\,m}\otimes\left[\omega_{J_1\,m_1}(p_1)x_{i,\,J_2\,m_2}(p_2)-
(-1)^{J_1+J_2+J}x_{i,\,J_2\,m_2}(p_2)\omega_{J_1\,m_1}(p_1)\right], 
\label{gaugetr}
\end{align}
where we have used (\ref{VC_FSH_2}) and (\ref{Chat_23_32}) with the notation (\ref{symbols_sum}). 
The expression does not vanish by taking the partial trace $\tr_k$, but does under the total trace $\Tr=\tr_n\tr_k$. 
The total trace $\Tr\,X_i(x)$ yields the observables $\tr_k\, x_{i,\,0\,0}(p)$. 
For the case of external angular momenta sufficiently smaller than the cutoff, i.e. $J_1,\,J_2\ll j$ and thus $J\ll j$, 
the contribution of $\hat{C}^{J\,m\,(jj)}_{J_1\,m_1\,(jj)\,J_2\,m_2\,(jj)}$, which includes the $6j$ symbol as in (\ref{VC_FSH}), 
is greatly suppressed when $J_1+J_2+J$ is odd, 
according to the formula (\ref{6j_RI}). This allows us to replace 
the sign factor $(-1)^{J_1+J_2+J}$ with unity. Hence, we can effectively consider the partial trace $\tr_k\,X_i(x)$ or 
equivalently $\tr_k\, x_{i,\,J\,m}(p)$ as 
gauge invariant observables~\footnote{
Nevertheless, the two-point function of $\tr_k\, x_{i,\,J\,m}(p)$ yields the $SO(4)$ breaking term $\Delta\Gamma$ 
in (\ref{Seff2}). There we used the ordinary renormalization prescription of local field theory, i.e., 
subtraction by local counter terms, which would not respect the full gauge invariance in noncommutative gauge theory. 
In fact, under the gauge transformation (\ref{gaugetr}), the local counter term corresponding to (\ref{wf-ren_U(1)}) 
varies by the amount of the product of the divergent factor ($\ln M$) and the suppression for $J_1+J_2+J$ odd, which could 
be nonvanishing. 
We expect that finite quantities free from the renormalization factors such as (\ref{observables}) 
will avoid the issue and show restoration of the $SO(4)$ symmetry.
}. By repeating a similar argument, $\tr_k\,X_i(x)^\ell$ ($\ell=1,2,\cdots$) 
can be regarded as gauge invariant observables. 

Since the overall $U(1)$ part is uninteresting in the target theory (the ordinary $\cN=4$ SYM on $\R^4$), it is better 
to consider the observables subtracted by that part. For example, with use of the field variables 
\be
X_i^{({\rm TL})}(x)\equiv \sum_{J=0}^{2j} \sum_{m=-J}^{J} \int \frac{d^2 p}{(2\pi)^2} \,e^{ip\cdot x} \ 
 \hat{Y}^{(jj)}_{J\, m} \otimes x^{({\rm TL})}_{i,\, J \, m}(p)
\ee
with
\be
 x^{({\rm TL})}_{i,\, J \, m}(p)\equiv x_{i,\, J \, m}(p)-\frac{1}{k}\left(\tr_k\,x_{i,\, J \, m}(p)\right)\id_k, 
\ee
it would be worth seeing the restoration of the $SO(4)$ symmetry at the nonperturbative level 
by numerical simulation of the quantities ($L=2,3,\cdots$):
\be
\frac{\vev{\prod_{\ell=1}^L\tr_k\left(X_{i_\ell}^{({\rm TL})}(x_\ell)^2\right)}_{\rm conn.}}{\vev{\tr_k\left(X_i^{({\rm TL})}(x)^2\right)}^{L}}. 
\label{observables}
\ee
The subscript ``conn.'' in the numerator means that the connected part of the $L$-point correlation function is taken. 
The denominator is introduced as having finite quantities independent of possible wave function renormalizations.  
The point $x$ and the index $i\in\{3,\cdots, 7\}$ in the denominator can be freely chosen as a reference.  

Finally, we comment on the nonperturbative stability of the $k$-coincident fuzzy $S^2$ solution (\ref{FS sln0}) with (\ref{FS sln}), 
which will be relevant in numerical simulation. 
As an illustration, tunneling amplitudes from the fuzzy $S^2$ solution 
\begin{enumerate}
\item
to the trivial vacuum ($X_a=0$) 
\item     
to the $(k-1)$-coincident fuzzy $S^2$ solution
\be
X_a(x)=\frac{M}{3}\,L'_a, \qquad L'_a\equiv \begin{pmatrix}
L^{(n)}_a\otimes \id_{k-1} &   \\
                        & {\bf 0}_n
\end{pmatrix} 
\label{k-1_FS}
\ee
\item
to the solutions 
\be
X_a(x)=\frac{M}{3}\,L''_a, \qquad L''_a\equiv \begin{pmatrix}
L^{(n)}_a\otimes \id_{k-2} &   &   \\
           & L^{(n+\ell)}_a &      \\
           &           & L^{(n-\ell)}_a 
\end{pmatrix} 
\qquad (\ell=1,2,\cdots, n-1) 
\label{k-2_FS}
\ee
\item
to the solutions 
\be
X_a(x)=\frac{M}{3}\,L'''_a, \qquad L'''_a\equiv \begin{pmatrix}
L^{(n)}_a\otimes \id_{k-1} &   &   \\
           & L^{(\ell)}_a &      \\
           &           & L^{(n-\ell)}_a 
\end{pmatrix} 
\qquad (\ell=1,2,\cdots, n-1) 
\label{k-3_FS}
\ee
\end{enumerate}
are evaluated in appendix~\ref{app:tunneling}. 
Although the tunneling amplitudes are expected to be suppressed due to the infinite volume of the space-time $\R^2$~\cite{Hanada:2010kt}, 
we will see this in more detail. 
Indeed, by scaling the length of the spatial direction 
in the two-dimensional space-time faster than $1/M$, 
all the results are shown to 
become zero in the successive limits (Step 1 and Step 2 in section~\ref{sec:loop}). 
This supports the nonperturbative stability of the solution (\ref{FS sln0}) with (\ref{FS sln}) in taking the successive limits.

\section*{Acknowledgements}
The authors would like to thank Masanori Hanada and Hiroshi Suzuki for collaboration during the early stages of this work, 
and Issaku Kanamori, Hidehiko Shimada and Hiroshi Suzuki for useful discussions. 
They would also like to express their gratitude to the KITP Santa Barbara, Kyushu University, University of Belgrade and 
the YITP Kyoto University, 
where various stages of this work were undertaken. 
The work of S.~M. is supported in part by  
Grant-in-Aid for Scientific Research (C), 15K05060. 
The work of F.~S. is supported in part by 
Grant-in-Aid for Scientific Research (C), 25400289. 
 
\appendix
\section{Deformed action in BTFT form}
\label{app:action}

The deformed action in the BTFT description $S=S_b+S_f$ ($S_b$ and $S_f$ denote its bosonic and fermionic parts, respectively) is explicitly 
given by
\begin{align}
S_b=&\frac{1}{g_{2d}^2}\int d^2x\ {\rm Tr}\Biggl\{
F_{12}^2 
+ \left( D_\mu B_{\bA} \right)^2 
+ \left( D_\mu X_{\ui} \right)^2 
+ \frac{1}{4} \left( D_\mu C \right)^2
+D_\mu \phi_+ D_\mu \phi_- \nn \\
&-[X_{\ui},B_{\bA}]^2 
-\left[ X_3, X_4 \right]^2 
-\left[ B_1, B_2 \right]^2 
-\left[ B_2, B_3 \right]^2 
-\left[ B_3, B_1 \right]^2 \nn \\
&+\frac{1}{4}\left[ \phi_+, \phi_- \right]^2
+\frac{1}{4}\left[ \phi_+, C \right] \left[ C, \phi_- \right]
-\frac{1}{4}\left[ C, X_{\ui} \right]^2
-\frac{1}{4}\left[ C, B_{\bA} \right]^2 \nn \\
&+\left[ \phi_+, X_{\ui} \right] \left[ X_{\ui}, \phi_- \right]
+\left[ \phi_+,B_{\bA} \right] \left[ B_{\bA}, \phi_- \right] \nn \\
&+\frac{2M^2}{81}\left(B_{\bA}^2 + X_{\ui}^2\right)
+\frac{M^2}{9}
\left(\frac{C^2}{4}+\phi_+ \phi_- \right)
-\frac{M}{2}C[\phi_+,\phi_-] \nn \\
&-\frac{4iM}{9}B_3\left(F_{12}+i[X_3,X_4]\right)
\Biggr\}, \\
S_f=&\frac{1}{g_{2d}^2}\int d^2x\ {\rm Tr}\Biggl\{
i\eta_+ D_\mu \psi_{-\mu}
+i\eta_- D_\mu \psi_{+\mu}
-\eta_+ \left[ X_{\ui}, \rho_{-\ui} \right]
-\eta_- \left[ X_{\ui}, \rho_{+\ui} \right] \nn \\
&+\psi_{+\mu} \left[ \phi_-, \psi_{+\mu} \right]
-\psi_{-\mu}\left[ \phi_+, \psi_{-\mu} \right]
+\psi_{+\mu}\left[ C, \psi_{-\mu} \right]  \nn \\
&+\rho_{+\ui}\left[ \phi_-, \rho_{+\ui} \right]
-\rho_{-\ui}\left[ \phi_+, \rho_{-\ui} \right]
+\rho_{+\ui}\left[ C, \rho_{-\ui} \right] \nn \\
&+\chi_{+\bA}\left[ \phi_-, \chi_{+\bA} \right] 
-\chi_{-\bA}\left[ \phi_+, \chi_{-\bA} \right]
+\chi_{+\bA}\left[ C, \chi_{-\bA} \right] \nn \\
&+\frac{1}{4}\eta_+\left[ \phi_-, \eta_+ \right] 
-\frac{1}{4}\eta_-\left[ \phi_+, \eta_- \right]
-\frac{1}{4}\eta_+\left[ C, \eta_- \right] \nn \\
&-\eta_+\left[ B_{\bA}, \chi_{-\bA} \right] 
-\eta_-\left[ B_{\bA}, \chi_{+\bA} \right]
-2 \epsilon_{\bA\bB\bC} \chi_{\bA}\left[ B_{\bB}, \chi_{-{\bC}} \right] \nn \\
&+2i \chi_{+1}\left( D_1 \rho_{-3} + D_2 \rho_{-4} 
            -i \left[ X_3, \psi_{-1} \right] -i\left[ X_4, \psi_{-2} \right] \right) \nn \\
&+2i \chi_{+2}\left( D_1 \rho_{-4} - D_2 \rho_{-3} 
            -i \left[ X_4, \psi_{-1} \right] +i\left[ X_4, \psi_{-2} \right] \right) \nn \\
&+2i \chi_{+3}\left( D_1 \psi_{-2} - D_2 \psi_{-1} 
            +i \left[ X_4, \rho_{-3} \right] -i\left[ X_3, \rho_{-4} \right] \right) \nn \\
&-2i \chi_{-1}\left( D_1 \rho_{+3} + D_2 \rho_{+4} 
            -i \left[ X_3, \psi_{+1} \right] -i\left[ X_4, \psi_{+2} \right] \right) \nn \\
&-2i \chi_{-2}\left( D_1 \rho_{+4} - D_2 \rho_{+3} 
            -i \left[ X_4, \psi_{+1} \right] +i\left[ X_3, \psi_{+2} \right] \right) \nn \\
&-2i \chi_{-3}\left( D_1 \psi_{+2} - D_2 \psi_{+1} 
            +i \left[ X_4, \rho_{+3} \right] -i\left[ X_3, \rho_{+4} \right] \right) \nn \\ 
&+2\psi_{+1}\left( \left[ B_1, \rho_{-3} \right] + \left[B_2, \rho_{-4} \right] + \left[ B_3, \psi_{-2} \right] \right) \nn \\
&+2\psi_{+2}\left( \left[ B_1, \rho_{-4} \right] - \left[B_2, \rho_{-3} \right] - \left[ B_3, \psi_{-1} \right] \right) \nn \\
&+2\rho_{+3}\left(- \left[ B_1, \psi_{-1} \right] + \left[B_2, \psi_{-2} \right] - \left[ B_3, \rho_{-4} \right] \right) \nn \\
&+2\rho_{+4}\left( -\left[ B_1, \psi_{-2} \right] - \left[B_2, \psi_{-1} \right] + \left[ B_3, \rho_{-3} \right] \right) \nn \\
&+\frac{2M}{3}\psi_{+\mu}\psi_{-\mu}
+\frac{2M}{9}\rho_{+\ui}\rho_{-\ui}
+\frac{4M}{9}\chi_{+A}\chi_{-A}
-\frac{M}{6}\eta_+\eta_-
\Biggr\}. 
\end{align}

\section{Fuzzy spherical harmonics}
\label{app:FSH}

In this appendix, we give definitions and properties 
of various fuzzy spherical harmonics \cite{Ishiki:2006yr,Ishii:2008ib} that are relevant in the text. 

Let $|j\,r\ket$ ($r=-j, -j+1, \cdots, j$) an orthonormal basis of $n(=2j+1)$-dimensional space of a spin-$j$ representation of $SU(2)$ 
normalized by 
\be
\bfra j\,r|j'\,r'\ket = \delta_{j\,j'} \,\delta_{r\,r'}.
\label{SU2basis_norm}
\ee
Here $j$ is assumed to take a non-negative integer or half-integer value. 
The $SU(2)$ generators $L_a$ ($a=1,2,3$) satisfying 
$[L_a, L_b ] = i\epsilon_{abc} L_c$ act on the basis as
\bea
L_3 |j\,r\ket & = & r|j\,r\ket. \nn \\
L_+ |j\,r\ket & = & \sqrt{(j-r)(j+r+1)} \,|j\,r+1\ket, \nn \\
L_- |j\,r\ket & = & \sqrt{(j+r)(j-r+1)}\, |j\,r-1\ket, 
\label{SU2 basis}
\eea
where $L_\pm = L_1\pm iL_2$. 

By expressing $\{ |j\,r\ket \}$ as $n$-dimensional unit vectors as 
\be
|j\,-j\ket = \begin{pmatrix}1 \\ 0 \\ 0 \\ \vdots \\ 0 \\ 0 \end{pmatrix},  \quad 
|j\,-j+1\ket = \begin{pmatrix}0 \\ 1 \\ 0 \\ \vdots \\ 0 \\ 0 \end{pmatrix},  \cdots, \quad
|j\, j\ket = \begin{pmatrix}0 \\ 0 \\ 0 \\ \vdots \\ 0 \\ 1 \end{pmatrix},
\label{hyouji}
\ee
any $n\times n$ matrix $M$ can be written as 
\be
M = \begin{pmatrix} M_{-j, \, -j} & M_{-j, \,-j+1} & \cdots & M_{-j,\,j} \\
                 M_{-j+1, \, -j} & M_{-j+1,\,-j+1} & \cdots & M_{-j+1, \,j} \\
                   \vdots        &     \vdots      &        & \vdots        \\
                   M_{j,\,-j}    & M_{j,\,-j+1}    & \cdots & M_{j,\, j}  \end{pmatrix} 
 = \sum_{r,r'=-j}^j M_{r,r'} |j\, r\ket\bfra j\,r'|. 
\ee                 
 
The adjoint action of $L_a$ to $M$ is defined as 
\be
L_a\circ M \equiv [L_a, M] = \sum_{r,r'}M_{r,r'}\Bigl(L_a|j\,r\ket\bfra j\,r'| -|j\,r\ket\bfra j\,r'|L_a\Bigr). 
\label{app:FSH_adj}
\ee
Then, it is easy to see that 
\be
[L_a\circ, L_b\circ] = i\epsilon_{abc} L_c\circ. 
\ee

\subsection{(Scalar) fuzzy spherical harmonics}
(Scalar) fuzzy spherical harmonics is defined by  
\be
\hat{Y}^{(jj)}_{J\,m}\equiv \sqrt{n} \sum_{r,r'=-j}^j (-1)^{-j+r'} C^{J\,m}_{j\,r \,j\,-r'} |j\,r\ket\bfra j \,r'|,
\label{app:FSH_def}
\ee
where $C^{J\,m}_{j\,r \, j \, -r'}\equiv \bfra j\,j\,\,r\,-r'|J\,m\ket$ is a Clebsch-Gordan (C-G) coefficient 
vanishing unless $m=r-r'$. 
In the basis (\ref{hyouji}), $\hat{Y}^{(jj)}_{J\,m}$ is an $n\times n$ matrix whose $(r, r')$ component is given by 
$\sqrt{n} \,(-1)^{-j+r'} C^{J\,m}_{j\,r \,j\,-r'}$. 
Note that $J$ and $m=r-r'$ take integer values as seen from the C-G coefficient. 

{}From the definition (\ref{app:FSH_adj}),
\be
L_3\circ \hat{Y}^{(jj)}_{J\,m} = m\hat{Y}^{(jj)}_{J\,m}, 
\ee
and the recursion relation for C-G coefficients~\footnote{See, e.g.,  
eq. (4) in Chapter 8.6.2 of \cite{QTAM}.},  
\bea
& & \sqrt{(c\pm\gamma)(c\mp\gamma +1)}\, C^{c\,\gamma\mp 1}_{a\,\alpha \, b \, \beta} \nn \\
& & \hspace{7mm}= \sqrt{(a\mp\alpha)(a\pm\alpha +1)}\,C^{c\,\gamma}_{a\,\alpha\pm 1\,b\,\beta} 
+\sqrt{(b\mp\beta)(b\pm\beta +1)}\,C^{c\,\gamma}_{a\,\alpha\,b\,\beta\pm 1}
\eea
leads to
\bea
L_+\circ \hat{Y}^{(jj)}_{J\,m} & = & \sqrt{(J-m)(J+m+1)}\,\hat{Y}^{(jj)}_{J\,m+1}, \nn \\ 
L_-\circ \hat{Y}^{(jj)}_{J\,m} & = & \sqrt{(J+m)(J-m+1)}\,\hat{Y}^{(jj)}_{J\,m-1} .
\eea
Therefore, we have
\be
\left(L_a\circ\right)^2\hat{Y}^{(jj)}_{J\,m} = J(J+1)\,\hat{Y}^{(jj)}_{J\,m}. 
\ee
 
$C^{J\,m}_{j\, r\, j\, -r'}$ is real, and the relation
\be
C^{J\,m}_{j\,r\,j\,-r'}= C^{J\,-m}_{j\,r'\,j\,-r}
\ee
obeys the identities \cite{QTAM}
\be
C^{J_3\,m_3}_{J_1\,m_1\,J_2\,m_2}= (-1)^{J_1+J_2-J_3}\,C^{J_3\,m_3}_{J_2\,m_2\,J_1\,m_1} 
= (-1)^{J_1+J_2-J_3}\,C^{J_3\,-m_3}_{J_1\,-m_1\,J_2\,-m_2}. 
\label{CG_id}
\ee
Then, the hermitian conjugate of $\hat{Y}^{(jj)}_{J\,m}$ becomes 
\be
\left(\hat{Y}^{(jj)}_{J\,m}\right)^\dagger = (-1)^m \,\hat{Y}^{(jj)}_{J\,-m}. 
\label{HC_FSH}
\ee 
For the $n$-dimensional trace ``$\tr_n$'', the orthonormality 
\be
\frac{1}{n}\,\tr_n\left\{\left(\hat{Y}^{(jj)}_{J\,m}\right)^\dagger \hat{Y}^{(jj)}_{J'\,m'}\right\} 
=\delta_{J\,J'}\,\delta_{m\,m'},
\label{orthon_FSH}
\ee
or equivalently 
\be
\frac{1}{n}\,\tr_n\left\{\hat{Y}^{(jj)}_{J\,m} \,\hat{Y}^{(jj)}_{J'\,m'}\right\} 
=(-1)^m \delta_{J\,J'}\,\delta_{m+m'\, 0},
\ee
follows from the orthogonality of the C-G coefficients
\be
\sum_{\alpha,\beta} \,C^{c \,\gamma}_{a\,\alpha \, b\,\beta} C^{c'\,\gamma'}_{a\,\alpha\, b\,\beta} 
= \delta_{c \,c'} \,\delta_{\gamma \,\gamma'}. 
\label{CG_ortho}
\ee

Next, let us compute the trace of the product of three fuzzy spherical harmonics given by (\ref{VC_FSH_def}) in the text, 
which is equivalent to
\be
\hat{Y}^{(jj)}_{J_1\,m_1} \hat{Y}^{(jj)}_{J_2\,m_2}
=\sum_{J=0}^{2j} \sum_{m=-J}^J \hat{C}^{J\,m\,(jj)}_{J_1\,m_1\,(jj)\,J_2\,m_2\,(jj)} \hat{Y}^{(jj)}_{J\,m}.
\label{VC_FSH_2}
\ee 
{}From the definition (\ref{app:FSH_def}) and the identity~\footnote{See, e.g., eq. (10) in Chapter 8.4.3 of \cite{QTAM}.} 
\be
C^{J_3\,m_3}_{J_1\,m_1\,J_2\,m_2} 
= (-1)^{J_2+m_2} \sqrt{\frac{2J_3+1}{2J_1+1}}\,C^{J_1\,m_1}_{J_2\,-m_2\,J_3\,m_3}, 
\label{app:FSH_CG_id}
\ee
we have
\be
\hat{C}^{J_1\,m_1\,(jj)}_{J_2\,m_2\,(jj)\,J_3\,m_3\,(jj)} 
= \sqrt{2J_2+1}\sum_{r, r',r''}C^{J_3\,m_3}_{j\,r''\,j\,r'}
C^{J_1\,m_1}_{j\,r\,j\,r'} C^{j\,r}_{j\,r''\,J_2\,m_2}. 
\label{app:FSH_Chat}
\ee
Furthermore, 
\be
\sum_{\alpha, \beta, \delta}C^{c\,\gamma}_{a\,\alpha\,b\,\beta}C^{e\,\epsilon}_{d\,\delta\,b\,\beta} 
C^{d\,\delta}_{a\,\alpha\,f\,\varphi}= (-1)^{b+c+d+f}\sqrt{(2c+1)(2d+1)}\, C^{e\,\epsilon}_{c\,\gamma\,f\,\varphi} 
\begin{Bmatrix} a & b & c \\ e & f & d \end{Bmatrix}
\ee
(See, e.g., eq.~(12) in Chapter 8.7.3 in~\cite{QTAM}) and the property of the $6j$ symbol, 
\be
\begin{Bmatrix} a & b & C \\ A & B & c \end{Bmatrix} = 
\begin{Bmatrix} A & B & C \\ a & b & c \end{Bmatrix} 
\label{6j_property1}
\ee
recast (\ref{app:FSH_Chat}) as 
\be
\hat{C}^{J_1\,m_1\,(jj)}_{J_2\,m_2\,(jj)\,J_3\,m_3\,(jj)} 
= (-1)^{J_1+2j} \sqrt{n(2J_2+1)(2J_3+1)}\,C^{J_1\,m_1}_{J_2\,m_2\,J_3\,m_3} 
\begin{Bmatrix} J_1 & J_2 & J_3 \\ j & j & j \end{Bmatrix}.
\label{VC_FSH}
\ee
The first equality of (\ref{CG_id}) leads to 
\be
\hat{C}^{J_1\,m_1\,(jj)}_{J_2\,m_2\,(jj)\,J_3\,m_3\,(jj)}=(-1)^{J_2+J_3-J_1}\,\hat{C}^{J_1\,m_1\,(jj)}_{J_3\,m_3\,(jj)\,J_2\,m_2\,(jj)}\,.
\label{Chat_23_32}
\ee
%

\subsection{Spin-$S$ fuzzy spherical harmonics}
Spin-$S$ fuzzy spherical harmonics is defined by 
\be
\hat{\cY}^{S\,n'}_{J\,m,\,\tilde{J}\,(jj)}\equiv 
\sum_{p=-\tilde{J}}^{\tilde{J}}C^{J\,m}_{\tilde{J}\,p\,S \,n'} \hat{Y}^{(jj)}_{\tilde{J}\,p} 
= C^{J\,m}_{\tilde{J}\,m-n'\,S\,n'} \hat{Y}^{(jj)}_{\tilde{J}\,m-n'}, 
\label{SpinS_FSH}
\ee
where $\hat{Y}^{(jj)}_{\tilde{J}\,m-n'}$ stands for a harmonics of 
the orbital angular momentum $(\tilde{J}, m-n')$ 
on the fuzzy $S^2$.  
Combined with a wave function with spin $S$, $\chi_{S\,n'}$, which satisfies
\be
\vec{S}^2 \chi_{S\,n'}= S(S+1) \,\chi_{S\, n'}, \qquad S_z \chi_{S\, n'}=n'\,\chi_{S\,n'}, 
\ee
$\hat{\cY}^{S\,n'}_{J\,m,\,\tilde{J}\,(jj)}\chi_{S\,n'}$ represents the irreducible representation 
of the total angular momentum $(J, m)$ 
obtained from the tensor product $(\tilde{J}, m-n')\otimes (S, n')$. 

In the text, $S_\pm, S_3$ are related to the $SU(2)_R$ generators (\ref{SU2R_gen}) as 
\be
S_{\pm}=J_{\pm\pm}, \qquad S_z=\frac12\,J_0.
\ee   
$\chi_{S\,n'}$ comes from the wave functions of the doublets 
$(\psi_{+\mu},\psi_{-\mu})$, 
$(\rho_{+\ui}, \rho_{-\ui})$, 
$(\chi_{+\bA},\chi_{-\bA})$, 
$(\eta_+,-\eta_-)$ for the $S=\frac12$ case, 
and from the wave functions of the triplet $(\frac{1}{\sqrt{2}}\phi_+, \frac12 C, \frac{1}{\sqrt{2}}\phi_-)$  for the $S=1$ case.   

\paragraph{Scalar fuzzy spherical harmonics} 
For $S=0$, (\ref{SpinS_FSH}) reduces to the scalar fuzzy spherical harmonics previously discussed: 
\be
\hat{\cY}^{00}_{J\,m, \,\tilde{J}(jj)} = C^{J\, m}_{\tilde{J}\,m\,0\,0}\hat{Y}^{(jj)}_{J\,m} 
= \delta_{J\,\tilde{J}} \hat{Y}^{(jj)}_{J\,m}. 
\ee

\paragraph{Vector fuzzy spherical harmonics}
For $S=1$, $\hat{\cY}^{1\,n'}_{J\,m,\,\tilde{J}\,(jj)}$ ($n'=1, 0,-1$) are used for the 
mode expansion of $\frac{1}{\sqrt{2}}\phi_+, \frac12 C, -\frac{1}{\sqrt{2}}\phi_-$, respectively. 

On the other hand, from (\ref{rename_scalar}) 
the following basis $\vec{\hat{Y}}^\rho_{J\,m\,(jj)}$ ($\rho=1, 0, -1$) 
is convenient to expand $\vec{Y}\equiv (X_9, X_{10}, X_8)^T$ in modes: 
\be
\vec{\hat{Y}}^{\rho=1}_{J\,m\,(jj)}=i\vec{\hat{\cY}}_{J+1\, m,\,J\,(jj)}, \qquad
\vec{\hat{Y}}^{\rho=-1}_{J\,m\,(jj)}=-i\vec{\hat{\cY}}_{J\, m,\,J+1\,(jj)}, \qquad 
\vec{\hat{Y}}^{\rho=0}_{J\,m\,(jj)} = \vec{\hat{\cY}}_{J\,m,\,J\,(jj)}
\ee
with 
\be
\vec{\hat{\cY}}_{J\, m,\,\tilde{J}\,(jj)}  = 
\begin{pmatrix} \hat{\cY}^{i=1}_{J\,m,\,\tilde{J}\,(jj)} \\ 
\hat{\cY}^{i=2}_{J\,m,\,\tilde{J}\,(jj)} \\ 
\hat{\cY}^{i=3}_{J\,m,\,\tilde{J}\,(jj)} \end{pmatrix} \\
\equiv  \frac{1}{\sqrt{2}} \begin{pmatrix} 
-\hat{\cY}^{1\,1}_{J\,m,\,\tilde{J}\,(jj)} + \hat{\cY}^{1\,-1}_{J\,m,\,\tilde{J}\,(jj)} \\
-i\hat{\cY}^{1\,1}_{J\,m,\,\tilde{J}\,(jj)} -i \hat{\cY}^{1\,-1}_{J\,m,\,\tilde{J}\,(jj)} \\
\sqrt{2} \,\hat{\cY}^{1\,0}_{J\, m,\,\tilde{J}\,(jj)} \end{pmatrix}. 
\ee
 
More explicitly, 
\bea
\vec{\hat{Y}}^{\rho=1}_{J\,m\,(jj)} & = & iV \begin{pmatrix} \hat{\cY}^{1\,1}_{J+1\,m,\,J\,(jj)} \\
 \hat{\cY}^{1\,0}_{J+1\,m,\,J\,(jj)} \\
\hat{\cY}^{1\,-1}_{J+1\,m,\,J\,(jj)} \end{pmatrix} 
= iV\begin{pmatrix} C^{J+1\,m}_{J\,m-1\,1\,1} \hat{Y}^{(jj)}_{J\,m-1} \\
C^{J+1\,m}_{J\,m\,1\,0} \hat{Y}^{(jj)}_{J\,m} \\
C^{J+1\,m}_{J\,m+1\,1\,-1} \hat{Y}^{(jj)}_{J\,m+1} \end{pmatrix}  ,
\label{vec_FSH_1}\\
\vec{\hat{Y}}^{\rho=0}_{J\,m\,(jj)} & = & V \begin{pmatrix} \hat{\cY}^{1\,1}_{J\,m,\,J\,(jj)} \\
 \hat{\cY}^{1\,0}_{J\,m,\,J\,(jj)} \\
\hat{\cY}^{1\,-1}_{J\,m,\,J\,(jj)} \end{pmatrix} 
= V\begin{pmatrix} C^{J\,m}_{J\,m-1\,1\,1} \hat{Y}^{(jj)}_{J\,m-1} \\
C^{J\,m}_{J\,m\,1\,0} \hat{Y}^{(jj)}_{J\,m} \\
C^{J\,m}_{J\,m+1\,1\,-1} \hat{Y}^{(jj)}_{J\,m+1} \end{pmatrix}  ,
\label{vec_FSH_2}\\
\vec{\hat{Y}}^{\rho=-1}_{J\,m\,(jj)} & = & -iV \begin{pmatrix} \hat{\cY}^{1\,1}_{J\,m,\,J+1\,(jj)} \\
 \hat{\cY}^{1\,0}_{J\,m,\,J+1\,(jj)} \\
\hat{\cY}^{1\,-1}_{J\,m,\,J+1\,(jj)} \end{pmatrix} 
= -iV\begin{pmatrix} C^{J\,m}_{J+1\,m-1\,1\,1} \hat{Y}^{(jj)}_{J+1\,m-1} \\
C^{J\,m}_{J+1\,m\,1\,0} \hat{Y}^{(jj)}_{J+1\,m} \\
C^{J\,m}_{J+1\,m+1\,1\,-1} \hat{Y}^{(jj)}_{J+1\,m+1} \end{pmatrix},  
\label{vec_FSH_3}
\eea 
where $V$ is a unitary matrix  
\be
V\equiv \frac{1}{\sqrt{2}} \begin{pmatrix} -1 & 0 & 1 \\ -i & 0 & -i \\ 0 & \sqrt{2} & 0 \end{pmatrix} 
\ee
with $\det V=-i$. 
 
We can see that $J, m\in \Z$ in $\vec{\hat{Y}}^\rho_{J\,m\,(jj)}$ from (\ref{vec_FSH_1})--(\ref{vec_FSH_3}). 
Also, for the $J=0$ case, $\hat{\cY}^{1\,n'}_{0\,m,\,0\,(jj)}=0$ because of $C^{0\,m}_{0\,m\,1\,n'}=0$. 
Thus, 
\be
\vec{\hat{Y}}^{\rho=0}_{0\,m\,(jj)}=0.
\label{vec_FSH_J=0}
\ee

\paragraph{Spinor fuzzy spherical harmonics}
For $S=\frac12$, the spinor fuzzy spherical harmonics 
\be
\hat{Y}^\kappa_{J\,m\,(jj)\,\alpha}\qquad \left(\kappa=1, -1, \quad \alpha=\frac12, -\frac12\right) 
\ee
is defined as 
\bea
\hat{Y}^{\kappa=1}_{J\,m\,(jj)\,\alpha} & \equiv & \hat{\cY}^{S=\frac12\,\alpha}_{J+\frac12\,m,\,J\,(jj)} 
= C^{J+\frac12\,m}_{J\,m-\alpha\,\frac12\,\alpha}\hat{Y}^{(jj)}_{J\,m-\alpha}, 
\label{sp_FSH_1}\\
\hat{Y}^{\kappa=-1}_{J\,m\,(jj)\,\alpha} & \equiv & \hat{\cY}^{S=\frac12\,\alpha}_{J\,m,\,J+\frac12\,(jj)} 
= C^{J\,m}_{J+\frac12\,m-\alpha\,\frac12\,\alpha}\hat{Y}^{(jj)}_{J+\frac12\,m-\alpha}.
\label{sp_FSH_2}
\eea
Here, $\kappa$ labels the spinor basis, and $\alpha$ labels the spinor components for each basis.      

Note that 
$m\in \Z +\frac12$ for $\hat{Y}^{\kappa=1}_{J\,m\,(jj)\,\alpha}$, because $J$ runs integers in $\hat{Y}^{(jj)}_{J\,m-\alpha}$. 
On the other hand, 
$J\in \Z +\frac12$ and  
$m\in \Z+\frac12$ for $\hat{Y}^{\kappa=-1}_{J\,m\,(jj)\,\alpha}$.

\subsection{Hermitian conjugates} 
{}From (\ref{SpinS_FSH}), the hermitian conjugate of the spin-$S$ fuzzy spherical harmonics 
$\hat{\cY}^{S\,n'}_{J\,m,\,\tilde{J}\,(jj)}$ 
reads 
\bea
\left(\hat{\cY}^{S\,n'}_{J\,m,\,\tilde{J}\,(jj)}\right)^\dagger 
& = & (-1)^{-J+\tilde{J}+S+m-n'} C^{J \,-m}_{\tilde{J}\,-m+n'\,S\,-n'} \hat{Y}^{(jj)}_{\tilde{J}\,-m+n'} \nn \\
& = & (-1)^{-J+\tilde{J}-S+m+n'} \hat{\cY}^{S\,-n'}_{J\,-m,\,\tilde{J}\,(jj)}, 
\label{HC_SpinS_FSH}
\eea
where we have used (\ref{CG_id}), (\ref{HC_FSH}) and $S-n'\in \Z$.

The vector fuzzy spherical harmonics with $\rho=1$ (\ref{vec_FSH_1}), 
\be
\hat{Y}^{\rho=1}_{J\,m\,(jj)\,i} = i\sum_{n'=-1}^{1}V_{i\,n'}\hat{\cY}^{1\,n'}_{J+1\,m,\,J\,(jj)}, 
\ee
has the hermitian conjugate as 
\be
\left(\hat{Y}^{\rho=1}_{J\,m\,(jj)\,i}\right)^\dagger = (-1)^{m+1}\,\hat{Y}^{\rho=1}_{J\,-m\,(jj)\,i},  
\ee
which can be seen from (\ref{HC_SpinS_FSH}) and the identity 
\be
V_{i\,n'}^*(-1)^{-n'}= V_{i\,-n'} .
\label{V_id}
\ee
Repeating a similar argument for (\ref{vec_FSH_2}) and (\ref{vec_FSH_3}), 
we conclude that
\be
\left(\hat{Y}^{\rho}_{J\,m\,(jj)\,i}\right)^\dagger = (-1)^{m+1}\,\hat{Y}^{\rho}_{J\,-m\,(jj)\,i}. 
\label{HC_vec_FSH}
\ee 

For the spinor fuzzy spherical harmonics (\ref{sp_FSH_1}) and (\ref{sp_FSH_2}), their hermitian conjugates 
turn out to be 
\be
\left(\hat{Y}^\kappa_{J\,m\,(jj)\,\alpha}\right)^\dagger = (-1)^{m+1+\kappa\alpha}\,
\hat{Y}^\kappa_{J\,-m\,(jj)\,-\alpha}.
\ee

\subsection{Orthonormality}
{}From (\ref{SpinS_FSH}) and (\ref{orthon_FSH}), we have 
\be
\frac{1}{n}\,\tr_n\left\{\left(\hat{\cY}^{S\,n'}_{J'\,m_1,\,\tilde{J}_1\,(jj)}\right)^\dagger 
\hat{\cY}^{S\,n'}_{J''\,m_2,\,\tilde{J}_2\,(jj)}\right\} 
= \delta_{\tilde{J}_1\,\tilde{J}_2}\delta_{m_1\,m_2}\,
C^{J'\,m_1}_{\tilde{J}_1\,m_1-n'\,S\,n'} C^{J''\,m_1}_{\tilde{J}_1\,m_1-n'\,S\,n'}.
\ee
Taking the sum over $n'$ leads to 
\be
\sum_{n'=-S}^S \frac{1}{n}\,\tr_n\left\{\left(\hat{\cY}^{S\,n'}_{J'\,m_1,\,\tilde{J}_1\,(jj)}\right)^\dagger 
\hat{\cY}^{S\,n'}_{J''\,m_2,\,\tilde{J}_2\,(jj)}\right\} 
= \delta_{J'\,J''}\,\delta_{\tilde{J}_1\,\tilde{J}_2}\delta_{m_1\,m_2},   
\label{orthon_SpinS_FSH}
\ee
where we have used~\footnote{Note that the sum over $m'$ in the middle of (\ref{app:FSH_CG_ortho}) is trivial because of the momentum conservation 
$m_1=m'+n'$. The second equality is nothing but (\ref{CG_ortho}).}   
\be
\sum_{n'=-S}^S C^{J'\,m_1}_{\tilde{J}_1\,m_1-n'\,S\,n'} C^{J''\,m_1}_{\tilde{J}_1\,m_1-n'\,S\,n'}
= \sum_{m'=-\tilde{J}_1}^{\tilde{J}_1}\sum_{n'=-S}^S
C^{J'\,m_1}_{\tilde{J}_1\,m'\,S\,n'} C^{J''\,m_1}_{\tilde{J}_1\,m'\,S\,n'}= \delta_{J'\,J''}. 
\label{app:FSH_CG_ortho}
\ee

For the vector fuzzy spherical harmonics (\ref{vec_FSH_1}), the identity 
\be
\sum_{i=1}^3V_{i\,n_1'}^*V_{i\,n_2'}=\delta_{n_1'\,n_2'}  \qquad (\mbox{$V$ is unitary}) 
\ee
and (\ref{orthon_SpinS_FSH}) imply 
\be
\sum_{i=1}^3 \frac{1}{n}\,\tr_n\left\{\left(\hat{Y}^{\rho=1}_{J'\,m_1\,(jj)\,i}\right)^\dagger 
\hat{Y}^{\rho=1}_{J''\,m_2\,(jj)\,i}\right\} = \delta_{J'\,J''}\,\delta_{m_1\,m_2}.
\ee
For cases including (\ref{vec_FSH_2}) and (\ref{vec_FSH_3}), 
similar formulas are obtained, and we conclude that 
\be
\sum_{i=1}^3 \frac{1}{n}\,\tr_n\left\{\left(\hat{Y}^{\rho_1}_{J'\,m_1\,(jj)\,i}\right)^\dagger 
\hat{Y}^{\rho_2}_{J''\,m_2\,(jj)\,i}\right\} = \delta_{\rho_1\,\rho_2}\delta_{J'\,J''}\,\delta_{m_1\,m_2}.
\label{orthon_vec_FSH}
\ee 

For the spinor fuzzy spherical harmonics (\ref{sp_FSH_1}) and (\ref{sp_FSH_2}), 
\be
\sum_{\alpha=-\frac12}^{\frac12}\frac{1}{n} \,\tr_n\left\{
\left(\hat{Y}^{\kappa_1}_{J'\,m_1\,(jj)\,\alpha}\right)^\dagger
\hat{Y}^{\kappa_2}_{J''\,m_2\,(jj)\,\alpha}\right\} = \delta_{\kappa_1\,\kappa_2}\delta_{J'\,J''}
\delta_{m_1\,m_2}
\label{orthon_sp_FSH}
\ee
holds.

\subsection{Some C-G coefficients}
\label{sec:CG}
The C-G coefficient is related to the $3j$ symbol as 
\be
C^{J\,m}_{J'\,m_1\,J''\,m_2} = (-1)^{-J'+J''-m}\sqrt{2J+1}\begin{pmatrix} J' & J'' & J \\ m_1 & m_2 & -m
\end{pmatrix}. 
\label{CG_3j}
\ee

Here we present the explicit form of some C-G coefficients that will be used later. 
\bea
C^{J+1\,m}_{J\,m-1\,1\,1} & = & \sqrt{\frac{(J+m+1)(J+m)}{2(2J+1)(J+1)}}, \label{2nd_1}\\
C^{J+1\,m}_{J\,m\,1\,0} & = & \sqrt{\frac{(J+m+1)(J-m+1)}{(2J+1)(J+1)}}, \label{2nd_2}\\
C^{J+1 \,m}_{J\,m+1\,1\,-1} & = & \sqrt{\frac{(J-m+1)(J-m)}{2(2J+1)(J+1)}}, \label{2nd_3}
\eea
\bea
C^{J\,m}_{J+1\,m-1\,1\,1} & = & \sqrt{\frac{(J-m+2)(J-m+1)}{2(2J+3)(J+1)}},  \\
C^{J\,m}_{J+1\,m\,1\,0} & = & -\sqrt{\frac{(J-m+1)(J+m+1)}{(2J+3)(J+1)}},  \\
C^{J\,m}_{J+1\,m+1\,1\,-1} & = & \sqrt{\frac{(J+m+2)(J+m+1)}{2(2J+3)(J+1)}}, 
\eea
\bea
C^{J\,m}_{J\,m-1\,1\,1} & = & -\sqrt{\frac{(J+m)(J-m+1)}{2(J+1)J}},  \\
C^{J\,m}_{J\,m\,1\,0} & = & \frac{m}{\sqrt{(J+1)J}},  \\
C^{J\,m}_{J\,m+1\,1\,-1} & = & \sqrt{\frac{(J-m)(J+m+1)}{2(J+1)J}}, \label{1st}
\eea
\bea
C^{J+\frac12 \,m}_{J\,m-\frac12 \,\frac12 \,\frac12} = \sqrt{\frac{J+m+\frac12}{2J+1}},  
& & 
C^{J+\frac12 \,m}_{J\,m+\frac12 \,\frac12 \,-\frac12} = \sqrt{\frac{J-m+\frac12}{2J+1}}, 
\label{4th} \\
C^{J\,m}_{J+\frac12 \,m-\frac12 \,\frac12 \,\frac12} = -\sqrt{\frac{J-m+1}{2(J+1)}}, & & 
C^{J\,m}_{J+\frac12 \,m+\frac12 \,\frac12 \,-\frac12} = \sqrt{\frac{J+m+1}{2(J+1)}}.
\eea

\subsection{$\vec{L}\circ$-actions to scalar, vector and spinor fuzzy spherical harmonics}
In this subsection, we will show that the following four relations hold:
\bea
& & \vec{L}\circ\hat{Y}^{(jj)}_{J\,m} = \sqrt{J(J+1)}\,\vec{\hat{Y}}^{\rho=0}_{J\,m\,(jj)}, 
\label{L_FSH} \\
& & \vec{L}\circ \cdot \vec{\hat{Y}}^\rho_{J\,m\,(jj)} = \sqrt{J(J+1)}\,\delta_{\rho\,0}\hat{Y}^{(jj)}_{J\,m}, 
\label{L_vec_FSH_1}\\
& & i\vec{L}\circ \times \vec{\hat{Y}}^\rho_{J\,m\,(jj)} + \vec{\hat{Y}}^\rho_{J\,m\,(jj)} = 
\rho (J+1) \,\vec{\hat{Y}}^\rho_{J\,m\,(jj)}, 
\label{L_vec_FSH_2}\\
& & \left(\vec{\sigma}\cdot\vec{L}\circ +\frac34\right) \hat{Y}^\kappa_{J\,m\,(jj)} 
= \kappa\left(J+\frac34\right)\hat{Y}^\kappa_{J\,m\,(jj)}, 
\label{L_sp_FSH}
\eea
where 
\be
\hat{Y}^\kappa_{J\,m\,(jj)}=\begin{pmatrix} \hat{Y}^\kappa_{J\,m\,(jj)\,\alpha=\frac12} \\
\hat{Y}^\kappa_{J\,m\,(jj)\,\alpha=-\frac12} \end{pmatrix}, 
\ee
$\vec{\sigma}$ are the Pauli matrices, and 
\be
\vec{\sigma}\cdot\vec{L} = \begin{pmatrix} L_3 & L_1-iL_2 \\ L_1+iL_2 & -L_3\end{pmatrix} 
= \begin{pmatrix} L_3 & L_- \\ L_+ & -L_3 \end{pmatrix}. 
\ee

\paragraph{Proof of (\ref{L_FSH})} 
Note that 
\bea
& & \hspace{-7mm}\sqrt{J(J+1)}\left(\hat{Y}^{\rho=0}_{J\,m\,(jj)\,i=1}+i\hat{Y}^{\rho=0}_{J\,m\,(jj)\,i=2}\right) \nn \\
& & = \sqrt{2J(J+1)}\,\hat{\cY}^{1\,-1}_{J\,m,\,J\,(jj)} 
= \sqrt{2J(J+1)}\,C^{J\,m}_{J\,m+1\,1\,-1} \,\hat{Y}^{(jj)}_{J\,m+1}.
\eea
Using (\ref{1st}), we can see 
\bea
\sqrt{J(J+1)}\left(\hat{Y}^{\rho=0}_{J\,m\,(jj)\,i=1}+i\hat{Y}^{\rho=0}_{J\,m\,(jj)\,i=2}\right) & = & 
\sqrt{(J-m)(J+m+1)}\,\hat{Y}^{(jj)}_{J\,m+1} \nn \\
& = & L_+\circ\hat{Y}^{(jj)}_{J\,m}. 
\label{pr_L_FSH_1}
\eea

Similarly, we obtain 
\bea
& & \sqrt{J(J+1)}\left(\hat{Y}^{\rho=0}_{J\,m\,(jj)\,i=1}-i\hat{Y}^{\rho=0}_{J\,m\,(jj)\,i=2}\right) 
= L_-\circ\hat{Y}^{(jj)}_{J\,m}, 
\label{pr_L_FSH_2}\\
& & \sqrt{J(J+1)}\,\hat{Y}^{\rho=0}_{J\,m\,(jj)\,i=3} = L_3\circ\hat{Y}^{(jj)}_{J\,m}. 
\label{pr_L_FSH_3}
\eea

(\ref{pr_L_FSH_1}), (\ref{pr_L_FSH_2}) and (\ref{pr_L_FSH_3}) mean (\ref{L_FSH}).

\paragraph{Proof of (\ref{L_vec_FSH_1})} 
For $\rho=0$, acting $\vec{L}\circ\cdot$ on (\ref{L_FSH}) leads to 
\be
J(J+1)\,\hat{Y}^{(jj)}_{J\,m} = \sqrt{J(J+1)}\,\vec{L}\circ \cdot \vec{\hat{Y}}^{\rho=0}_{J\,m\,(jj)}, 
\ee
which is nothing but (\ref{L_vec_FSH_1}) when $J\neq 0$. 
At $J=0$, (\ref{L_vec_FSH_1}) trivially holds because of (\ref{vec_FSH_J=0}).  

For $\rho=1$, 
\bea
& & \hspace{-4mm}\vec{L}\circ\cdot \vec{\hat{Y}}^{\rho=1}_{J\,m\,(jj)}  
= \frac12 L_+\circ\left(\hat{Y}^{\rho=1}_{J\,m\,(jj)\,i=1}-i\hat{Y}^{\rho=1}_{J\,m\,(jj)\,i=2}\right) 
+\frac12 L_-\circ\left(\hat{Y}^{\rho=1}_{J\,m\,(jj)\,i=1}+i\hat{Y}^{\rho=1}_{J\,m\,(jj)\,i=2}\right) \nn \\
& & \hspace{22mm}+L_3\circ \hat{Y}^{\rho=1}_{J\,m\,(jj)\,i=3} \nn \\
& & = -\frac{i}{\sqrt{2}}\, L_+\circ \hat{\cY}^{1\,1}_{J+1\,m,\,J\,(jj)} 
+\frac{i}{\sqrt{2}} \,L_-\circ \hat{\cY}^{1\,-1}_{J+1\,m,\,J\,(jj)} +iL_3\circ \hat{\cY}^{1\,0}_{J+1\,m,\,J\,(jj)}
\nn \\
& & = -\frac{i}{\sqrt{2}}\,C^{J+1\,m}_{J\,m-1\,1\,1} \,L_+\circ\hat{Y}^{(jj)}_{J\,m-1} 
+\frac{i}{\sqrt{2}} \,C^{J+1\,m}_{J\,m+1\,1\,-1} \,L_-\circ \hat{Y}^{(jj)}_{J\,m+1} 
+iC^{J+1\,m}_{J\,m\,1\,0} \,L_3\circ \hat{Y}^{(jj)}_{J\,m} \nn \\
& & = \left[-\frac{i}{\sqrt{2}}\,C^{J+1\,m}_{J\,m-1\,1\,1}\,\sqrt{(J-m+1)(J+m)}\right. \nn \\
& & \hspace{7mm}\left. +\frac{i}{\sqrt{2}}\,C^{J+1\,m}_{J\,m+1\,1\,-1}\,\sqrt{(J+m+1)(J-m)} 
+iC^{J+1\,m}_{J\,m\,1\,0}\,m\right] \hat{Y}^{(jj)}_{J\,m} .
\label{pr_L_vec_FSH_1}
\eea
By using (\ref{2nd_1})--(\ref{2nd_3}), it turns out that the r.h.s. of (\ref{pr_L_vec_FSH_1}) 
vanishes. 

Finally, $\vec{L}\circ\cdot\vec{\hat{Y}}^{\rho=-1}_{J\,m\,(jj)} =0$ is similarly shown, which completes the proof of (\ref{L_vec_FSH_1}). 

\paragraph{Proof of (\ref{L_vec_FSH_2})} 
When $J=0$, (\ref{L_vec_FSH_2}) trivially holds from (\ref{vec_FSH_J=0}). 
Let us focus on the case $J\neq 0$.  

For $\rho=0$, acting $\vec{L}\circ\times$ on (\ref{L_FSH}) yields  
\be
i\vec{L}\circ\times\vec{\hat{Y}}^{\rho=0}_{J\,m\,(jj)} = 
\frac{i}{\sqrt{J(J+1)}}\,\vec{L}\circ\times\vec{L}\circ\hat{Y}^{(jj)}_{J\,m} 
=-\frac{1}{\sqrt{J(J+1)}}\,\vec{L}\circ \hat{Y}^{(jj)}_{J\,m} 
= -\vec{\hat{Y}}^{\rho=0}_{J\,m\,(jj)}.
\label{pr_L_vec_FSH_2}
\ee
Since
\be
\vec{L}\circ\times\vec{L}\circ = \begin{pmatrix} L_2\circ L_3\circ -L_3\circ L_2\circ \\
L_3\circ L_1\circ -L_1\circ L_3\circ \\ L_1\circ L_2\circ -L_2\circ L_1\circ \end{pmatrix} 
=i\begin{pmatrix} L_1\circ \\ L_2 \circ \\ L_3\circ \end{pmatrix} = i\vec{L}\circ , 
\ee
(\ref{pr_L_vec_FSH_2}) proves (\ref{L_vec_FSH_2}) for $\rho=0$. 

For $\rho=1$, let us consider, e.g., the $i=3$ component: 
\bea
& & i\left(L_1\circ\hat{Y}^{\rho=1}_{J\,m\,(jj)\,i=2}-L_2\circ\hat{Y}^{\rho=1}_{J\,m\,(jj)\,i=1}\right) \nn \\
& & = -\frac{1}{\sqrt{2}}\,\left[L_1\circ\left(-i\hat{\cY}^{1\,1}_{J+1\,m,\,J\,(jj)}
-i\hat{\cY}^{1\,-1}_{J+1\,m,\,J\,(jj)}\right) 
-L_2\circ\left(-\hat{\cY}^{1\,1}_{J+1\,m,\,J\,(jj)}
+\hat{\cY}^{1\,-1}_{J+1\,m,\,J\,(jj)}\right)\right] \nn \\
& & = \frac{i}{\sqrt{2}}\,\left(L_+\circ\hat{\cY}^{1\,1}_{J+1\,m,\,J\,(jj)} 
+L_-\circ\hat{\cY}^{1\,-1}_{J+1\,m,\,J\,(jj)}\right) \nn \\
& & = \frac{i}{\sqrt{2}}\,\left( C^{J+1\,m}_{J\,m-1\,1\,1}\,L_+\circ\hat{Y}^{(jj)}_{J\,m-1}
+C^{J+1\,m}_{J\,m+1\,1\,-1}\,L_-\circ\hat{Y}^{(jj)}_{J\,m+1}\right).
\label{pr_L_vec_FSH_3}
\eea
With use of (\ref{2nd_1})--(\ref{2nd_3}), the r.h.s.  of (\ref{pr_L_vec_FSH_3}) becomes 
\be
(\mbox{r.h.s. of (\ref{pr_L_vec_FSH_3})}) 
= iJC^{J+1\,m}_{J\,m\,1\,0}\,\hat{Y}^{(jj)}_{J\,m} =  iJ\hat{\cY}^{1\,0}_{J+1\,m,\,J\,(jj)} = J\hat{Y}^{\rho=1}_{J\,m\,(jj)\,i=3},
\ee
which shows (\ref{L_vec_FSH_2}) for $\rho=1$ and $i=3$. 

We can similarly show (\ref{L_vec_FSH_2}) for the remaining cases including $\rho=-1$.

\paragraph{Proof of (\ref{L_sp_FSH})} 
For $\kappa=1$, let us consider, e.g., the $\alpha=\frac12$ component: 
\bea
& & \left(\left(\vec{\sigma}\cdot\vec{L}\circ +\frac34\right)
\hat{Y}^{\kappa=1}_{J\,m\,(jj)}\right)_{\alpha=\frac12} 
= \left(L_3\circ +\frac34\right) \hat{Y}^{\kappa=1}_{J\,m\,(jj)\,\alpha=\frac12} 
+L_-\circ\hat{Y}^{\kappa=1}_{J\,m\,(jj)\,\alpha=-\frac12} \nn \\
& & \hspace{24mm}= C^{J+\frac12\,m}_{J\,m-\frac12\,\frac12\,\frac12}\,\left(L_3\circ +\frac34\right) 
\hat{Y}^{(jj)}_{J\,m-\frac12} +C^{J+\frac12\,m}_{J\,m+\frac12\, \frac12\,-\frac12} \,L_-\circ
\hat{Y}^{(jj)}_{J\,m+\frac12}. 
\label{pr_L_sp_FSH}
\eea
By using (\ref{4th}), the r.h.s. of (\ref{pr_L_sp_FSH}) can be expressed as 
\be
(\mbox{r.h.s. of (\ref{pr_L_sp_FSH})}) 
= \left(J+\frac34\right) C^{J+\frac12\,m}_{J\,m-\frac12\,\frac12\,\frac12}\,\hat{Y}^{(jj)}_{J\,m-\frac12} 
=\left(J+\frac34\right)\,\hat{Y}^{\kappa=1}_{J\,m\,(jj)\,\alpha=\frac12},  
\ee
showing (\ref{L_sp_FSH}) for $\kappa=1$ and $\alpha=\frac12$.  

We can similarly prove (\ref{L_sp_FSH}) for all the other cases.

\subsection{Vertex coefficients}
We start by showing the formula for the trace of the product of three kinds of spin-$S$ fuzzy spherical harmonics: 
\bea
& & \hspace{-7mm}\sum_{n_1, n_2, n_3}\frac{1}{n}\,\tr_n\left\{
\left(\hat{\cY}^{S_1\,n_1}_{J_1\,m_1,\,\tilde{J}_1\,(jj)}\right)^\dagger
\hat{\cY}^{S_2\,n_2}_{J_2\,m_2,\,\tilde{J}_2\,(jj)}\hat{\cY}^{S_3\,n_3}_{J_3\,m_3,\,\tilde{J}_3\,(jj)}\right\}
C^{S_1\,n_1}_{S_2\,n_2\,S_3\,n_3} \nn \\
& & = (-1)^{\tilde{J}_1+2j}\sqrt{n(2S_1+1)(2\tilde{J}_1+1)(2J''+1)(2\tilde{J}_2+1)(2J_3+1)(2\tilde{J}_3+1)} \nn \\
& & \hspace{7mm}\times \begin{Bmatrix} J_1 & \tilde{J}_1 & S_1 \\ J_2 & \tilde{J}_2 & S_2 \\
J_3 & \tilde{J}_3 & S_3 \end{Bmatrix} C^{J_1\,m_1}_{J_2\, m_2\, J_3\,m_3} 
\begin{Bmatrix} \tilde{J}_1 & \tilde{J}_2 & \tilde{J}_3 \\ j & j & j \end{Bmatrix}.
\label{VC_SpinS_FSH}
\eea

{}From the definition (\ref{SpinS_FSH}) and (\ref{VC_FSH}), 
\bea
& & \hspace{-7mm}(\mbox{l.h.s. of (\ref{VC_SpinS_FSH})}) \nn \\
& & =\sum_{n_1,n_2,n_3} C^{J_1\,m_1}_{\tilde{J}_1\,m_1-n_1\,S_1\,n_1} 
C^{J_2\,m_2}_{\tilde{J}_2\,m_2-n_2\,S_2\,n_2} C^{J_3\,m_3}_{\tilde{J}_3\,m_3-n_3\,S_3\,n_3} 
C^{S_1\,n_1}_{S_2\,n_2\,S_3\,n_3} 
\hat{C}^{\tilde{J}_1\,m_1-n_1\,(jj)}_{\tilde{J}_2\,m_2-n_2\,(jj)\,\tilde{J}_3\,m_3-n_3\,(jj)} \nn \\
& & = \sum_{p_1,p_2,p_3}\sum_{n_1,n_2,n_3} C^{J_1\,m_1}_{\tilde{J}_1\,p_1\,S_1\,n_1} 
C^{J_2\,m_2}_{\tilde{J}_2\,p_2\,S_2\,n_2} C^{J_3\,m_3}_{\tilde{J}_3\,p_3\,S_3\,n_3} 
 C^{S_1\,n_1}_{S_2\,n_2\,S_3\,n_3} 
\hat{C}^{\tilde{J}_1\,p_1\,(jj)}_{\tilde{J}_2\,p_2\,(jj)\,\tilde{J}_3\,p_3\,(jj)} \nn \\
& & = \sum_{p_3,n_3}\left[\sum_{p_1,n_1,p_2,n_2}C^{J_1\,m_1}_{\tilde{J}_1\,p_1\,S_1\,n_1} 
C^{J_2\,m_2}_{\tilde{J}_2\,p_2\,S_2\,n_2} C^{\tilde{J}_1\,p_1}_{\tilde{J}_2\,p_2\,\tilde{J}_3\,p_3} 
C^{S_1\,n_1}_{S_2\,n_2\,S_3\,n_3}\right] \nn \\
& & \hspace{14mm} \times C^{J_3\,m_3}_{\tilde{J}_3\,p_3\,S_3\,n_3} (-1)^{\tilde{J}_1+2j}
\sqrt{n(2\tilde{J}_2+1)(2\tilde{J}_3+1)}\,\begin{Bmatrix} \tilde{J}_1 & \tilde{J}_2 & \tilde{J}_3 \\ j & j & j 
\end{Bmatrix}.
\label{app:FSH_Vcoeff}
\eea
Applying the identity~\footnote{See, e.g., eq.~(26) in Chapter 8.7.4 of \cite{QTAM}.} 
\bea
& & \hspace{-7mm}\sum_{\beta, \gamma, \epsilon, \varphi} C^{a\,\alpha}_{b\,\beta\,c\,\gamma} 
C^{d\,\delta}_{e\,\epsilon\,f\,\varphi} C^{b\,\beta}_{e\,\epsilon\,g\,\eta} 
C^{c\,\gamma}_{f\,\varphi\,j\,\mu}  
\nn \\
& & =\sum_{k,\kappa}\sqrt{(2b+1)(2c+1)(2d+1)(2k+1)}\,C^{k\,\kappa}_{g\,\eta\,j\,\mu} 
C^{a\,\alpha}_{d\,\delta\,k\,\kappa} \,\begin{Bmatrix} a & b & c \\ d & e & f  \\ k & g & j \end{Bmatrix} 
\eea
to $[\sum_{p_1,n_1,p_2,n_2} \cdots]$ in (\ref{app:FSH_Vcoeff}) together with the orthogonality (\ref{CG_ortho}), 
we obtain (\ref{VC_SpinS_FSH}). 

 To derive formulas for the traces including the vector and spinor fuzzy spherical harmonics, 
it is convenient to recast the expressions (\ref{vec_FSH_1})--(\ref{vec_FSH_3}) as well as 
(\ref{sp_FSH_1}) and (\ref{sp_FSH_2}) into the concise form: 
\be
\hat{Y}^\rho_{J\,m\,(jj)\,i} = i^{\rho}\sum_{n'=-1}^{1} V_{i\,n'}\,\hat{\cY}^{1\,n'}_{Q\,m,\,\tilde{Q}\,(jj)}
\label{vec_FSH}
\ee
with $Q\equiv J + \delta_{\rho,\, 1}$, $\tilde{Q}\equiv J + \delta_{\rho, \,-1}$, and   
\be
\hat{Y}^\kappa_{J\,m\,(jj)\,\alpha} = \hat{\cY}^{\frac12\,\alpha}_{U\,m,\,\tilde{U}\,(jj)} 
\label{sp_FSH}
\ee 
with $U\equiv J+\frac12\delta_{\kappa, \,1}$, $\tilde{U}\equiv J+\frac12\delta_{\kappa, \,-1}$.  
In the remaining part of this section, we compute the vertex coefficients defined by (\ref{VC_D_def})--(\ref{VC_G_def}) 
in the text.

\subsubsection{$\hat{\cD}$} 
We plug (\ref{vec_FSH}) into (\ref{VC_D_def}) and note that 
\be
\sum_{i=1}^3V_{i\,n_1}V_{i\,-n_2} = \sum_{i=1}^3 V_{i\,n_1}V_{i\,n_2}^*(-1)^{-n_2} = \delta_{n_1\,n_2}(-1)^{-n_2},
\ee
which follows from (\ref{V_id}), 
so that the $\hat{\cD}$ coefficient is expressed as 
\bea
& & \hspace{-7mm}\hat{\cD}^{J\,m\,(jj)}_{J_1\,m_1\,(jj)\,\rho_1\,J_2\,m_2\,(jj)\,\rho_2} 
\nn \\
& & = i^{\rho_1+\rho_2}\sum_{n_1=-1}^{1}(-1)^{-n_1} \frac{1}{n}\,\tr_n\left\{ 
\left(\hat{\cY}^{00}_{J\,m,\,J\,(jj)}\right)^\dagger \hat{\cY}^{1\,n_1}_{Q_1\,m_1,\,\tilde{Q}_1\,(jj)} 
\hat{\cY}^{1\,-n_1}_{Q_2\,m_2,\,\tilde{Q}_2\,(jj)}\right\} 
\eea
with $Q_a=J_a+\delta_{\rho_a,\,1}$, $\tilde{Q}_a=J_a+\delta_{\rho_a,\,-1}$ ($a=1,2$).  

Next, we use (\ref{HC_SpinS_FSH}) to rewrite it as 
\bea
& & \hspace{-7mm}\hat{\cD}^{J\,m\,(jj)}_{J_1\,m_1\,(jj)\,\rho_1\,J_2\,m_2\,(jj)\,\rho_2} \nn \\
& & = i^{\rho_1+\rho_2}(-1)^{Q_1-\tilde{Q}_1+1+m_1+m}
\sum_{n_1=-1}^{1}\frac{1}{n}\,\tr_n\left\{ \left(\hat{\cY}^{1\,n_1}_{Q_1\,-m_1,\,\tilde{Q}_1\,(jj)}\right)^\dagger 
\hat{\cY}^{1\,n_1}_{Q_2\,m_2,\,\tilde{Q}_2\,(jj)} \hat{\cY}^{00}_{J\,-m,\,J\,(jj)}\right\} \nn \\
& & = i^{\rho_1+\rho_2}(-1)^{Q_1-\tilde{Q}_1+1+m_1+m}\nn \\
& & \hspace{4mm} \times\sum_{n_1,n_2=-1}^{1}
\frac{1}{n}\,\tr_n\left\{ \left(\hat{\cY}^{1\,n_1}_{Q_1\,-m_1,\,\tilde{Q}_1\,(jj)}\right)^\dagger 
\hat{\cY}^{1\,n_2}_{Q_2\,m_2,\,\tilde{Q}_2\,(jj)} \hat{\cY}^{00}_{J\,-m,\,J\,(jj)}\right\} 
C^{1\,n_1}_{1\,n_2\,0\,0}. 
\eea
In the last equality, we have inserted $1=\sum_{n_2} \delta_{n_1\,n_2} = \sum_{n_2}C^{1\,n_1}_{1\,n_2\,0\,0}$. 

Then, applying the formula (\ref{VC_SpinS_FSH}) and 
\bea
C^{Q_1\,-m_1}_{Q_2\,m_2\,J\,-m} & = &  (-1)^{Q_2-m_2}\sqrt{\frac{2Q_1+1}{2J+1}}\,C^{J\,-m}_{Q_1\,-m_1\,Q_2\,-m_2} 
\nn \\
& = & (-1)^{Q_2-m_2}\sqrt{\frac{2Q_1+1}{2J+1}} \cdot (-1)^{Q_1+Q_2-J}C^{J\,m}_{Q_1\,m_1\,Q_2\,m_2} 
\eea
yields 
\bea
\hat{\cD}^{J\,m\,(jj)}_{J_1\,m_1\,(jj)\,\rho_1\,J_2\,m_2\,(jj)\,\rho_2} 
& = & i^{\rho_1+\rho_2}(-1)^{2Q_1+2Q_2+2j-J+1+m_1-m_2+m} \nn \\
& & \times \sqrt{3n(2Q_1+1)(2\tilde{Q}_1+1)(2Q_2+1)(2\tilde{Q}_2+1)(2J+1)} \nn \\
& & \times \begin{Bmatrix} Q_1 & \tilde{Q}_1 & 1 \\ Q_2 & \tilde{Q}_2 & 1 \\ J & J & 0 
\end{Bmatrix} C^{J\,m}_{Q_1\,m_1\,Q_2\,m_2} 
\begin{Bmatrix} \tilde{Q}_1 & \tilde{Q}_2 & J \\ j & j & j \end{Bmatrix}.
\label{app:FSH_Dhat}
\eea

We here note that $J, J_1, J_2, m, m_1, m_2$ are integers, $m=m_1+m_2$ from 
the C-G coefficient $C^{J\,m}_{Q_1\,m_1\,Q_2\,m_2}$, 
\be
(2Q_a+1)(2\tilde{Q}_a+1) = (2J_a+1)(2J_a+2\rho_a^2+1) \qquad (a=1,2) ,
\label{Q_Qtilde}
\ee
the symmetric property of the $6j$ symbol 
\be
\begin{Bmatrix} a & b & c \\ A & B & C \end{Bmatrix} = 
\begin{Bmatrix} b & a & c \\ B & A & C \end{Bmatrix} =
\begin{Bmatrix} a & c & b \\ A & C & B \end{Bmatrix},  
\label{6j_property2}
\ee
and the relation~\footnote{See, e.g., eq.~(1) in Chapter 10.9.1 of \cite{QTAM}.}
\be
\begin{Bmatrix} a & b & c \\ d & e & f \\ g & h & 0 \end{Bmatrix} = \delta_{c\,f}\,\delta_{g\,h}\, 
\frac{(-1)^{b + c + d + g}}{\sqrt{(2c+1)(2g+1)}}\, \begin{Bmatrix} a & b & c \\ e & d & g\end{Bmatrix}. 
\label{9j_to_6j}
\ee
These reduce (\ref{app:FSH_Dhat}) to 
\bea
& & \hspace{-7mm}\hat{\cD}^{J\,m\,(jj)}_{J_1\,m_1\,(jj)\,\rho_1\,J_2\,m_2\,(jj)\,\rho_2} \nn \\
& & = i^{\rho_1+\rho_2}(-1)^{2j+\tilde{Q}_1+Q_2}
\sqrt{n(2J_1+1)(2J_1+2\rho_1^2+1)(2J_2+1)(2J_2+2\rho_2^2+1)} \nn \\
& & \hspace{4mm}\times  C^{J\,m}_{Q_1\,m_1\,Q_2\,m_2} 
\begin{Bmatrix} Q_1 & \tilde{Q}_1 & 1 \\ \tilde{Q}_2 & Q_2 & J \end{Bmatrix} \,
\begin{Bmatrix} J & \tilde{Q}_1 & \tilde{Q}_2 \\ j & j & j \end{Bmatrix}.
\label{VC_D}
\eea

Equivalently, we can rewrite (\ref{VC_D_def}) as 
\be
\sum_{i=1}^3\hat{Y}^{\rho_1}_{J_1\,m_1\,(jj)\,i} \hat{Y}^{\rho_2}_{J_2\,m_2\,(jj)\,i} 
= \sum_{J=0}^{2j}\sum_{m=-J}^J \hat{\cD}^{J\,m\,(jj)}_{J_1\,m_1\,(jj)\,\rho_1\,J_2\,m_2\,(jj)\,\rho_2} 
\hat{Y}^{(jj)}_{J\,m}, 
\label{VC_D_2}
\ee
and 
\bea
\left(\hat{Y}^{(jj)}_{J\,m}\right)^\dagger \hat{Y}^{\rho_1}_{J_1\,m_1\,(jj)\,i} & = & 
\sum_{\rho_2, J_2, m_2} \hat{\cD}^{J\,m\,(jj)}_{J_1\,m_1\,(jj)\,\rho_1\,J_2\,m_2\,(jj)\,\rho_2} 
\left(\hat{Y}^{\rho_2}_{J_2\,m_2\,(jj)\,i}\right)^\dagger,
\label{VC_D_3} \\
\hat{Y}^{\rho_2}_{J_2\,m_2\,(jj)\,i}\left(\hat{Y}^{(jj)}_{J\,m}\right)^\dagger & = & 
\sum_{\rho_1, J_1, m_1} \hat{\cD}^{J\,m\,(jj)}_{J_1\,m_1\,(jj)\,\rho_1\,J_2\,m_2\,(jj)\,\rho_2} 
\left(\hat{Y}^{\rho_1}_{J_1\,m_1\,(jj)\,i}\right)^\dagger. 
\label{VC_D_4}
\eea

Finally, the relation
\be
\hat{\cD}^{J\,m\,(jj)}_{J_1\,m_1\,(jj)\,\rho_1\,J_2\,m_2\,(jj)\,\rho_2} =(-1)^{\tilde{Q}_1+\tilde{Q}_2-J}\,
\hat{\cD}^{J\,m\,(jj)}_{J_2\,m_2\,(jj)\,\rho_2\,J_1\,m_1\,(jj)\,\rho_1}
\label{cD_12_21}
\ee
holds from the first equality in (\ref{CG_id}). 
 
\subsubsection{$\hat{\cE}$} 
We plug (\ref{vec_FSH}) into (\ref{VC_E_def}) and note that 
\be
\sum_{i',j',k'=1}^3\epsilon_{i'j'k'}V_{i'\,n_1}V_{j'\,n_2}V_{k'\,n_3} = \epsilon_{n_1 n_2 n_3}\det V 
= -i\epsilon_{n_1 n_2 n_3}
\ee
with $\epsilon_{n_1=1\,n_2=0\,n_3=-1}=+1$ and 
\be
\hat{\cY}^{1\,n_1}_{Q_1\,m_1,\,\tilde{Q}_1\,(jj)}= (-1)^{Q_1-\tilde{Q}_1+1+m_1+n_1} 
\left(\hat{\cY}^{1\,-n_1}_{Q_1\, -m_1, \,\tilde{Q}_1\,(jj)}\right)^\dagger, 
\ee
so that the $\hat{\cE}$ coefficient is expressed as 
\bea
& & \hat{\cE}_{J_1\,m_1\,(jj)\,\rho_1\,J_2\,m_2\,(jj)\,\rho_2\,J_3\,m_3\,(jj)\,\rho_3} 
= i^{\rho_1+\rho_2+\rho_3-1}\sum_{n_1, n_2,n_3=-1}^{1} \epsilon_{-n_1\,n_2\,n_3} 
(-1)^{Q_1-\tilde{Q}_1+1+m_1-n_1} \nn \\
& & \hspace{24mm}\times \frac{1}{n} \,\tr_n \left\{ 
\left(\hat{\cY}^{1\,n_1}_{Q_1\, -m_1, \,\tilde{Q}_1\,(jj)}\right)^\dagger
\hat{\cY}^{1\,n_2}_{Q_2\, m_2, \,\tilde{Q}_2\,(jj)} \hat{\cY}^{1\,n_3}_{Q_3\, m_3, \,\tilde{Q}_3\,(jj)}\right\}. 
\eea

Here, it can be seen from the formulas in section~\ref{sec:CG} that 
\be
\epsilon_{-n_1\,n_2\,n_3}(-1)^{-n_1} = -\sqrt{2}\,C^{1\,n_1}_{1\,n_2\,1\,n_3}  
\label{epsilon_CG}
\ee
holds. This and (\ref{VC_SpinS_FSH}) lead to 
\bea
 & & \hat{\cE}_{J'\,m_1\,(jj)\,\rho_1\,J''\,m_2\,(jj)\,\rho_2\,J_3\,m_3\,(jj)\,\rho_3}
= i^{\rho_1+\rho_2+\rho_3-1}(-1)^{Q_1+m_1+2j} \nn \\
& & \hspace{24mm} \times \sqrt{6n(2\tilde{Q}_1+1)(2Q_2+1)(2\tilde{Q}_2+1)(2Q_3+1)(2\tilde{Q}_3+1)} \nn \\
& & \hspace{24mm} \times \begin{Bmatrix} Q_1 & \tilde{Q}_1 & 1 \\ Q_2 & \tilde{Q}_2 & 1 \\ Q_3 & \tilde{Q}_3 & 1 
\end{Bmatrix} C^{Q_1\,-m_1}_{Q_2\,m_2\,Q_3\,m_3} 
\begin{Bmatrix} \tilde{Q}_1 & \tilde{Q}_2 & \tilde{Q}_3 \\ j & j & j \end{Bmatrix}.
\eea

Finally, we use 
\be
C^{Q_1\,-m_1}_{Q_2\,m_2\,Q_3\,m_3} 
 = (-1)^{-Q_2+Q_3+m_1}\sqrt{2Q_1+1}\,\begin{pmatrix} 
Q_1 & Q_2 & Q_3 \\ m_1 & m_2 & m_3  \end{pmatrix},
\ee
which follows from (\ref{CG_3j}),
and note $m_1\in \Z$,  
\bea
-Q_1-\tilde{Q}_1-\rho_1 & = & -2J_1-\delta_{\rho_1,\, 1}-\delta_{\rho_1, \,-1}-\rho_1 \in 2\Z, \nn \\
Q_2-\tilde{Q}_2-\rho_2 & = & \delta_{\rho_2,\,1}-\delta_{\rho_2,\,-1}-\rho_2 = 0, \nn\\
-Q_3-\tilde{Q}_3-\rho_3 & \in & 2\Z
\eea
and (\ref{Q_Qtilde}), to arrive at the formula 
\bea
& & \hat{\cE}_{J_1\,m_1\,(jj)\,\rho_1\,J_2\,m_2\,(jj)\,\rho_2\,J_3\,m_3\,(jj)\,\rho_3} 
= i^{-\rho_1-\rho_2-\rho_3-1} (-1)^{-\tilde{Q}_1-\tilde{Q}_2-\tilde{Q}_3+2j} \nn \\
& & \hspace{7mm} \times  \sqrt{6n(2J_1+1)(2J_1+2\rho_1^2+1)(2J_2+1)(2J_2+2\rho_2^2+1)(2J_3+1)(2J_3+2\rho_3^2+1)} \nn \\
& & \hspace{7mm} \times \begin{Bmatrix} Q_1 & \tilde{Q}_1 & 1 \\ Q_2 & \tilde{Q}_2 & 1 \\ Q_3 & \tilde{Q}_3 & 1 
\end{Bmatrix} \begin{pmatrix} 
Q_1 & Q_2 & Q_3 \\ m_1 & m_2 & m_3  \end{pmatrix}
\begin{Bmatrix} \tilde{Q}_1 & \tilde{Q}_2 & \tilde{Q}_3 \\ j & j & j \end{Bmatrix}.
\label{VC_E}
\eea 

Equivalently, we can rewrite (\ref{VC_E_def}) as 
\bea
& & \hspace{-7mm}\sum_{i',j'=1}^3\epsilon_{i'j'k'}\hat{Y}^{\rho_1}_{J_1\,m_1\,(jj)\,i'} \hat{Y}^{\rho_2}_{J_2\,m_2\,(jj)\,j'} 
\nn \\
& & = \sum_{\rho_3, J_3, m_3} \hat{\cE}_{J_1\,m_1\,(jj)\,\rho_1\,J_2\,m_2\,(jj)\,\rho_2\,J_3\,m_3\,(jj)\,\rho_3} 
\left(\hat{Y}^{\rho_3}_{J_3\,m_3\,(jj)\,k'}\right)^\dagger. 
\label{VC_E_2}
\eea

\subsubsection{$\hat{\cF}$} 
In the expression 
\be
\hat{\cF}^{J_1\,m_1\,(jj)\,\kappa_1}_{J_2\,m_2\,(jj)\,\kappa_2\,J\,m\,(jj)} 
= \sum_{\alpha_1=-\frac12}^{\frac12}\frac{1}{n}\,\tr_n\left\{ 
\left(\hat{\cY}^{\frac12\,\alpha_1}_{U_1\,m_1,\,\tilde{U}_1\,(jj)}\right)^\dagger 
 \hat{\cY}^{\frac12\,\alpha_1}_{U_2\,m_2,\,\tilde{U}_2\,(jj)} \hat{\cY}^{00}_{J\,m,\,J\,(jj)}\right\} 
\ee
with $U_a=J_a+\frac12\delta_{\kappa_a,\,1}$, $\tilde{U}_a = J_a+\frac12\delta_{\kappa_a, \,-1}$ ($a=1,2$), 
which is obtained by plugging (\ref{vec_FSH}) and (\ref{sp_FSH}) into (\ref{VC_F_def}), we insert 
$1=\sum_{\alpha_2=-\frac12}^{\frac12}\delta_{\alpha_1\,\alpha_2} 
= \sum_{\alpha_2=-\frac12}^{\frac12}C^{\frac12\,\alpha_1}_{\frac12\,\alpha_2\, 0\,0}$ and apply 
the formula (\ref{VC_SpinS_FSH}). Then, the $\hat{\cF}$ coefficient becomes 
\bea
\hat{\cF}^{J_1\,m_1\,(jj)\,\kappa_1}_{J_2\,m_2\,(jj)\,\kappa_2\,J\,m\,(jj)} 
& = & (-1)^{\tilde{U}_1+2j} \sqrt{2n(2\tilde{U}_1+1)(2U_2+1)(2\tilde{U}_2+1)}\,(2J+1) \nn \\
& &  \times \begin{Bmatrix} U_1 & \tilde{U}_1 & \frac12 \\ U_2 & \tilde{U}_2 & \frac12 \\ J & J & 0 
\end{Bmatrix} C^{U_1\, m_1}_{U_2\,m_2\,J\,m} 
\begin{Bmatrix} \tilde{U}_1 & \tilde{U}_2 & J \\ j & j & j \end{Bmatrix}.
\label{app:FSH_Fhat}
\eea

Finally, noting 
\be
(2U_2+1)(2\tilde{U}_2+1) = (2J''+1)(2J''+2) 
\label{U_Utilde}
\ee
and (\ref{9j_to_6j}) reduces (\ref{app:FSH_Fhat}) to 
\bea
\hat{\cF}^{J_1\,m_1\,(jj)\,\kappa_1}_{J_2\,m_2\,(jj)\,\kappa_2\,J\,m\,(jj)} 
& = & (-1)^{2j+U_2+J+\frac12}\sqrt{n(2\tilde{U}_1+1)(2J_2+1)(2J_2+2)(2J+1)} \nn \\
& & \times C^{U_1\, m_1}_{U_2\,m_2\,J\,m} \begin{Bmatrix} U_1 & \tilde{U}_1 & \frac12 \\ \tilde{U}_2 & U_2 & J\end{Bmatrix}
\begin{Bmatrix} \tilde{U}_1 & \tilde{U}_2 & J \\ j & j & j \end{Bmatrix}.
\label{VC_F}
\eea
More explicitly, 
for $(\kappa_1, \kappa_2)=(1,1)$, 
\bea
\hat{\cF}^{J_1\,m_1\,(jj)\,1}_{J_2\,m_2\,(jj)\,1\,J\,m\,(jj)}& = & (-1)^{2j+J_1+J_2+\frac12-m_2}\,\sqrt{n(J+J_1+J_2+2)(-J+J_1+J_2+1)} \nn \\
& & \times C^{J\,m}_{J_1+\frac12\,\,m_1\,J_2+\frac12\,\,-m_2}
\begin{Bmatrix} J_1& J_2& J \\ j & j & j\end{Bmatrix},
\eea
for $(\kappa_1, \kappa_2)=(-1,-1)$, 
\bea
\hat{\cF}^{J_1\,m_1\,(jj)\,-1}_{J_2\,m_2\,(jj)\,-1\,J\,m\,(jj)}& = & (-1)^{2j+J_1+J_2+\frac12-m_2}\,\sqrt{n(J+J_1+J_2+2)(-J+J_1+J_2+1)} \nn \\
& & \times C^{J\,m}_{J_1\,m_1\,J_2\,\,-m_2}
\begin{Bmatrix} J_1+\frac12 & J_2+\frac12 & J \\ j & j & j\end{Bmatrix},
\eea
for $(\kappa_1, \kappa_2)=(1,-1)$, 
\bea
\hat{\cF}^{J_1\,m_1\,(jj)\,1}_{J_2\,m_2\,(jj)\,-1\,J\,m\,(jj)}& = & (-1)^{2j+J_1+J_2-m_2}\,
\sqrt{n\left(J+J_1-J_2+\frac12\right)\left(J-J_1+J_2+\frac12\right)} \nn \\
& & \times C^{J\,m}_{J_1+\frac12\,\,m_1\,J_2\,\,-m_2}
\begin{Bmatrix} J_1& J_2+\frac12 & J \\ j & j & j\end{Bmatrix},
\eea
and, for $(\kappa_1, \kappa_2)=(-1,1)$, 
\bea
\hat{\cF}^{J_1\,m_1\,(jj)\,-1}_{J_2\,m_2\,(jj)\,1\,J\,m\,(jj)}& = & (-1)^{2j+J_1+J_2-m_2}\,
\sqrt{n\left(J+J_1-J_2+\frac12\right)\left(J-J_1+J_2+\frac12\right)} \nn \\
& & \times C^{J\,m}_{J_1\,m_1\,J_2+\frac12\,\,\,-m_2}
\begin{Bmatrix} J_1+\frac12 & J_2& J \\ j & j & j\end{Bmatrix}.
\eea

The property 
\be
C^{U_1\,m_1}_{U_2\,m_2\,J\,m}=(-1)^{J+m}\sqrt{\frac{2U_1+1}{2U_2+1}}\,C^{U_2\,-m_2}_{U_1\,-m_1\,J\,m}
\ee
together with (\ref{6j_property1}) and (\ref{6j_property2}) leads to 
\be
\hat{\cF}^{J_1\,m_1\,(jj)\,\kappa_1}_{J_2\,m_2\,(jj)\,\kappa_2\,J\,m\,(jj)}=
(-1)^{J+\tilde{U}_1-\tilde{U}_2+m_1-m_2-\frac12\kappa_1+\frac12\kappa_2}\,
\hat{\cF}^{J_2\,-m_2\,(jj)\,\kappa_2}_{J_1\,-m_1\,(jj)\,\kappa_1\,J\,m\,(jj)}.
\label{cF_12_21}
\ee

\subsubsection{$\hat{\cG}$} 
We substitute (\ref{vec_FSH}) and (\ref{sp_FSH}) in (\ref{VC_G_def}) to obtain
\bea
& & \hat{\cG}^{J_1\,m_1\,(jj)\,\kappa_1}_{J_2\,m_2\,(jj)\,\kappa_2\,J\,m\,(jj)\,\rho} 
= i^\rho \sum_{\alpha,\beta=-\frac12}^{\frac12} \sum_{n'=-1}^{1} 
\left(\sum_{i=1}^3 \sigma^i_{\alpha\beta} V_{i\,n'} \right) \nn \\
& & \hspace{24mm}\times \frac{1}{n}\, \tr_n \left\{ 
\left(\hat{\cY}^{\frac12\, \alpha}_{U_1\,m_1,\,\tilde{U}_1\,(jj)}\right)^\dagger 
\hat{\cY}^{\frac12\, \beta}_{U_2\,m_2,\,\tilde{U}_2\,(jj)} \hat{\cY}^{1\,n'}_{Q\,m,\,\tilde{Q}\,(jj)}\right\}. 
\eea

Note that 
\be
\sum_{i=1}^3 \sigma^i_{\alpha\beta}V_{i\,n'}= \sqrt{3}\,C^{\frac12\,\alpha}_{\frac12\,\beta\,1\,n'} 
\label{sigmaV_CG}
\ee
holds from 
\bea
\sum_{i=1}^3\sigma^i_{\alpha\beta}V_{i\,n'} &= & 
-\frac{n'}{\sqrt{2}} 
\left(\delta_{\alpha,\,\frac12}\delta_{\beta,\,-\frac12}+\delta_{\alpha,\,-\frac12}\delta_{\beta,\,\frac12}\right) 
-\frac{|n'|}{\sqrt{2}}
\left(\delta_{\alpha,\,\frac12}\delta_{\beta,\,-\frac12}-\delta_{\alpha,\,-\frac12}\delta_{\beta,\,\frac12}\right) 
\nn \\
& & +\delta_{n',\, 0}
\left(\delta_{\alpha,\,\frac12}\delta_{\beta,\,\frac12}-\delta_{\alpha,\,-\frac12}\delta_{\beta,\,-\frac12}\right), 
\eea
and formulas in section~\ref{sec:CG}. 
Together with this, (\ref{VC_SpinS_FSH}), 
(\ref{Q_Qtilde}), and (\ref{U_Utilde}) lead to 
\bea
& & \hspace{-7mm}\hat{\cG}^{J_1\,m_1\,(jj)\,\kappa_1}_{J_2\,m_2\,(jj)\,\kappa_2\,J\,m\,(jj)\,\rho}
= i^\rho (-1)^{\tilde{U}_1+2j}\sqrt{6n(2\tilde{U}_1+1)(2J_2+1)(2J_2+2)(2J+1)(2J+2\rho^2+1)} \nn \\
& & \hspace{30mm} \times 
\begin{Bmatrix} U_1 & \tilde{U}_1 & \frac12 \\ U_2 & \tilde{U}_2 & \frac12 \\ Q & \tilde{Q} & 1 
\end{Bmatrix} C^{U_1\, m_1}_{U_2\,m_2\,Q\,m} 
\begin{Bmatrix} \tilde{U}_1 & \tilde{U}_2 & \tilde{Q} \\ j & j & j \end{Bmatrix}.
\label{VC_G}
\eea

\section{Computational details of one-point functions}
\label{app:one-point_func}
\setcounter{equation}{0}
In this appendix, the details of the computation of (\ref{y_E}), (\ref{y_D}), and (\ref{y_G}) are presented. 
The gauge transformation property of the one-point function $\vev{y_{J\,m,\,\rho}(p)}$ is also discussed. 

\subsection{(\ref{y_E})}  
\label{app:sum_E}
Here, we calculate the sum of $\hat{\cE}$ with respect to $m'$ in (\ref{y_E}). 
{}From the expression of $\hat{\cE}$ (\ref{VC_E}) and $2\tilde{Q}'=2(J'+\delta_{\rho',\,-1})\in 2\Z$, it reduces to 
the sum of $3j$ symbols:
\bea
& & \hspace{-7mm}\sum_{m'=-Q'}^{Q'} (-1)^{-m'}
\hat{\cE}_{\bar{J}\,-\bar{m}\,(jj)\,\bar{\rho}\,J'\,m'\,(jj) \rho'\,J'\,-m'\,(jj)\,\rho'}
= i^{-\bar{\rho}-2\rho'-1}\,(-1)^{-\tilde{\bar{Q}}+2j}\,\nn \\
& & \times\sqrt{6n(2\bar{J}+1)(2\bar{J}+2\bar{\rho}^2+1)} \, (2J'+1)(2J'+2\rho'^2+1) \nn \\
& & \times \begin{Bmatrix} \bar{Q} & \tilde{\bar{Q}} & 1 \\
                 Q'   &  \tilde{Q}'    &  1  \\
                 Q'   & \tilde{Q}'     & 1  \end{Bmatrix} 
\begin{Bmatrix} \tilde{\bar{Q}} & \tilde{Q}'  &  \tilde{Q}'  \\  j   &   j  &   j  \end{Bmatrix}
\sum_{m'=-Q'}^{Q'} (-1)^{-m'}
\begin{pmatrix} \bar{Q}  &  Q'  &  Q'  \\  -\bar{m}  & m'  &  -m' \end{pmatrix} 
\label{sum_E}
\eea  
with $\tilde{\bar{Q}}= \bar{J}+\delta_{\bar{\rho},\,-1}$.

The relation~\footnote{See, e.g., eq.~(13) in Chapter 8.1.3 of \cite{QTAM}.}
\be
\sum_{m'=-Q'}^{Q'} (-1)^{-m'}
\begin{pmatrix} \bar{Q}  &  Q'  &  Q'  \\  -\bar{m}  & m'  &  -m' \end{pmatrix} 
= \sum_{m'=-Q'}^{Q'} (-1)^{-m' +\bar{Q}} \sum_{J=|\bar{Q}-Q'|}^{\bar{Q}+Q'} \sum_{m=-J}^J 
C^{J\,m}_{\bar{Q}\,-\bar{m}\,Q'\,m'}\,C^{0\,0}_{J\,m\,Q'\,-m'}
\ee
together with
\bea
C^{J\,m}_{\bar{Q}\,-\bar{m}\,Q'\,m'} & = & (-1)^{Q'+m'}\sqrt{\frac{2J+1}{2\bar{Q}+1}}\,
C^{\bar{Q}\,\bar{m}}_{J\,-m\,Q'\,m'}, \nn \\
C^{\bar{Q}\,\bar{m}}_{J\,-m\,Q'\,m'} & = & (-1)^{J+Q'-\bar{Q}} \,C^{\bar{Q}\, -\bar{m}}_{J\,m\,Q'\,-m'}
\eea
and the orthogonality (\ref{CG_ortho}) 
leads to 
\bea
\sum_{m'=-Q'}^{Q'} (-1)^{-m'}
\begin{pmatrix} \bar{Q}  &  Q'  &  Q'  \\  -\bar{m}  & m'  &  -m' \end{pmatrix} 
 & = & (-1)^{Q'} \sqrt{2Q'+1}\,\delta_{\bar{Q}\,0}\,\delta_{\bar{m}\,0} \nn \\
& = & (-1)^{Q'} \sqrt{2Q'+1}\,\delta_{\bar{\rho},\,-1}\,\delta_{\bar{J}\, 0}\,\delta_{\bar{m}\,0}. 
\label{sum_E:sum_3j}
\eea
In addition, 
\be
\begin{Bmatrix} 1  &  \tilde{Q}'  &  \tilde{Q}'  \\  j  & j  &  j  \end{Bmatrix}  = 
\begin{Bmatrix} j  &  \tilde{Q}'  &  j  \\  1  & j  & \tilde{Q}'  \end{Bmatrix}  = (-1)^{\tilde{Q}'+2j+1}\,\frac{1}{2\sqrt{j(j+1)n}}\,
\sqrt{\frac{\tilde{Q}'(\tilde{Q}'+1)}{2\tilde{Q}'+1}} 
\label{sum_E:6j}
\ee
and 
\be
\begin{Bmatrix} 0  &  1  &  1  \\
                          Q'  & \tilde{Q}' & 1  \\
                          Q'  & \tilde{Q}' & 1 \end{Bmatrix} 
 =  \frac{(-1)^{Q'+\tilde{Q}'}}{\sqrt{3(2Q'+1)}}\,
\begin{Bmatrix} 1  &  1  &  1  \\  \tilde{Q}'  &  \tilde{Q}'  &  Q'  \end{Bmatrix} 
= \frac{-Q'(Q'+1)+\tilde{Q}'(\tilde{Q'}+1)+2}{6\sqrt{2(2Q'+1)\tilde{Q}'(\tilde{Q}'+1)(2\tilde{Q}'+1)}},
\label{sum_E:9j}
\ee                          
which are seen from, e.g., Table 9.2 in Chapter 9.12 and eq.~(2) in Chapter 10.9.1 of \cite{QTAM}.  
Plugging these into (\ref{sum_E}), we end up with 
\bea
& & \sum_{m'=-Q'}^{Q'} (-1)^{-m'}
\hat{\cE}_{\bar{J}\,-\bar{m}\,(jj)\,\bar{\rho}\,J'\,m'\,(jj) \rho'\,J'\,-m'\,(jj)\,\rho'} \nn \\
& & \hspace{7mm}= \frac{-1}{2\sqrt{j(j+1)}}\,(2J'+1+2\delta_{\rho',\,1})\,[\rho'(J'+1)-1]\,
\delta_{\bar{\rho},\,-1}\,\delta_{\bar{J}\, 0}\,\delta_{\bar{m}\,0}.
\label{sum_E_3}
\eea
Here, we have used 
\bea
& & 2j\in \Z, \qquad Q'+\tilde{Q}'+\rho'=2(J'+\delta_{\rho',\,1})\in 2\Z, \nn \\
& & -Q'(Q'+1) +\tilde{Q}'(\tilde{Q}'+1)+2 = -2[\rho'(J'+1)-1].
\eea

\subsection{(\ref{y_D})}  
\label{app:sum_D}
Let us calculate the sum of $\hat{\cD}$ with respect to $m$ in (\ref{y_D}). 

For $9j$ symbols, each elementary permutation of rows or columns gives a multiplicative sign factor, 
\be
(-1)^{\mbox{sum of all the angular momenta}},  
\ee
as seen from Chapter 10.4.1 of \cite{QTAM}. 
Namely, 
\begin{align}
\begin{Bmatrix} a & b & c \\ d & e & f \\ g & h & k \end{Bmatrix}  
&= (-1)^{a+b+\cdots+k}\begin{Bmatrix} d & e & f \\ a & b & c \\  g & h & k \end{Bmatrix}
 = (-1)^{a+b+\cdots+k} \begin{Bmatrix} a & b & c \\ g & h & k \\ d & e & f  \end{Bmatrix} \nn \\
& = (-1)^{a+b+\cdots+k} \begin{Bmatrix} b & a & c \\ e & d & f \\ h & g & k \end{Bmatrix} 
= (-1)^{a+b+\cdots+k} \begin{Bmatrix} a & c & b \\ d & f & e \\ g & k & h \end{Bmatrix}.
\label{9j_perm}
\end{align}

{}From (\ref{cD_12_21}), 
\bea
& & \sum_{m=-J}^J
\left(\hat{\cD}^{J\,m\,(jj)}_{\bar{J}\,-\bar{m}\,(jj)\,\bar{\rho}\,J\,m\,(jj)\,0}
-\hat{\cD}^{J\,m\,(jj)}_{J\,m\,(jj)\,0\, \bar{J}\,-\bar{m}\,(jj)\,\bar{\rho}}\right) 
= \left(1-(-1)^{\tilde{\bar{Q}}}\right)\sum_{m=-J}^J
\hat{\cD}^{J\,m\,(jj)}_{\bar{J}\,-\bar{m}\,(jj)\,\bar{\rho}\,J\,m\,(jj)\,0} \nn \\
& & \hspace{7mm}= \left(1-(-1)^{\tilde{\bar{Q}}}\right)\,i^{\bar{\rho}}\,(-1)^{J+2j+1}
\sqrt{3n(2\bar{J}+1)(2\bar{J}+2\bar{\rho}^2+1)}\,(2J+1)^{3/2} \nn \\
& & \hspace{10mm} \times \begin{Bmatrix} \bar{Q} & \tilde{\bar{Q}} & 1 \\ J & J & 1 \\ J & J & 0 \end{Bmatrix} 
\begin{Bmatrix} J & \tilde{\bar{Q}} & J \\ j & j & j \end{Bmatrix} 
\left(\sum_{m=-J}^J C^{J\,m}_{\bar{Q}\,-\bar{m}\,J\,m}\right). 
\label{sum_D}
\eea
Because the sum of the C-G coefficient in the last line can be written as the sum of a $3j$ symbol, 
\be
\sum_{m=-J}^J C^{J\,m}_{\bar{Q}\,-\bar{m}\,J\,m}
= \sum_{m=-J}^J (-1)^{\bar{Q}-J+m}\,\sqrt{2J+1}\,
\begin{pmatrix} \bar{Q} & J & J \\ -\bar{m} & m & -m \end{pmatrix}, 
\ee
it can be computed similarly to the derivation of (\ref{sum_E:sum_3j}): 
\be
\sum_{m=-J}^J C^{J\,m}_{\bar{Q}\,-\bar{m}\,J\,m} = (2J+1)\,\delta_{\bar{Q}\,0}\,\delta_{\bar{m}\,0} 
= (2J+1)\,\delta_{\bar{\rho},\,-1}\,\delta_{\bar{J}\,0}\,\delta_{\bar{m}\,0}  .
\label{sum_D:sum_CG}                              
\ee
Then, (\ref{sum_D}) reduces to 
\bea
& & \hspace{-7mm}\sum_{m=-J}^J
\left(\hat{\cD}^{J\,m\,(jj)}_{\bar{J}\,-\bar{m}\,(jj)\,\bar{\rho}\,J\,m\,(jj)\,0}
-\hat{\cD}^{J\,m\,(jj)}_{J\,m\,(jj)\,0\, \bar{J}\,-\bar{m}\,(jj)\,\bar{\rho}}\right) \nn \\
& & = -6i(-1)^{J+2j+1}\,\sqrt{n}\,(2J+1)^{5/2}\,
\delta_{\bar{\rho},\,-1}\,\delta_{\bar{J}\,0}\,\delta_{\bar{m}\,0} \begin{Bmatrix} 0 & 1 & 1 \\ J & J & 1 \\ J & J & 0 \end{Bmatrix} 
\begin{Bmatrix} J & 1 & J \\ j & j & j \end{Bmatrix}. 
\label{sum_D_2}
\eea

As in (\ref{sum_E:6j}), we have 
\be
\begin{Bmatrix} J & 1 & J \\ j & j & j \end{Bmatrix} = 
(-1)^{J+2j+1}\,\frac{1}{2\sqrt{j(j+1)n}}\,\sqrt{\frac{J(J+1)}{2J+1}}.  
\label{sum_D:6j}
\ee
Equation (\ref{9j_perm}) and the formula~\footnote{See, e.g., eq.~(3) in Chapter 10.9.1 of \cite{QTAM}.} 
\be
\begin{Bmatrix}
a & b & c \\ d & 0 & f \\ g & h & 0 \end{Bmatrix} =\delta_{df}\delta_{bh}\delta_{cf}\delta_{gh}\,\frac{(-1)^{a-b-c}}{(2b+1)(2c+1)}
\ee
lead to 
\be
\begin{Bmatrix} 0 & 1 & 1 \\ J & J & 1 \\ J & J & 0 \end{Bmatrix} = 
-\begin{Bmatrix} 1 & 0 & 1 \\ J & J & 1 \\ J & J & 0 \end{Bmatrix} = 
\begin{Bmatrix} J & J & 1 \\ 1 & 0 & 1 \\ J & J & 0 \end{Bmatrix} = -\frac{1}{3(2J+1)}. 
\label{sum_D:9j}
\ee
Plugging (\ref{sum_D:6j}) and (\ref{sum_D:9j}) into (\ref{sum_D_2}), we end up with 
\be
\sum_{m=-J}^J
\left(\hat{\cD}^{J\,m\,(jj)}_{\bar{J}\,-\bar{m}\,(jj)\,\bar{\rho}\,J\,m\,(jj)\,0}
-\hat{\cD}^{J\,m\,(jj)}_{J\,m\,(jj)\,0\, \bar{J}\,-\bar{m}\,(jj)\,\bar{\rho}}\right) 
= i\frac{\sqrt{J(J+1)}\,(2J+1)}{\sqrt{j(j+1)}}\,
\delta_{\bar{\rho},\,-1}\,\delta_{\bar{J}\,0}\,\delta_{\bar{m}\,0}.
\label{sum_D_3}
\ee

\subsection{(\ref{y_G})}  
\label{app:sum_G}
Here, we compute the sum of $\hat{\cG}$ with respect to $m$ in (\ref{y_G}). From the expression of 
$\hat{\cG}$ (\ref{VC_G}), the sum reduces to the sum of the C-G coefficient $C^{U\,-m}_{U\,-m\,\bar{Q}\,-\bar{m}}$. 
It can be obtained by using (\ref{sum_D:sum_CG}) as 
\be
\sum_{m=-U}^UC^{U\,-m}_{U\,-m\,\bar{Q}\,-\bar{m}} 
= (-1)^{\bar{Q}}\sum_{m=-U}^UC^{U\,-m}_{\bar{Q}\,-\bar{m}\,U\,-m}
= (-1)^{\bar{Q}}\,(2U+1)\,\delta_{\bar{Q}\,0}\,\delta_{\bar{m}\,0} 
= (2U+1)\,\delta_{\bar{\rho},\,-1}\,\delta_{\bar{J}\,0}\,\delta_{\bar{m}\,0}. 
\ee
Then, we have 
\begin{align}
\sum_{m=-U}^U\hat{\cG}^{J\,-m \,(jj) \,\kappa}_{J \,-m\,(jj)\,\kappa\,\bar{J}\,-\bar{m}\,(jj)\,\bar{\rho}}
=& -i(-1)^{\tilde{U}+2j}\,3\sqrt{2n(2\tilde{U}+1)(2J+1)(2J+2)}\,(2U+1)
 \nn \\
&\times \delta_{\bar{\rho},\,-1}\,\delta_{\bar{J}\,0}\,\delta_{\bar{m}\,0}\,
\begin{Bmatrix} U & \tilde{U} & \frac12 \\ U & \tilde{U} & \frac12 \\ 0 & 1 & 1 \end{Bmatrix}
\begin{Bmatrix} \tilde{U} & \tilde{U} & 1 \\ j & j & j \end{Bmatrix}. 
\label{sum_G}
\end{align}

By a similar derivation of (\ref{sum_E:6j}) and (\ref{sum_E:9j}), 
\bea
& & \begin{Bmatrix} \tilde{U} & \tilde{U} & 1 \\ j & j & j \end{Bmatrix} 
= (-1)^{\tilde{U}+2j+1}\,\frac{1}{2\sqrt{j(j+1)n}}\,\sqrt{\frac{\tilde{U}(\tilde{U}+1)}{2\tilde{U}+1}},  
\label{sum_G:6j}
\\
& & \begin{Bmatrix} U & \tilde{U} & \frac12 \\ U & \tilde{U} & \frac12 \\ 0 & 1 & 1 \end{Bmatrix}
= \frac{(-1)^{U+\tilde{U}+\frac32}}{\sqrt{3(2U+1)}}\,
\begin{Bmatrix} \frac12 & \tilde{U} & U \\ \tilde{U} & \frac12 & 1 \end{Bmatrix}
= \frac{\frac34 -U(U+1)+\tilde{U}(\tilde{U}+1)}{3\sqrt{2\tilde{U}(\tilde{U}+1)(2U+1)(2\tilde{U}+1)}}.\nn \\
\label{sum_G:9j}
\eea
Note that $U+\tilde{U}+\frac32 = 2J+2 \in \Z$ since $J$ is an integer or a half-integer.

Plugging (\ref{sum_G:6j}) and (\ref{sum_G:9j}) into (\ref{sum_G}), we end up with
\bea
\sum_{m=-U}^U\hat{\cG}^{J\,-m \,(jj) \,\kappa}_{J \,-m\,(jj)\,\kappa\,\bar{J}\,-\bar{m}\,(jj)\,\bar{\rho}}
& = & i\frac{1}{2\sqrt{j(j+1)}}\,\sqrt{\frac{2U+1}{2\tilde{U}+1}}\,\sqrt{(2J+1)(2J+2)}\,
 \delta_{\bar{\rho},\,-1}\,\delta_{\bar{J}\,0}\,\delta_{\bar{m}\,0} \nn \\
& & \times \left[\frac34 -U(U+1) +\tilde{U}(\tilde{U}+1)\right] \nn \\
& = &i\frac{1}{2\sqrt{j(j+1)}}\,(2U+1)\,\left[\frac34-\kappa\left(J+\frac34\right)\right] 
\,\delta_{\bar{\rho},\,-1}\,\delta_{\bar{J}\,0}\,\delta_{\bar{m}\,0}.\nn \\
\label{sum_G_3}
\eea

\subsection{Gauge transformation property of $y_{J\,m,\,\rho}(p)$}
\label{app:gauge_y}
Here, we consider the gauge transformation of $y_{J\,m,\,\rho}(p)$. 
The gauge transformations of $\tilde{X}_{8,9,10}$ are
\bea
\delta \tilde{X}_9 & = & i\frac{M}{3}[L_1, \,\Omega] + i[\tilde{X}_9,\,\Omega], \nn \\
\delta \tilde{X}_{10} & = & i\frac{M}{3}[L_2, \,\Omega] + i[\tilde{X}_{10},\,\Omega], \nn \\
\delta \tilde{X}_8 & = & i\frac{M}{3}[L_3, \,\Omega] + i[\tilde{X}_8,\,\Omega], 
\eea 
where $\Omega$ is a gauge transformation parameter. 
Then the gauge transformation of $\vec{\tilde{Y}}=(\tilde{X}_9, \tilde{X}_{10}, \tilde{X}_8)^T$ is expressed as  
\bea
\delta\vec{\tilde{Y}}(p) = i\frac{M}{3}\vec{L}\circ\tilde{\Omega}(p) +
i\int\frac{d^2q}{(2\pi)^2}\,[\vec{\tilde{Y}}(q),\,\tilde{\Omega}(p-q)]
\label{gauge_Y}
\eea
in the momentum representation in $\R^2$. 
Furthermore, in terms of modes on fuzzy $S^2$, 
\bea
\vec{\tilde{Y}}(p) & = & \sum_{\rho=-1}^1\sum_{J=\delta_{\rho,\,0}}^{2j-\delta_{\rho,\,-1}} \sum_{m=-Q}^Q 
\vec{\hat{Y}}^\rho_{J\,m\,(jj)}\otimes y_{J\,m,\,\rho}(p), \label{mode_y} \\
\tilde{\Omega}(p) & = & \sum_{J=0}^{2j} \sum_{m=-J}^J \hat{Y}^{(jj)}_{J\,m}\otimes \omega_{J\,m}(p), 
\label{gauge_omega}
\eea  
(\ref{gauge_Y}) reads 
\bea
& &\hspace{-7mm}\delta y_{J\,m,\,\rho}(p) = \delta_{\rho,\,0}\,i\frac{M}{3}\sqrt{J(J+1)}\,\omega_{J\,m}(p) \nn \\
& & + \sum_{\rho'=-1}^1\sum_{J'=\delta_{\rho',\,0}}^{2j-\delta_{\rho',\,-1}} \sum_{m=-Q'}^{Q'}\sum_{J''=0}^{2j} 
\sum_{m''=-J''}^{J''}(-1)^{m''-m+1}\hat{\cD}^{J''\,-m''\,(jj)}_{J\,-m\,(jj)\,\rho\,J'\,m'\,(jj)\,\rho'} 
\nn \\
& & \hspace{7mm} \times i\int\frac{d^2q}{(2\pi)^2}\,\left[y_{J'\,m',\,\rho'}(q)\,\omega_{J''\,m''}(p-q) 
-(-1)^{\tilde{Q}'+\tilde{Q}-J''}\omega_{J''\,m''}(p-q)\,y_{J'\,m',\,\rho'}(q)\right], \nn \\
& & 
\label{gauge_y}
\eea
where we have used (\ref{cD_12_21}). 
   
Let us focus on the case of $J=m=0$ where the first term of (\ref{gauge_y}) vanishes:
\bea
& & \hspace{-7mm}\delta y_{0\,0,\,\rho}(p) =  
\sum_{\rho'=-1}^1\sum_{J'=\delta_{\rho',\,0}}^{2j-\delta_{\rho',\,-1}} \sum_{m=-Q'}^{Q'}\sum_{J''=0}^{2j} 
\sum_{m''=-J''}^{J''}(-1)^{m''+1}\hat{\cD}^{J''\,-m''\,(jj)}_{0\,0\,(jj)\,\rho\,J'\,m'\,(jj)\,\rho'} 
\nn \\
& & \times i\int\frac{d^2q}{(2\pi)^2}\,\left[y_{J'\,m',\,\rho'}(q)\,\omega_{J''\,m''}(p-q) 
-(-1)^{\tilde{Q}'+\delta_{\rho,\,-1}-J''}\omega_{J''\,m''}(p-q)\,y_{J'\,m',\,\rho'}(q)\right] . \nn \\
& & 
\label{gauge_y00}
\eea
Since $y_{0\,0,\,\rho=0}(p)$ does not exist, we consider the cases $\rho=\pm 1$. 
Note that $\tr_k \,y_{0\,0,\,\rho}(p)$ is gauge invariant when each term of the summand satisfies 
\be
\tilde{Q}'+\delta_{\rho,\,-1}-J''=J'-J''+\delta_{\rho,\,-1}+\delta_{\rho',\,-1} \in 2\Z.
\label{gauge_inv_cond}
\ee  

For $\rho=1$, $\hat{\cD}$ can be computed as 
\bea
\lefteqn{\hat{\cD}^{J''\,-m''\,(jj)}_{0\,0\,(jj)\,1\,J'\,m'\,(jj)\,\rho'} 
 = \delta_{\rho',\,0}\left[i\frac{m'}{\sqrt{J'(J'+1)}}\,\delta_{J'\,J''}\,\delta_{m'+m'',\,0}\right]} \nn \\
 &  & \hspace{7mm}+ \delta_{\rho',\,1}\left[-\sqrt{\frac{(J'-m'+1)(J'+m'+1)}{(J'+1)(2J'+1)}}\,
\delta_{J'\,J''}\,\delta_{m'+m'',\,0}\right] \nn \\ 
& &\hspace{7mm} + \delta_{\rho',\,-1}\left[-\sqrt{\frac{(J'-m'+1)(J'+m'+1)}{(J'+1)(2J'+3)}}\,
\delta_{J'+1,\,J''}\,\delta_{m'+m'',\,0}\right] ,
\eea
from which we find that (\ref{gauge_inv_cond}) is satisfied for each case of $\rho'=0,\pm 1$. 
Hence, $\tr_k \,y_{0\,0,\,\rho=1}(p)$ is gauge invariant.  

On the other hand, for $\rho=-1$, $\hat{\cD}$ becomes 
\bea
\lefteqn{\hat{\cD}^{J''\,-m''\,(jj)}_{0\,0\,(jj)\,-1\,J'\,m'\,(jj)\,\rho'} 
 =  \delta_{\rho',\,0}\left[\frac{i}{2\sqrt{j(j+1)}}\,\sqrt{J'(J'+1)}\,\delta_{J'\,J''}\,\delta_{m'+m'',\,0}
 \right]} \nn \\
 & & + \delta_{\rho',\,1}\left[\frac{-1}{2\sqrt{j(j+1)}}\,
 \sqrt{\frac{(J'+2j+2)(2j-J')(J'+1)}{2J'+3}}\,\delta_{J'+1,\,J''}\,\delta_{m'+m'',\,0}\right] \nn \\
 & & + \delta_{\rho',\,-1}\left[\frac{-1}{2\sqrt{j(j+1)}}\,
 \sqrt{\frac{(J'+2j+2)(2j-J')(J'+1)}{2J'+1}}\,\delta_{J'\,J''}\,\delta_{m'+m'',\,0}\right].  
\eea
Thus, (\ref{gauge_inv_cond}) is satisfied for $\rho'=\pm 1$, but not for $\rho'=0$. 
This shows that $\tr_k \,y_{0\,0,\,\rho=-1}(p)$ is not gauge invariant.

\subsection{$\tr_n\,\vec{\hat{Y}}^{\rho=-1}_{0\,0\,(jj)}$}
\label{app:tr_vecY}
Here, we show that $\tr_n\,\vec{\hat{Y}}^{\rho=-1}_{0\,0\,(jj)}$ vanishes, which is consistent with the gauge invariance.   

{}From the definition of vector and scalar fuzzy spherical harmonics, 
\bea
& & \vec{\hat{Y}}^{\rho=-1}_{0\,0\,(jj)} = \begin{pmatrix} 
\frac{-i}{\sqrt{6}}\,\left(\hat{Y}^{(jj)}_{1\,1}-\hat{Y}^{(jj)}_{1\,-1}\right) \\ 
\frac{-1}{\sqrt{6}}\,\left(\hat{Y}^{(jj)}_{1\,1}+\hat{Y}^{(jj)}_{1\,-1}\right) \\ 
\frac{i}{\sqrt{3}}\,\hat{Y}^{(jj)}_{1\,0}
\end{pmatrix}, 
\\
& & \tr_n\,\hat{Y}^{(jj)}_{J\,m} = \sqrt{n}\sum_{r=-j}^j (-1)^{-j+r}\,C^{J\,m}_{j\,r\,j\,-r}. 
\label{tr_Y}
\eea
Due to the angular momentum conservation, (\ref{tr_Y}) vanishes for nonzero $m$: 
\be
\tr_n \,\hat{Y}^{(jj)}_{1\,1} =  \tr_n \,\hat{Y}^{(jj)}_{1\,-1} = 0.
\label{tr_Y_11_1-1}
\ee
We note that $C^{1\,0}_{j\,r\,j\,-r} = (-1)^{2j-1}\,C^{1\,0}_{j\,-r\,j\,r}$ and $j-r\in \Z$ to show that 
\bea
\tr_n\,\hat{Y}^{(jj)}_{1\,0} & = & \sqrt{n}\sum_{r=-j}^j (-1)^{-j+r}\,C^{1\,0}_{j\,r\,j\,-r} 
= \sqrt{n}\sum_{r=-j}^j (-1)^{-j+r}\,(-1)^{2j-1}\,C^{1\,0}_{j\,-r\,j\,r}\nn \\
 & = & -\sqrt{n}\sum_{r=-j}^j (-1)^{j-r}\,C^{1\,0}_{j\,r\,j\,-r} = - \tr_n\,\hat{Y}^{(jj)}_{1\,0}=0. 
\label{tr_Y_10} 
\eea
Thus, (\ref{tr_Y_11_1-1}) and (\ref{tr_Y_10}) lead to 
\be
\tr_n\,\vec{\hat{Y}}^{\rho=-1}_{0\,0\,(jj)} = 0. 
\ee

\section{Proof of Theorems 1 and 2}
\label{app:proof}
In this appendix, we give proofs of Theorems 1 and 2, which are claimed in the text. 

\subsection{Theorem 1}
\label{app:proof R2}
Here, in order to show Theorem 1, we evaluate $\cA_{II}^{\rm UV}$ in (\ref{cAUV_II}): 
\bea
 \cA^{\rm UV}_{II} & = & \left(\frac{M}{3}\right)^2 \sum_{J',J''\in \text{Region II}} 
n \,f(J',J'',J)\, 
\Wsj^2 
\, I_{a,b}, 
 \label{AII} \\
 I_{a,b} & \equiv & \int \frac{d^2\tq}{(2\pi)^2}
\frac{g(\tp,\tq)}{M_{a,b}\left(\tP_i(J') ; \tQ_k(J'') ; \tq,\tp
 \right)}, 
 \label{tptJ}
\eea
where $g(\tp, \tq)$ and $f(J', J''; J)$ are supposed to have properties (\ref{asympt g}) and (\ref{form_f}). 
We also assume that the integration $I_{a,b}$ converges.  
Recall that we are considering the situation of
$\tp_\mu = \frac{3}{M} p_\mu$ and $J=\frac{3}{M} u$ typically of the order of $\cO(M^{-1})$. 
In Region II defined by (\ref{total_region}) and (\ref{regions}), 
it is convenient to change the summation variables 
from $J'$ and $J''$ to 
\begin{equation}
J_+\equiv J'+J'', \qquad 
\Delta \equiv J'-J''. 
\label{J+-}
\end{equation}
Then, $J_+ \gg \Delta$ for a sufficiently small $M$. 
By noting that both $J_+$ and $\Delta$ take even integers or odd integers since 
$J'=\frac{1}{2} \left( J_+ + \Delta \right)$ and 
$J''=\frac{1}{2} \left( J_+ - \Delta \right)$ should be integers, 
the summation can be rewritten as
\begin{equation}
\sum_{J',J''\in {\rm Region\ II}}  =
\sum_{J_B \le J_+ \le 4j}^{\rm even} \sum_{|\Delta|\le J}^{\rm even}
+ \sum_{J_B \le J_+ \le 4j}^{\rm odd} \sum_{|\Delta|\le J}^{\rm odd}. 
\end{equation}
{}From $|\Delta| \le J$, 
it can be seen that there is a constant $C_1$ such that 
\begin{equation}
|f(J',J'';J)| \le C_1 M^{-N_J-N_\Delta} J_+^{N_1+N_2}. 
\label{bound1}
\end{equation}

In this region, we can use Edmonds' formula, which holds for 
$a,b,c \gg f,m,n$ ($f,m,n\in \Z$ or $\Z+\frac{1}{2}$)~\footnote{See, e.g., 
eq. (14) in Chapter 9.9.1 of \cite{QTAM}.}: 
\begin{equation}
\left\{ \begin{matrix} a & a+n & f \\ b+m & b & c \end{matrix} \right\}
\simeq \frac{ (-1)^{a+b+c+f+m}}{\sqrt{ (2a+1)(2b+1) }}
d^{f}_{mn}(\beta),
\label{Edmonds}
\end{equation}
where 
\begin{equation}
\cos\beta = \frac{ a(a+1) + b(b+1) - c(c+1) }{2\sqrt{a(a+1)b(b+1)}}, 
\end{equation}
and $d^f_{mn}(\beta)$ is a real function related to the Wigner $D$-function. It is explicitly given as
\bea
d_{m\,n}^f(\beta) &=& (-1)^{f-n}\frac{1}{2^f}\sqrt{\frac{(f+m)!}{(f-m)!(f+n)!(f-n)!}}\,
(1-X)^{-\frac{m-n}{2}}\,(1+X)^{-\frac{m+n}{2}}\nn \\
& & \times \,\frac{d^{f-m}}{dX^{f-m}}\left[(1-X)^{f-n}(1+X)^{f+n}\right]
\label{d-function}
\eea
with $X= \cos \beta$ and $0<\beta<\frac{\pi}{2}$. 
By putting $a=J''$, $n=\Delta$, $f=J$, $m=0$ and $b=c=j$ in this formula, 
we obtain 
\begin{equation}
n\Wsj^2 \simeq \frac{1}{2J''+1} \left( d^J_{0\Delta}(\beta) \right)^2  
\label{edmonds}
\end{equation}
and
\begin{align}
\cos\beta = \frac{1}{2} \sqrt{ \frac{J''(J''+1)}{j(j+1)}} 
= \frac{J_+}{4j}\left(1+\cO\left(\frac{\Delta}{J+},\,\frac{1}{J_+} \right)\right), 
\end{align}
which means that there is a constant $C_2$ such that 
\begin{equation}
n\Wsj^2 \le \frac{C_2}{J_+}\left( d^J_{0\Delta}(\beta_+) \right)^2 \qquad \mbox{with} \qquad \cos\beta_+ \equiv \frac{J_+}{4j}. 
\label{bound2}
\end{equation}

Next, let us evaluate the integral $I_{a,b}$ in (\ref{tptJ}). 
Recall that 
the polynomials $\tP_i(J)$ and $\tQ_k(J)$ in the denominator of 
the integrand are actually of the form (\ref{tPtQ_actual}). 
Because of $|\tp| \ll J''$ and $|\Delta| \ll J_+$ in Region II, 
we see that 
$\frac{1}{\tq^2+\tP_k(J')}$ and  
$\frac{1}{(\tq+\tp)^2+\tQ_k(J'')}$ 
are bounded from the above as 
\begin{equation}
\frac{1}{\tq^2+\tP_i(J')}< \frac{c'}{\tq^2+J_+^2}, \qquad 
\frac{1}{(\tq+\tp)^2+\tQ_k(J'')}< \frac{c''}{\tq^2+J_+^2},
\end{equation}
where $c'$ and $c''$ are some constants of the order of $\cO(1)$. 
Therefore, with use of (\ref{asympt g}), 
$I_{a,b}$ is bounded as
\begin{align}
|I_{a,b}| &\le {C'}_3 \int \frac{d^2\tq}{(2\pi)^2}
\frac{|g(\tp,\tq)|}{
\left( \tq^2 + J_+^2 \right)^{a+b}
}
\le C_3 M^{-n_p} J_+^{2+n_q-2(a+b)}. 
\label{bound3}
\end{align}
$C'_3$ and $C_3$ are constants of the order of $\cO(1)$.  

Note that the identity 
\begin{equation}
\sum_{\Delta=-J}^J \left( d^J_{0\Delta}(\beta_+)\right)^2  = 1 
\label{d2formula1}
\end{equation}
follows from 
\bea
& & d^J_{MM'}(\beta)=(-1)^{M-M'}\,d^J_{M'M}(\beta), \label{formula-d1}\\
& & \sum_{M''=-J}^J (-1)^{M''-M'}\,d_{MM''}^J(\beta)\,d^J_{M''M'}(\beta)=\delta_{MM'} \label{formula-d2}.  
\eea
These are seen, e.g., in eq. (1) in Chapter 4.4 and eq. (10) in Chapter 4.7.2 of \cite{QTAM}. 
Combining (\ref{bound1}), (\ref{bound2}), (\ref{bound3}) and (\ref{d2formula1}), 
we have
\be
|A_{II}^{UV}| 
=C_{II} M^{2-n_p-N_J-N_\Delta}
\sum_{J_B \le J_+ \le 4j}
J_+^{1+n_q+N_1+N_2-2(a+b)} 
\label{AII_bound}
\ee
with an $\cO(1)$ constant $C_{II}$. 

Recalling that $j=\cO(M^{-2})$ and $J_B=\cO(M^{-2\alpha})$, we end up with 
\begin{align}
|\cA_{II}^{UV}| \le 
\begin{cases}
C_+ M^{W^+} & (w>0) \\
C_0M^{W^-}|\ln M| & (w=0) \\
C_-M^{-2\alpha w-n_p-N_J-N_\Delta+2} 
< 
C_-M^{W^-}  & (w<0), 
\end{cases}
\end{align}
where 
$w \equiv N_1+N_2-2(a+b)+n_q+2$, 
$W^+ \equiv -N_J-N_\Delta -n_p + 2 -2w$ and  
$W^- \equiv -N_J-N_\Delta -n_p + 2 -w$. 
$C_+$, $C_0$ and $C_-$ are $M$-independent constants. 
That immediately proves the theorem.

\subsection{Theorem 2}
\label{app:proof R1}
In Region I, we express 
$\cA_I^{\rm UV}$ in (\ref{cAUV_I}) 
in terms of rescaled variables of the order of $\cO(1)$: 
\begin{equation}
 u=\frac{M}{3}\,J, \qquad 
 u'=\frac{M}{3}\,J', \qquad 
 u''=\frac{M}{3}\,J''.  
\end{equation}
 Then $f(J',J'';J)$ becomes
 \begin{equation}
 f(J',J'';J) = \hat{C}\ \left(\frac{M}{3}\right)^{-(N_1+N_2)-N_J-N_\Delta} 
 (u'-u'')^{N_\Delta} (u')^{N_1} (u'')^{N_2}, \quad (\hat{C} = C u^{N_J}
 ). 
 \label{fRI}
 \end{equation}
Furthermore, by noting $J,J',J'' \ll j$, the Wigner $6j$ symbol can be approximated as 
\bea
 \left\{ \begin{matrix} J' & J'' & J \\ j & j & j  \end{matrix} \right\}^2 
& \simeq & \frac{1+(-1)^{J'+J''+J}}{2} \frac{1}{n}
\frac{ \left[ \left( \frac{J+J'+J''}{2}\right)! \right]^2}{\left[ \left(  \frac{-J+J'+J''}{2}\right)!  
 \left(  \frac{J-J'+J''}{2} \right)!\left(  \frac{J+J'-J''}{2} \right)!  \right]^2}\nn \\
& & \times \frac{  (J+J'+J'')!  (J-J'+J'')!  (J+J'-J'')!}{ (J+J'+J''+1)!} 
\label{6j_RI}
\eea
with $n=2j+1$, e.g., according to eq. (4) in Chapter 9.9.1 and eq.~(32) in Chapter 8.5.2 of \cite{QTAM}. 
Applying the Stirling formula to the factorials in the r.h.s.~\footnote{
Since $J, J', J'' \geq \cO(M^{-1})\gg 1$, this is justified for a generic point in Region I 
except for in the vicinity of the boundary. We can see that the contribution from such exceptional points is negligible 
in the Moyal limit.}
leads to 
\begin{align}
 \left\{ \begin{matrix} J' & J'' & J \\ j & j & j  \end{matrix} \right\}^2 
 &\simeq  \frac{2}{\pi n} \frac{1}{\sqrt{ (J+J'+J'')(-J+J'+J'')(J-J'+J'')(J+J'-J'') }}  
\label{6j in RI} \\
&\simeq  \left( \frac{M}{3} \right)^2 \frac{2}{\pi n} 
\frac{1}{\sqrt{\left[ (u'+u'')^2 - u^2 \right]  \left[ u^2 - (u'-u'')^2  \right] } } 
\label{WRI}
\end{align}
for $J'+J''+J$ even. 

In the meantime, we assume that $a, \,b\geq 1$. 
The other cases, $a=0$ or $b=0$, will be considered separately later.  
The integral $I_{a,b}$ in (\ref{tptJ}) becomes~\footnote{
Note that $g(\tp,\tq)=\left(\frac{M}{3}\right)^{-n_p-n_q}g(p,q)$ as $g$ is a homogeneous polynomial.} 
\begin{align}
I_{a,b} \sim &
\left(\frac{M}{3}\right)^{2(a+b)-n_p-n_q-2}
\int \frac{d^2q}{(2\pi)^2}
\frac{g(p,q)}{\left(q^2+(u')^2\right)^a\left((q+p)^2+(u'')^2\right)^b} \nn \\
=&\left(\frac{M}{3}\right)^{2(a+b)-n_p-n_q-2}
\frac{\Gamma(a+b)}{\Gamma(a)\Gamma(b)}
\int_0^1 dt\  t^{b-1}(1-t)^{a-1}  \int  \frac{d^2q}{(2\pi)^2} 
\frac{ g(p,q) }{ \left[ (q+tp)^2 + X(t) \right]^{a+b} } \nn \\
=&\left(\frac{M}{3}\right)^{2(a+b)-n_p-n_q-2}
\frac{\Gamma(a+b)}{\Gamma(a)\Gamma(b)}
\int_0^1 dt\  t^{b-1}(1-t)^{a-1}  \int  \frac{d^2q}{(2\pi)^2} 
\frac{ g(p,q-tp) }{ \left[ q^2 + X(t) \right]^{a+b} }, 
\label{IRI}
\end{align}
where we have used the Feynman integral formula, 
\begin{equation}
 \frac{1}{A^aB^b}= \frac{\Gamma(a+b)}{\Gamma(a)\Gamma(b)} \int_0^1 dt
\frac{t^{a-1}(1-t)^{b-1}}{\left(At+B(1-t)\right)^{a+b}},  
\label{Feynman_int}
\end{equation} 
and 
\begin{equation}
X(t) \equiv t(1-t)p^2 + (1-t) (u')^2 + t (u'')^2. 
\end{equation}

Combining (\ref{fRI}), (\ref{WRI}), and (\ref{IRI}), we have
\bea
\cA^{\rm UV}_I & = & C_I
 \left(\frac{M}{3}\right)^{-(N_1+N_2)-N_J-N_\Delta+2(a+b)-n_p-n_q} \nn \\
& &\times \int_{D_u} du' du'' 
 \frac{(u')^{N_1} (u'')^{N_2} (u'-u'')^{N_\Delta}} 
  {\sqrt{\left[ (u'+u'')^2 - u^2 \right]  \left[ u^2 - (u'-u'')^2  \right]}} \nn \\
& &\times \int_0^1 dt\  t^{b-1}(1-t)^{a-1}  \int  \frac{d^2q}{(2\pi)^2} 
\frac{ g(p,q-tp) }{ \left[ q^2 + X(t) \right]^{a+b} },
\label{SI}
\eea
where the integration region $D_u$ stands for 
\be
D_u\equiv \{(u',u'')| \,u \le u'+u'' \le u_B, \,|u'-u''| \le u\} 
\label{Du}
\ee 
with $u_B\equiv\frac{M}{3} J_B=\cO(M^{1-2\alpha})\gg 1$, and $C_I$ is an $\cO(1)$ constant.

Note that the exponent of $\frac{M}{3}$ is nothing but $W^-$
and is always positive. 
In fact, 
\begin{description}
\item $w>0$: \, 
{}From Theorem 1, $W^+>0$ and $W^- = W^+ + w > W^+ > 0$; 
\item $w\leq 0$:\,
$W^- > 0$ from Theorem 1. 
\end{description}

Similarly to the case of Theorem 1, let us consider the situation that the $q$-integrals in $I_{a,b}$ converge in the UV region, 
because the assumption of Theorem 2, that $\cA_{II}^{\rm UV}$ vanishes, implies that there is no UV divergence in $\cA_I^{\rm UV}$.  
The only possible divergence in the integration in (\ref{SI}) is from the IR region: 
$u'\sim 0$ or $u''\sim 0$. 

Note that $g(p,q-tp)$ can be expanded as 
\begin{equation}
 g(p,q-tp) = \sum_{\ell\geq 0}\alpha_{\ell}(t) q^{2\ell},  
 \label{ell-expand}
\end{equation}
because parity-odd terms with respect to $q_\mu \to -q_\mu$ trivially vanish in the integration. 
The coefficient $\alpha_{\ell}(t)$ that depends on $p^2$ is a polynomial of $t$.  
After the rescaling $q_\mu = X(t)^{1/2}\hat{q}_\mu$, we obtain 
\begin{align}
\cA_I^{\rm UV}
\sim&\left(\frac{M}{3}\right)^{W^-} \sum_{\ell\geq 0}
\int \frac{d^2 \hq}{(2\pi)^2}
 \frac{\hq^{2\ell}}{\left(\hq^2+1\right)^{a+b}} \nn \\
&\times 
\int_{D_u} du' du'' 
 \frac{(u')^{N_1} (u'')^{N_2} (u'-u'')^{N_\Delta}} 
  {\sqrt{\left[ (u'+u'')^2 - u^2 \right]  \left[ u^2 - (u'-u'')^2  \right]}}\,I_\ell(u',u'')  , 
\\
I_\ell(u',u'')\equiv & 
\int_0^1 dt\  \frac{t^{b-1}(1-t)^{a-1}\alpha_\ell(t)}  
{\left[t(1-t)p^2+(1-t)(u')^2+t(u'')^2\right]^{a+b-\ell-1}}.
\end{align}
As mentioned above, we are considering the situation that the $\hat{q}$-integrals UV converge, i.e.  
\begin{equation}
 a+b-\ell-1 >0
\end{equation}
for each $\ell$, and $a, \,b\geq 1$ in the computation here. 

For arbitrary $\ell$ such that $\alpha_\ell(t)\neq 0$, the most singular behaviors as $u'\sim 0$ and $u''\sim 0$ come 
from $t\sim 0$ and $t\sim 1$ in the integral $I_\ell(u',u'')$, 
respectively. 
We evaluate them with the assumption that $\alpha_\ell(t)$ behaves as (\ref{alpha_l(t)}). 
For $u'\sim 0$, the dominant contributions to $I_\ell$ are 
\bea
I_\ell(u'\sim 0,\, u'') & \sim & \int^{(u')^2}_0dt
\,\frac{t^{b-1+a_\ell^{(0)}}}{\left[(u')^2 + t((u'')^2+p^2-(u')^2)-t^2p^2\right]^{a+b-\ell-1}}  
\nn \\
& = & (u')^{-2(a-\ell-a^{(0)}_\ell-1)}\int^1_0dt \,
\frac{t^{b-1+a_\ell^{(0)}}}{\left[1 + t((u'')^2+p^2-(u')^2)-t^2(u')^2p^2 \right]^{a+b-\ell-1}} \nn \\
& \sim & (u')^{-2(a-\ell-a^{(0)}_\ell-1)}. 
\label{I_l(0,u)}
\eea 
The integral in the second line is finite since $u''\gtrsim u=\cO(1)$ for $u'\sim 0$ in $D_u$.  
Similarly, we have 
\be
I_\ell(u', u''\sim 0)\sim (u'')^{-2(b-\ell-a^{(1)}_\ell-1)}.
\label{I_l(u,0)}
\ee

The IR contribution in $\cA_I^{\rm UV}$ around $(u', u'')=(0,u)$ comes from (\ref{I_l(0,u)}):  
\be
\cA_{I\,(0,u)}^{\rm UV}\equiv M^{W^-}
\int_{(0,u)}du'du''\frac{(u')^{N_1-2(a-\ell-a^{(0)}_\ell-1)}}{\sqrt{\left[ (u'+u'')^2 - u^2 \right]  \left[ u^2 - (u'-u'')^2  \right]}},
\label{cAUV_I(0,u)_def}
\ee
which is evaluated by changing the variables as 
\be
u'=x\cos\theta, \qquad u''=u+x\sin\theta
\ee
with $x\gtrsim \frac{M}{3}$ and $-\frac{\pi}{4}<\theta<\frac{\pi}{4}$. Then, 
\bea
\cA_{I\,(0,u)}^{\rm UV} & = & M^{W^-}\left(\int_{M/3}dx\,x^{N_1-2(a-\ell-a^{(0)}_\ell-1)}\right)\frac{1}{u}
\int^{\pi/4}_0d\theta\frac{(\cos\theta)^{N_1-2(a-\ell-a^{(0)}_\ell-1)}}{\sqrt{\cos(2\theta)}} \nn \\
& \sim & M^{W^- +N_1-2(a-\ell-a^{(0)}_\ell-1)+1}. 
\label{cAUV_I(0,u)}
\eea
Note that the $\theta$-integral is finite. The remaining IR contribution around $(u',u'')=(u,0)$, 
\be
\cA_{I\,(u,0)}^{\rm UV}\equiv M^{W^-}
\int_{(u,0)}du'du''\frac{(u'')^{N_2-2(b-\ell-a^{(1)}_\ell-1)}}{\sqrt{\left[ (u'+u'')^2 - u^2 \right]  \left[ u^2 - (u'-u'')^2  \right]}},
\ee
can be obtained in the same manner as 
\be
\cA_{I\,(u,0)}^{\rm UV}\sim M^{W^- +N_2-2(b-\ell-a^{(1)}_\ell-1)+1}. 
\label{cAUV_I(u,0)}
\ee

{}From (\ref{cAUV_I(0,u)}) and (\ref{cAUV_I(u,0)}), we can see that $\cA_I^{\rm UV}$ vanishes in the Moyal limit 
if both 
\begin{equation}
	W^- + \left(N_1-2(a-\ell-a_\ell^{(0)}-1)\right)+1 
\end{equation}
and 
\begin{equation}
	W^- +  \left(N_2-2(b-\ell-a_\ell^{(1)}-1)\right)+1 
\end{equation}
are positive for arbitrary $\ell$ such that $\alpha_\ell(t)\neq 0$. 
This is equivalent to the statement of Theorem 2 for $a, \,b\geq 1$. 

Next, let us consider the case that one of $a$ and $b$ is zero (say $b=0$). 
Without using (\ref{Feynman_int}), we have
\bea
\left.\cA_I^{\rm UV}\right|_{b=0} & = & C_I
 \left(\frac{M}{3}\right)^{W^-} 
\int_{D_u} du' du'' 
 \frac{(u')^{N_1} (u'')^{N_2} (u'-u'')^{N_\Delta}} 
  {\sqrt{\left[ (u'+u'')^2 - u^2 \right]  \left[ u^2 - (u'-u'')^2  \right]}} \nn \\
& & \times \int  \frac{d^2q}{(2\pi)^2} 
\frac{ g(p,q) }{ \left[ q^2 + (u')^2 \right]^{a} }.
\label{SI_b=0}
\eea
The rescaling $q_\mu = u'\hat{q}_\mu$ and the expansion (\ref{ell-expand}) with $t=0$ lead to 
\begin{align}
\left.\cA_I^{\rm UV}\right|_{b=0} \sim & \left(\frac{M}{3}\right)^{W^-} \sum_{\ell\geq 0}\alpha_\ell(0)
\int \frac{d^2\hq}{(2\pi)^2}\frac{\hq^{2\ell}}{\left(\hq^2+1\right)^{a}} \nn \\
&\times \int_{D_u} du' du'' 
 \frac{(u')^{N_1+2(-a+\ell+1)} (u'')^{N_2} (u'-u'')^{N_\Delta}} 
  {\sqrt{\left[ (u'+u'')^2 - u^2 \right]  \left[ u^2 - (u'-u'')^2  \right]}}.
\end{align}
Here, IR singular behavior around $(u',u'')= (0, u)$ has the same form as (\ref{cAUV_I(0,u)_def}) with $a_\ell^{(0)}=0$: 
\bea
\left.\cA_{I\,(0,u)}^{\rm UV}\right|_{b=0} & = & M^{W^-}
\int_{(0,u)}du'du''\frac{(u')^{N_1-2(a-\ell-1)}}{\sqrt{\left[ (u'+u'')^2 - u^2 \right]  \left[ u^2 - (u'-u'')^2  \right]}} \nn \\
& \sim & M^{W^-+N_1-2(a-\ell-1)+1},
\eea
while the contribution around $(u',u'')=(u,0)$ becomes proportional to $M^{W^-+N_2+1}$, which clearly vanishes as $M\to 0$. 
This proves Theorem 2 for the $b=0$ case.

\section{Flat limit of fuzzy spherical harmonics}
\label{app:FSH flat limit}
In this appendix, we express the fuzzy spherical harmonics in terms of plane waves on the Moyal plane in order to 
prepare to obtain the one-loop effective action in the Moyal limit (the limit at Step 1 in section~\ref{sec:loop}). 
Once the expression is obtained, taking the commutative limit (at Step 2) is straightforward. 

\subsection{$|j\,r\ket$ near $r=j$ in large $j$}
Let us consider the spin-$(j, r)$ eigenstates of the $SU(2)$ generators $\vec{L}$ given 
in (\ref{SU2basis_norm}) and (\ref{SU2 basis}). 
For large $j$, we focus on a region near the north pole ($r=j$) to put 
\be
r =j-s  \qquad (s=0,1,2,\cdots) \qquad \mbox{with}\qquad s\ll \cO(j).
\ee
Then, the second and third equations in (\ref{SU2 basis}) become 
\bea
L_+|j\,j-s\ket & = & \sqrt{2j}\sqrt{s}\left[1+\cO(s/j)\right]|j\,j-s+1\ket, \nn \\
L_-|j\,j-s\ket & = & \sqrt{2j}\sqrt{s+1}\left[1+\cO(s/j)\right]|j\,j-s-1\ket, 
\eea
which suggests the definition 
\be
|s\kett \equiv |j\,j-s\ket, \quad a\equiv \frac{1}{\sqrt{2j}}L_+,\quad 
a^\dagger \equiv \frac{1}{\sqrt{2j}}L_- 
\ee
so that $a$ ($a^\dagger$) acts on $|s\kett$ as an annihilation (creation) operator:   
\bea
& & a|s\kett = \sqrt{s}|s-1\kett, \qquad a^\dagger |s\kett = \sqrt{s+1} |s+1\kett, 
\eea 
satisfying the algebra 
\begin{equation}
[a,a^\dagger] = 1
\end{equation}
on the Hilbert space spanned by $|s\kett$. 
These operators  
$a$ and $a^\dagger$ can be regarded as coordinates of the Moyal plane $(\hat{\xi},\hat{\eta})$: 
\be
[\hat{\xi},\hat{\eta}]=i\Theta
\ee
with 
\be
\hat{\zeta}\equiv \hat{\xi}+i\hat{\eta} = \sqrt{2\Theta}\,a, \qquad
\hat{\zeta}^\dagger\equiv \hat{\xi}-i\hat{\eta} = \sqrt{2\Theta} \,a^\dagger.
\label{Moyal_zeta}
\ee

\subsection{$\hat{Y}^{(jj)}_{J\,m}$ in large $j$}
Let us substitute $r=j-s$, $r'=j-s'$ in the definition of
$\hat{Y}^{(jj)}_{J\,m}$ (\ref{app:FSH_def}) 
and consider the case $s, s'\ll\cO(j)$: 
\be
\hat{Y}^{(jj)}_{J\,m} = \sqrt{2J+1}\sum_{s,s'} C^{j\,j-s}_{j\,j-s'\,J\,m}|s\kett\bfraa s'| . 
\label{FSH_flat1}
\ee
Since the C-G coefficient $C^{c\,\gamma}_{c\,\alpha\,b\,\beta}$ for $c\gg b$ 
can be evaluated as~\footnote{See, e.g., eq.~(7) in section 8.9.1 of \cite{QTAM}.} 
\be
C^{c\,\gamma}_{c\,\alpha\, b\,\beta} \simeq 
\delta_{\beta,\gamma-\alpha}\sqrt{\frac{4\pi}{2b+1}}\,Y_{b\,\beta}(\vartheta, 0)
\ee
with $Y_{b\,\beta}$ being a spherical harmonics and 
\be
\cos\frac{\vartheta}{2}=\sqrt{\frac{c+\gamma+\frac12}{2c+1}},\qquad 
\sin \frac{\vartheta}{2}=\sqrt{\frac{c-\gamma+\frac12}{2c+1}},
\ee
we have 
\be
C^{j\,j-s}_{j\,j-s'\,J\,m} \simeq \delta_{m,\,s'-s}\sqrt{\frac{4\pi}{2J+1}}\,Y_{J\,m}(\vartheta,0)
\ee
with
\be
\cos\frac{\vartheta}{2}=\sqrt{1-\frac{s+\frac12}{n}},\qquad 
\sin \frac{\vartheta}{2}=\sqrt{\frac{s+\frac12}{n}}. 
\ee
Note that 
$\vartheta \simeq 2\sqrt{\frac{s+\frac12}{n}}$ is actually small for $s\ll j$, 
corresponding to the fact that we are looking at the neighborhood of the north pole. 
Then the fuzzy spherical harmonics (\ref{FSH_flat1}) can be evaluated as
\be
\hat{Y}^{(jj)}_{J\,m} \simeq \sqrt{4\pi} \sum_{s,s'}\delta_{m,\,s'-s}Y_{J\,m}(\vartheta, 0)|s\kett \bfraa s'|. 
\label{FSH_flat2}
\ee

\subsection{Stereographic transformation} 
Next, we consider a stereographic transformation from $S^2\backslash \mbox{(south pole})$ with the radius $R=3/M$ 
to $\R^2$. As in Fig.~\ref{fig:stereographic}, the plane $\R^2$ is tangent to the $S^2$ at the north pole, 
and the north pole coincides with the origin of $\R^2$. 
\begin{figure}
\begin{center}
\includegraphics[height=7cm, width=10cm, clip]{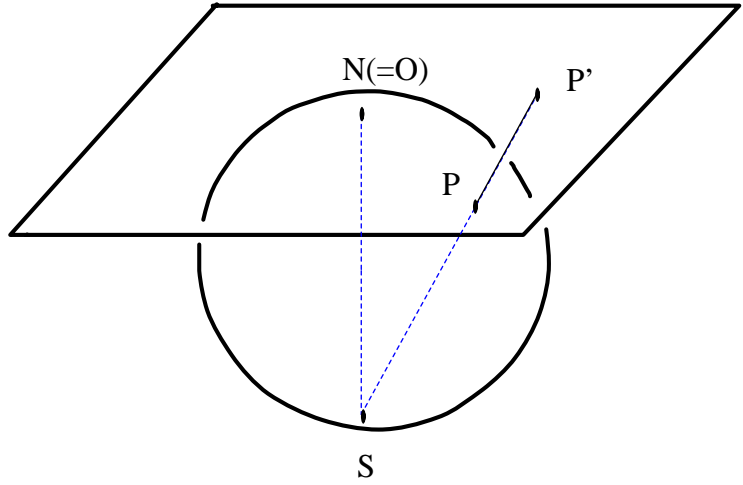}
\end{center}
\caption{\small The stereographic transformation from $S^2\backslash \mbox{(south pole})$ to $\R^2$. 
The ray from the south pole (S) maps the point P on $S^2$ to the point P' on $\R^2$. 
The north pole (N) coincides with the origin (O) on the plane.}
\label{fig:stereographic}
\end{figure}
%
The coordinates $(\vartheta, \varphi)$ of $S^2$ are mapped to the coordinates $(\xi, \eta)$ of $\R^2$ as 
\be
\xi = 2R\tan\frac{\vartheta}{2}\cos\varphi, \qquad \eta = 2R\tan \frac{\vartheta}{2}\sin\varphi.  
\ee
Equivalently, in terms of the complex coordinates $\zeta=\xi+i\eta$, $\bar{\zeta}=\xi-i\eta$, 
\be
\zeta = 2R\tan\frac{\vartheta}{2}\,e^{i\varphi}, 
\qquad \bar{\zeta} = 2R\tan \frac{\vartheta}{2}\,e^{-i\varphi}. 
\label{stereo_zeta} 
\ee  
In the Moyal limit, we send $R\to \infty$ and $\vartheta\to 0$ with $R\vartheta$ kept fixed. 
In this limit, $\zeta$ and $\bar\zeta$ become 
\be
\zeta \to R\vartheta\,e^{i\varphi}, \qquad \bar{\zeta} \to R\vartheta\,e^{-i\varphi}.
\ee
Recall that the radius $R$ and 
the noncommutativity of the Moyal plane 
are given by $R=\frac{3}{M}$ and 
$\Theta = \frac{18}{M^2n}$, respectively. 
Then, when $\vartheta$ is small as $\vartheta\simeq 2\sqrt{\frac{s+\frac{1}{2}}{n}}$, 
the absolute value of $\zeta$ can be written as 
\be
|\zeta| = R\vartheta = \frac{6}{M\sqrt{n}}\sqrt{s+\frac12} = \sqrt{2\Theta}\sqrt{s+\frac12}, 
\ee
which corresponds to the ``radial length'' of the Moyal coordinates (\ref{Moyal_zeta}): 
\be
\frac12(\hat{\zeta}\hat{\zeta}^\dagger + \hat{\zeta}^\dagger\hat{\zeta})
= \Theta(aa^\dagger + a^\dagger a) = 2\Theta\left(s+\frac12\right)\qquad \mbox{on}\qquad |s\kett. 
\ee

Let us next rewrite the spherical harmonics in (\ref{FSH_flat2}) 
using these parametrizations. 
When $J\to\infty$ and $\vartheta\to 0$ with finite $J\vartheta$, 
the harmonics $Y_{J\,m}(\vartheta, 0)$ can be expressed by the Bessel function as~\footnote{For example, see eq.~(9) 
in section 5.12.3 of \cite{QTAM}.} 
\be
Y_{J\,m}(\vartheta, 0) \simeq \sqrt{\frac{J}{2\pi}}\,(-1)^m J_m(J\vartheta).  
\label{sh_bessel}
\ee
In our case, since the absolute value of the momentum in the $\R^2$
is kept finite as
\begin{equation}
u= \frac{1}{R}J = \frac{M}{3}J, 
\end{equation}
the combination $J\vartheta$ is evaluated as 
\be
J\vartheta = uR\vartheta = u\sqrt{2\Theta}\sqrt{s+\frac12}=u|\zeta|. 
\label{uzeta}
\ee
Then (\ref{FSH_flat2}) can be rewritten as
\be
\hat{Y}^{(jj)}_{J\,m} \simeq \sqrt{\frac{6u}{M}}\sum_{s,s'}\delta_{m,\,s'-s}(-1)^m J_m(u|\zeta|)|s\kett \bfraa s'|. 
\label{FSH_flat3}
\ee
More explicitly, 
\be
\hat{Y}^{(jj)}_{J\,0} \simeq \sqrt{\frac{6u}{M}}\sum_{s=0,1,2,\cdots}
J_0\left(u\sqrt{2\Theta}\sqrt{s+\frac12}\right) |s\kett\bfraa s|, 
\label{FSH_0_flat3}
\ee
and, for $m=1,2,\cdots$, 
\bea
\hat{Y}^{(jj)}_{J\,m} & \simeq & \sqrt{\frac{6u}{M}} \,(-1)^m\sum_{s=0,1,2,\cdots} 
J_m \left(u\sqrt{2\Theta}\sqrt{s+\frac12}\right) |s\kett\bfraa m+s|, 
\label{FSH_m_flat3}\\
\hat{Y}^{(jj)}_{J\,-m} & = & (-1)^{-m} \left(\hat{Y}^{(jj)}_{J\,m}\right)^\dagger \nn \\
 & \simeq & \sqrt{\frac{6u}{M}} \sum_{s=0,1,2,\cdots} 
J_m \left(u\sqrt{2\Theta}\sqrt{s+\frac12}\right) |m+s\kett\bfraa s|.
\label{FSH_-m_flat3}
\eea

Here we should note that
eq. (\ref{sh_bessel}) holds even for $J$ smaller than $\cO(1/M)$. More precisely, 
it is valid for $J \geq \cO(M^{-\varepsilon})$ with $0<\varepsilon \ll 1$. 
In fact, according to the formula (4) in section 5.12.2 of \cite{QTAM}, 
$Y_{J\,\pm m}(\vartheta, 0)$ can be evaluated as 
\be
Y_{J\,\pm m}(\vartheta, 0) \simeq  (\mp 1)^m \frac{1}{m!}\sqrt{\frac{2J+1}{4\pi}\frac{(J+m)!}{(J-m)!}} 
\left(\frac{\vartheta}{2}\right)^m 
\left[1-\frac{3J(J+1)-m(m+1)}{3(m+1)}\left(\frac{\vartheta}{2}\right)^2\right]
\label{YJm_theta_small}
\ee
for $0<\vartheta \ll 1$ and $m=0,1,\cdots ,J$. 
When $J\gg \cO(1)$ and $J\vartheta \ll \cO(1)$, it becomes 
\bea
Y_{J\,\pm m}(\vartheta, 0) & \simeq & (\mp 1)^m\frac{1}{m!} \sqrt{\frac{J}{2\pi}} \left(\frac{J\vartheta}{2}\right)^m
\left[1-\frac{1}{m+1}\left(\frac{J\vartheta}{2}\right)^2\right] 
\nn \\
 & = & (\mp 1)^m \sqrt{\frac{J}{2\pi}} J_m(J\vartheta) +\cO((J\vartheta)^{m+4}) ,
\label{remark_YJm_Jm}
\eea
which reproduces the first two terms of the expansion of (\ref{sh_bessel}) for small $J\vartheta$.   

Similarly, 
$Y_{J\,m}(\vartheta,0)$ for $J\leq \cO(M^{-\varepsilon})$ reads 
\bea
Y_{J\, 0}(\vartheta, 0) &\simeq & \sqrt{\frac{2J+1}{4\pi}} \times \left[1+\cO(\vartheta^2)\right]=\cO(1), \nn \\
Y_{J\, \pm m}(\vartheta, 0) & \simeq & \cO(\vartheta^m)=\cO(M^m)\sim 0\qquad (m=1,2, \cdots,J). 
\eea
Plugging this into (\ref{FSH_flat2}) leads to 
\be
\hat{Y}^{(jj)}_{J\, m} \simeq \delta_{m\,0} \sqrt{2J+1}\,\times \id_n  \quad \mbox{for} \quad J \leq \cO(M^{-\varepsilon}). 
\label{FSH_Jsmall}
\ee

\subsection{Plane-wave basis on the Moyal plane}
A field $f(x)$ on $\R^2$  
\be
f(x) = \int \frac{d^2p}{(2\pi)^2} \,e^{ip\cdot x} \tilde{f}(p)
\ee
with $x_\mu=(\xi, \eta)$ and $p\cdot x = p_1\xi + p_2\eta$ corresponds to the operator 
\be
\hat{f} = \int \frac{d^2p}{(2\pi)^2} \,e^{ip\cdot \hat{x}} \tilde{f}(p)
\ee
with $p\cdot\hat{x} = p_1\hat{\xi}+p_2\hat{\eta}$. 
The field $f(x)$ and the operator $\hat{f}$ are connected with each other 
through the operator 
\be
\hat{\Delta}(x)\equiv \int \frac{d^2p}{(2\pi)^2} \,e^{ip\cdot (\hat{x}-x)}
\ee
as
\be
\hat{f} = \int d^2x f(x) \hat{\Delta}(x), 
\qquad 
f(x) = 2\pi\Theta \Tr\left(\hat{f}\hat{\Delta}(x)\right). 
\label{fx_hatf}
\ee
The second formula follows from 
\be
\Tr\left(e^{ip\cdot \hat{x}}\,e^{iq\cdot \hat{x}}\right) = \frac{2\pi}{\Theta} \,\delta^2(p+q),
\ee
which can be shown, for instance, by using the eigenstate of $\hat{\xi}$: 
\be
\hat{\xi}|\xi\ket = \xi|\xi\ket, \qquad \bfra\xi|\hat{\eta} = -i\Theta \partial_\xi\bfra\xi|, \qquad 
\Tr(\cdots) = \int d\xi \bfra\xi|(\cdots)|\xi\ket.   
\ee
It can also be shown that 
\be
\Tr\left(\hat{\Delta}(x) \,\hat{\Delta}(x')\right) = \frac{1}{2\pi\Theta}\,\delta^2(x-x'). 
\ee

The product $\hat{f}_1\hat{f}_2$ corresponds to the Moyal product 
\be
(f_1*f_2)(x)\equiv \left.e^{i\frac{\Theta}{2}(\partial_\xi\partial_{\eta'}-\partial_\eta\partial_{\xi'})}
f_1(\xi,\eta)f_2(\xi',\eta')\right|_{\xi'=\xi,\,\eta'=\eta}
\label{Moyal}
\ee
through
\be
\hat{f}_1\hat{f}_2 = \int d^2x\,(f_1*f_2)(x)\,\hat{\Delta}(x), \qquad 
(f_1*f_2)(x) = 2\pi\Theta \Tr\left(\hat{f}_1\hat{f}_2\hat{\Delta}(x)\right).
\ee
{}From (\ref{fx_hatf}), $\hat{f}$ can be expanded by the plane-wave basis as  
\be
\hat{f} =2\pi\Theta \int d^2x\,\Tr\left(\hat{f}\hat{\Delta}(x)\right)\,\hat{\Delta}(x) 
= 2\pi\Theta\int\frac{d^2p}{(2\pi)^2}\,\Tr\left(\hat{f}\,e^{ip\cdot\hat{x}}\right)\,e^{-ip\cdot\hat{x}}. 
 \label{hatf_PW}
\ee

\subsection{Computation of $\bfraa s| e^{ip\cdot\hat{x}}|s'\kett$}
Let us compute the matrix elements of the plane-wave basis $e^{ip\cdot\hat{x}}$ with respect to the states $|s\kett$. 
Using the complex combination 
\be
p\equiv p_1+ip_2, \qquad \bar{p}\equiv p_1-ip_2, 
\ee
the basis can be expressed as 
\be
e^{ip\cdot\hat{x}} = e^{i\frac12\left(\bar{p}\hat{\zeta}+p\hat{\zeta}^\dagger\right)} 
= e^{-\frac{\Theta}{4}p\bar{p}}\,e^{i\sqrt{2\Theta}\,\frac{p}{2}a^\dagger}\,
e^{i\sqrt{2\Theta}\,\frac{\bar{p}}{2} a}. 
\ee
Then, 
\bea
\bfraa s| e^{ip\cdot\hat{x}}|s'\kett & = & e^{-\frac{\Theta}{4}p\bar{p}}
\sum_{t=0}^s\frac{\left(i\sqrt{2\Theta}\,\frac{p}{2}\right)^t}{t!}
\sum_{t'=0}^{s'}\frac{\left(i\sqrt{2\Theta}\,\frac{\bar{p}}{2}\right)^{t'}}{t'!} \,
\bfraa s|\left(a^\dagger\right)^t a^{t'}|s'\kett \nn \\
& = & e^{-\frac{\Theta}{4}|p|^2}\,e^{i(s-s')\varphi_p}\sum_{t=0}^s\sum_{t'=0}^{s'}
\delta_{t'-t,\,s'-s} \,\frac{\sqrt{s!s'!}}{(s-t)!}
\frac{\left(i\sqrt{2\Theta}\,\frac{|p|}{2}\right)^{t+t'}}{t!\,t'!}
\eea
with
$p=|p|\,e^{i\varphi_p}$ and $\bar{p}= |p|\,e^{-i\varphi_p}$, 
which leads to 
\begin{align}
e^{ip\cdot\hat{x}} = & \sum_{s,s'}|s\kett\bfraa s|e^{ip\cdot\hat{x}} |s'\kett\bfraa s'| \nn \\
= & \sum_{m\in \Z}e^{-\frac{\Theta}{4}|p|^2}\,e^{-im\varphi_p}\sum_{s,s'}\delta_{m,\,s'-s}  
 \sum_{t=0}^s\sum_{t'=0}^{s'}
\delta_{m,\,t'-t} \,\frac{\sqrt{s!s'!}}{(s-t)!} \,
\frac{\left(i\sqrt{2\Theta}\,\frac{|p|}{2}\right)^{t+t'}}{t!\,t'!} 
\,|s\kett\bfraa s'|. 
\label{PW_ss'}
\end{align}

Note that 
\be
\sum_{t=0}^s\frac{s!}{(s-t)!}\frac{\left(-\frac{\Theta}{2}\,|p|^2\right)^t}{(m+t)!\,t!} 
=\frac{1}{m!}\,F\left(-s, m+1;\frac{\Theta}{2}\,|p|^2\right)
\label{conf_HG}
\ee
for $m=0,1,2,\cdots$, 
where 
$F(a,b;z)$ is a confluent hypergeometric function defined by 
\be
F(a,b;z)\equiv\sum_{t=0}^\infty \frac{(a)_t}{(b)_t}\frac{z^t}{t!}
\ee
with $(a)_t\equiv a(a+1)\cdots(a+t-1)$ and $(a)_0\equiv 1$. 
Then, (\ref{PW_ss'}) can be expressed as
\bea
e^{ip\cdot\hat{x}} & = & e^{-\frac{\Theta}{4}\,|p|^2}\sum_{s=0}^\infty\Biggl[
F\left(-s,1;\frac{\Theta}{2}\,|p|^2\right)\,|s\kett\bfraa s| \nn \\
& & \hspace{20mm}+\sum_{m=1}^\infty \frac{\left(i\sqrt{2\Theta}\,\frac{|p|}{2}\right)^m}{m!}\sqrt{\frac{(m+s)!}{s!}}\,
F\left(-s,m+1;\frac{\Theta}{2}|p|^2\right) \nn \\
& & \hspace{30mm}\times\Bigl(e^{-im\varphi_p}\,|s\kett\bfraa m+s| +e^{im\varphi_p}\,|m+s\kett\bfraa s|\Bigr)
\Biggr]. 
\label{PW_sm}
\eea

\subsection{$\hat{Y}^{(jj)}_{J\,m}$ in terms of the plane-wave basis}
By using (\ref{hatf_PW}), $\hat{Y}^{(jj)}_{J\,m}$ can be expanded as 
\be
\hat{Y}^{(jj)}_{J\,m} = 2\pi\Theta\int\frac{d^2p}{(2\pi)^2}\,\Tr\left(\hat{Y}^{(jj)}_{J\,m}\,e^{ip\cdot\hat{x}}\right) \,e^{-ip\cdot\hat{x}}.
\label{FSH_PW}
\ee
Plugging (\ref{FSH_0_flat3}), (\ref{FSH_m_flat3}), (\ref{FSH_-m_flat3}), and (\ref{PW_sm}) into (\ref{FSH_PW}) 
leads to 
\be
\hat{Y}^{(jj)}_{J\,m}  =  2\pi\Theta\,\sqrt{\frac{6u}{M}}\int\frac{d^2p}{(2\pi)^2}\,e^{-\frac{\Theta}{4}|p|^2}
\,e^{im\varphi_p}\,\cK_m(\Theta, u, |p|)\,e^{-ip\cdot\hat{x}}
\label{FSH_PW1}
\ee
for $J\geq \cO(M^{-\varepsilon})$. 
The kernels are given by  
\begin{align}
&\cK_0(\Theta, u, |p|) \equiv \sum_{s=0}^\infty J_0\left(u\sqrt{2\Theta}\sqrt{s+\frac12}\right)\,
F\left(-s,1;\frac{\Theta}{2}\,|p|^2\right), 
\label{kernel_0}\\
&\cK_m(\Theta, u, |p|) \equiv \frac{\left(-i\sqrt{2\Theta}\,\frac{|p|}{2}\right)^m}{m!}
\sum_{s=0}^\infty \sqrt{\frac{(m+s)!}{s!}}\,J_m\left(u\sqrt{2\Theta}\sqrt{s+\frac12}\right)\,
 F\left(-s,m+1;\frac{\Theta}{2}\,|p|^2\right), 
\label{kernel_m}\\
&\cK_{-m}(\Theta, u, |p|) \equiv \frac{\left(i\sqrt{2\Theta}\,\frac{|p|}{2}\right)^m}{m!}
\sum_{s=0}^\infty \sqrt{\frac{(m+s)!}{s!}}\,J_m\left(u\sqrt{2\Theta}\sqrt{s+\frac12}\right)\,
F\left(-s,m+1;\frac{\Theta}{2}\,|p|^2\right) 
\label{kernel_-m}
\end{align}
for $m=1,2,\cdots$.

\subsection{Computation of the kernels}
We can explicitly compute the kernels (\ref{kernel_0}), (\ref{kernel_m}), and (\ref{kernel_-m}) in the 
case of small $\Theta$. 

\subsubsection{$\cK_0$}
It is convenient to insert 
\be
1 = 2\int_0^\infty dv\,v\,v^{2s}\,e^{-v^2}\,\frac{1}{s!}
\ee
into the sum in (\ref{kernel_0}) for the computation. 
Then $\cK_0(\Theta, u, |p|)$ becomes 
\be
\cK_0(\Theta, u, |p|) = 2\int^\infty_0 dv\,v\sum_{s=0}^\infty v^{2s}\,e^{-v^2}\,
J_0\left(u\sqrt{2\Theta}\sqrt{s+\frac12}\right) \sum_{t=0}^s
\frac{\left(-\frac{\Theta}{2}\,|p|^2\right)^t}{(s-t)!\,(t!)^2}. 
\ee
After changing the order of the sums as 
\be
\sum_{s=0}^\infty\sum_{t=0}^s\cdots = \sum_{t=0}^\infty\sum_{s=t}^\infty\cdots, 
\ee
and making a shift $s\to s+t$, we obtain 
\be
\cK_0(\Theta, u, |p|) = 2\int^\infty_0 dv\,v\,e^{-v^2}\sum_{t=0}^\infty 
\frac{\left(-\frac{\Theta}{2}\,|p|^2v^2\right)^t}{(t!)^2}\sum_{s=0}^\infty 
J_0\left(u\sqrt{2\Theta}\sqrt{s+t+\frac12}\right)\frac{v^{2s}}{s!}. 
\label{kernel_0_2}
\ee

First, let us evaluate the sum over $s$ by setting $x=2\Theta s$
and $y=2\Theta t$. 
Using the Stirling formula, we have 
\be
\frac{v^{2s}}{s!}\simeq \frac{1}{\sqrt{2\pi}}\sqrt{\frac{2\Theta}{x}}\,e^{\frac{1}{2\Theta}f(x)}, \qquad
f(x) \equiv 2x\ln v -x\ln\frac{x}{2\Theta} +x. 
\label{kernel_fx}
\ee
For small $\Theta$, the summation over $s$ can be evaluated as an integral: 
\be
\sum_{s=0}^\infty 
J_0\left(u\sqrt{2\Theta}\sqrt{s+t+\frac12}\right)\frac{v^{2s}}{s!} \nn \\
\simeq \frac{1}{\sqrt{2\pi}}\frac{1}{2\Theta}\int_0^\infty dx\,J_0\left(u\sqrt{x+y+\Theta}\right)\, 
\sqrt{\frac{2\Theta}{x}}\,e^{\frac{1}{2\Theta}f(x)}, 
\label{kernel_0_sum_s}
\ee
in which contributions to the integral localize to the saddle point $x_0$ of $f(x)$ 
with 
\be
x_0=2\Theta\,v^2, \qquad 
f(x_0)=2\Theta v^2, \qquad f''(x_0)= -\frac{1}{2\Theta v^2}. 
\label{kernel_saddle}
\ee
Including Gaussian fluctuations around the saddle point, we obtain 
\be
\sum_{s=0}^\infty 
J_0\left(u\sqrt{2\Theta}\sqrt{s+t+\frac12}\right)\frac{v^{2s}}{s!} 
\simeq J_0\left(u\sqrt{\Theta}\sqrt{2v^2+2t+1}\right)\,e^{v^2}. 
\label{app:sum_s}
\ee
 
Next, the region of $t$ giving a dominant contribution to the sum is 
$0\leq t \lesssim t_0$ with $t_0$ satisfying 
\be
(t_0!)^2 = \left(\frac{\Theta}{2}\,|p|^2v^2\right)^{t_0}, 
\ee  
because for $t\gtrsim t_0$, the denominator $(t!)^2$ grows much faster than the numerator, 
and the contribution becomes negligible. The value of $t_0$ is evaluated as 
\be
t_0\simeq \sqrt{\frac{\Theta}{2}}\,|p|v, 
\ee
which means that the $t$-dependence of $J_0\left(u\sqrt{\Theta}\sqrt{2v^2+2t+1}\right)$ in (\ref{app:sum_s}) can be dropped 
and replaced with $J_0\left(u\sqrt{\Theta}\sqrt{2v^2+1}\right)$ for small $\Theta$. 
Noting  
\be
\sum_{t=0}^\infty 
\frac{\left(-\frac{\Theta}{2}\,|p|^2v^2\right)^t}{(t!)^2}=J_0\left(\sqrt{2\Theta}\,|p|v\right), 
\ee  
we have 
\be
\cK_0(\Theta, u, |p|) \simeq  2\int^\infty_0 dv\,v\,J_0\left(\sqrt{2\Theta}\,|p|v\right)\,
J_0\left(u\sqrt{\Theta}\sqrt{2v^2+1}\right). 
\ee

Finally, the region of $v$ dominant to the integral is $v\gg 1$. 
The region $0<v\lesssim \cO(1)$ gives merely an $\cO(1)$ contribution to the integral. 
So, in the expansion 
\be
J_0\left(u\sqrt{\Theta}\sqrt{2v^2+1}\right)= J_0\left(u\sqrt{2\Theta}\,v\right) 
+ \frac{u\sqrt{2\Theta}}{4v}\,J_0'\left(u\sqrt{2\Theta}\,v\right)+\cdots, 
\ee
we can neglect the second- and higher-order terms for $v \gg 1$. 
Hence, 
\be
\cK_0(\Theta, u, |p|) \simeq 2\int^\infty_0 dv\,v\,J_0\left(\sqrt{2\Theta}\,|p|v\right)\,
J_0\left(\sqrt{2\Theta}\,uv\right) = \frac{1}{\Theta}\frac{1}{|p|}\,\delta(|p|-u), 
\label{kernel_0_small_Theta}
\ee
where the formula 
\be
 \int_0^\infty dr\,r\,J_m(\alpha r)\,J_m(\beta\,r) 
= \frac{1}{\alpha}\,\delta(\alpha-\beta)
\label{app:int_bessel3}
\ee
for $\alpha, \beta>0$ was used. Equation (\ref{app:int_bessel3}) is derived below. 

{}From (\ref{FSH_PW1}) and (\ref{kernel_0_small_Theta}), we obtain 
\be
\hat{Y}^{(jj)}_{J\,0} \simeq 2\pi\sqrt{\frac{6u}{M}}\,e^{-\frac{\Theta}{4}u^2}\int \frac{d^2p}{(2\pi)^2}\,
\frac{1}{|p|}\,\delta(|p|-u)\,e^{-ip\cdot\hat{x}}
\ee
for $J\geq \cO(M^{-\varepsilon})$ and small $\Theta$. 

\paragraph{Derivation of (\ref{app:int_bessel3})} 
Since $J_m(z) = (-1)^{-m}J_{-m}(z)$, it is sufficient to show (\ref{app:int_bessel3}) for $m=0,1,2, \cdots$. 
Let us start with the Bessel equations 
\bea
 & & \left(\partial^2_r+\frac{1}{r}\partial_r+\alpha^2-\frac{m^2}{r^2}\right)J_m(\alpha r)= 0, 
\label{app:bessel_alpha}\\
 & & \left(\partial^2_r+\frac{1}{r}\partial_r+\beta^2-\frac{m^2}{r^2}\right)J_m(\beta r)= 0.  
 \label{app:bessel_beta}
\eea
With use of the asymptotic form
\be
J_m(z) = \sqrt{\frac{2}{\pi z}}\,\cos\left(z-\frac{2m+1}{4}\pi\right)\times \left\{1+\cO(1/z)\right\},
\ee
$\int_0^\infty dr r\left[J_m(\beta r)\times {\rm (\ref{app:bessel_alpha})}- J_m(\alpha r)\times {\rm (\ref{app:bessel_beta})}\right]$ 
leads to 
\begin{align}
\int_0^\infty dr\,r\,J_m(\alpha r)\,J_m(\beta\,r) & =\frac{1}{\alpha^2-\beta^2}\frac{1}{\pi}\,\lim_{r\to \infty}\left[
\left(\sqrt{\frac{\alpha}{\beta}}+\sqrt{\frac{\beta}{\alpha}}\right)\sin((\alpha-\beta)r) 
\right. \nn \\
& \hspace{28mm}\left. 
-\left(\sqrt{\frac{\alpha}{\beta}}-\sqrt{\frac{\beta}{\alpha}}\right)(-1)^m \cos((\alpha+\beta)r) \right].
\label{app:int_bessel}
\end{align}

Let us pick 
a test function $f(\beta)$ such that $f(0)$ is finite and $f(\beta)$ rapidly decays as $\beta\to \infty$, 
and compute $\int_0^\infty d\beta \,\beta\, f(\beta)\times$ (\ref{app:int_bessel}). 
Noting that 
\begin{align}
\int_0^\infty d\beta\,\beta\,f(\beta)\,\frac{1}{\alpha^2-\beta^2}
\left(\sqrt{\frac{\alpha}{\beta}}-\sqrt{\frac{\beta}{\alpha}}\right)\cos((\alpha+\beta)r) 
& = \int_0^\infty d\beta\,\frac{f(\beta)}{\alpha+\beta}\sqrt{\frac{\beta}{\alpha}}\,\cos((\alpha+\beta)r) 
\nn \\
& \to 0 \qquad (r\to \infty) 
\end{align}
due to the rapid oscillation as $r\to \infty$, we obtain 
\be
\int_0^\infty d\beta\,\beta \,f(\beta)\int_0^\infty dr\,r\,J_m(\alpha r)\,J_m(\beta\,r) 
 = \lim_{r\to \infty} \frac{1}{\pi}\int_0^\infty d\beta\,f(\beta)\sqrt{\frac{\beta}{\alpha}}\,
\frac{\sin((\beta-\alpha)r)}{\beta-\alpha}. 
\label{app:int_bessel2}
\ee
Because the contribution to the integral on the r.h.s. localizes to the neighborhood of $\beta= \alpha$ due to the rapid oscillation, 
\be
(\mbox{r.h.s. of (\ref{app:int_bessel2})}) = \frac{f(\alpha)}{\pi} \lim_{r\to\infty}\int_0^\infty d\beta\, 
  \frac{\sin((\beta-\alpha)r)}{\beta-\alpha} = \frac{f(\alpha)}{\pi} \int^\infty_{-\infty}dy \,\frac{\sin y}{y} = f(\alpha). 
\ee
This shows that (\ref{app:int_bessel3}) holds.

\subsubsection{$\cK_m$ ($m=1,2,\cdots$)}
Following similar steps to the case of $\cK_0$, the expression for $\cK_m$ 
corresponding to (\ref{kernel_0_2}) becomes 
\bea
\cK_m(\Theta, u, |p|) & = & \left(-i\sqrt{2\Theta}\,\frac{|p|}{2}\right)^m\,2\int^\infty_0dv\,v\,e^{-v^2}
\sum_{t=0}^\infty\frac{\left(-\frac{\Theta}{2}\,|p|^2v^2\right)^t}{(m+t)!\,t!} \nn \\
& & \times \sum_{s=0}^\infty J_m\left(u\sqrt{2\Theta}\sqrt{s+t+\frac12}\right)\,
\sqrt{\frac{(m+s+t)!}{(s+t)!}}\,\frac{v^{2s}}{s!}. 
\label{kernel_m_2}
\eea
We treat the summation over $s$ as discussed before:
\bea
& & \sum_{s=0}^\infty J_m\left(u\sqrt{2\Theta}\sqrt{s+t+\frac12}\right)\,
\sqrt{\frac{(m+s+t)!}{(s+t)!}}\,\frac{v^{2s}}{s!} \nn \\
& &  \hspace{7mm}\simeq J_m\left(u\sqrt{\Theta}\sqrt{2v^2+2t+1}\right)\,(m+v^2+t)^{m/2}\,e^{v^2}. 
\eea 
Since the summation over $t$ is dominated 
by the region $0\leq t \lesssim \sqrt{\frac{\Theta}{2}}\,|p|v$, 
the $t$-dependence of $J_m\left(u\sqrt{\Theta}\sqrt{2v^2+2t+1}\right)$ 
can be discarded for small $\Theta$. 
Noting 
\be
\left(\sqrt{2\Theta}\,\frac{|p|}{2}\,v\right)^m\sum_{t=0}^\infty\frac{\left(-\frac{\Theta}{2}\,|p|^2v^2\right)^t}{(m+t)!\,t!} = J_m\left(\sqrt{2\Theta}\,|p|v\right), 
\ee
we have 
\be
\cK_m(\Theta, u, |p|) \simeq (-i)^m\,2\int^\infty_0dv\,v\,J_m\left(\sqrt{2\Theta}\,|p|v\right)\,
\left(1+\frac{m}{v^2}\right)^{m/2}J_m\left(u\sqrt{\Theta}\sqrt{2v^2+1}\right).
\ee
Because the region $v\gg 1$ dominates the integral, 
we can approximate as 
\be
 J_m\left(u\sqrt{\Theta}\sqrt{2v^2+1}\right)\simeq J_m\left(u\sqrt{2\Theta}\,v\right), 
\qquad  \left(1+\frac{m}{v^2}\right)^{m/2}\simeq 1
\ee
in evaluating the integral. Hence, use of (\ref{app:int_bessel3}) leads to 
\be
\cK_m(\Theta, u, |p|) \simeq (-i)^m\,2\int^\infty_0dv\,v\,J_m\left(\sqrt{2\Theta}\,|p|v\right)\,
J_m\left(\sqrt{2\Theta}\,uv\right)
= (-i)^m\,\frac{1}{\Theta}\frac{1}{|p|}\,\delta(|p|-u).
\label{kernel_m_small_Theta}
\ee

{}From (\ref{FSH_PW1}) and (\ref{kernel_m_small_Theta}), we obtain 
\be
\hat{Y}^{(jj)}_{J\,m}\simeq 2\pi\,\sqrt{\frac{6u}{M}}\,(-i)^m\,e^{-\frac{\Theta}{4}\,u^2}
\int\frac{d^2p}{(2\pi)^2}\,\frac{1}{|p|}\,\delta(|p|-u)\,e^{im\varphi_p}\,e^{-ip\cdot\hat{x}}
\ee
for$J\geq \cO(M^{-\varepsilon})$, small $\Theta$, and $m=1,2,\cdots$ fixed.

\subsubsection{$\cK_{-m}$ ($m=1,2,\cdots$)}
Since (\ref{kernel_-m}) is of the same form as (\ref{kernel_m}) with $-i$ to $i$ changed in the prefactor 
$\frac{\left(-i\sqrt{2\Theta}\,\frac{|p|}{2}\right)^m}{m!}$, 
the computation for $\cK_m$ is repeated to obtain $\cK_{-m}$. 
The result is 
\bea
& & \cK_{-m}(\Theta, u, |p|) \simeq  i^m\,\frac{1}{\Theta}\frac{1}{|p|}\,\delta(|p|-u), 
\label{kernel_-m_small_Theta} \\
& & 
\hat{Y}^{(jj)}_{J\,-m}\simeq 2\pi\,\sqrt{\frac{6u}{M}}\,i^m\,e^{-\frac{\Theta}{4}\,u^2}
\int\frac{d^2p}{(2\pi)^2}\,\frac{1}{|p|}\,\delta(|p|-u)\,e^{-im\varphi_p}\,e^{-ip\cdot\hat{x}}
\eea
for $J\geq \cO(M^{-\varepsilon})$, small $\Theta$, and $m=1,2,\cdots$ fixed. 

\subsection{Summary}
Summarizing the results obtained in the previous subsections, 
for $J\geq \cO(M^{-\varepsilon})$, small $\Theta$ and $m\in\Z$ fixed,  
the kernel and the fuzzy spherical harmonics are evaluated as 
\be
\cK_m(\Theta,u,|p|) \simeq (-i)^m\,\frac{1}{\Theta}\frac{1}{|p|}\,\delta(|p|-u)
\qquad (m\in \Z)
\label{Kernel_small_Theta}
\ee
and 
\be
\hat{Y}^{(jj)}_{J\,m}\simeq 2\pi\,\sqrt{\frac{6u}{M}}\,(-i)^m\,e^{-\frac{\Theta}{4}\,u^2}
\int\frac{d^2p}{(2\pi)^2}\,\frac{1}{|p|}\,\delta(|p|-u)\,e^{im\varphi_p}\,e^{-ip\cdot\hat{x}}. 
\label{FSH_small_Theta}
\ee 

Let us consider the commutative limit $\Theta\to 0$ at the final step 
(Step 2 in section~\ref{sec:loop}). 
The coordinates $\hat{x}_\mu=(\hat{\xi}, \hat{\eta})$ reduce to  
the c-numbers $x_\mu=(\xi, \eta)$. We call the polar angle of $x_\mu$ in $\R^2$  
$\varphi$. Then, 
\be
p\cdot\hat{x}\to |p||\zeta|\cos(\varphi_p-\varphi)   
\ee  
with $\zeta=\xi+i\eta$ and $\varphi=\arg(\zeta)$. 
Then (\ref{FSH_small_Theta}) becomes
\be
\hat{Y}^{(jj)}_{J\,m}\to \sqrt{\frac{6u}{M}}\,
\left[\frac{(-i)^m}{2\pi}\int^{2\pi}_0d\varphi_p\,e^{im\varphi_p-iu|\zeta|\cos\varphi_p}\right]
\,e^{im\varphi} =  \sqrt{\frac{6u}{M}}\,J_{-m}(u|\zeta|)\,e^{im\varphi}. 
\label{FSH_Theta_zero}
\ee

We can see that (\ref{FSH_Theta_zero}) is consistent with the orthonormality 
\be
\delta_{J\,J'}\,\delta_{m\,m'} = \frac{1}{n}\,\tr_n\left\{\left(\hat{Y}^{(jj)}_{J\,m}\right)^\dagger 
\hat{Y}^{(jj)}_{J'\,m'}\right\} 
\label{FSH_orthogonal}
\ee
as follows. 
In the commutative limit, 
\be
\frac{1}{n}\,\tr_n\left\{\cdots\right\} \to \int\frac{d\Omega}{4\pi}\left\{\cdots\right\},  \qquad 
d\Omega = \sin\vartheta \,d\vartheta\,d\varphi. 
\ee
{}From (\ref{uzeta}), 
\be
\sin\vartheta \simeq \vartheta = \frac{1}{R}\,|\zeta|, \qquad d\vartheta = \frac{1}{R}\,d|\zeta|, 
\ee
and thus 
\be
d\Omega \simeq \frac{1}{R^2}\,|\zeta|\,d|\zeta|\,d\varphi. 
\ee
Using (\ref{FSH_Theta_zero}), the r.h.s. of (\ref{FSH_orthogonal}) reads 
\bea
(\mbox{r.h.s. of (\ref{FSH_orthogonal})}) & \to & \int \frac{d\Omega}{4\pi}\,\frac{6}{M}\sqrt{uu'}\,
J_{-m}(u|\zeta|)\,J_{-m'}(u'|\zeta|)\,e^{-i(m-m')\varphi} \nn \\
& = & \frac{\sqrt{uu'}}{R}\,\delta_{m\,m'}\int^\infty_0d|\zeta|\,|\zeta|\,J_{-m}(u|\zeta|)\,J_{-m'}(u'|\zeta|). 
\eea
$J_{-m}(z)=(-1)^mJ_m(z)$ and the formula (\ref{app:int_bessel3}) leads to 
\be
 (\mbox{r.h.s. of (\ref{FSH_orthogonal})}) \to \frac{1}{R}\,\delta_{m\,m'}\,\delta(u-u'),
\ee 
which is consistent with the l.h.s. of (\ref{FSH_orthogonal}) because 
$
\delta_{J\,J'} \to \frac{1}{R}\,\delta(u-u')
$
in the limit.

\section{Nonperturbative stability of the fuzzy sphere solution}
\label{app:tunneling}
In this appendix, to examine the nonperturbative stablity of the $k$-coincident fuzzy $S^2$ solution (\ref{FS sln0}) with (\ref{FS sln}), 
we evaluate the tunneling amplitudes from the fuzzy $S^2$ solution 
\begin{enumerate}
\item
to the trivial vacuum ($X_a=0$) 
\item     
to the $(k-1)$-coincident fuzzy $S^2$ solution (\ref{k-1_FS})
\item
to the solutions (\ref{k-2_FS})
\item
to the solutions (\ref{k-3_FS}). 
\end{enumerate}

\subsection{Tunneling amplitude to the trivial vacuum}
\label{app:F-0}
We assume the form of a tunneling solution to the trivial vacuum to be  
\be
X_a(x)= f(x)\,\frac{M}{3}\,L_a
\label{inst_ansatz}
\ee
with $f(x)=1$ at $x_1=-\infty$ and $f(x)=0$ at $x_1=+\infty$. 
$x_1$ is regarded as Euclidean time.  

Plugging (\ref{inst_ansatz}) into  
the relevant part of the action,
\bea
S_{\rm inst}& = & \frac{1}{g_{2d}^2}\int d^2x\,\Tr\left[(\partial_\mu X_a)^2+\left(i[X_8,\,X_9]+\frac{M}{3}X_{10}\right)^2 \right. \nn \\
& & \left. + \left(i[X_9,\,X_{10}]+\frac{M}{3}X_{8}\right)^2 +\left(i[X_{10},\,X_8]+\frac{M}{3}X_{9}\right)^2 \right],
\label{inst_action}
\eea
leads to 
\be
S_{\rm inst} = \frac{k}{g_{2d}^2}j(j+1)n\int d^2x
\left[\frac{M^2}{9}(\partial_\mu f(x))^2 +\left(\frac{M^2}{9}\right)^2 f(x)^2(f(x)-1)^2\right].
\label{inst_action2}
\ee
Here, 
\be
\Tr\,(L_a^2) = \tr_n\left((L_a^{(n)})^2\right)\,\tr_k(\id_k) = k\,\tr_n(j(j+1)\id_n) = kj(j+1)n
\ee
has been used. 

By rescaling $\xi_\mu =\frac{M}{3}\,x_\mu$ with $f(x) = \tilde{f}(\xi)$, the expression becomes 
\bea
S_{\rm inst} & = & \frac{k}{g_{2d}^2}j(j+1)n\,\frac{M^2}{9}\,\tilde{S}_{\rm inst} = 
\frac{k}{g_{2d}^2}\frac{2}{\Theta}\,j(j+1)\,\tilde{S}_{\rm inst}, 
\label{inst_action3} \\
\tilde{S}_{\rm inst} & \equiv &\int d^2\xi
\left[(\partial_\mu \tilde{f}(\xi))^2 +\tilde{f}(\xi)^2(\tilde{f}(\xi)-1)^2\right].
\label{inst_action_tilde}
\eea
An $x_2$-independent solution (the so-called kink solution) is explicitly given by 
\be
\tilde{f}(\xi) = \frac12\left[1-\tanh\left(\frac12\xi_1+c\right)\right], 
\label{kink}
\ee
where $c$ is an integration constant. 

Since the value of $\tilde{S}_{\rm inst}$ at the solution (\ref{kink}) is proportional to the length of the $\xi_2$-direction, 
the tunneling rate ($\simeq e^{-S_{\rm inst}}$) vanishes when the $\xi_2$-direction is noncompact.   
In order to discuss this in more detail, 
suppose the extent of the $\xi_2$-direction is $0\leq \xi_2 \leq L$; namely, $x_2$ ranges over $0\leq x_2 \leq \tilde{L}(\equiv\frac{3}{M}L)$. 
Then, the value of the action at the kink solution becomes 
\be
S_{\rm inst} = \frac{k}{g_{2d}^2}\frac{2}{\Theta}\frac{L}{3}\,j(j+1)=\frac{4\pi k}{g_{4d}^2}\frac{\tilde{L}}{9}\,Mj(j+1)
\label{inst_action_f}
\ee
with $g_{4d}^2=2\pi \Theta g_{2d}^2$. 
At Step 1 (the Moyal limit) in the successive limits, which sends $j$ as $j \propto M^{-2}\to \infty$, 
the tunneling is completely suppressed ($S_{\rm inst} \to +\infty$) even for $\tilde{L}$ finite. 

\subsection{Tunneling amplitude to the $(k-1)$-coincident fuzzy $S^2$ solution}
\label{app:F-1}
We obtain a tunneling solution to the $(k-1)$-coincident fuzzy $S^2$ solution (\ref{k-1_FS}) by assuming the form  
\be
X_a = \frac{M}{3}\begin{pmatrix}
L^{(n)}_a\otimes \id_{k-1} &   \\
                        & f(x)L_a^{(n)}
\end{pmatrix} 
\ee
with $f(x)=1$ at $x_1=-\infty$ and $f(x)=0$ at $x_1=+\infty$. 

The computation is parallel with the previous subsection. 
The result is given by replacing the overall factor $k$ with $1$ in (\ref{inst_action3}) and (\ref{inst_action_f}). 
Thus, this tunneling is also completely suppressed at Step 1 in the successive limits. 

\subsection{Tunneling amplitude to the solutions (\ref{k-2_FS})}
\label{app:F-2}
In the process to (\ref{k-2_FS}), two of the $k$ $L_a^{(n)}$ in the fuzzy sphere solution (\ref{FS sln0}), (\ref{FS sln}) 
recombine to $L_a^{(n+\ell)}$ and $L_a^{(n-\ell)}$. The process for $\ell=1$ seems the most relevant in all of the tunneling processes 
and should be considered. 
We make a bound for the amplitude, since obtaining the explicit solution seems technically complicated. 
As discussed in~\cite{Yee:2003ge}, we recast the action as 
\bea
S_{\rm inst} & = & \frac{1}{g_{2d}^2}\int d^2x\,\Tr\left[\left(\pm\partial_1X_a+i\epsilon_{abc}X_bX_c+\frac{M}{3}X_a\right)^2+(\partial_2X_a)^2 \right. 
\nn \\
& & \hspace{27mm}\left.\mp\partial_1\left\{\frac23i\epsilon_{abc}X_aX_bX_c+\frac{M}{3}X_aX_a\right\}\right] \nn \\
& \geq & \mp \frac{1}{g_{2d}^2}\int dx_2\,\Tr\left[\frac23i\epsilon_{abc}X_aX_bX_c+\frac{M}{3}X_aX_a\right]_{x_1=-\infty}^{x_1=+\infty},  
\label{BPS_bound}
\eea
where the signs are chosen so that the right-hand side is nonnegative. 
The inequality is saturated by the solution of~\footnote{Interestingly, it can be seen that 
the solutions in appendices~\ref{app:F-0} and \ref{app:F-1} saturate the bound.} 
\be
\pm\partial_1X_a+i\epsilon_{abc}X_bX_c+\frac{M}{3}X_a=0, \qquad \partial_2X_a=0  
\label{BPS_eqs}
\ee
with (\ref{FS sln0}) at $x_1=-\infty$ and (\ref{k-2_FS}) at $x_2=+\infty$. 
The right-hand side of (\ref{BPS_bound}) becomes 
\be
\mp \frac{\tilde{L}}{g_{2d}^2}\left(\frac{M}{3}\right)^3\,\Tr\,\frac13\left\{(L_a'')^2-(L_a)^2\right\}
=\mp\frac{\tilde{L}}{g_{2d}^2}\left(\frac{M}{3}\right)^3\frac{n}{2}\,\ell^2.
\ee
Taking the lower sign, we eventually have the bound 
\be
S_{\rm inst} \geq \frac{2\pi}{g_{4d}^2}\frac{\tilde{L}M}{3}\,\ell^2.
\ee 
Thus, we see that sending $\tilde{L}$ to infinity faster than $1/M$ suppresses the tunneling rate and stabilizes the fuzzy sphere 
configuration (\ref{FS sln0}) with (\ref{FS sln}). 
 
This should be regarded as a sufficient condition, and there could be cases in which tunneling does not occur in milder conditions. 
For instance, suppose that tunneling solutions break $Q_+$ or $Q_-$ supersymmetry. Then, zero-modes of the associated Nambu-Goldstone fermions appear 
in the path integral, 
and annihilate amplitudes that do not soak up the zero-modes. 

\subsection{Tunneling amplitude to the solutions (\ref{k-3_FS})}
\label{app:F-3}
We can evaluate a similar bound for the process in which one of the $k$ $L_a^{(n)}$ in the fuzzy sphere solution 
splits into $L_a^{(\ell)}$ and $L_a^{(n-\ell)}$ with $\ell=1,2,\cdots, n-1$. 
Then, the upper sign is taken in (\ref{BPS_bound}). The result reads 
\be
S_{\rm inst} \geq \frac{2\pi}{g_{4d}^2}\frac{\tilde{L}M}{3}\,\frac{\ell(n-\ell)}{2}. 
\ee
The same procedure of sending $\tilde{L}$ as in appendix~\ref{app:F-2} maintains the stability.

\bibliographystyle{JHEP}
{\small 
\bibliography{refs}
}
\end{document}